\documentclass[aps,prd,showkeys,superscriptaddress,singlecolumn,nofootinbib,floatfix]{revtex4-2}

\usepackage{subfiles}
\usepackage{amsmath}
\usepackage{amssymb}
\usepackage{amsthm}
\usepackage{mathrsfs}
\usepackage{graphicx}
\usepackage{epstopdf}
\usepackage{fancyhdr}
\usepackage{array}
\usepackage[all]{xy}
\usepackage{eufrak}
\usepackage{euscript}
\usepackage{enumerate}
\usepackage{slashed}
\usepackage{hyperref}
\usepackage{subfigure} % NEEDED IN ORDER TO USE SUBFIG
\usepackage{epstopdf} % COMPILE EPS (AND PDF) FIGURES USING PDFLATEX!!!

\graphicspath{{./figures/}} %%% Path to include figures files
\hypersetup{pdftex,colorlinks=true,linkcolor=blue,citecolor=blue,menucolor=black,urlcolor=blue,filecolor=blue}

\begin{document}

\title{Holographic entropy production in a Bjorken expanding\\ hot and dense strongly coupled quantum fluid}

\author{Romulo Rougemont}
\email{rougemont@ufg.br}
\affiliation{Instituto de F\'{i}sica, Universidade Federal de Goi\'{a}s, Av. Esperan\c{c}a - Campus Samambaia, CEP 74690-900, Goi\^{a}nia, Goi\'{a}s, Brazil}

\author{Willians Barreto}
\email{willians.barreto@ufabc.edu.br}
\affiliation{Centro de Ci\^{e}ncias Naturais e Humanas, Universidade Federal do ABC, Av. dos Estados 5001, 09210-580 Santo Andr\'{e}, S\~{a}o Paulo, Brazil}
\affiliation{Centro de F\'{i}sica Fundamental, Universidad de Los Andes, M\'{e}rida 5101, Venezuela}

\begin{abstract}
We analyze the time evolution of several physical observables, namely the pressure anisotropy, the scalar condensate, the charge density, and also, for the first time, the non-equilibrium entropy for a Bjorken expanding strongly coupled $\mathcal{N}=4$ Supersymmetric Yang-Mills plasma charged under an Abelian $U(1)$ subgroup of the global $SU(4)$ R-symmetry. This represents a far-from-equilibrium, hot and dense strongly coupled quantum fluid with a critical point in its phase diagram. For some sets of initial data preserving all the energy conditions, dynamically driven transient violations of the dominant and the weak energy conditions are observed when the plasma is still far from the hydrodynamic regime. The energy conditions violations get stronger at larger values of the chemical potential to temperature ratio, $\mu/T$, indicating that those violations become more relevant as the strongly coupled quantum fluid approaches its critical regime. For some of those energy conditions violations, it is observed a clear correlation with different plateau structures formed in the far from equilibrium entropy, indicating the presence of transient, early time windows where the Bjorken expanding plasma has zero entropy production even while being far from equilibrium. The hydrodynamization of the pressure anisotropy and also the much later thermalization of the scalar condensate are generally found to be delayed, within small relative tolerances, as $\mu/T$ is increased towards criticality. The value of $\mu/T$ in the medium is enhanced by increasing its initial charge density, and/or also by reducing its initial energy density.
\end{abstract}

\maketitle
\tableofcontents

%\newpage

%%%%%%%%%%%%%%%%%%%%%%%%%%%%%%%%%
\section{Introduction}
\label{sec:intro}

The analysis of the evolution of the medium produced in ultrarelativistic heavy ion collisions \cite{Arsene:2004fa,Adcox:2004mh,Back:2004je,Adams:2005dq,Aad:2013xma}, which leads to the formation of tiny droplets of a strongly coupled quark-gluon plasma \cite{Gyulassy:2004zy,Heinz:2013th,Luzum:2013yya,Shuryak:2014zxa} in high energy particle colliders, constitutes one of the main contemporary motivations to study the real time physics of far from equilibrium media, and also how hydrodynamization and thermalization are achieved at later times \cite{Florkowski:2017olj,Romatschke:2017ejr}.

Although QCD is the fundamental microscopic theory describing the strong nuclear interaction, a self-consistent, first principles microscopic description of heavy ion collisions is currently unavailable. This is mainly due to the fact that during the evolution of the medium created in such collisions, QCD matter passes through several different regimes which include strongly coupled physics \cite{Shuryak:2014zxa}, requiring therefore a non-perturbative treatment of QCD. The main of such approaches is lattice QCD, which is greatly successful in the description of QCD spectroscopy \cite{Durr:2008zz,Borsanyi:2014jba} and QCD thermodynamics at zero \cite{Aoki:2006we,Borsanyi:2010bp,Borsanyi:2013bia,Borsanyi:2016ksw} and moderate values of baryon chemical potential \cite{Borsanyi:2013hza,Borsanyi:2014ewa,Bazavov:2017dus,Borsanyi:2021sxv}. However, mainly due to the so-called sign problem \cite{Philipsen:2012nu,Meyer:2011gj,Ratti:2018ksb,Alexandru:2020wrj}, current lattice QCD calculations face severe difficulties regarding the calculation of physical observables at large values of baryon density, which are relevant for the search of a putative critical point in the QCD phase diagram in the plane of temperature and baryon chemical potential, and also concerning the calculation of physical observables evaluated at real time, which are important in the analysis of non-equilibrium phenomena.

In face of the aforementioned limitations, several different alternative and simpler approaches based on phenomenological models, effective theories, and also toy models are usually employed to make either quantitative predictions which may be tested against experimental heavy ion data, or to obtain possible qualitative insights on some general aspects of the quark-gluon plasma.

One such alternative or complementary approach is the holographic gauge/gravity duality \cite{Maldacena:1997re,Gubser:1998bc,Witten:1998qj,Witten:1998zw}. Indeed, different holographic models can be devised, with either a more phenomenological and quantitative approach to QCD \cite{Gursoy:2007cb,Gursoy:2007er,Gubser:2008ny,Gubser:2008yx,DeWolfe:2010he,DeWolfe:2011ts,Finazzo:2014cna,Rougemont:2015wca,Rougemont:2015ona,Finazzo:2015xwa,Rougemont:2017tlu,Critelli:2017oub,Rougemont:2018ivt,Grefa:2021qvt,Grefa:2022sav,Rougemont:2015oea,Finazzo:2016mhm,Critelli:2016cvq,Rougemont:2020had,Jarvinen:2021jbd,Demircik:2021zll,Hoyos:2021njg,Knaute:2017opk,Cai:2022omk}, or with a more qualitative focus on possible broad and general properties of strongly coupled media \cite{Casalderrey-Solana:2011dxg,DeWolfe:2013cua} (which may not be necessarily tied to specific quantitative results in QCD). In this very regard, probably the most important and general outcome of the holographic gauge/gravity duality is the \emph{almost} universal result for the shear viscosity to entropy density ratio in strongly coupled quantum fluids, $\eta/s=1/4\pi$, which is valid for any isotropic and translationally invariant gauge/gravity dual with two derivatives of the metric field in the gravity action \cite{Kovtun:2003wp,Buchel:2003tz,Kovtun:2004de}.

Realistically emulating all the stages of heavy ion collisions in holography is something currently unfeasible, mainly because the holographic system in the gauge/gravity approach is always strongly coupled, therefore missing the correct time evolution of the effective coupling of the medium in actual heavy ion collisions. However, an interesting and important line of investigation can be established on the grounds of understanding the physical possibilities, and also putative general properties, arising from the real time dynamics of far from equilibrium strongly coupled quantum fluids, in contrast to the scenarios realized in classical weakly coupled approaches like kinetic theory \cite{Arnold:2002zm,Kurkela:2014tea,Almaalol:2020rnu}. Contrasting such different approaches may lead e.g. to a better qualitative understanding on the essential physical properties behind the several different stages comprised in actual heavy ion collisions. Particularly, in the context of far from equilibrium holography, different physical observables have been studied under several different kinds of dynamic gauge/gravity models, see e.g. \cite{Chesler:2008hg,Chesler:2009cy,Chesler:2010bi,Heller:2011ju,Heller:2012je,vanderSchee:2012qj,Casalderrey-Solana:2013aba,Chesler:2013lia,vanderSchee:2014qwa,Jankowski:2014lna,Fuini:2015hba,Chesler:2015bba,Bellantuono:2015hxa,Pedraza:2014moa,DiNunno:2017obv,Attems:2016tby,Casalderrey-Solana:2016xfq,Grozdanov:2016zjj,Attems:2017zam,Romatschke:2017vte,Spalinski:2017mel,Critelli:2017euk,Casalderrey-Solana:2017zyh,Critelli:2018osu,Attems:2018gou,Kurkela:2019set,Cartwright:2019opv,Cartwright:2020qov,Rougemont:2021qyk,Rougemont:2021gjm,Cartwright:2021maz,Ecker:2021ukv,Ghosh:2021naw,Cartwright:2022hlg} for a non-exhaustive list.

In the present work, we proceed with our analyses of far from equilibrium holographic dynamics in the context of a strongly coupled conformal plasma expanding according to the inhomogeneous boost invariant Bjorken flow \cite{Bjorken:1982qr}. We investigate how several physical observables of a top-down holographic Einstein-Maxwell-Dilaton toy model for a hot and dense plasma with a critical point in its phase diagram, called `One R-Charged Black Hole' (1RCBH) model \cite{Gubser:1998jb,Behrndt:1998jd,Kraus:1998hv,Cai:1998ji,Cvetic:1999ne,Cvetic:1999rb} (see also \cite{DeWolfe:2011ts}), are affected by different variations of the far from equilibrium initial data (like the initial charge and energy densities), leading to different values of the chemical potential to temperature ratio of the medium, $\mu/T$. In a previous work \cite{Critelli:2018osu}, the Bjorken flow for this model was analyzed mainly in terms of how the hydrodynamization time associated to the convergence of the pressure anisotropy to the Navier-Stokes result is affected by increasing the value of $\mu/T$ towards criticality. By considering variations of the initial charge density of the medium, while keeping fixed both the initial profile for the metric anisotropy and the initial energy density, it was observed that, on average, the fluid takes more time to converge to the Navier-Stokes hydrodynamic constitutive relation as $\mu/T$ increases, with the hydrodynamization time of the pressure anisotropy displaying a particularly sharp rise close to the critical point of the model.

On the other hand, there are also several important aspects of the 1RCBH model undergoing Bjorken flow which have not been analyzed in \cite{Critelli:2018osu}, and whose study constitutes the specific focus of the present work. Namely: a) in order to see the late time effective thermalization of the scalar condensate, which is associated to its convergence to the corresponding thermodynamic equilibrium result, in the present work we carry out our numerical simulations for much longer times (what clearly demands a numerical code implemented using efficient programming languages for numerical purposes, as we will discuss in appendix \ref{sec:app3}); b) in order to analyze several qualitatively different physical possibilities for the time evolution of the system, we deploy a more general and qualitatively varied set of initial data, now including also the analysis for variations of the initial energy density of the medium and variations of the initial profile for the bulk dilaton field (besides variations of the initial charge density and of the initial profile for the metric anisotropy); c) more importantly, for the first time, we also analyze the holographic entropy production \cite{Chesler:2009cy,Tian:2014goa} in the 1RCBH model expanding according to the Bjorken flow dynamics.

Regarding the holographic entropy of a fluid in thermodynamic equilibrium, it is well known that it is dual to the Bekenstein-Hawking black hole entropy \cite{Bekenstein:1973ur,Hawking:1974sw} measured by the area of the event horizon of a static black hole in equilibrium within the higher dimensional bulk.\footnote{For other notions of the concept of entropy also employed in the holographic gauge/gravity duality in different contexts, see e.g. \cite{Ryu:2006bv,Ryu:2006ef,Nishioka:2009un,Rangamani:2016dms,Faulkner:2013ana,Engelhardt:2014gca,Braga:2016wzx,Braga:2020opg}. For a recent non-holographic calculation of entropy production in Bjorken flow within the context of the QCD phase diagram, see \cite{Dore:2022qyz}.} However, when the fluid is far from equilibrium, there are clear indications that the area of a dynamic event horizon is no longer an adequate measure of the entropy of the medium. In fact, as argued e.g. in \cite{Figueras:2009iu}, one expects that entropy production is local in time, therefore, associating the far from equilibrium entropy to the area of a dynamic event horizon seems unnatural, since in such a context the event horizon can only be determined by knowing the entire future evolution of the black hole geometry and, consequently, it is a global rather than a local observable. Moreover, in the case of the holographic conformal soliton flow \cite{Friess:2006kw}, which corresponds to an ideal fluid and, thus, has zero entropy production at all times, the area of the event horizon diverges \cite{Figueras:2009iu}, providing an explicit example that it is not an adequate measure of holographic entropy production. On the other hand, the area of the apparent horizon is calculated locally in time, as expected for an entropy function. Furthermore, for the conformal soliton flow, the area of the apparent horizon is constant in time \cite{Figueras:2009iu}, as expected for the entropy of an ideal fluid.\footnote{As discussed in \cite{Figueras:2009iu}, for the conformal soliton flow the system does not settle down to a stationary state at late times, and the apparent horizon does not converge to the event horizon in this case, contrary to what happen in cases where dissipation drives the evolution of the system.} Indeed, in many other works in the holographic literature the non-equilibrium entropy has been associated to the area of a dynamic apparent horizon \cite{Chesler:2009cy,Heller:2011ju,Heller:2012je,Jankowski:2014lna,vanderSchee:2014qwa,Grozdanov:2016zjj,Buchel:2016cbj,Muller:2020ziz,Engelhardt:2017aux}, and we also follow the same approach. Since here the apparent horizon later converges to the event horizon close to equilibrium, the holographic non-equilibrium entropy so defined is assured to converge to the usual Bekenstein-Hawking thermodynamic entropy in equilibrium, what should be observed at late times in the evolution of the dynamic system.

In two previous works \cite{Rougemont:2021qyk,Rougemont:2021gjm}, by analyzing the dynamical evolution of some initial data, we pointed out some notable correlations between far from equilibrium plateau structures observed in the entropy, indicating the presence of transient, early time windows with zero entropy production in the Bjorken expanding $\mathcal{N}=4$ Supersymmetric Yang-Mills (SYM) plasma at zero density, and some later transient violations of classical energy conditions \cite{HawkingEllisBook,WaldBookGR1984} in the medium. Although violations of classical energy conditions by quantum effects are well known (see e.g. the discussions in \cite{Visser:1999de,Costa:2021hpu}), the physical possibility that such violations maybe can also happen during the dynamical evolution of the QCD medium produced in heavy ion collisions is something not currently contemplated in the literature, since phenomenological models of pre-hydrodynamic stage of heavy ion collisions are commonly modeled using classical kinetic theory approaches \cite{Kurkela:2018wud}, where such violations do not take place due to the positiveness of the single particle distribution function \cite{degroot}.\footnote{It would be important to check, within transport approaches, whether such violations can be observed outside the classical regime due to quantum effects, what requires moving from classical kinetic theory to the quantum Kadanoff-Baym equations \cite{Baym:1961zz}.} More generally, as shown in \cite{Rougemont:2021qyk,Rougemont:2021gjm} in the context of the holographic Bjorken flow, and as also known from some other far from equilibrium holographic dynamics \cite{Arnold:2014jva}, it is important to recognize that strongly coupled quantum fluids may generally display dynamically driven violations of classical energy conditions, even for some initial data satisfying all such conditions.

In the present work, we shall see that such violations are generally enhanced by increasing the value of $\mu/T$ in the hot and dense medium described by the holographic 1RCBH model. We will also discuss the correlations observed when plateau structures are produced in the non-equilibrium entropy and later violations of energy conditions, and how the value of $\mu/T$ affects such correlations.

This work is organized as follows. In section \ref{sec:2}, we review the holographic formulation of the Bjorken flow for the 1RCBH model, providing more details than in the previous work \cite{Critelli:2018osu}, what will be useful for the reader interested in reproducing our results, and we also briefly present some specific results for the thermodynamics of the model which will be used in the analysis of the effective thermalization of the scalar condensate and in some analytical consistency checks for the late time numerics of our calculations, involving also the non-equilibrium entropy. In section \ref{sec:3}, we discuss the form of the set of initial data which will be analyzed in the present work and how we shall organize their different variations, besides also defining the set of normalized observables we will consider to study different physical aspects of the dynamical evolution of the selected initial data; we also briefly review the formulas for the holographic non-equilibrium entropy and the dominant and weak energy conditions. In sections \ref{sec:4}, \ref{sec:5}, and \ref{sec:extra} we analyze, respectively, the effects associated to variations in the initial charge density, the initial energy density, and the initial dilaton profile, while keeping the remaining initial data fixed. We shall work with a subset of initial metric anisotropies originally considered for the SYM plasma in \cite{Rougemont:2021qyk,Rougemont:2021gjm}, which will imply several qualitatively different physical possibilities for the dynamic evolution of the system, allowing us to extract possible general conclusions from the analysis presented. In section \ref{sec:conc} we present a summary with our main conclusions and future perspectives, while the appendices are devoted to discussions on numerical error analysis, further physical consistency checks of our numerical solutions, and also an analysis of our numerical code's performance, clearly showing the need for using efficient programming languages for numerical purposes in the present work.

We use a mostly plus metric signature and natural units where $c=\hbar=k_B=1$.

%%%%%%%%%%%%%%%%%%%%%%%%%%%%%%%%%
\section{Holographic Bjorken flow for the 1RCBH plasma}
\label{sec:2}

The 1RCBH model \cite{Gubser:1998jb,Behrndt:1998jd,Kraus:1998hv,Cai:1998ji,Cvetic:1999ne,Cvetic:1999rb} (see also \cite{DeWolfe:2011ts,Finazzo:2016psx,Critelli:2017euk}) is a solution of the five dimensional $\mathcal{N}=8$ gauged supergravity action \cite{Behrndt:1998jd} which, in turn, corresponds to a consistent truncation of type IIB superstrings on AdS$_5\times S^5$. This was shown to lie within a class of solutions equivalent to near-extremal spinning D3-branes on AdS$_5\times S^5$ \cite{Cvetic:1999ne}. The Kaluza-Klein compactification of $S^5$ on the spinning D3-branes solutions leads to a global $SU(4)$ R-charge symmetry, which has three independent Cartan subgroups $U(1)_{a}\times U(1)_{b}\times U(1)_{c}$ associated with three distinct conserved charges $(Q_{a},Q_{b},Q_{c})$ of the black hole background \cite{Behrndt:1998jd}. The general solution is known as the STU model, while the particular case of the 1RCBH model is obtained by considering only one charge by setting $Q \equiv Q_{a}$ and $Q_{b}=Q_{c}=0$. The effective five dimensional theory so obtained is holographycally dual to a $\mathcal{N}=4$ SYM plasma at finite temperature with a chemical potential associated to the conserved R-charge transforming under the Abelian $U(1)_{a}$ subgroup of the global $SU(4)$ R-charge symmetry. The effective five dimensional bulk action comprises the metric field, a scalar dilaton field, and an Abelian Maxwell field,
\begin{align}
S = \frac{1}{2 \kappa_5^2} \int d^5x \sqrt{-g} \left[ R - \frac{f(\phi)}{4} F_{\mu \nu} F^{\mu \nu} - \frac{1}{2} (\partial_{\mu} \phi)^2 - V(\phi) \right],
\label{eq:action}
\end{align}
where $\kappa_5^2\equiv 8 \pi G_5$, with $G_5$ being the five dimensional Newton's constant, and the dilaton potential $V(\phi)$ and the coupling function $f(\phi)$ between the dilaton and Maxwell fields are given by,
\begin{align}
V(\phi) = -\frac{1}{L^2} \left(8 e^{\phi/\sqrt{6}} + 4 e^{-\sqrt{2/3}\,\phi} \right), \qquad f(\phi) = e^{- 2\sqrt{2/3}\,\phi},
\label{eq:Vandf}
\end{align}
with $L$ being the asymptotic AdS$_5$ radius (which from now on we set to unity for simplicity). The bulk action \eqref{eq:action} is supplemented by two boundary terms: the Gibbons-Hawking-York action \cite{York:1972sj,Gibbons:1976ue}, needed for the well-posedness of the boundary Dirichlet problem; and the counterterm action \cite{Critelli:2017euk}, required in the holographic renormalization procedure \cite{Bianchi:2001kw,Skenderis:2002wp,deHaro:2000vlm} in order to consistently remove the boundary divergences of the on-shell action.

Several different aspects of the 1RCBH model have been already studied in the literature. For instance, the thermodynamics of the model was detailed discussed in \cite{DeWolfe:2011ts,Finazzo:2016psx}, where it was shown that this model has a critical point at $\mu/T=\pi/\sqrt{2}$, which is identified by the divergences of the second (and higher order) derivatives of the pressure, like the heat capacity and the charge susceptibility. Since the model has vanishing trace anomaly, it is conformal and all the dimensionless combinations of physical observables in equilibrium are functions of the dimensionless ratio $\mu/T$ (instead of functions of independent $T$ and $\mu$, since in the conformal case there is no dimensionful scale in the vacuum of the theory, contrary to what happens e.g. in QCD, where $\Lambda_\textrm{QCD}$ sets the scale with respect to which other dimensionful quantities, like $T$ and $\mu$, can be measured). For each possible value of $\mu/T$ there are two competing branches of solutions, one corresponding to thermodynamically stable black hole backgrounds, and another one corresponding to unstable black holes. The stable branch of black hole backgrounds constitutes the set of physically relevant solutions of the 1RCBH model at finite $\mu/T$. Due to its conformal nature, the phase diagram of the 1RCBH model is a line in the $\mu/T$ axis (instead of a plane with independent $T$ and $\mu$ directions), which starts at $\mu/T=0$, corresponding to the AdS$_5$-Schwarzschild solution, dual to the SYM plasma at finite temperature and zero chemical potential, and ends at the critical point $\mu/T=\pi/\sqrt{2}$ --- due to this peculiar feature, although there is a critical point in the phase diagram of the 1RCBH model (where several observables diverge with characteristic critical exponents \cite{DeWolfe:2011ts,Finazzo:2016psx}), there is no phase transition, since in this model $\mu/T$ can not be increased past its critical value.

Also the conductivity and charge diffusion in the 1RCBH model were analyzed in \cite{DeWolfe:2011ts}. The spectra of quasinormal modes and their critical behavior were discussed in \cite{Finazzo:2016psx,Critelli:2017euk}. The holographic renormalization and the far from equilibrium homogeneous isotropization dynamics were presented in \cite{Critelli:2017euk}. The evolution of holographic complexity \cite{Ebrahim:2018uky}, the holographic entanglement entropy and mutual information \cite{Ebrahim:2020qif}, the entanglement of purification \cite{Amrahi:2021lgh}, and the critical behavior of the hydrodynamic derivative series \cite{Asadi:2021hds} were also analyzed for the 1RCBH model. The far from equilibrium Bjorken flow dynamics and the analysis of the hydrodynamization times of the medium, including the vicinity of the critical point, were presented in \cite{Critelli:2018osu}.

Having discussed the above introductory remarks, the main purposes of the present section of the paper are the following: first, we briefly present below some specific results for the thermodynamics of the model which will be used in the analysis of the effective thermalization of the scalar condensate and in some analytical consistency checks for the late time numerics of our calculations, involving also the non-equilibrium entropy. Next, as the main focus of the present section, we will review the holographic formulation of the Bjorken flow dynamics of the 1RCBH model in more detail than in the previous short paper of Ref. \cite{Critelli:2018osu}.

Our new results will be mainly presented in sections \ref{sec:4}, \ref{sec:5}, and \ref{sec:extra}, after a discussion on the organization of the initial data and the normalization of the relevant physical observables in section \ref{sec:3}.

\subsection{Some important thermodynamic results}
\label{sec:2.1}

\begin{figure*}%[h]
\center
\subfigure[]{\includegraphics[width=0.49\textwidth]{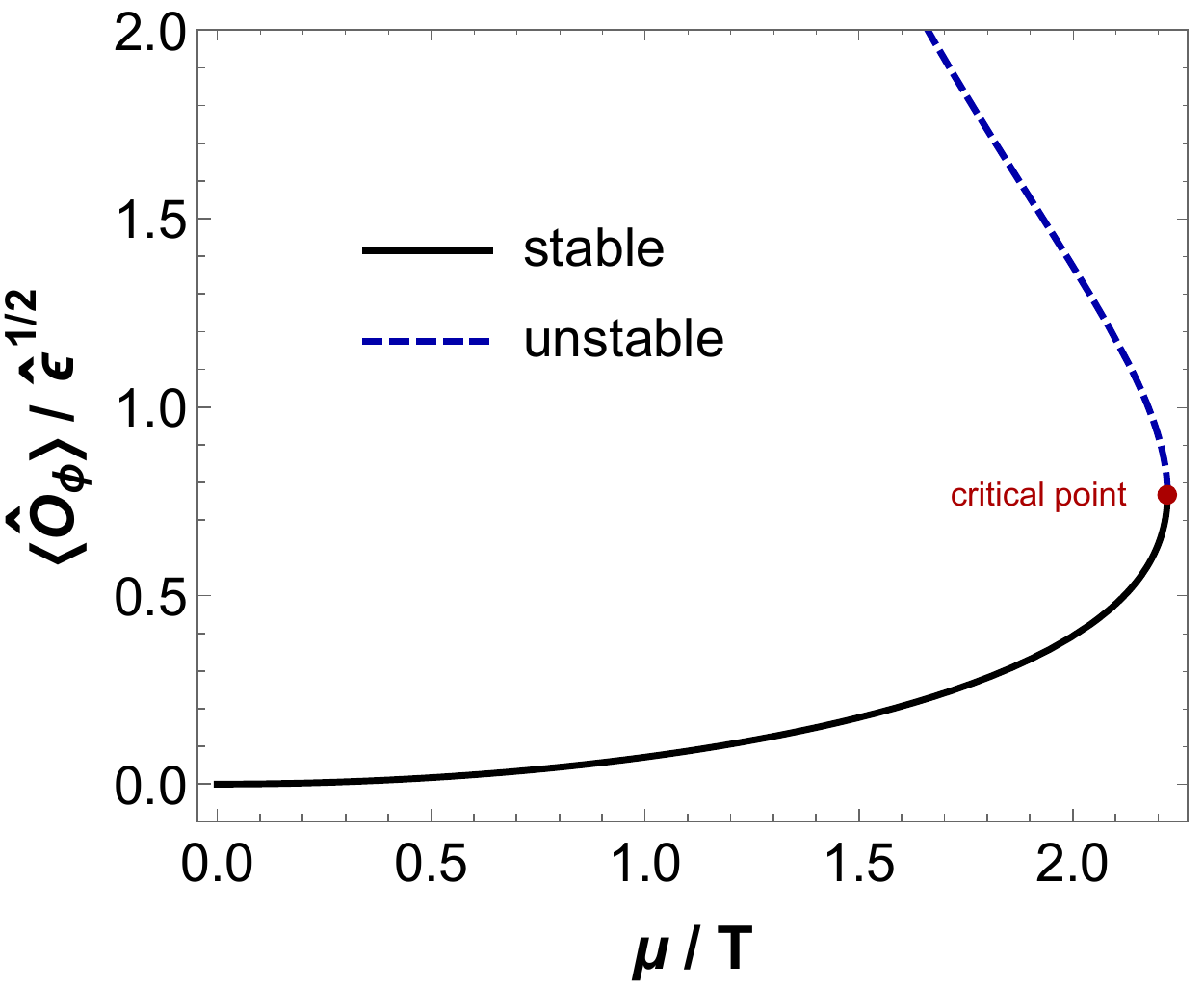}}
\subfigure[]{\includegraphics[width=0.47\textwidth]{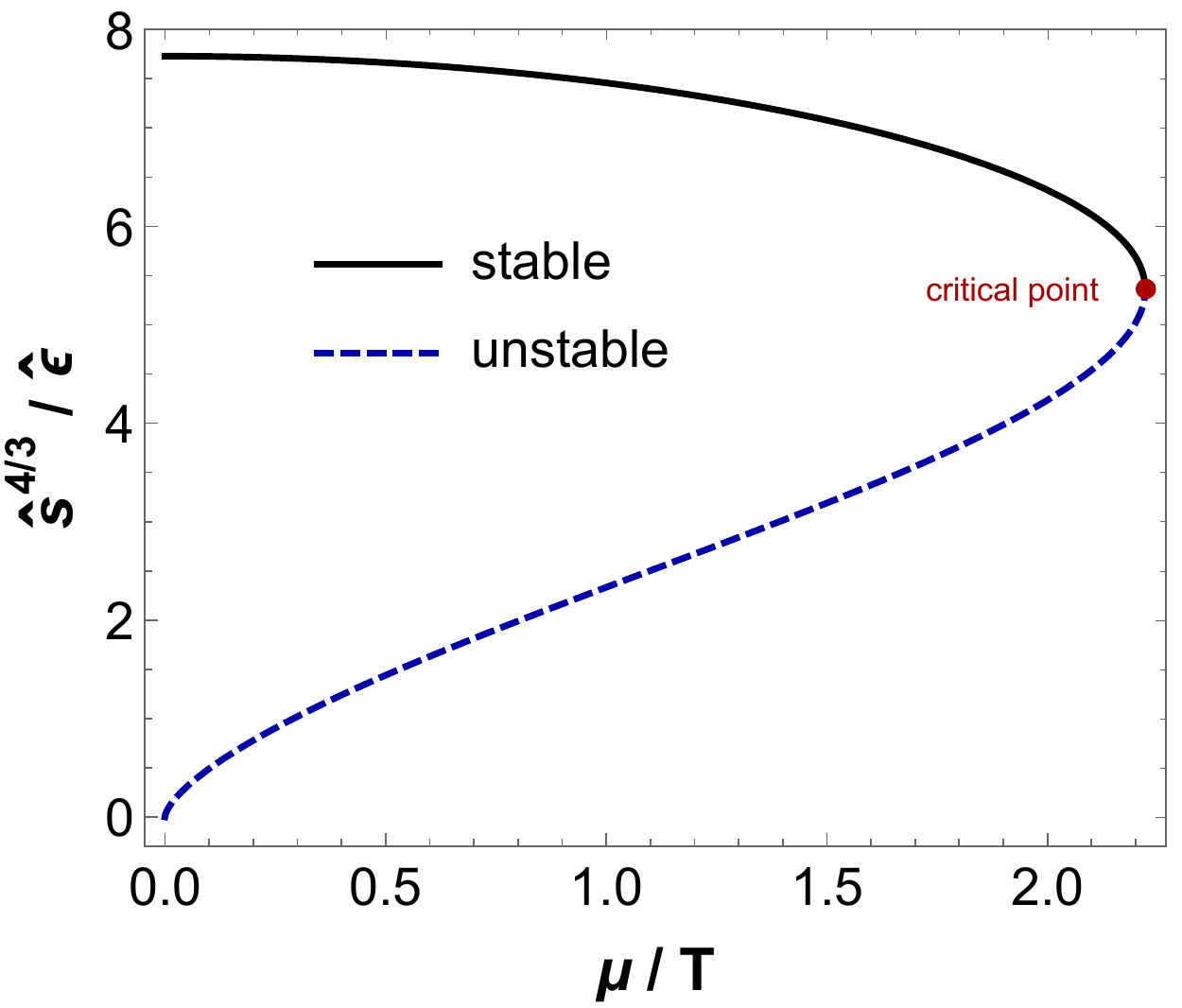}}
\subfigure[]{\includegraphics[width=0.47\textwidth]{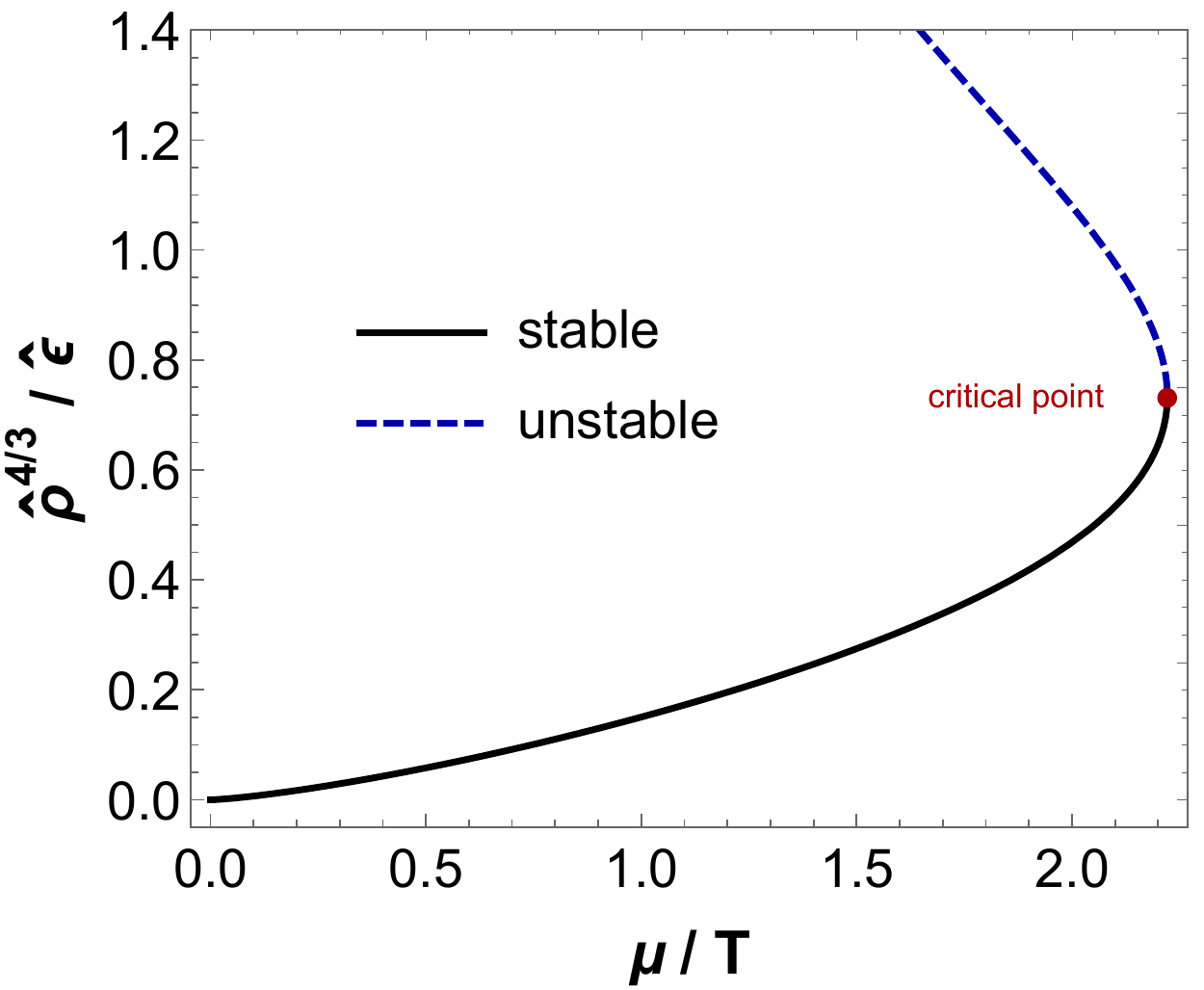}}
\caption{Analytical equilibrium results for dimensionless ratios involving (a) the scalar condensate, (b) the entropy density, and (c) the charge density, all of them normalized by the energy density. We display the thermodynamically stable and unstable branches of equilibrium 1RCBH solutions. In the stable branch, at $\mu/T=0$, one recovers the pure thermal SYM plasma results.}
\label{fig:eqfigs}
\end{figure*}

Let $X$ be any physical observable and let us define $\hat{X}\equiv\kappa_5^2 X = 4\pi^2 X/N_c^2$. By considering Eqs. (4.21) and (4.24) of \cite{Critelli:2017euk}, one obtains for the dual quantum field theory in thermal equilibrium the following result for the dimensionless ratio of the scalar condensate $\langle\hat{O}_\phi\rangle$ normalized by the square root of the energy density $\hat{\epsilon}$ of the medium,
\begin{align}
\frac{\langle\hat{O}_\phi\rangle}{\hat{\epsilon}^{1/2}}\biggr|_{\textrm{(eq.)}} = \kappa_5\, \frac{\langle O_\phi\rangle}{\epsilon^{1/2}}\biggr|_{\textrm{(eq.)}} = \frac{2Q^2}{3M},
\label{eqse1}
\end{align}
where $Q$ and $M$ are the charge and the mass of the equilibrium black hole solution. On the other hand, from Eq. (2.9) of \cite{Finazzo:2016psx}, one gets,
\begin{align}
M = \tilde{r}_\textrm{EH} \sqrt{\tilde{r}_\textrm{EH}^2+Q^2},
\label{eqse2}
\end{align}
where $\tilde{r}_\textrm{EH}$ is the radial location of the event horizon in equilibrium. By substituting Eq. \eqref{eqse2} into Eq. \eqref{eqse1}, it follows that,
\begin{align}
\frac{\langle\hat{O}_\phi\rangle}{\hat{\epsilon}^{1/2}}\biggr|_{\textrm{(eq.)}} = \frac{2 \left(Q/\tilde{r}_\textrm{EH}\right)^2}{3\, \sqrt{1+\left(Q/\tilde{r}_\textrm{EH}\right)^2}}.
\label{eqse3}
\end{align}
By using Eq. (2.12) of \cite{Finazzo:2016psx}, the result above can be written as below,
\begin{align}
\frac{\langle\hat{O}_\phi\rangle}{\hat{\epsilon}^{1/2}}\biggr|_{\textrm{(eq.)}} = \frac{4\left[\displaystyle{\frac{1\pm\sqrt{1-\left(x/x_c\right)^2}}{x/x_c}}\right]^2}{3\, \sqrt{1+2 \left[\displaystyle{\frac{1\pm\sqrt{1-\left(x/x_c\right)^2}}{x/x_c}}\right]^2}},
\label{eqse4}
\end{align}
where we defined the notation $x/x_c\equiv(\mu/T)/(\pi/\sqrt{2})$ as in \cite{Critelli:2018osu},\footnote{Note that the notation $x\equiv\mu/T$ has nothing to do with the spatial coordinate $x$ (which is not actually relevant in the present work). In particular, $x_c\equiv\pi/\sqrt{2}$ denotes the critical point.} and the lower (upper) signs specify the thermodynamically stable (unstable) branch of equilibrium 1RCBH solutions. The corresponding results are displayed in Fig. \ref{fig:eqfigs} (a).

Also, by using Eqs. (2.15) and (2.17) of \cite{Finazzo:2016psx}, and by taking into account the fact that in a four dimensional conformal theory, as in the case of the 1RCBH plasma, $\epsilon=3p$, one obtains that,
\begin{align}
\frac{\hat{s}^{4/3}}{\hat{\epsilon}}\biggr|_{\textrm{(eq.)}} = \kappa_5^{2/3}\,\, \frac{s^{4/3}}{\epsilon}\biggr|_{\textrm{(eq.)}} &= (4\pi^2)^{1/3}\,\, \frac{(s/N_c^2T^3)^{4/3}}{3p/N_c^2T^4}\biggr|_{\textrm{(eq.)}} \nonumber\\
&= (4\pi^2)^{1/3}\,\, \frac{\left[ \displaystyle{\frac{\pi^2}{16}} \left(3\pm \sqrt{1-\left(x/x_c\right)^2}\right)^2 \left(1\mp \sqrt{1-\left(x/x_c\right)^2}\right) \right]^{4/3}}{\displaystyle{\frac{3\pi^2}{128}} \left( 3\pm\sqrt{1-\left(x/x_c\right)^2} \right)^3 \left( 1\mp\sqrt{1-\left(x/x_c\right)^2} \right)},
\label{eqse5}
\end{align}
where, as before, the lower (upper) signs specify the thermodynamically stable (unstable) branch of equilibrium 1RCBH solutions. The corresponding results are shown in Fig. \ref{fig:eqfigs} (b). Moreover, in a completely analogous way, one can also easily obtain from Eqs. (2.16) and (2.17) of \cite{Finazzo:2016psx} the following results for the dimensionless ratio,
\begin{align}
\frac{\hat{\rho}^{4/3}}{\hat{\epsilon}}\biggr|_{\textrm{(eq.)}} = \kappa_5^{2/3}\,\, \frac{\rho^{4/3}}{\epsilon}\biggr|_{\textrm{(eq.)}} &= (4\pi^2)^{1/3}\,\, \frac{(\rho/N_c^2T^3)^{4/3}}{3p/N_c^2T^4}\biggr|_{\textrm{(eq.)}} \nonumber\\
&= (4\pi^2)^{1/3}\,\, \frac{\left[ \displaystyle{\frac{x}{16}} \left(3\pm \sqrt{1-\left(x/x_c\right)^2}\right)^2 \right]^{4/3}}{\displaystyle{\frac{3\pi^2}{128}} \left( 3\pm\sqrt{1-\left(x/x_c\right)^2} \right)^3 \left( 1\mp\sqrt{1-\left(x/x_c\right)^2} \right)},
\label{eqse6}
\end{align}
which are displayed in Fig. \ref{fig:eqfigs} (c).

The thermodynamically stable equilibrium results derived above will be used to obtain the value of $\mu/T$ in the medium from the late time evolution of each far from equilibrium initial data and also to analyze the late time effective thermalization of the scalar condensate in the Bjorken flow of the 1RCBH model. Since these are analytical results, they will be also employed to make important physical consistency checks of our numerical results in the late time evolution of the system.

\subsection{Bjorken flow dynamics}
\label{sec2.2}

Now we review in more details the holographic formulation of the Bjorken flow for the 1RCBH model \cite{Critelli:2018osu}.

Let us begin by writing down the general Einstein-Maxwell-Dilaton equations of motion coming from the extremization of the bulk action \eqref{eq:action},
\begin{subequations}
\begin{align}
R_{\mu\nu}-\frac{g_{\mu\nu}}{3}\left(V(\phi)-\frac{f(\phi)}{4}F_{\alpha\beta}^2\right) -\frac{1}{2}\partial_\mu\phi\partial_\nu\phi-\frac{f(\phi)}{2}F_{\mu\rho}F_\nu^\rho &=0, \label{eq:Einstein}\\
\nabla_\mu(f(\phi)F^{\mu\nu})=\frac{1}{\sqrt{-g}}\partial_\mu(f(\phi)F^{\mu\nu}) &=0,
\label{eq:Maxwell}\\
\frac{1}{\sqrt{-g}}\partial_\mu(\sqrt{-g}g^{\mu\nu}\partial_\nu\phi) -\partial_\phi V(\phi) - \frac{\partial_\phi f(\phi)}{4}F_{\mu\nu}^2 &=0. \label{eq:dilaton}
\end{align}
\end{subequations}

Now we particularize the above results to the Bjorken flow symmetry \cite{Bjorken:1982qr}, which is a geometry commonly used in the context of heavy ion collisions to provide a simplified modeling of the expansion of the medium near the beam axis. The Bjorken symmetry corresponds to boost invariance along the beam line, which we take as the $z$ axis, where the fluid expands at the speed of light, plus translation and $O(2)$ rotation invariance in the transverse $xy$ plane. In practice, this means that in the Bjorken flow one is completely neglecting the dynamics of the medium in the transverse plane, which is the reason why this flow is usually employed only near mid-rapidity in the context of heavy ion collisions. The Bjorken symmetry is more naturally described in terms of the Milne coordinates $(\tau,\xi,x,y)$, where the proper time $\tau$ and the spacetime rapidity $\xi$ are defined by,
\begin{align}
\tau\equiv\sqrt{t^2-z^2},\qquad \xi\equiv \textrm{arctanh}(v)= \textrm{arctanh}\left(\frac{z}{t}\right) = \frac{1}{2} \ln\left(\frac{t+z}{t-z}\right),
\label{eq:milne}
\end{align}
in terms of which the boundary four dimensional Minkowski metric is written as,
\begin{align}
ds^2_{\textrm{Mink$_4$}} = -d\tau^2+\tau^2 d\xi^2+dx^2+dy^2.
\label{eq:mink}
\end{align}
The expectation value of the energy-momentum tensor is written as $\langle T^{\mu}_\nu\rangle = \mathrm{diag}(-\varepsilon,p_L,p_T,p_T)$, where for a conformal system, as in the case of the 1RCBH model, the longitudinal and transverse pressures are written as functions of the energy density and its time derivative as follows,
\begin{align}
p_L(\tau)=-\epsilon(\tau)-\tau\partial_\tau\epsilon(\tau),\qquad p_T(\tau)=\epsilon(\tau)+\frac{\tau}{2}\partial_\tau\epsilon(\tau),
\label{eq:pLpT}
\end{align}
so that the pressure anisotropy of a conformal fluid undergoing Bjorken flow is given by,
\begin{align}
\frac{\Delta p}{\epsilon}\equiv \frac{p_T-p_L}{\epsilon} = 2+\frac{3}{2}\,\tau\frac{\partial_\tau\epsilon}{\epsilon}.
\label{eq:dPE}
\end{align}
Using five dimensional generalized infalling Eddington-Finkelstein coordinates suited to the holographic implementation of the characteristic formulation of general relativity \cite{Chesler:2009cy}, the ansatz for the bulk Einstein-Maxwell-Dilaton fields compatible with the Bjorken symmetry can be written as follows \cite{Critelli:2018osu},
\begin{align}
& ds^2 = 2d\tau\left[dr-A(\tau,r) d\tau \right]+\Sigma(\tau,r)^2\left[e^{-2B(\tau,r)}d\xi^2 + e^{B(\tau,r)}(dx^2+dy^2)\right], \ \ A_\mu dx^\mu = \Phi(\tau,r)d\tau, \ \ \phi = \phi(\tau,r),
\label{eq:lineElement}
\end{align}
where $r$ is the radial coordinate of the asymptotically AdS$_5$ spacetime, whose boundary is at $r\rightarrow\infty$, where $\tau$ becomes the usual four dimensional proper time of the Bjorken flow. In these coordinates, infalling radial null geodesics satisfy $\tau = \textrm{constant}$, while outgoing radial null geodesics satisfy $dr/d\tau=A(\tau,r)$. The metric in Eq. \eqref{eq:lineElement} is also invariant under radial diffeomorphism shifts of the form $r\to r+\lambda(\tau)$, with arbitrary $\lambda(\tau)$.

By substituting the ansatz \eqref{eq:lineElement} into the general Einstein-Maxwell-Dilaton equations of motion \eqref{eq:Einstein} --- \eqref{eq:dilaton}, one obtains the following set of coupled $1+1$ partial differential equations \cite{Critelli:2018osu},\footnote{In writing down the constraint Eq. \eqref{eq:Max-Const} coming from one component of Maxwell's equations, we made use of the other nontrivial component of Maxwell's equations, given by Eq. \eqref{eq:pde1}, taking also into account that, from the definitions of $d_+$ and $\mathcal{E}$, one may rewrite $(d_+\Phi)' = -\partial_\tau\mathcal{E} - A'\mathcal{E} - A\mathcal{E}'$. Moreover, in writing down the constraint Eq. \eqref{eq:EH-Const2} coming from Einstein's equations, we made use of the Hamiltonian constraint \eqref{eq:EH-HamiltConst} also following from Einstein's equations.}
\begin{subequations}
\begin{align}
\partial_\tau\mathcal{E}+A\mathcal{E}'+\left(3\frac{d_+\Sigma}{\Sigma}+\frac{\partial_\phi f}{f}d_+\phi\right)\mathcal{E} &=0,\label{eq:Max-Const}\\
\frac{3 \Sigma '}{\Sigma } + \frac{\partial_\phi f}{f}\phi ' + \frac{\mathcal{E}'}{\mathcal{E}} &=0,\label{eq:pde1}\\
4 \Sigma (d_+\phi)'+6 \phi'd_+\Sigma + 6 \Sigma'd_+\phi+\Sigma\mathcal{E}^2 \partial_\phi f - 2 \Sigma  \partial_\phi V &=0,\label{eq:pde2}\\
(d_+\Sigma)' +\frac{2 \Sigma '}{\Sigma }d_+\Sigma+\frac{\Sigma}{12} \left(2V + f \mathcal{E}^2\right) &=0,\label{eq:pde3}\\
\Sigma \, (d_{+}B)'+\frac{3}{2}(B' d_+\Sigma + \Sigma'd_+ B) &=0,\label{eq:pde4}\\
A'' + \frac{1}{12} \left(18 B'd_{+}B-\frac{72 \Sigma'd_{+}\Sigma}{\Sigma^2} + 6 \phi'd_{+}\phi-7 f \mathcal{E}^2-2 V\right) &=0,\label{eq:pde5}\\
\Sigma'' + \frac{\Sigma}{6} \left(3 \left(B'\right)^2+\left(\phi '\right)^2\right) &=0,\label{eq:EH-HamiltConst}\\
d_+(d_+\Sigma)+\frac{\Sigma}{2}(d_+B)^2-A'd_+\Sigma+\frac{\Sigma}{6}(d_+\phi)^2 &=0,\label{eq:EH-Const2}
\end{align}
\end{subequations}
where $X'\equiv \partial_r X$ is the directional derivative along infalling radial null geodesics, while $d_+X\equiv[\partial_\tau+A(\tau,r)\partial_r]X$ is the directional derivative along outgoing radial null geodesics, and we also defined $\mathcal{E}\equiv-\Phi'$. Notice there are five background functions to be determined, $\{A(\tau,r),\Sigma(\tau,r),B(\tau,r),\phi(\tau,r),\mathcal{E}(\tau,r)\}$, and there are also five dynamical equations \eqref{eq:pde1} --- \eqref{eq:pde5}, besides three constraint equations corresponding to Eqs. \eqref{eq:Max-Const}, \eqref{eq:EH-HamiltConst}, and \eqref{eq:EH-Const2}. As we are going to discuss in a moment, Eqs. \eqref{eq:pde1} --- \eqref{eq:EH-HamiltConst} constitute a nested set of equations of motion which can be systematically integrated using numerical techniques, while the constraint Eqs. \eqref{eq:Max-Const} and \eqref{eq:EH-Const2} can be used to monitor the accuracy of the numerical solutions so obtained, as it will be discussed in the numerical error analysis to be presented in appendix \ref{sec:app2}.

Let us now work out the near-boundary ultraviolet expansions of the bulk fields taking into account the boundary conditions associated to the holographic Bjorken flow. The bulk metric given in Eq. \eqref{eq:lineElement} must flow to the AdS$_5$ geometry in the ultraviolet, which is conformally equivalent to the four dimensional Minkowski metric \eqref{eq:mink} at the boundary. In order to satisfy this condition, one must impose the following boundary conditions for the metric coefficients,
\begin{align}
A(\tau,r\to\infty)\sim\frac{r^2}{2},\qquad B(\tau,r\to\infty)\sim-\frac{2}{3}\ln(\tau), \qquad \Sigma(\tau,r\to\infty)\sim\tau^{1/3}r.
\label{eq:metricbc}
\end{align}
Indeed, by substituting \eqref{eq:metricbc} into the five dimensional metric in Eq. \eqref{eq:lineElement} one obtains in the generalized infalling Eddington-Finkelstein coordinates,
\begin{align}
\lim_{r\to\infty} ds^2 = ds^2_{\textrm{AdS$_5$}} = 2d\tau dr + r^2 ds^2_{\textrm{Mink$_4$}},
\label{eq:metricbcheck}
\end{align}
where $r^2$ is the global conformal factor of AdS$_5$. Moreover, regarding the bulk dilaton field in the 1RCBH model, one notices from the dilaton potential in Eq. \eqref{eq:Vandf} that the conformal dimension of the associated scalar condensate in the dual quantum field theory at the boundary is $\Delta_\phi=2$, consequently, the dilaton field must flow to zero in the ultraviolet as implemented by the following boundary condition,
\begin{align}
\phi(\tau,r\to\infty)\sim \frac{ \phi_{4-\Delta_\phi}(\tau)}{r^{4-\Delta_\phi}} = \frac{\phi_2(\tau)}{r^2}.
\label{eq:dilatonbc}
\end{align}
Finally, concerning the nontrivial component of the Maxwell field, it is related to the R-charge chemical potential of the quantum fluid at the boundary according to the following boundary condition,
\begin{align}
\Phi(\tau\to\infty,r\to\infty)=\mu.
\label{eq:Maxbc}
\end{align}

Taking into account the above boundary conditions, the near-boundary ultraviolet expansions of the bulk fields are given by \cite{Bianchi:2001kw,Critelli:2017euk,Critelli:2018osu},
\begin{subequations}
\begin{align}
A(\tau,r) &= \frac{1}{2}[r+\lambda(\tau)]^2 -\partial_\tau\lambda(\tau) +\sum_{n=1}^{\infty} \frac{a_n(\tau)}{r^n}, \label{eq:expA}\\
B(\tau,r) &= -\frac{2}{3}\ln(\tau) +\sum_{n=1}^{\infty} \frac{b_n(\tau)}{r^n}, \label{eq:expB}\\
\Sigma(\tau,r) &= \tau^{1/3}[r+\lambda(\tau)] +\sum_{n=0}^{\infty} \frac{s_n(\tau)}{r^n}, \label{eq:expS}\\
\phi(\tau,r) &= \sum_{n=2}^{\infty} \frac{\phi_n(\tau)}{r^n}, \label{eq:expDil}\\
\Phi(\tau,r) &= \Phi_0(\tau) + \sum_{n=2}^{\infty} \frac{\Phi_n(\tau)}{r^n}, \label{eq:expMax}
\end{align}
\end{subequations}
where the vast majority of the ultraviolet expansion coefficients $\{a_n(\tau),b_n(\tau),s_n(\tau),\phi_n(\tau),\Phi_n(\tau)\}$ may be fixed as functions of just a few undetermined coefficients and their time derivatives. This can be accomplished by substituting the above ultraviolet expansions into the Einstein-Maxwell-Dilaton equations of motion and them solving the resulting algebraic equations order by order in powers of $r$. By working with the ultraviolet expansions up to order $n=8$ one identifies four undetermined coefficients in the aforementioned procedure: $\{a_2(\tau),\phi_2(\tau),\Phi_0(\tau),\Phi_2(\tau)\}$. The values of the ultraviolet coefficients $\{a_2(\tau),\phi_2(\tau)\}$ at the initial time slice $\tau_0$ can be freely chosen since they correspond to two of the four initial data of the 1RCBH model undergoing Bjorken flow, as we are going to discuss in a moment;\footnote{More precisely, as we are going to discuss, $\phi_2(\tau)$ is the boundary value of the subtracted dilaton field, $\phi_2(\tau)=\phi_s(\tau,u=0)$, with the initial profile of the subtracted dilaton field being one of the four initial data of the system. Once this profile is chosen, one automatically has the specification of the initial value $\phi_2(\tau_0)=\phi_s(\tau_0,u=0)$.} once their initial values are chosen, their subsequent time evolutions are determined by numerically solving the nested set of partial differential equations previously derived, as it will be discussed in more details afterwards in the text. Moreover, at order $n=6$ in the aforementioned algebraic procedure, one fixes that $\Phi_2(\tau)=c/\tau$, where $c$ is an undetermined constant; however, from the Bjorken flow hydrodynamic analysis of the local charge conservation equation, $\nabla_\mu \langle\hat{J}^\mu\rangle=0$, one concludes that for a conformal fluid, like the 1RCBH model, the time evolution of the R-charge density is given by $\hat{\rho}(\tau)\equiv\langle \hat{J}^\tau\rangle =\rho_0/\tau$, where $\rho_0$ is a parameter setting the initial charge density of the medium \cite{Critelli:2018osu}. Since the holographic renormalization procedure fixes that $\hat{\rho}(\tau)\equiv\langle\hat{J}^\tau\rangle=-\langle\hat{J}_\tau\rangle=-\Phi_2(\tau)$ \cite{Critelli:2017euk}, one then concludes that $\Phi_2(\tau)=c/\tau=-\rho_0/\tau$, where $\rho_0$ is also one of the freely chosen initial data of the system \cite{Critelli:2018osu}. Finally, none of the other ultraviolet expansion coefficients of the bulk fields directly depend on $\Phi_0(\tau)$, which may be formally identified with the value of the R-charge chemical potential through the boundary condition \eqref{eq:Maxbc} applied to the series expansion in \eqref{eq:expMax}. However, as it is clear from Eq. \eqref{eq:Maxbc}, the value of $\mu/T$ in the Bjorken flow of the 1RCBH model is not an initial data, and it can only be estimated in the late time evolution of the system \cite{Critelli:2018osu}, when the fluid approaches the equilibrium regime, as it will be explained afterwards in the text.

Before discussing the general schematics to integrate the nested set of $1+1$ partial differential equations of motion of the system, let us consider the radial integration of the Maxwell equation \eqref{eq:pde1}, where we make use of the specific form of Maxwell-dilaton coupling function given in \eqref{eq:Vandf},
\begin{align}
3\int \frac{d\Sigma}{\Sigma} -2\sqrt{\frac{2}{3}}\int d\phi + \int \frac{d\mathcal{E}}{\mathcal{E}} = 0 \,\,\,\Rightarrow\,\,\, \mathcal{E}=-\Phi'=e^{\bar{c}}\,\Sigma^{-3}\,e^{2\sqrt{2/3}\,\phi},
\label{eq:MaxField1}
\end{align}
where $\bar{c}$ is a radial integration constant, which can be fixed by evaluating \eqref{eq:MaxField1} close to the boundary at $r\to\infty$. In order to do it, we consider the near-boundary expansions for $\mathcal{E}(\tau,r\to\infty)\sim -\partial_r(\Phi_0+\Phi_2 r^{-2})=2\Phi_2 r^{-3}$ and for $e^{\bar{c}}\,\Sigma(\tau,r\to\infty)^{-3}\,e^{2\sqrt{2/3}\,\phi(\tau,r\to\infty)}\sim e^{\bar{c}}\,\tau^{-1}r^{-3}$, in which case one concludes that $e^{\bar{c}}=2\Phi_2(\tau)\tau=-2\rho_0$, where we used the aforementioned relation $\Phi_2(\tau)=-\rho_0/\tau$. Consequently,
\begin{align}
\mathcal{E}(\tau,r)=-2\rho_0\,\Sigma(\tau,r)^{-3}\,e^{2\sqrt{2/3}\,\phi(\tau,r)},
\label{eq:MaxField2}
\end{align}
so that the bulk `radial electric field' $\mathcal{E}=-\Phi'$ is completely determined in terms of the metric coefficient $\Sigma$ and the dilaton field $\phi$.

At this point we may lay down the general reasoning used to numerically integrate the nested set of $1+1$ partial differential equations of motion describing the Bjorken flow dynamics of the 1RCBH model:
\begin{enumerate}[i.]
\item On the hypersurface defined at the initial time slice $\tau_0$, one chooses the initial profiles for the metric anisotropy coefficient $B(\tau_0,r)$ and for the dilaton field $\phi(\tau_0,r)$, besides also the initial values for the charge density $\rho_0$ and for the dynamical ultraviolet coefficient $a_2(\tau_0)$;\footnote{As we will discuss in details in the next subsections \ref{sec:2.3} and \ref{sec:2.4}, one also has to choose the initial value for the radial shift function $\lambda(\tau_0)$, while its time evolution may be conveniently obtained by requiring that the radial position of the apparent horizon remains fixed during the time evolution of the system \cite{Chesler:2013lia}.}
\item Next one radially solves the Hamiltonian constraint \eqref{eq:EH-HamiltConst} to obtain $\Sigma(\tau_0,r)$, which at this step automatically fixes the value of $\mathcal{E}(\tau_0,r)$ through Eq. \eqref{eq:MaxField2};
\item Next one radially solves Eq. \eqref{eq:pde3} to obtain $d_+\Sigma(\tau_0,r)$;
\item Next one radially solves Eq. \eqref{eq:pde4} to obtain $d_+B(\tau_0,r)$;
\item Next one radially solves the dilaton Eq. \eqref{eq:pde2} to obtain $d_+\phi(\tau_0,r)$;
\item Next one radially solves Eq. \eqref{eq:pde5} to obtain $A(\tau_0,r)$;
\item At this step, from the definition of the directional derivative along outgoing radial null geodesics, $d_+\equiv\partial_\tau+A(\tau,r)\partial_r$, one already has $\{\partial_\tau B(\tau_0,r),\partial_\tau\phi(\tau_0,r)\}$, besides also $\partial_\tau a_2(\tau_0)$ (as we will discuss in details in the next subsections \ref{sec:2.3} and \ref{sec:2.4}), which together with the initial profiles chosen for the metric anisotropy and the dilaton field, and the initial value chosen for $a_2(\tau)$, comprise the set of initial conditions required to evolve $\{B(\tau_0,r),\phi(\tau_0,r),a_2(\tau_0)\}$ to the next time slice $\tau_0+\Delta\tau$ using some discrete integration technique (here we employ the fourth order Adams-Bashforth method to integrate in time, while the radial integrations are performed using the pseudospectral method \cite{boyd01});
\item Repeat the previous steps to obtain all the fields in the current time slice and iteratively evolve the system to the next time slices until reaching any desired end time $\tau_\textrm{end}$ for the numerical simulations.
\end{enumerate}
Furthermore, as aforementioned, the constraint Eqs. \eqref{eq:Max-Const} and \eqref{eq:EH-Const2} are used to check the numerical accuracy of the solutions obtained with the above general algorithm, as we shall discuss in appendix \ref{sec:app2}.

\subsection{Renormalized 1-point functions, field redefinitions and the apparent horizon}
\label{sec:2.3}

The first few terms in the ultraviolet near-boundary expansions of the bulk fields \eqref{eq:expA} --- \eqref{eq:expMax} explicitly read as below,
\begin{subequations}
\begin{align}
A(\tau,r) &= \frac{r^2}{2}+\lambda r+\frac{\lambda^2-2\partial_\tau\lambda}{2} +\frac{a_2}{r^2} +\frac{-4\lambda a_2+\partial_\tau a_2}{2r^3} +\mathcal{O}\!\left(r^{-4}\right),\label{eq:expA2}\\
\Sigma(\tau,r) &= \tau^{1/3}r+\frac{1+3\tau\lambda}{3\tau^{2/3}}-\frac{1}{9\tau^{5/3}r} +\frac{5+9\tau\lambda}{81\tau^{8/3}r^2} - \frac{20+60\tau\lambda+54\tau^2\lambda^2+27\tau^4\phi_2^2}{486\tau^{11/3}r^3} +\mathcal{O}\!\left(r^{-4}\right),\label{eq:expS2}\\
B(\tau,r) &= -\frac{2}{3}\ln(\tau) - \frac{2}{3\tau r} + \frac{1+2\tau\lambda}{3\tau^2r^2} - \frac{2+6\tau\lambda+6\tau^2\lambda^2}{9\tau^3r^3}\nonumber\\
& +\frac{6+24\tau\lambda+36\tau^2\lambda^2+24\tau^3\lambda^3-36\tau^4a_2 -2\tau^4\phi_2^2-27\tau^5\partial_\tau a_2-3\tau^5\phi_2\partial_\tau\phi_2}{36\tau^4r^4} \nonumber\\
& -\frac{1}{180\tau^5r^5}\left[24+120\tau^2\lambda^2(2+2\tau\lambda+\tau^2\lambda^2) -48\tau^4a_2+315\tau^5\partial_\tau a_2+35\tau^5\phi_2\partial_\tau\phi_2+15\tau^6 (\partial_\tau\phi_2)^2 \right. \nonumber\\
&\left. -\,20\tau\lambda(-6+36\tau^4a_2+2\tau^4\phi_2^2+27\tau^5\partial_\tau a_2+3\tau^5 \phi_2\partial_\tau\phi_2)+135\tau^6\partial^2_\tau a_2 +15\tau^6\phi_2\partial^2_\tau\phi_2\right]+\mathcal{O}\!\left(r^{-6}\right),\label{eq:expB2}\\
\phi(\tau,r) &= \frac{\phi_2}{r^2} +\frac{-2\lambda\phi_2+\partial_\tau\phi_2}{r^3} +\frac{36 \tau\lambda^2\phi_2+\sqrt{6}\tau\phi_2^2+(3-36\tau\lambda)\partial_\tau\phi_2 +9\tau\partial_\tau^2\phi_2}{12\tau r^4} +\mathcal{O}\!\left(r^{-5}\right),\label{eq:expphi2}\\
[d_+B](\tau,r) &= -\frac{1}{3\tau}+\frac{1}{3\tau^2r}-\frac{1+\tau\lambda}{3\tau^3r^2} +\frac{6+12\tau\lambda+6\tau^2\lambda^2+36\tau^4a_2+2\tau^4\phi_2^2 +27\tau^5\partial_\tau a_2+3\tau^5\phi_2\partial_\tau\phi_2}{18\tau^4r^3} +\mathcal{O}\!\left(r^{-4}\right),\label{eq:expdB2}\\
[d_+\Sigma](\tau,r) &= \frac{\tau^{1/3}r^2}{2}+\frac{(1+3\tau\lambda)r}{3\tau^{2/3}} +\frac{-1+2\tau\lambda+3\tau^2\lambda^2}{6\tau^{5/3}}+\frac{10}{81\tau^{8/3}r} +\frac{-100-120\tau\lambda+972\tau^4a_2+81\tau^4\phi_2^2}{972\tau^{11/3}r^2} +\mathcal{O}\!\left(r^{-3}\right),\label{eq:expdS2}\\
[d_+\phi](\tau,r) &= -\frac{\phi_2}{r} +\mathcal{O}\!\left(r^{-2}\right).\label{eq:expdphi2}
\end{align}
\end{subequations}

The holographic renormalization procedure for the 1RCBH model \cite{Critelli:2017euk} allows one to write down the renormalized 1-point functions of the dual quantum field theory, namely, the expectation values of the boundary energy-momentum tensor $\langle T_{\mu\nu}\rangle$, the $U(1)$ vector current $\langle J_\mu\rangle$, and the scalar condensate $\langle O_\phi\rangle$, in terms of the bulk gravity ultraviolet coefficients $\{a_2(\tau),\phi_2(\tau),\Phi_2(\tau)=-\rho_0/\tau\}$ and their time derivatives. In order to obtain those relations, one first needs to relate the holographic radial coordinates $r$ and $\bar{\rho}$ written,\footnote{We use here the bar to denote the radial Fefferman-Graham coordinate $\bar{\rho}$ in order to avoid confusing its notation with the charge density $\rho(\tau)$.} respectively, in the generalized infalling Eddington-Finkelstein coordinates, where we developed our previously discussed analysis of the holographic Bjorken flow dynamics, and in the Fefferman-Graham coordinates used in the holographic renormalization procedure \cite{Critelli:2017euk}. For the holographic Bjorken flow this relation is explicitly given by \cite{Rougemont:2021gjm},
\begin{align}
r(\bar{\rho})=\frac{1}{\sqrt{\bar{\rho}}}-\frac{a_2(\tau)\bar{\rho}^{3/2}}{4}-\frac{\partial_\tau a_2(\tau)\bar{\rho}^2}{10} + \mathcal{O}(\bar{\rho}^{5/2}).
\label{eq:rrho}
\end{align}
By substituting Eq. \eqref{eq:rrho} into the near-boundary expansions \eqref{eq:expA2} --- \eqref{eq:expdphi2}, one identifies the relevant ultraviolet coefficients in Fefferman-Graham coordinates entering the holographic renormalization formulas for the 1-point functions of the the dual quantum field theory at the boundary \cite{Critelli:2017euk}. The final results for the renormalized 1-point functions of the 1RCBH plasma undergoing Bjorken flow read as follows \cite{Critelli:2018osu},
\begin{subequations}
\begin{align}
\hat{\epsilon}(\tau) &\equiv \kappa_{5}^{2}\langle T_{\tau\tau} \rangle(\tau) = -3a_2(\tau)-\frac{1}{6}\phi_{2}(\tau)^{2},\label{eq:hatE}\\
\hat{p}_T(\tau) &\equiv \kappa_{5}^{2}\langle T_x^x \rangle(\tau) =\kappa_{5}^{2}\langle T_y^y \rangle(\tau) = -3a_2(\tau)-\frac{1}{6}\phi_{2}(\tau)^{2}-\frac{3}{2}\tau \partial_\tau a_2(\tau) -\frac{1}{6}\tau\phi_{2}(\tau)\partial_\tau \phi_2(\tau), \label{eq:hatpT}\\
\hat{p}_L(\tau) &\equiv \kappa_{5}^{2}\langle T_\xi^\xi \rangle(\tau) =  3a_2(\tau)+\frac{1}{6}\phi_{2}(\tau)^{2}+3\tau \partial_\tau a_2(\tau) +\frac{1}{3}\tau\phi_{2}(\tau)\partial_\tau \phi_2(\tau), \label{eq:hatpL}\\
\hat{\rho}(\tau) &\equiv \kappa_{5}^{2}\langle J^{\tau} \rangle(\tau) = -\Phi_2(\tau) =  \frac{\rho_0}{\tau}, \label{eq:hatrho}\\
\langle\hat{O}_\phi\rangle(\tau) &\equiv \kappa_{5}^{2}\langle \mathcal{O}_{\phi} \rangle(\tau) = -\phi_2(\tau). \label{eq:hatOphi}
\end{align}
\end{subequations}
From the above holographic formulas for the physical observables of the 1RCBH plasma, one can immediately confirm that the boundary system is conformal, since it has vanishing trace anomaly: $\langle \hat{T}_\mu^\mu\rangle = -\hat{\epsilon}+\hat{p}_L+2\hat{p}_T=0$.

In order to facilitate the numerical implementation of the general steps previously discussed to integrate the equations of motions of the system, we make use of field redefinitions corresponding to subtractions of the leading order near-boundary terms in the ultraviolet expansions of the bulk fields \eqref{eq:expA2} --- \eqref{eq:expdphi2}, such that the boundary values of the subtracted fields go to radial constants at the boundary,
\begin{subequations}
\begin{align}
u^2A_{s}(\tau,u) &\equiv A(\tau,u) - \frac{1}{2u^2} -\frac{\lambda}{u} -\frac{\lambda^2}{2}+\partial_\tau\lambda, \label{eq:subA}\\
u^3\Sigma_{s}(\tau,u) &\equiv \Sigma(\tau,u) - \frac{\tau^{1/3}}{u} -\frac{1+3 \tau  \lambda }{3 \tau ^{2/3}}+\frac{u}{9\tau^{5/3}} -\frac{(5+9\tau\lambda)u^2}{81 \tau^{8/3}}, \label{eq:subS}\\
u^4B_{s}(\tau,u) &\equiv B(\tau, u) +\frac{2 \ln(\tau )}{3}+\frac{2u}{3 \tau }-\frac{(1+2 \tau  \lambda)u^2}{3 \tau ^2} +\frac{(2+6 \tau  \lambda+6 \tau ^2 \lambda^2)u^3}{9 \tau ^3}, \label{eq:subB}\\
u^2 \phi_{s}(\tau,u) &\equiv \phi(\tau,u), \label{eq:subphi}\\
u^3[d_{+}B]_{s}(\tau,u) &\equiv [d_{+}B](\tau,u) + \frac{1}{3 \tau }-\frac{u}{3 \tau ^2}+\frac{(1+\tau\lambda)u^2}{3 \tau ^3}, \label{eq:subdB}\\
u^2 [d_{+}\Sigma]_{s}(\tau,u) &\equiv [d_{+}\Sigma](\tau,u) -\frac{\tau^{1/3}}{2 u^2} -\frac{1+3 \tau  \lambda}{3 \tau ^{2/3} u} -\frac{-1+2\tau\lambda+3\tau^2\lambda^2}{6 \tau^{5/3}}-\frac{10 u}{81 \tau ^{8/3}}, \label{eq:subdS}\\
u [d_{+}\phi]_{s}(\tau,u) &\equiv [d_{+}\phi](\tau,u), \label{eq:subdphi}\\
\mathcal{E}_s(\tau,u) & = \mathcal{E}(\tau,u), \label{eq:subE}
\end{align}
\end{subequations}
where we defined the new radial coordinate $u\equiv 1/r$, in terms of which the boundary lies at $u=0$. In terms of the above subtracted fields one can easily see that the corresponding boundary values are given by the following radial constants \cite{Critelli:2018osu},
\begin{subequations}
\begin{align}
A_s(\tau,u=0) &= a_2, \label{eq:bcAs}\\
\Sigma_s(\tau,u=0) &= - \frac{20+60\tau\lambda+54\tau^2\lambda^2+27\tau^4\phi_2^2}{486\tau^{11/3}}, \label{eq:bcSs}\\
B_s(\tau,u=0) &= \frac{6+24\tau\lambda+36\tau^2\lambda^2+24\tau^3\lambda^3-36\tau^4a_2 -2\tau^4\phi_2^2-27\tau^5\partial_\tau a_2-3\tau^5\phi_2\partial_\tau\phi_2}{36\tau^4}, \label{eq:bcBs}\\
\phi_s(\tau,u=0) &= \phi_2, \label{eq:bcphis}\\
[d_+B]_s(\tau,u=0) &= \frac{6+12\tau\lambda+6\tau^2\lambda^2+36\tau^4a_2+2\tau^4\phi_2^2 +27\tau^5\partial_\tau a_2+3\tau^5\phi_2\partial_\tau\phi_2}{18\tau^4}, \label{eq:bcdBs}\\
[d_+\Sigma]_s(\tau,u=0) &= \frac{-100-120\tau\lambda+972\tau^4a_2+81\tau^4\phi_2^2}{972\tau^{11/3}}, \label{eq:bcdSs}\\
[d_+\phi]_s(\tau,u=0) &= -\phi_2. \label{eq:bcdphis}
\end{align}
\end{subequations}
The equations of motion to be numerically solved as functions of $(\tau,u)$ are then given by the original equations of motion rewritten in terms of the subtracted fields, whose boundary values were obtained above.

Before proceeding to a discussion on the numerics in the next subsection \ref{sec:2.4}, we briefly discuss the calculation of the apparent horizon of the far from equilibrium black hole solutions.

The apparent horizon is the outermost trapped null surface inside the event horizon, separating a spacetime region where the geodesics are directed outward with light rays moving outward and a region where the light rays along the same geodesics move inward. Consequently, inside the apparent horizon all light rays move inward --- see e.g. Fig. 2 of Ref. \cite{Chesler:2009cy} for a clear illustration. The apparent horizon converges to the event horizon at late times, when the black hole geometry approaches equilibrium. Since the apparent horizon lies inside the event horizon, by cutting off the radial integration of the bulk equations of motion at some position inside the apparent horizon, one assures that the radial domain of the bulk geometry causally connected to observers at the boundary is adequately taken into account and no physical information is lost in such integration procedure.

In the holographic Bjorken flow, the radial position of the apparent horizon within the bulk generally presents wide fluctuations during the time evolution of the system. This is clearly inconvenient if one wants to cutoff the radial integration at some value of the holographic radial coordinate and hold this cutoff fixed for any value of time, since these wide fluctuations may eventually lead the radial cutoff to lie beyond, instead of behind the apparent horizon for some values of time, what could lead to loss of information and inaccurate physical results. Fortunately, it is possible to fix this issue by using the radial diffeomorphism shift invariance mentioned below Eq. \eqref{eq:lineElement}, which involves the function $\lambda(\tau)$. Indeed, since it is an arbitrary function of time, $\lambda(\tau)$ may be chosen in a way such as to keep the radial position of the apparent horizon fixed during the time evolution of the system \cite{Chesler:2013lia}.

For any metric field of the form shown in Eq. \eqref{eq:lineElement}, by assuming that the radial position of the apparent horizon $r_{\textrm{AH}}$ remains constant in time, one calculates it as the value of the radial coordinate which satisfies the following condition \cite{Chesler:2013lia},
\begin{align}
[d_+\Sigma](\tau,r_{\textrm{AH}})=0.
\label{eq:AppHor}
\end{align}
By requiring that $\partial_\tau r_{\textrm{AH}}(\tau)=0$ and that Eq. \eqref{eq:AppHor} remain valid at all times, it follows that $\partial_\tau [d_+\Sigma](\tau,r_{\textrm{AH}})=0$, implying that $d_+[d_+\Sigma](\tau,r_{\textrm{AH}})=A(\tau,r_{\textrm{AH}})\partial_r [d_+\Sigma](\tau,r_{\textrm{AH}})$. Substituting this result into the constraint Eq. \eqref{eq:EH-Const2}, and then combining it with other components of Einstein's equations, one obtains that,
\begin{align}
A(\tau,u_{\textrm{AH}}) = \frac{6([d_+B](\tau,u_{\textrm{AH}}))^2 + 2([d_+\phi](\tau,u_{\textrm{AH}}))^2}{2V+f\mathcal{E}^2},
\label{eq:Astar}
\end{align}
where we used Eq. \eqref{eq:AppHor}. By using now Eq. \eqref{eq:subA} evaluated at the apparent horizon position $u=u_{\textrm{AH}}$, it follows that,
\begin{align}
\partial_\tau\lambda(\tau) &= u_{\textrm{AH}}^2 A_s(\tau,u_{\textrm{AH}})+\frac{1}{2u_{\textrm{AH}}^2}+\frac{\lambda(\tau)}{u_{\textrm{AH}}}+\frac{\lambda^2(\tau)}{2} -A(\tau,u_{\textrm{AH}}).
\label{eq:dlambda}
\end{align}
Once one chooses the initial value $\lambda(\tau_0)$, the function $\lambda(\tau)$ is evolved in time by using Eq. \eqref{eq:dlambda}, which keeps the radial position of the apparent horizon fixed during the time evolution. In practice, as done in \cite{Rougemont:2021qyk,Rougemont:2021gjm}, we choose $\lambda(\tau_0)=0$ and solve Eq. \eqref{eq:AppHor} using Eq. \eqref{eq:subS} with the Newton-Raphson method. In this way, the apparent horizon remains fixed, within some numerical tolerance, in its initial position calculated with the initial condition $\lambda(\tau_0)=0$.

\subsection{Numerical solver}
\label{sec:2.4}

In order to numerically integrate the nested set of $1+1$ partial differential equations of motion of the system, we discretize the radial and time directions. We employ the pseudospectral method \cite{boyd01} to integrate in the radial direction and make use of the fourth order Adams-Bashforth method to integrate in the time direction. In both cases, the general algorithmic steps are the same discussed e.g. in \cite{Critelli:2017euk,Rougemont:2021gjm}, and we refer the interested reader to those works for the details.

Here we supplement the information concerning the extra steps required to implement these algorithms in the case of the 1RCBH model undergoing Bjorken flow. The complete set of freely chosen initial data needed to be specified on the initial time slice $\tau_0$ is given by $\{B_s(\tau_0,u),\phi_s(\tau_0,u),a_2(\tau_0),\rho_0;\lambda(\tau_0)\}$. In order to evolve the system to the next time slices, one also needs to obtain the expressions for the time derivatives of these data. As discussed at the end of the previous subsection \ref{sec:2.3}, we set here $\lambda(\tau_0)=0$ and $\partial_\tau\lambda(\tau)$ is given by Eq. \eqref{eq:dlambda}. On the other hand, the physically different initial data will be determined by different choices for the subset $\{B_s(\tau_0,u),\phi_s(\tau_0,u),a_2(\tau_0),\rho_0\}$, which will be discussed in the next section \ref{sec:3}. It remains, therefore, the task of obtaining the expressions for the time derivatives $\{\partial_\tau B_s(\tau,u),\partial_\tau\phi_s(\tau,u),\partial_\tau a_2(\tau)\}$ required to evolve the system in time.

The expressions for $\partial_\tau B_s$ and $\partial_\tau\phi_s$ can be derived from the expressions for $d_+B=\partial_\tau B+A\partial_r B$ and $d_+\phi=\partial_\tau \phi+A\partial_r \phi$ rewritten in terms of the subtracted fields and the radial coordinate $u=1/r$. After some tedious algebraic manipulations, one arrives at the following results,
\begin{align}
\partial_\tau B_s(\tau,u) &= \frac{[d_+B]_s}{u} -\frac{2}{3\tau^4 u} - \frac{2A_s}{3\tau} + \frac{2uA_s}{3\tau^2} - \frac{2u^2A_s}{3\tau^3} + 4u^3A_sB_s + \frac{2B_s}{u} + \frac{B_s'}{2} + u^4A_sB_s' \nonumber\\
& +\, \left(4B_s-\frac{2}{u\tau^3}+\frac{4u A_s}{3\tau} - \frac{2u^2 A_s}{\tau^2}+uB_s'\right)\lambda+\left(\! - \frac{1}{3\tau^3} \!-\! \frac{7}{3\tau^2 u} \!+\! 2uB_s \!-\! \frac{2u^2A_s}{\tau} +\frac{u^2B_s'}{2} \right)\lambda^2 \nonumber\\
& -\, \left( \frac{1}{\tau^2} + \frac{4}{3\tau u} \right)\lambda^3 - \frac{\lambda^4}{\tau} + \left( \frac{2}{3\tau^3} - 4uB_s - u^2B_s' + \frac{2\lambda}{\tau^2} + \frac{2\lambda^2}{\tau} \right)\partial_\tau\lambda,\label{eq:dtBs}\\
\partial_\tau\phi_s(\tau,u) & = \frac{[d_+\phi]_s}{u} +\frac{\phi_s'}{2}+u^4A_s\phi_s' +\frac{\phi_s}{u}+2u^3A_s\phi_s+\left(u\phi_s'+2\phi_s\right)\lambda +\left(\frac{u^2\phi_s'}{2}+u\phi_s\right)\lambda^2-\left(2u\phi_s+u^2\phi_s'\right)\partial_\tau\lambda,\label{eq:dtphis}
\end{align}
where $X_s'(\tau,u)\equiv \partial_u X_s(\tau,u)$ is evaluated at any constant time slice by applying the pseudospectral finite differentiation matrix \cite{boyd01} to the numerically known solution $X_s(\tau,u)$, which is expressed as a vector with $N$ components corresponding to the values of $X_s(\tau,u)$ on top of the $N$ collocation points of the Chebyshev-Gauss-Lobatto radial grid \cite{Critelli:2017euk,Rougemont:2021gjm}.

Regarding the expression for $\partial_\tau a_2(\tau)$, it can be determined directly from Eq. \eqref{eq:bcBs} once $\partial_\tau \phi_2(\tau)$ is known, which in turn follows from Eqs. \eqref{eq:expphi2} and \eqref{eq:subphi},
\begin{align}
\partial_\tau\phi_2(\tau) = \phi_s'(\tau,u=0)+2\lambda(\tau)\phi_2(\tau).
\label{eq:dtphi2}
\end{align}

In the algorithms employed in \cite{Critelli:2017euk,Critelli:2018osu,Rougemont:2021qyk,Rougemont:2021gjm} to integrate in the radial direction, the boundary values of the fields are calculated separately, which can be done here by using Eqs. \eqref{eq:bcAs} --- \eqref{eq:bcdphis} and the following results also coming from the near-boundary expansions \eqref{eq:expA2} --- \eqref{eq:expdphi2},
\begin{align}
\partial_\tau B_s(\tau,u=0) &= -\frac{2}{3\tau^5}-\frac{2\lambda}{\tau^4}-\frac{2\lambda^2}{\tau^3}-\frac{2\lambda^3}{3\tau^2}+\frac{2\partial_\tau\lambda}{3\tau^3}+\frac{2\lambda\partial_\tau\lambda}{\tau^2}+\frac{2\lambda^2\partial_\tau\lambda}{\tau}-\frac{7\partial_\tau a_2}{4}-\frac{7\phi_2 \partial_\tau\phi_2}{36}-\frac{\tau(\partial_\tau\phi_2)^2}{12}\nonumber\\
&-\,\frac{3\tau\partial^2_\tau a_2}{4}-\frac{\tau\phi_2\partial^2_\tau\phi_2}{12},\label{eq:dtBs0}\\
\partial_\tau\phi_s(\tau,u=0) &= \partial_\tau\phi_2,\label{eq:dtphis0}
\end{align}
where, in particular, the following second derivatives are obtained from Eqs. \eqref{eq:expB2} and \eqref{eq:expphi2}, respectively,
\begin{align}
\partial^2_\tau a_2(\tau) &= -\frac{4 B'_s(\tau,u=0)}{3\tau}-\frac{8}{45\tau^6} -\frac{8\lambda}{9\tau^5}-\frac{16\lambda^2}{9\tau^4}-\frac{16\lambda^3}{9\tau^3} -\frac{8\lambda^4}{9\tau^2}+\frac{16a_2}{45\tau^2}+\frac{16\lambda a_2}{3\tau}+\frac{8\lambda\phi_2^2}{27\tau}\nonumber\\
&-\, \frac{7\partial_\tau a_2}{3\tau}+4\lambda\partial_\tau a_2 -\frac{7\phi_2\partial_\tau\phi_2}{27\tau}+\frac{4\lambda\phi_2\partial_\tau\phi_2}{9} -\frac{(\partial_\tau\phi_2)^2}{9}-\frac{\phi_2\partial^2_\tau\phi_2}{9},\label{eq:dt2a2}\\
\partial^2_\tau \phi_2(\tau) &= \frac{2\phi''_s(\tau,u=0)}{3}-4\lambda^2\phi_2-\sqrt{\frac{2}{27}}\,\phi_2^2-\frac{\partial_\tau\phi_2}{3\tau} +4\lambda\partial_\tau\phi_2.\label{eq:dt2phi2}
\end{align}

We close this section by remarking that for the numerical calculations carried out in the present work we used $N=21$ collocation points in the radial grid and a time step size of $\Delta\tau=12\times10^{-5}$. However, for three specific initial conditions to be discussed next we needed to increase the number of collocation points to $N\sim 30$ in order to eliminate numerically spurious oscillations in the normalized scalar condensate at early times.

%%%%%%%%%%%%%%%%%%%%%%%%%%%%%%%%%
\section{Non-equilibrium entropy, initial data, physical observables and energy conditions}
\label{sec:3}

In this section, we discuss the holographic formula for the non-equilibrium entropy density, the form of the initial data $\{B_s(\tau_0,u),\phi_s(\tau_0,u),a_2(\tau_0),\rho_0\}$ which will be analyzed in the present work, besides presenting the conventions we will use to plot the dimensionless ratios for the physical observables, and also the form of the dominant and weak energy conditions for a conformal fluid undergoing Bjorken flow.

Concerning the holographic calculation of the non-equilibrium entropy density, it proceeds as follows \cite{Chesler:2009cy,Rougemont:2021qyk,Rougemont:2021gjm}. The area of the apparent horizon reads,
\begin{align}
A_{\textrm{AH}}(\tau) = \int d^3x \sqrt{-g}\biggr|_{u=u_{\textrm{AH}}} = \int dx dy d\xi \sqrt{-g}\biggr|_{u=u_{\textrm{AH}}} = \sqrt{-g}\biggr|_{u=u_{\textrm{AH}}} \mathcal{A} = |\Sigma(\tau,u_{\textrm{AH}})|^3 \mathcal{A},
\label{eq:AreaHor}
\end{align}
where $\mathcal{V}(\tau) = \tau \mathcal{A} = \tau \int dx dy d\xi$ is the expanding volume of the medium in Bjorken flow. Analogously to the Bekenstein-Hawking relation, one obtains for the non-equilibrium holographic entropy the following relation, written in terms of the area of the apparent horizon,
\begin{align}
S_{\textrm{AH}}(\tau) = \frac{A_{\textrm{AH}}(\tau)}{4G_5} = \frac{2\pi |\Sigma(\tau,u_{\textrm{AH}})|^3 \mathcal{A}}{\kappa_5^2},
\label{eq:BHrel}
\end{align}
and the entropy density is, therefore,
\begin{align}
\hat{s}_{\textrm{AH}}(\tau) \equiv \kappa_5^2\, s_{\textrm{AH}}(\tau) = \kappa_5^2\, \frac{S_{\textrm{AH}}(\tau)}{\mathcal{V}(\tau)} = \frac{2\pi|\Sigma(\tau,u_{\textrm{AH}})|^3}{\tau},
\label{eq:hats}
\end{align}
where $\Sigma(\tau,u_{\textrm{AH}})$ is calculated in terms of the numerical result for $\Sigma_s(\tau,u_{\textrm{AH}})$ through Eq. \eqref{eq:subS} evaluated at the radial location of the apparent horizon.

Regarding the initial profile for the subtracted metric anisotropy coefficient, we are going to work with the following general form \cite{Rougemont:2021qyk,Rougemont:2021gjm} (which is considerably broader than the forms considered e.g. in \cite{Critelli:2018osu,wilkaodamassa}),
\begin{equation}
B_s(\tau_0,u) = \Omega_1 \cos(\gamma_1 u) + \Omega_2 \tan(\gamma_2 u) + \Omega_3 \sin(\gamma_3 u) + \sum_{i=0}^{5}\beta_i u^i  +\, \frac{\alpha}{u^4} \left[-\frac{2}{3} \ln\left(1+ \frac{u}{\tau _0}\right) + \frac{2 u^3}{9 \tau_0^3} - \frac{u^2}{3 \tau _0^2}+\frac{2 u}{3 \tau _0}\right],
\label{eq:Bs0}
\end{equation}
such that one needs to choose the values of the parameters $\{\Omega_i,\gamma_i,\beta_i,\alpha\}$ in \eqref{eq:Bs0}, and also the values of the initial data $\{\phi_s(\tau_0,u),a_2(\tau_0),\rho_0\}$ in order to fully specify a given initial condition. We take $\tau_0=0.2$ as the initial time of our numerical simulations and evolve each initial condition up to $\tau_\textrm{end}=35$. This end time used in our present numerical simulations is about five times larger than considered in previous works \cite{Critelli:2018osu,Rougemont:2021qyk,Rougemont:2021gjm}, what is needed in order to be able to see the effective thermalization of the scalar condensate at late times in the evolution of the system. The set of parameters $\{\Omega_i,\gamma_i,\beta_i,\alpha\}$ in \eqref{eq:Bs0} which we are going to explore in this paper is provided in Table \ref{tabICs}. With these different profiles for $B_s(\tau_0,u)$ it is possible to generate many physically different possibilities for the time evolution of the 1RCBH plasma expanding according to the Bjorken flow dynamics.
\begin{table}[h]
\centering
\begin{tabular}{|c||c|c|c|c|c|c|c|c|c|c|c|c|c|}
\hline
$B_s\#$ & $\Omega_1$ & $\gamma_1$ & $\Omega_2$ & $\gamma_2$ & $\Omega_3$ & $\gamma_3$ & $\beta_0$ & $\beta_1$ & $\beta_2$ & $\beta_3$ & $\beta_4$ & $\beta_5$ & $\alpha$ \\
\hline
\hline
1 & 0 & 0 & 0 & 0 & 0 & 0 & 0.5 & -0.5 & 0.4 & 0.2 & -0.3 & 0.1 & 1 \\
\hline
2 & 0 & 0 & 0 & 0 & 0 & 0 & -0.2 & -0.5 & 0.3 & 0.1 & -0.2 & 0.4 & 1 \\
\hline
3 & 0 & 0 & 0 & 0 & 0 & 0 & 0.1 & -0.4 & 0.3 & 0 & -0.1 & 0 & 1 \\
\hline
4 & 0 & 0 & 1 & 1 & 0 & 0 & 0 & 0 & 0 & 0 & 0 & 0 & 1 \\
\hline
5 & 0 & 0 & 0 & 0 & 0 & 0 & -0.2 & -0.5 & 0 & 0 & 0 & 0 & 1 \\
\hline
6 & 0 & 0 & 0 & 0 & 0 & 0 & -0.2 & -0.4 & 0 & 0 & 0 & 0 & 1 \\
\hline
7 & 0 & 0 & 0 & 0 & 0 & 0 & -0.2 & -0.6 & 0 & 0 & 0 & 0 & 1 \\
\hline
8 & 0 & 0 & 0 & 0 & 0 & 0 & -0.3 & -0.5 & 0 & 0 & 0 & 0 & 1 \\
\hline
9 & 0 & 0 & 0 & 0 & 1 & 8 & 0 & 0 & 0 & 0 & 0 & 0 & 1 \\
\hline
10 & 1 & 8 & 0 & 0 & 0 & 0 & -0.2 & -0.5 & 0 & 0 & 0 & 0 & 1 \\
\hline
11 & 0.5 & 8 & 0 & 0 & 0 & 0 & -0.2 & -0.5 & 0 & 0 & 0 & 0 & 1 \\
\hline
\end{tabular}
\caption{Set of parameters for the initial profile of the subtracted metric anisotropy \eqref{eq:Bs0} analyzed in this work.}
\label{tabICs}
\end{table}

In what concerns the initial profile for the subtracted dilaton field, we are going to consider the following general form,
\begin{equation}
\phi_s(\tau_0,u) = \sum_{i=0}^{3}a_i u^i + \alpha_0\, e^{-u^2/\sigma} + \alpha_1 \cos(\omega_1 u) + \alpha_2 \sin(\omega_2 u),
\label{eq:phis0}
\end{equation}
The set of parameters $\{a_i,\alpha_i,\sigma,\omega_i\}$ in \eqref{eq:phis0} which we are going to explore in this paper is provided in Table \ref{tabphis}. \footnote{Notice that $\phi_s1$ in Table \ref{tabphis} is the trivial dilaton profile, $\phi_s(\tau_0,u)=0$.}
\begin{table}[h]
\centering
\begin{tabular}{|c||c|c|c|c|c|c|c|c|c|c|}
\hline
$\phi_s\#$ & $a_0$ & $a_1$ & $a_2$ & $a_3$ & $\alpha_0$ & $\alpha_1$ & $\alpha_2$ & $\sigma$ & $\omega_1$ & $\omega_2$ \\
\hline
\hline
1 & 0 & 0 & 0 & 0 & 0 & 0 & 0 & 2 & 0 & 0 \\
\hline
2 & 0 & 1 & 1 & 0 & 0 & 0 & 0 & 2 & 0 & 0 \\
\hline
3 & -0.8 & 1.1 & 0.6 & 0 & 0 & 0 & 0 & 2 & 0 & 0 \\
\hline
4 & 0.7 & -0.9 & 0 & 0.4 & 0 & 0 & 0 & 2 & 0 & 0 \\
\hline
5 & 0 & 0 & 0 & 0 & 0.5 & 0 & 0 & 2 & 0 & 0 \\
\hline
6 & 0 & 0 & 0 & 0 & 0 & 1 & 0 & 2 & 1 & 0 \\
\hline
7 & 0 & 0 & 0 & 0 & 0 & 0 & 1 & 2 & 0 & 5 \\
\hline
\end{tabular}
\caption{Set of parameters for the initial profile of the subtracted dilaton field \eqref{eq:phis0} analyzed in this work.}
\label{tabphis}
\end{table}

As discussed before, the initial data $\rho_0$ sets the initial value of the charge density of the medium, $\hat{\rho}(\tau_0)=\rho_0/\tau_0$. On the other hand, $a_2(\tau_0)$ together with $\phi_2(\tau_0)=\phi_s(\tau_0,u=0)$ specify the initial value of the energy density of the fluid, $\hat{\epsilon}(\tau_0)$, according to Eq. \eqref{eq:hatE}. In section \ref{sec:4}, we will present our results for the physical observables of the 1RCBH plasma undergoing Bjorken flow, using as initial data the profiles for $B_s(\tau_0,u)$ given in Table \ref{tabICs} with variations on the initial charge density of the medium, keeping fixed its initial energy density and setting $\phi_s(\tau_0,u)=0$. In section \ref{sec:5}, we vary instead the initial energy density of the medium, while keeping fixed its initial charge density and setting again $\phi_s(\tau_0,u)=0$. In section \ref{sec:extra}, we shall analyze the results obtained with the different profiles for $\phi_s(\tau_0,u)$ given in Table \ref{tabphis}, while keeping fixed the remaining initial data.

Regarding the physical observables we shall consider in this work, they will be expressed through the following dimensionless ratios: $\Delta \hat{p}/\hat{\epsilon}=(\hat{p}_T-\hat{p}_L)/\hat{\epsilon}$, which is obtained from Eqs. \eqref{eq:hatE} --- \eqref{eq:hatpL}; $\langle\hat{O}_\phi\rangle/\hat{\epsilon}^{1/2}$, which is obtained from Eqs. \eqref{eq:hatE} and \eqref{eq:hatOphi}; $\hat{\rho}^{4/3}/\hat{\epsilon}$, which is obtained from Eqs. \eqref{eq:hatE} and \eqref{eq:hatrho}; while for the entropy, we will use the ratio $\hat{s}_\textrm{AH}^{4/3}/\hat{\epsilon}$, which is obtained from Eqs. \eqref{eq:hatE} and \eqref{eq:hats}, only in the late time physical consistency analysis of the numerical solutions to be discussed in appendix \ref{sec:app1}. Indeed, whilst $\hat{s}_\textrm{AH}^{4/3}/\hat{\epsilon}$ is important to check the late time convergence of the entropy density to the thermodynamically stable branch of equilibrium black hole solutions previously discussed in section \ref{sec:2.1}, since the energy density is a nontrivial function of time, one needs a different dimensionless normalization for the entropy density $s_{\textrm{AH}}(\tau)$ in order to be able to follow the actual time evolution of the entropy function $S_{\textrm{AH}}(\tau)$, which is directly related to the area of the apparent horizon $A_\textrm{AH}(\tau)$ through Eq. \eqref{eq:BHrel}. This is important in order to check the validity of the second law of thermodynamics during the time evolution of the fluid, which is associated here with the fact that the area of the apparent horizon should not decrease as a function of time during the evolution of the system; moreover, it is also important to directly track the time evolution of the area of the apparent horizon because flat regions for this observable are a direct measure of regions in the time evolution of the fluid with zero entropy production \cite{Rougemont:2021qyk,Rougemont:2021gjm}. Thus, for the main physical analyses present in this paper we are going to describe the time evolution of the entropy of the medium through the following dimensionless ratio,
\begin{align}
\frac{\tau\hat{s}_\textrm{AH}(\tau)}{\Lambda^2} = \frac{\hat{S}_\textrm{AH}(\tau)}{\mathcal{A}\Lambda^2} = \frac{2\pi A_\textrm{AH}(\tau)}{\mathcal{A}\Lambda^2} = \frac{2\pi|\Sigma(\tau,u_\textrm{AH})|^3}{\Lambda^2},
\label{eq:normS}
\end{align}
where $\Lambda$ is an energy scale, extracted for each initial condition, through a fit of the late time result for the full numerical energy density to its corresponding analytical hydrodynamic Navier-Stokes result \cite{Critelli:2018osu}. This energy scale shall also be used to define the following effective dimensionless time measure, $\omega_\Lambda(\tau) \equiv \tau T_{\textrm{eff}}(\tau)$, where we follow \cite{Florkowski:2017olj,Rougemont:2021qyk,Rougemont:2021gjm} and take the `effective non-equilibrium temperature' (at zero density), $T_{\textrm{eff}}(\tau)$, to be given by the third-order hydrodynamic truncation for the energy density of the pure thermal SYM plasma \cite{Booth:2009ct},
\begin{equation}
T_{\textrm{3rd}}^{(0)}(\tau) = \frac{\Lambda}{(\Lambda\tau)^{1/3}} \left[ 1 - \frac{1}{6\pi(\Lambda\tau)^{2/3}} + \frac{-1+\ln 2}{36\pi^2(\Lambda\tau)^{4/3}} +\, \frac{-21 + 2\pi^2 + 51\ln 2 - 24\ln^2 2}{1944\pi^3 (\Lambda\tau)^2} \right].
\label{eq:time}
\end{equation}
All the aforementioned dimensionless ratios for the physical observables of the 1RCBH plasma undergoing Bjorken flow will be numerically interpolated as functions of the dimensionless time measure $\omega_\Lambda(\tau)\equiv\tau T_{\textrm{3rd}}^{(0)}(\tau)$ in order to plot our main results in sections \ref{sec:4} and \ref{sec:5}.

Regarding the value of $\mu/T$ for each initial condition, as discussed before, it is not an initial data in the Bjorken flow of the 1RCBH plasma, instead it is extracted from the latest time result for $[\hat{\rho}^{4/3}/\hat{\epsilon}](\tau)$, which is matched to the corresponding thermodynamically stable equilibrium result for $[\hat{\rho}^{4/3}/\hat{\epsilon}]_\textrm{(eq.)}$ discussed in section \ref{sec:2.1}. Clearly, for a reliable estimate of the value of $\mu/T$, this matching procedure needs to be done when $[\hat{\rho}^{4/3}/\hat{\epsilon}](\tau)$ is almost stabilized in a constant value at late times, and the result will be more precise the larger is the end time $\tau_\textrm{end}$ of the numerical simulations.

Closely related to the above discussion, several important analytical consistency checks of our numerical results can be performed. Since all the near-equilibrium hydrodynamic results and also all the equilibrium thermodynamic results are functions of $\mu/T$, and not of the specific initial conditions of the far from equilibrium fluid, once the value of $\mu/T$ for a given initial condition, extracted from the late time analysis of $[\hat{\rho}^{4/3}/\hat{\epsilon}](\tau)$, is plugged into the analytical expressions for $[\langle\hat{O}_\phi\rangle/\hat{\epsilon}^{1/2}]_\textrm{(eq.)}$ and $[\hat{s}_\textrm{AH}^{4/3}/\hat{\epsilon}]_\textrm{(eq.)}$ in the thermodynamically stable branches of equilibrium black hole solutions, one has the asymptotic values that should be attained at late times by the dynamical observables $[\langle\hat{O}_\phi\rangle/\hat{\epsilon}^{1/2}](\tau)$ and $[\hat{s}_\textrm{AH}^{4/3}/\hat{\epsilon}](\tau)$; moreover, for $[\Delta \hat{p}/\hat{\epsilon}](\tau)$, one should obtain at late times convergence of the numerical results to the corresponding analytical hydrodynamic Navier-Stokes (NS) results coming from Eq. \eqref{eq:dPE},
\begin{align}
\left[\frac{\Delta\hat{p}}{\hat{\epsilon}}\right]_\textrm{NS}(\tau,\mu/T)=2+\frac{3}{2}\,\tau\frac{\partial_\tau\hat{\epsilon}_\textrm{NS}(\tau,\mu/T)}{\hat{\epsilon}_\textrm{NS}(\tau,\mu/T)},
\label{eq:NSpressure}
\end{align}
with $\hat{\epsilon}_\textrm{NS}(\tau,\mu/T)$ given by Eq. (12) of \cite{Critelli:2018osu} evaluated at the same value of $\mu/T$ obtained from the late time analysis of $ [\hat{\rho}^{4/3}/\hat{\epsilon}](\tau)$. We indeed consistently see, for all the initial conditions, the late time convergence of all these physical observables to their corresponding analytical hydrodynamic or thermodynamic results evaluated at the same value of $\mu/T$. These outcomes constitute independent and highly nontrivial physical consistency checks of our numerical simulations by analytical results in the late time regime of the system.

We close this section by briefly reviewing the weak energy condition (WEC) and the dominant enegy condition (DEC) for a conformal fluid expanding according to the Bjorken flow dynamics, which will be important in our physical analyses in the course of the next sections. These classical energy conditions are commonly postulated in general relativity in order to constrain the content of matter's energy-momentum tensor used in Einstein's equations and ensure energy positiveness, even though some quantum effects are known to violate such classical energy conditions \cite{Visser:1999de,Costa:2021hpu}. The WEC states that $\langle \hat{T}_{\mu\nu}\rangle t^\mu t^\nu\ge 0$ for any timelike vector $t^\mu$. It implies the following inequalities for a conformal fluid undergoing Bjorken flow \cite{Janik:2005zt,Rougemont:2021qyk,Rougemont:2021gjm},
\begin{align}
\hat{\epsilon}(\tau)\ge 0\,;\qquad \{\partial_\tau\hat{\epsilon}(\tau)\le 0,\,\, \tau\partial_\tau\ln[\hat{\epsilon}(\tau)]\ge -4\}\,\,\,\Rightarrow\,\,\, -4 \le \left[\frac{\Delta\hat{p}}{\hat{\epsilon}}\right](\tau)\le 2.
\label{eq:WEC}
\end{align}
Therefore, one may violate the WEC by either having a negative energy density and/or by having a normalized pressure anisotropy assuming values outside the aformentioned interval. All the cases with transient, far from equilibrium violations of the WEC analyzed here, and also in Refs. \cite{Rougemont:2021qyk,Rougemont:2021gjm}, were related to violations on the bounds for the pressure anisotropy in \eqref{eq:WEC}, while the energy density was always positive during the time evolution of the medium. We further remark that for a conformal fluid, the strong energy condition (SEC), stating that $\langle\hat{T}_{\mu\nu}\rangle t^\mu t^\nu\ge -\langle \hat{T}_\mu^\mu\rangle /2$, is equivalent to the WEC, since $\langle\hat{T}_\mu^\mu\rangle=0$ in a conformal system. On the other hand, the DEC states that for any future-directed timelike vector $t^\mu$, $X^\mu \equiv -\langle\hat{T}^{\mu\nu}\rangle t_\nu$ must also be a future-directed timelike or null vector. This is a sufficient but not a necessary condition to establish causal propagation of matter \cite{WaldBookGR1984}. It was shown in \cite{Rougemont:2021qyk,Rougemont:2021gjm} that for a conformal fluid the DEC implies the following inequalities, which are more restrictive than the WEC \eqref{eq:WEC},
\begin{align}
\hat{\epsilon}(\tau) \ge 0\,;\qquad -1 \le\left[\frac{\Delta\hat{p}}{\hat{\epsilon}}\right](\tau)\le 2.
\label{eq:DEC}
\end{align}

%%%%%%%%%%%%%%%%%%%%%%%%%%%%%%%%%
\section{Results for variations of the initial charge density}
\label{sec:4}

\begin{figure*}%[h]
\center
\subfigure[]{\includegraphics[width=0.49\textwidth]{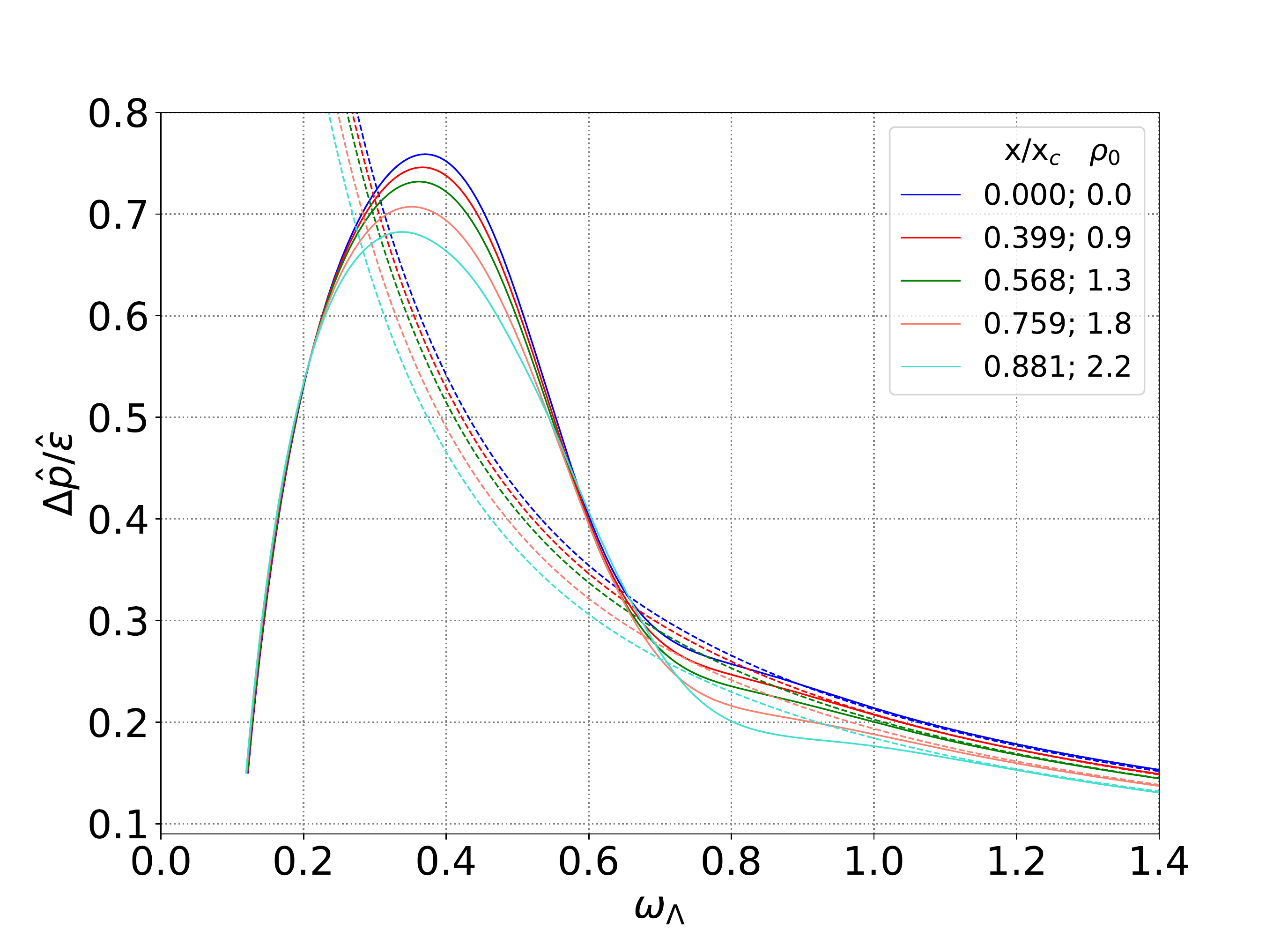}}
\subfigure[]{\includegraphics[width=0.49\textwidth]{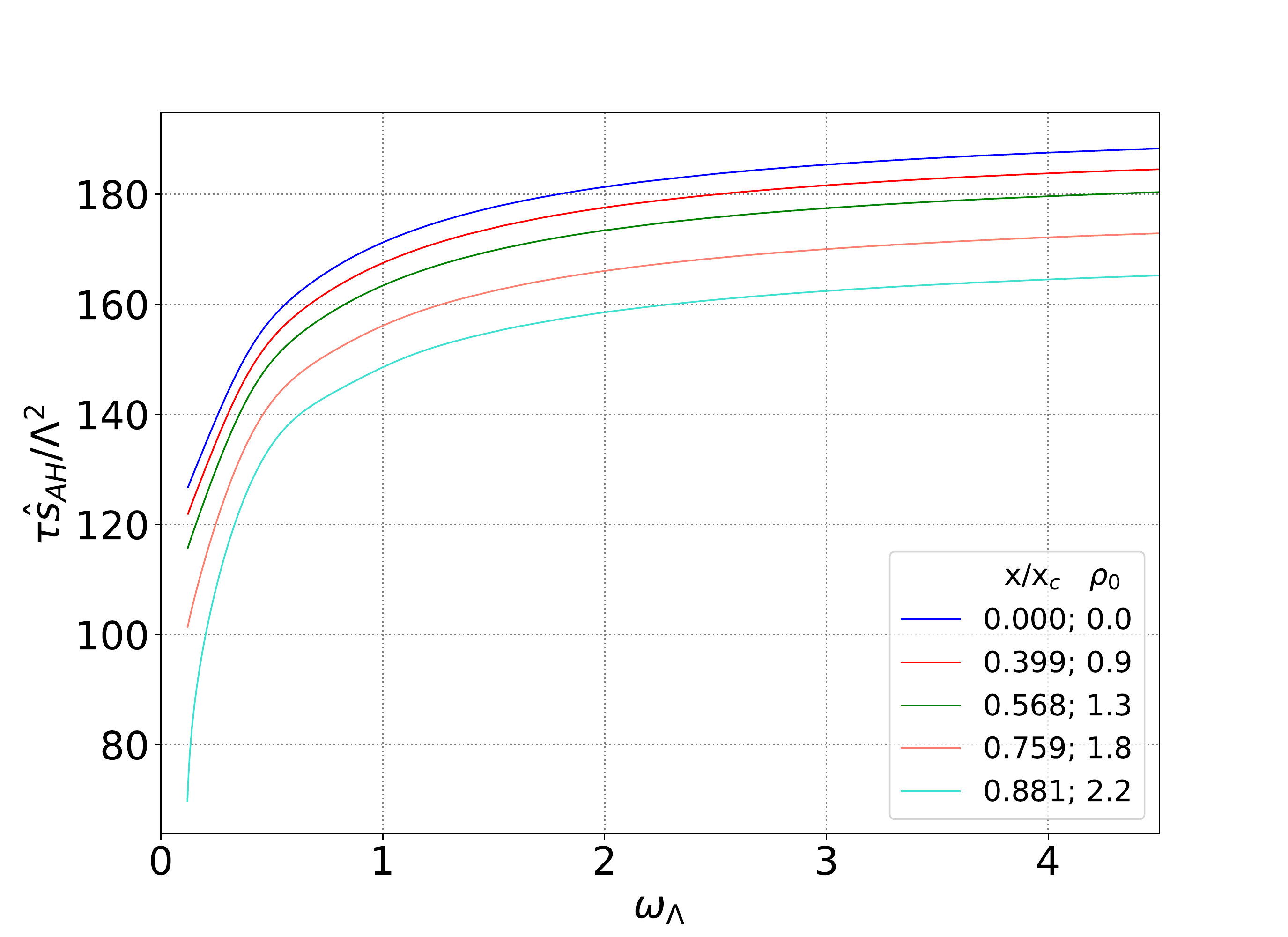}}
\subfigure[]{\includegraphics[width=0.49\textwidth]{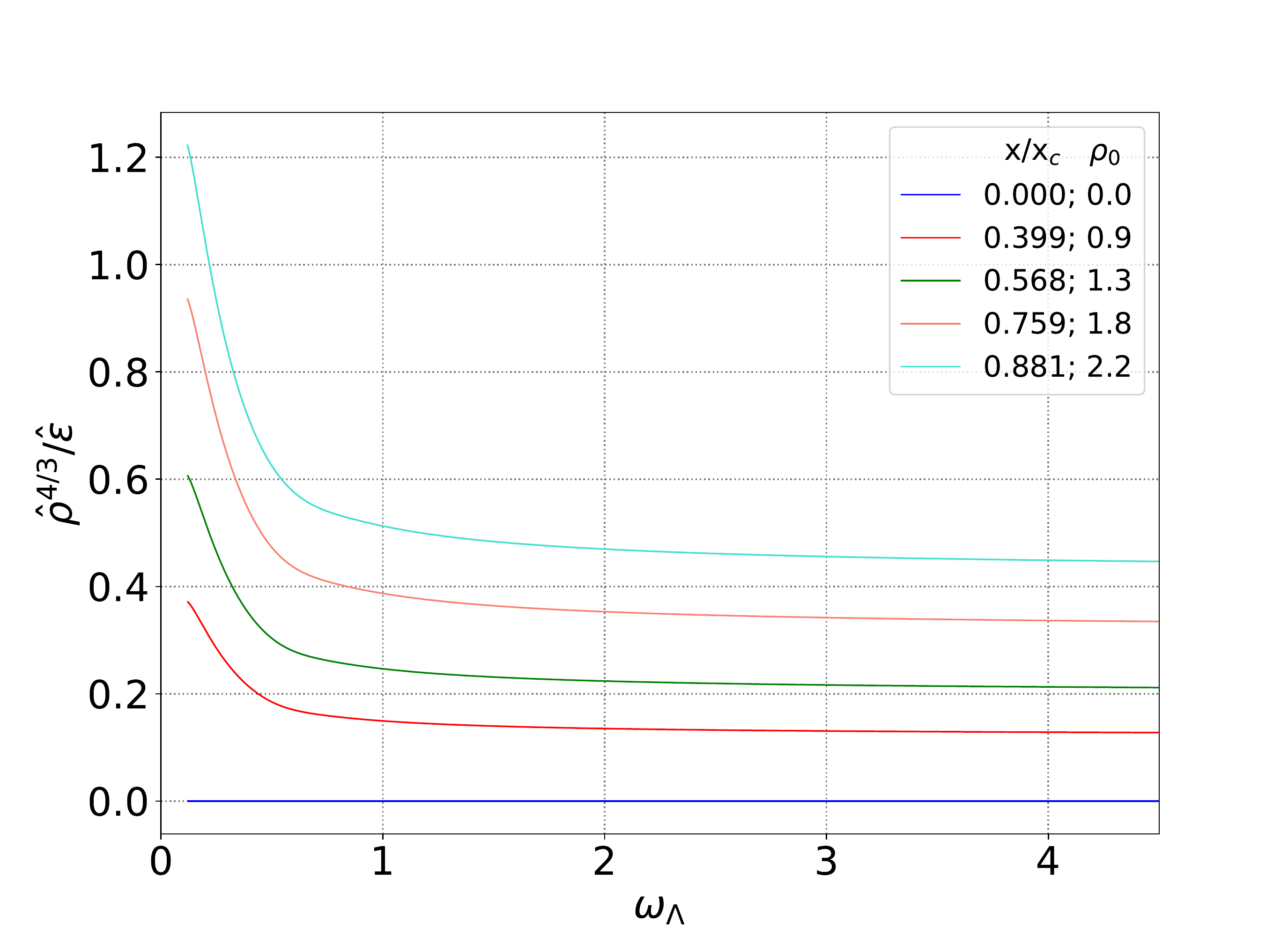}}
\subfigure[]{\includegraphics[width=0.49\textwidth]{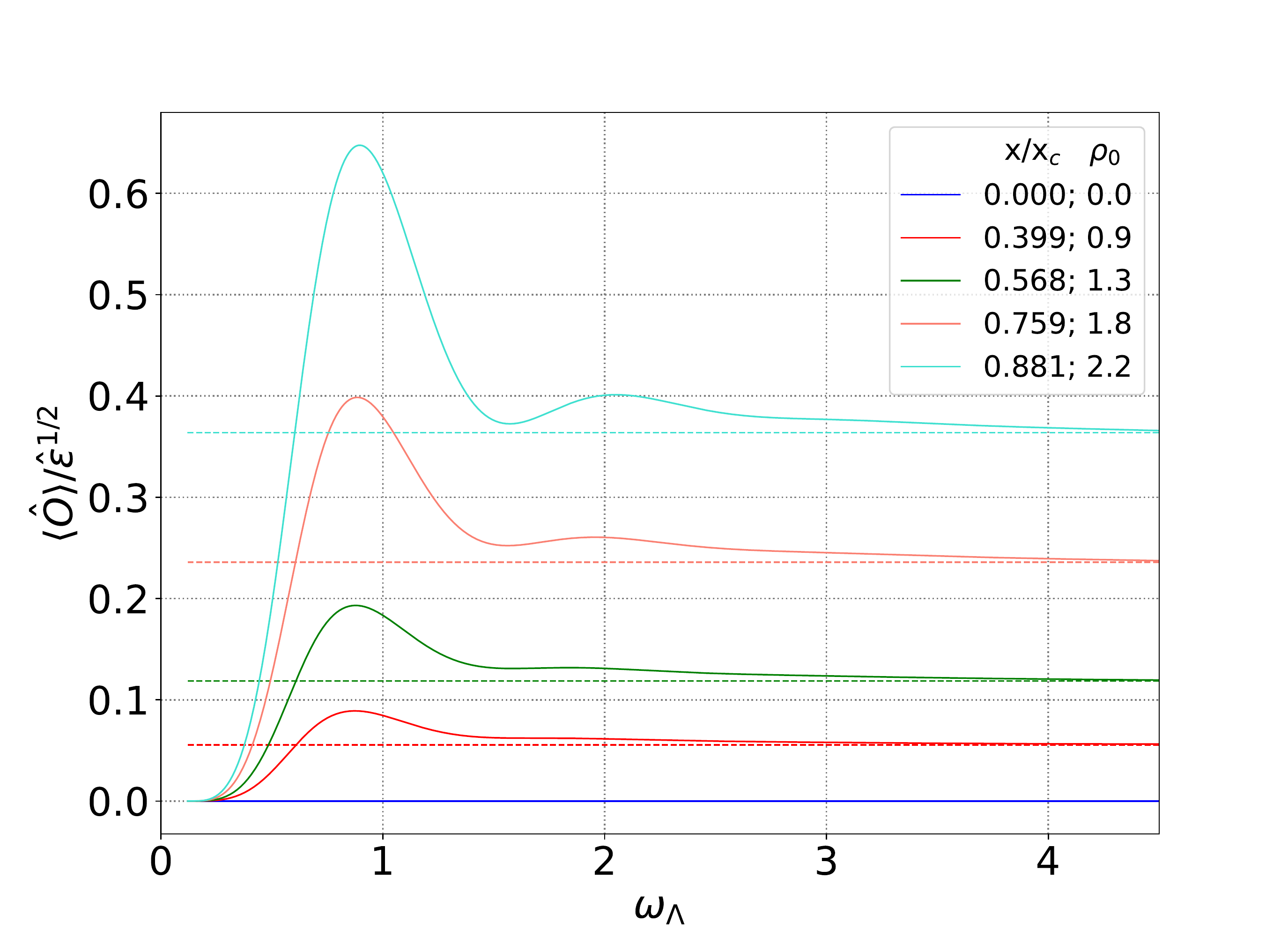}}
\caption{(a) Normalized pressure anisotropy (solid lines) and the corresponding hydrodynamic Navier-Stokes result (dashed lines), (b) normalized non-equilibrium entropy $\hat{S}_\textrm{AH}/\mathcal{A}\Lambda^2=\tau\hat{s}_\textrm{AH}/\Lambda^2$, (c) normalized charge density, and (d) normalized scalar condensate (solid lines) and the corresponding thermodynamic stable equilibrium result (dashed lines). Results obtained for variations of $\rho_0$ keeping fixed $B_s1$ in Table \ref{tabICs} with $a_2(\tau_0)=-6.67$. Note that $x_c\equiv\left(\mu/T\right)_c=\pi/\sqrt{2}$ is the critical point.}
\label{fig:result1}
\end{figure*}

\begin{figure*}%[h]
\center
\subfigure[]{\includegraphics[width=0.49\textwidth]{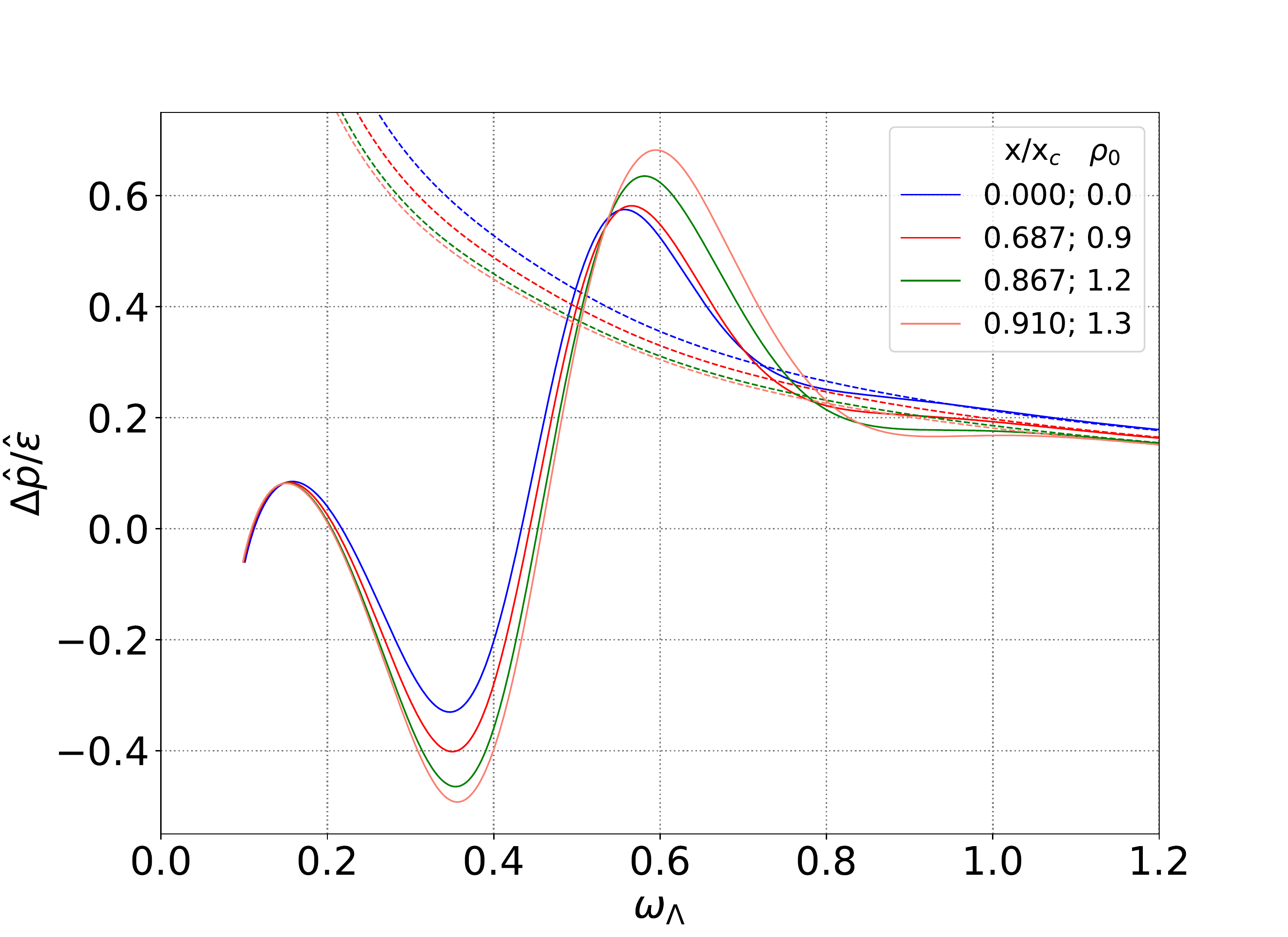}}
\subfigure[]{\includegraphics[width=0.49\textwidth]{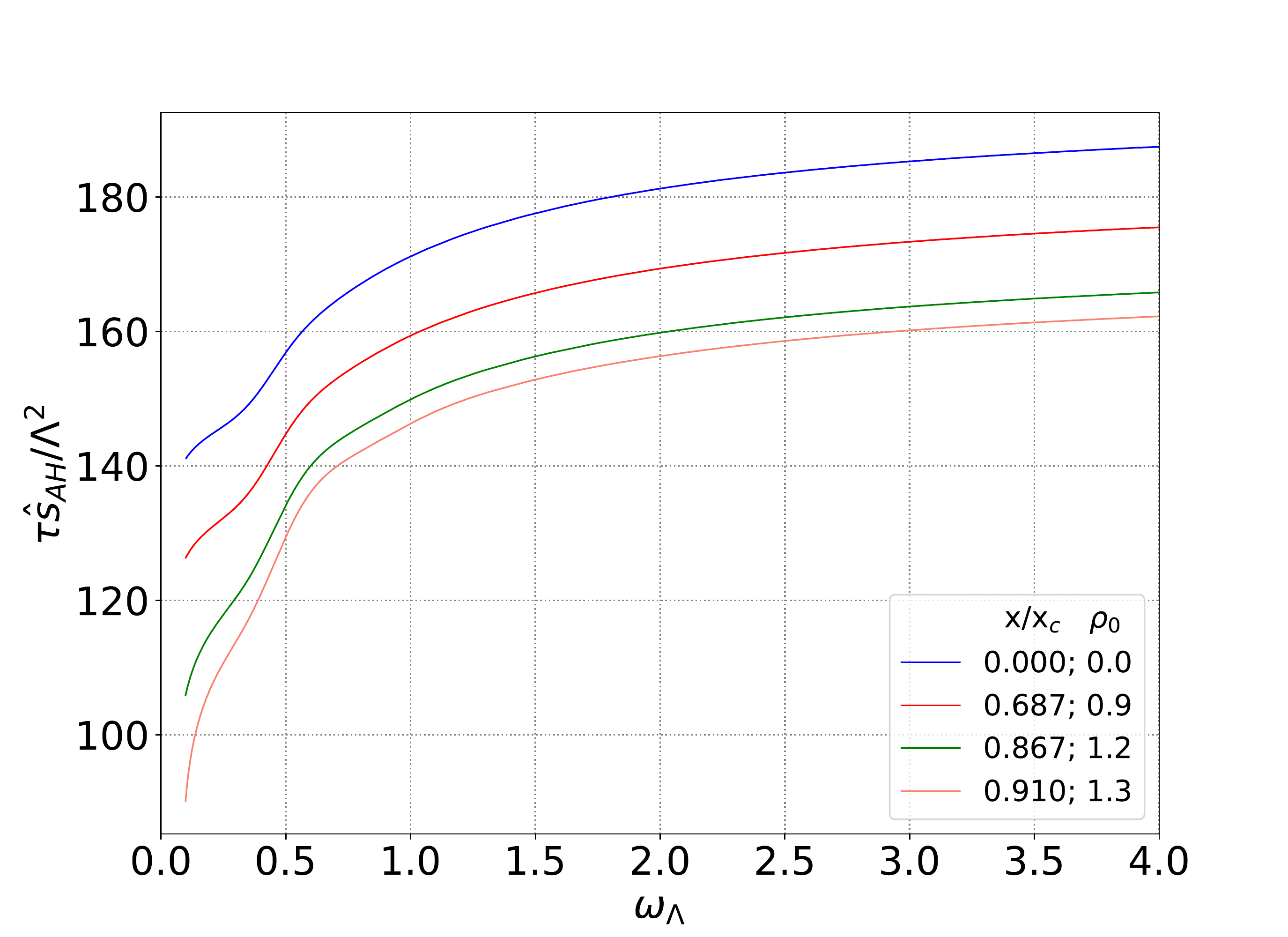}}
\subfigure[]{\includegraphics[width=0.49\textwidth]{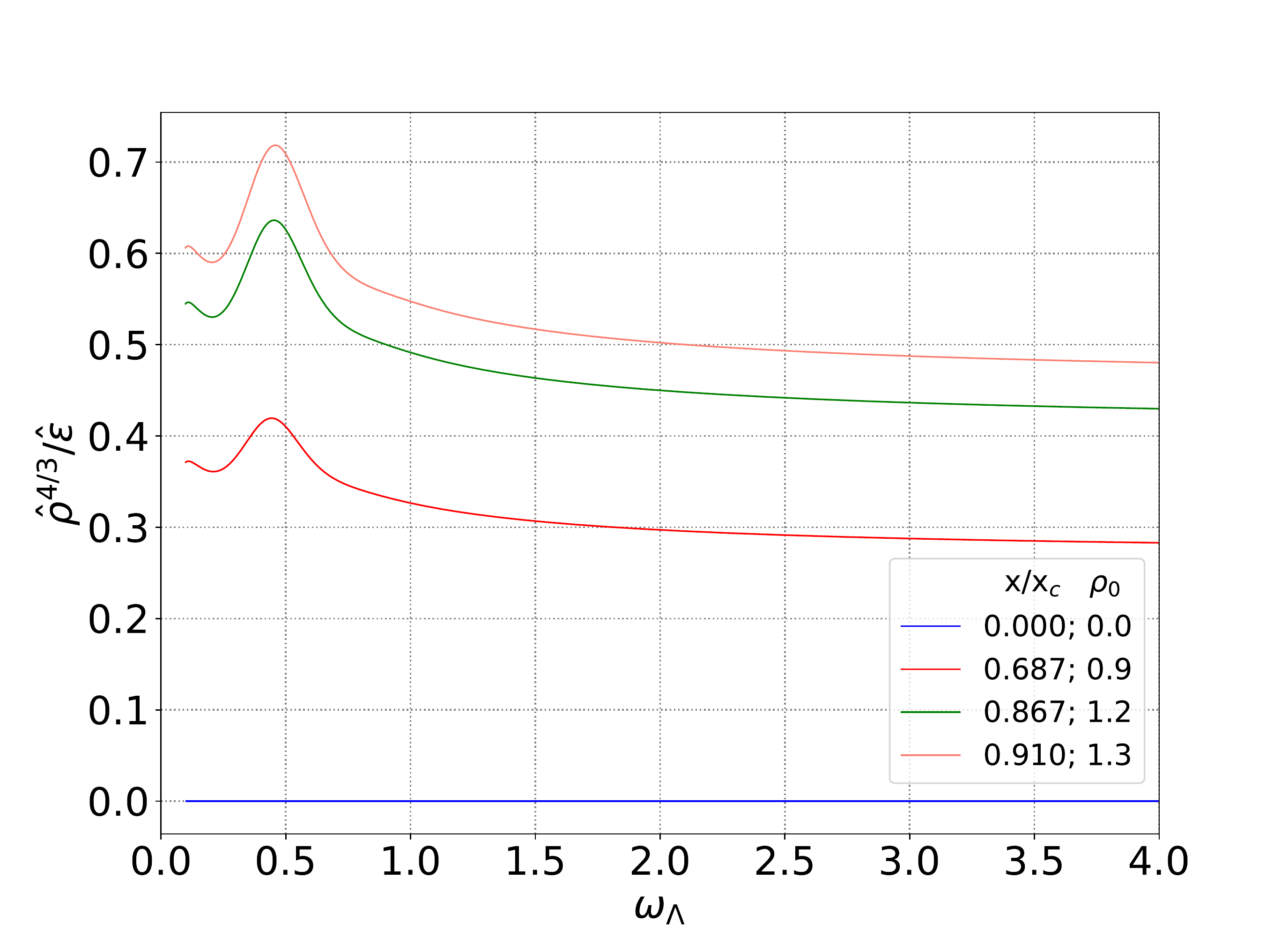}}
\subfigure[]{\includegraphics[width=0.49\textwidth]{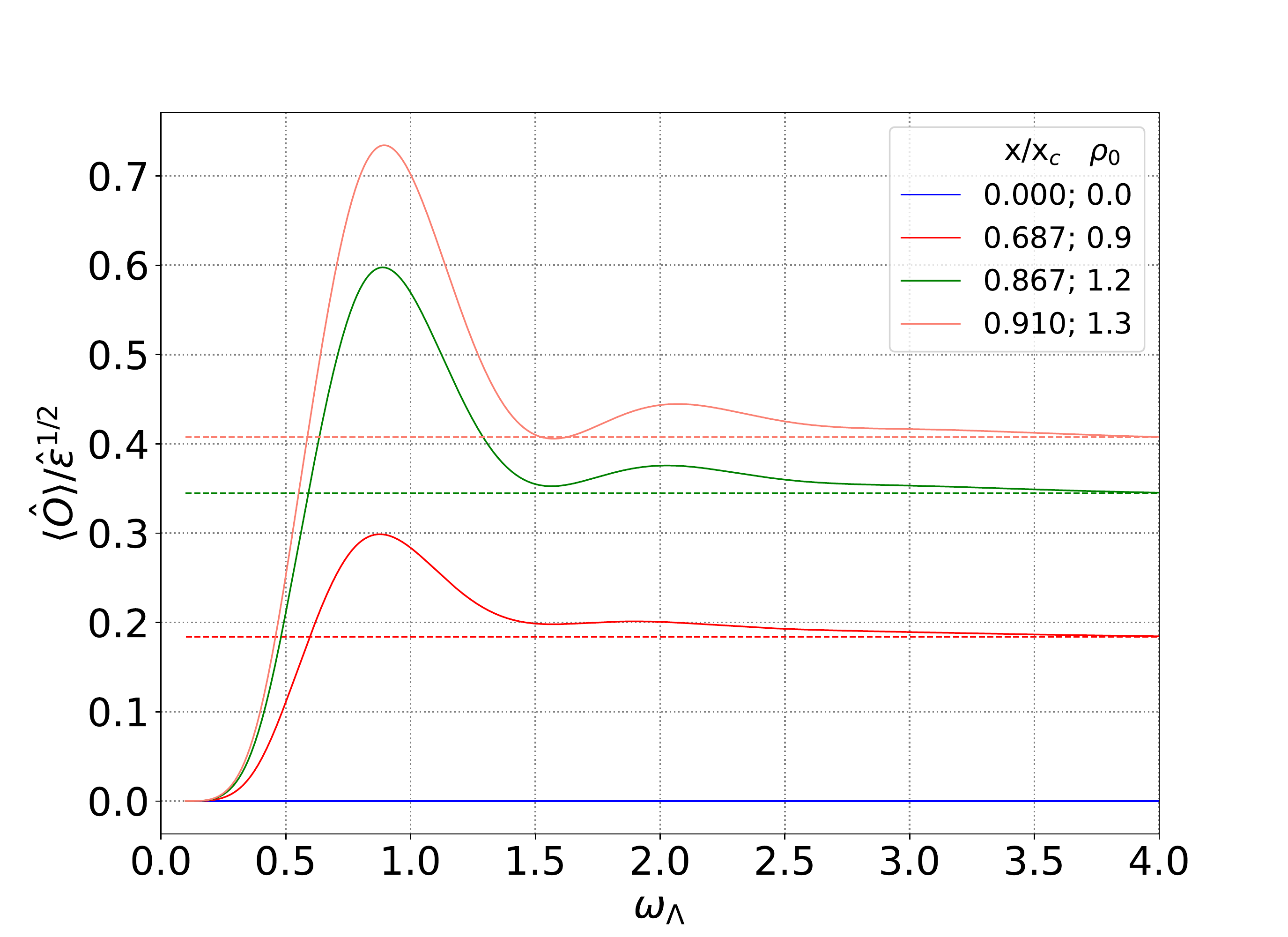}}
\caption{(a) Normalized pressure anisotropy (solid lines) and the corresponding hydrodynamic Navier-Stokes result (dashed lines), (b) normalized non-equilibrium entropy $\hat{S}_\textrm{AH}/\mathcal{A}\Lambda^2=\tau\hat{s}_\textrm{AH}/\Lambda^2$, (c) normalized charge density, and (d) normalized scalar condensate (solid lines) and the corresponding thermodynamic stable equilibrium result (dashed lines). Results obtained for variations of $\rho_0$ keeping fixed $B_s2$ in Table \ref{tabICs} with $a_2(\tau_0)=-6.67$. Note that $x_c\equiv\left(\mu/T\right)_c=\pi/\sqrt{2}$ is the critical point.}
\label{fig:result2}
\end{figure*}

\begin{figure*}%[h]
\center
\subfigure[]{\includegraphics[width=0.49\textwidth]{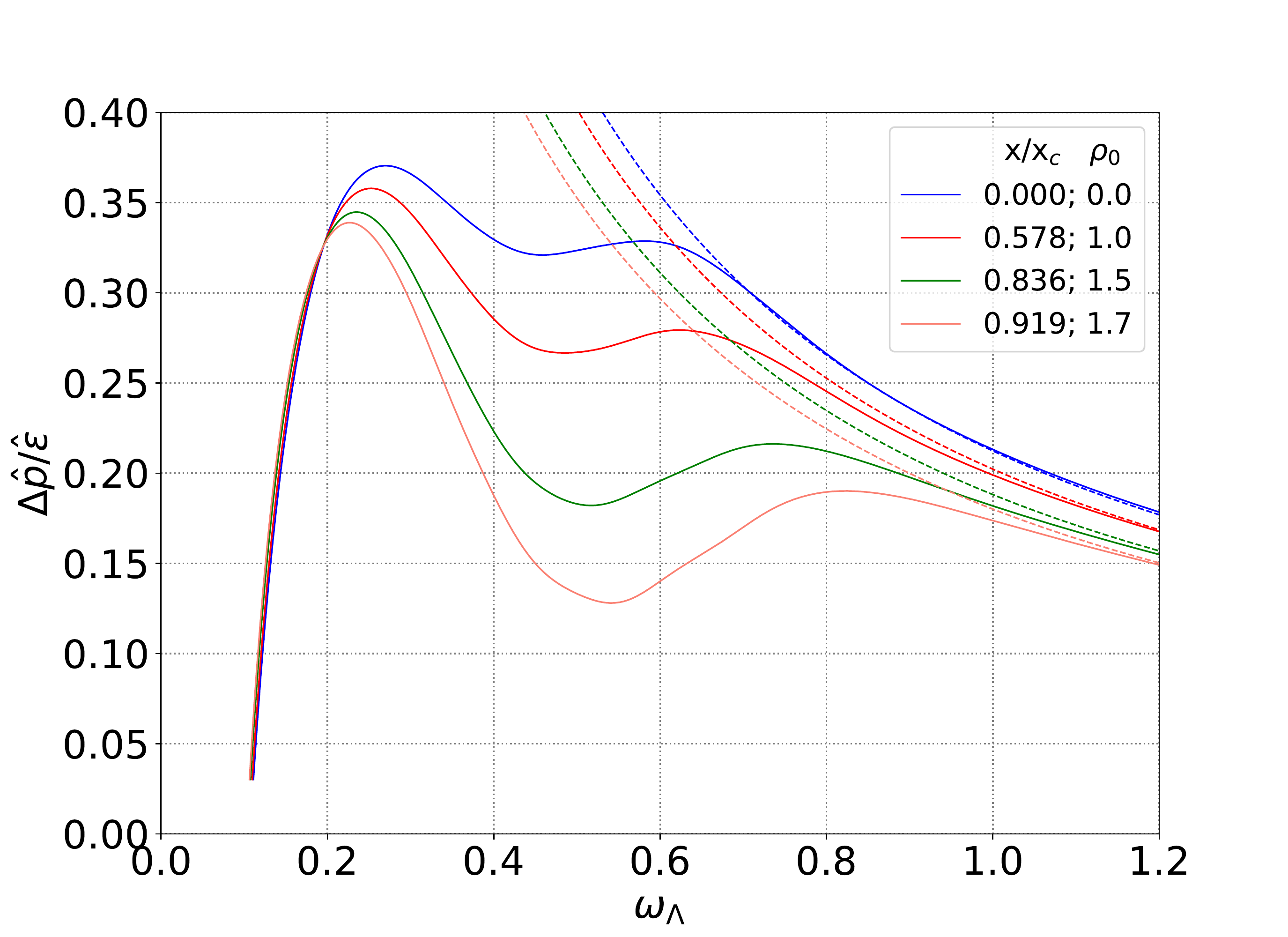}}
\subfigure[]{\includegraphics[width=0.49\textwidth]{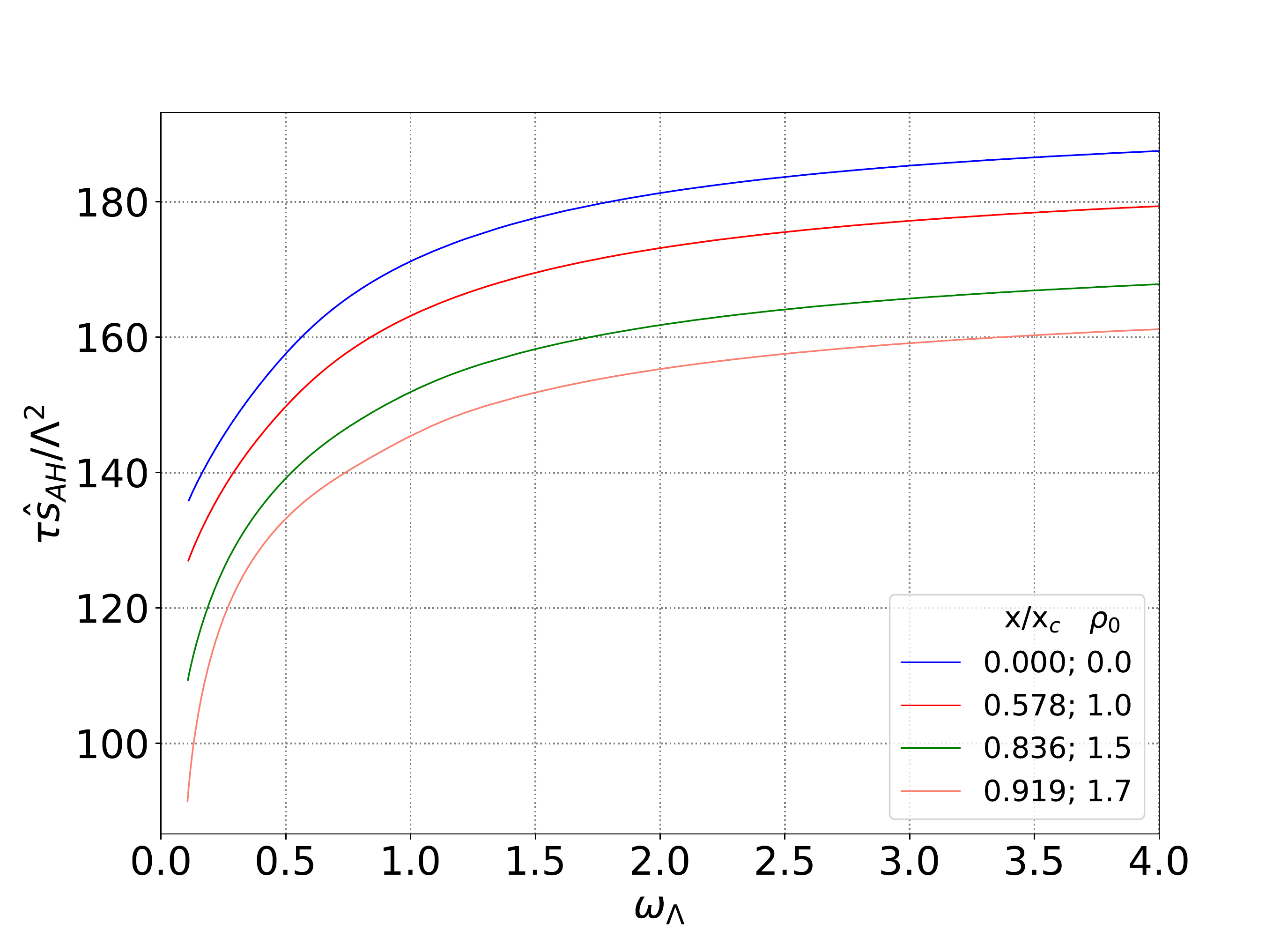}}
\subfigure[]{\includegraphics[width=0.49\textwidth]{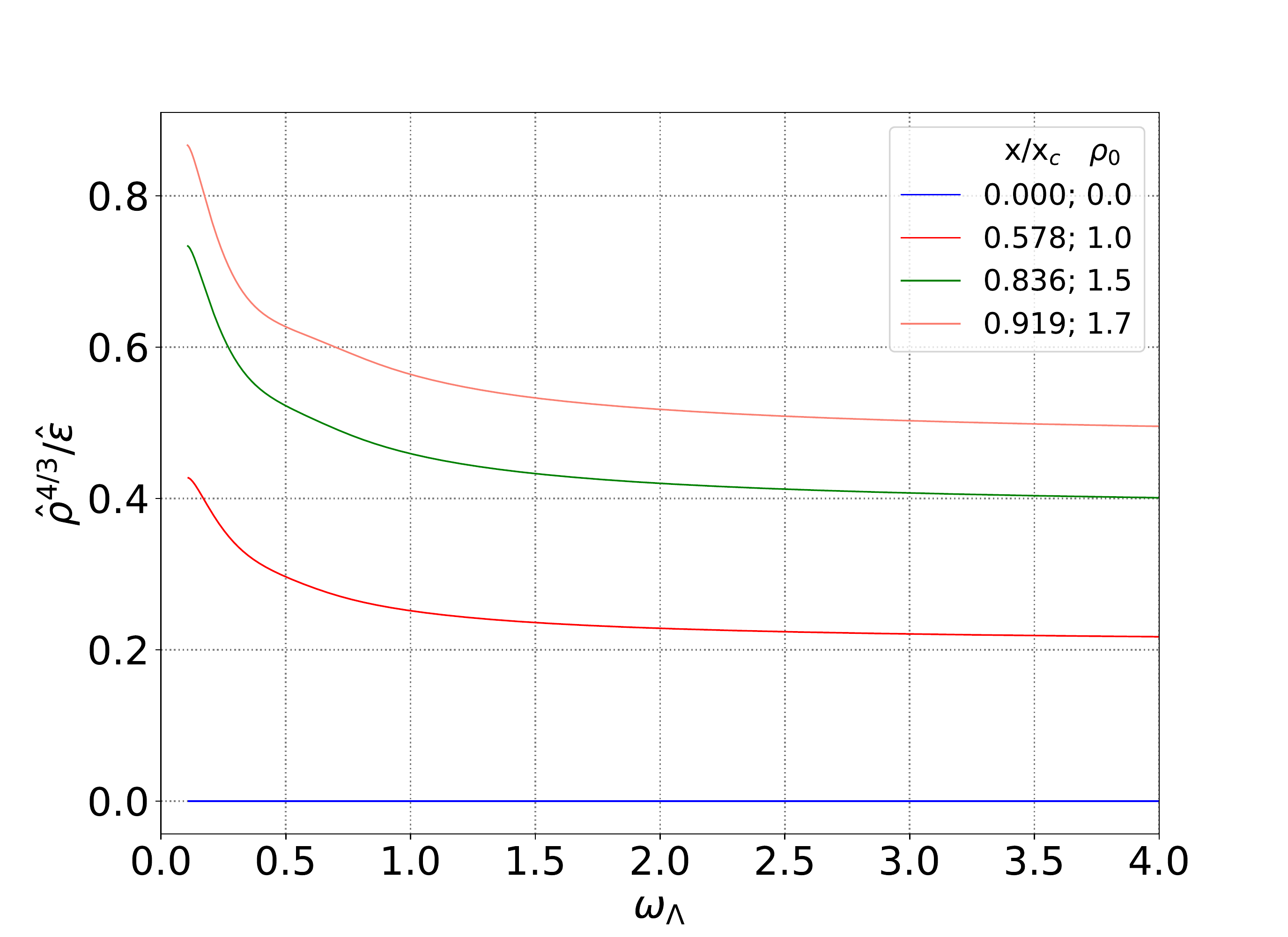}}
\subfigure[]{\includegraphics[width=0.49\textwidth]{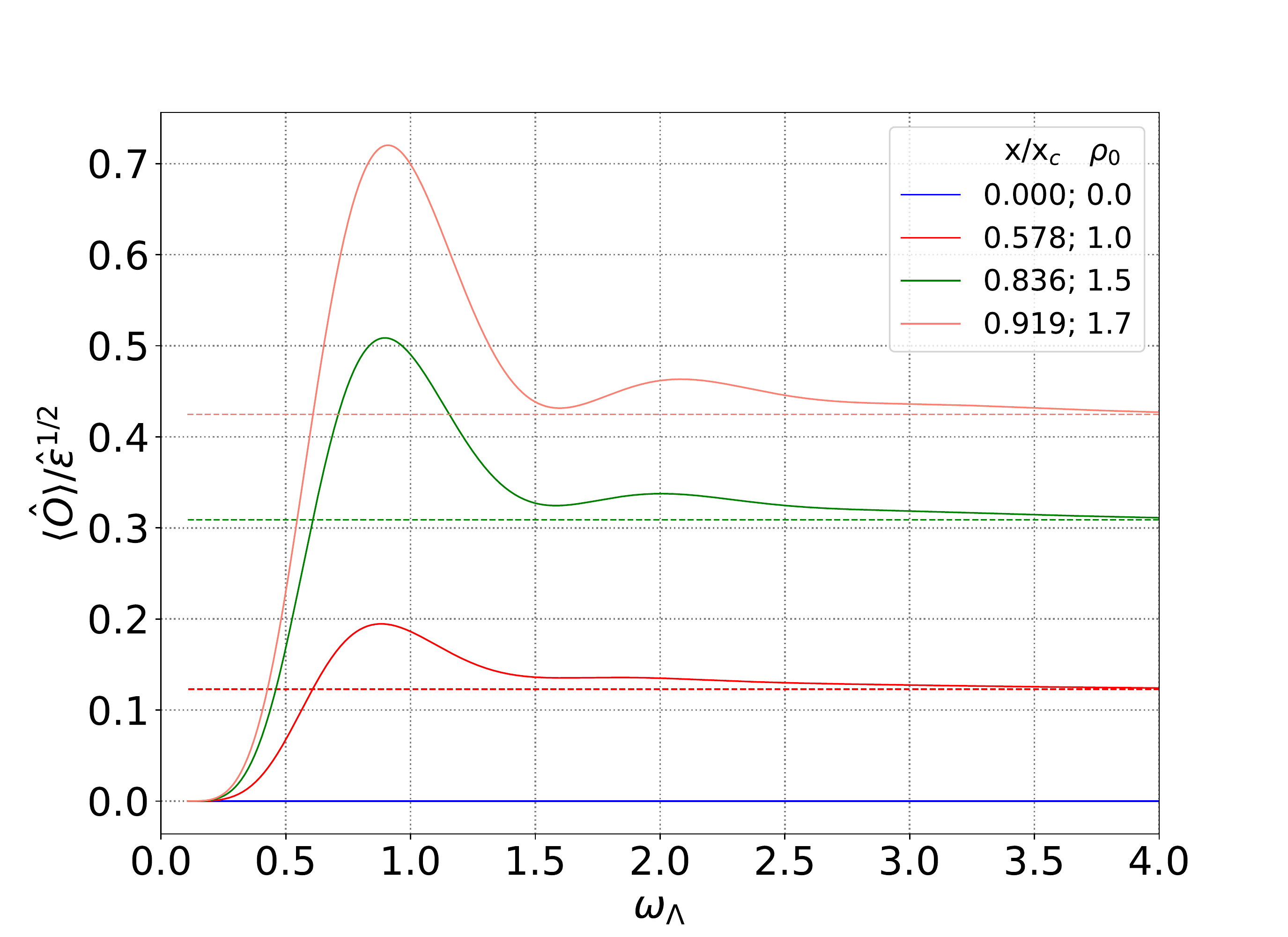}}
\caption{(a) Normalized pressure anisotropy (solid lines) and the corresponding hydrodynamic Navier-Stokes result (dashed lines), (b) normalized non-equilibrium entropy $\hat{S}_\textrm{AH}/\mathcal{A}\Lambda^2=\tau\hat{s}_\textrm{AH}/\Lambda^2$, (c) normalized charge density, and (d) normalized scalar condensate (solid lines) and the corresponding thermodynamic stable equilibrium result (dashed lines). Results obtained for variations of $\rho_0$ keeping fixed $B_s3$ in Table \ref{tabICs} with $a_2(\tau_0)=-6.67$. Note that $x_c\equiv\left(\mu/T\right)_c=\pi/\sqrt{2}$ is the critical point.}
\label{fig:result3}
\end{figure*}

\begin{figure*}%[h]
\center
\subfigure[]{\includegraphics[width=0.49\textwidth]{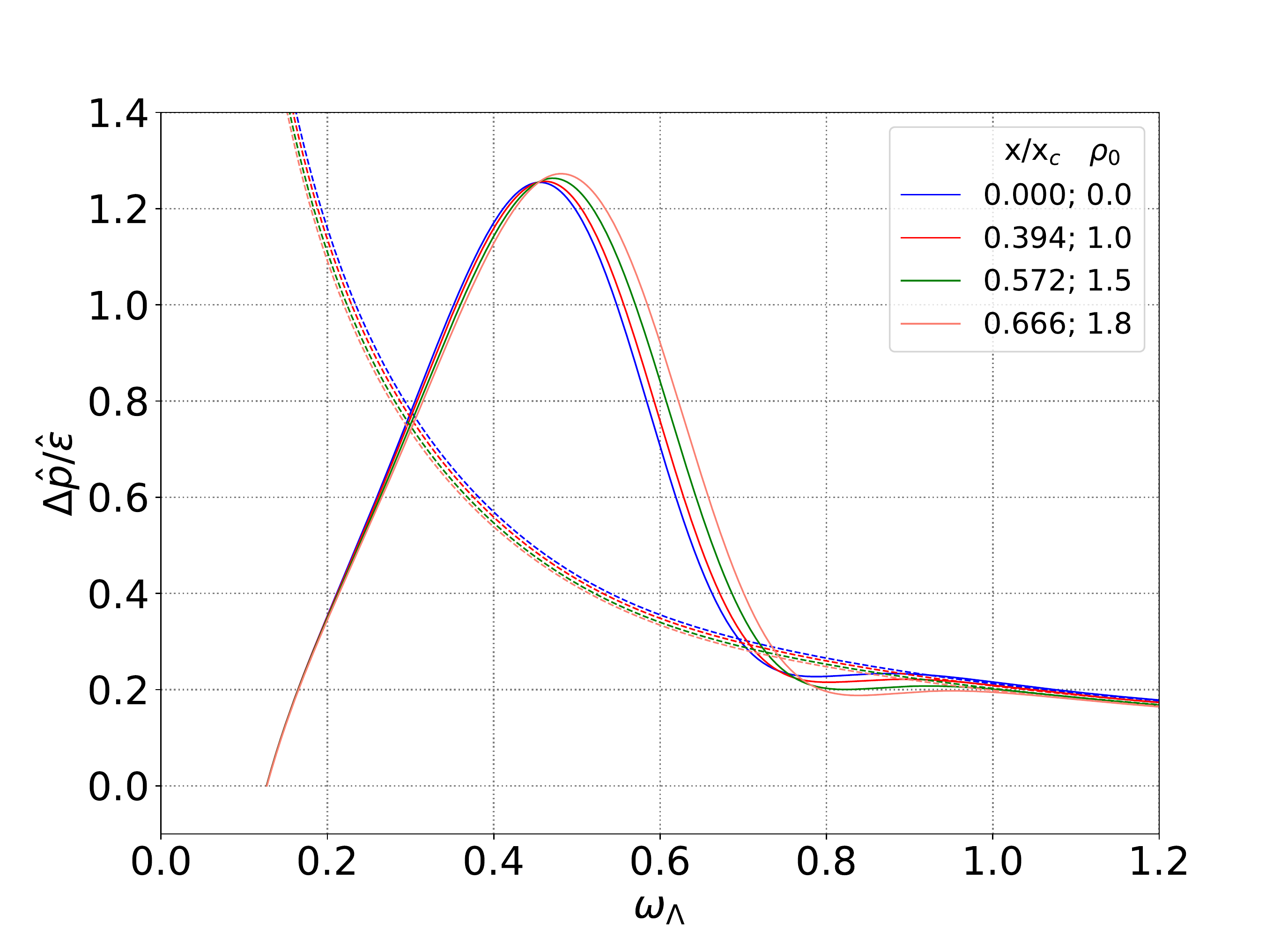}}
\subfigure[]{\includegraphics[width=0.49\textwidth]{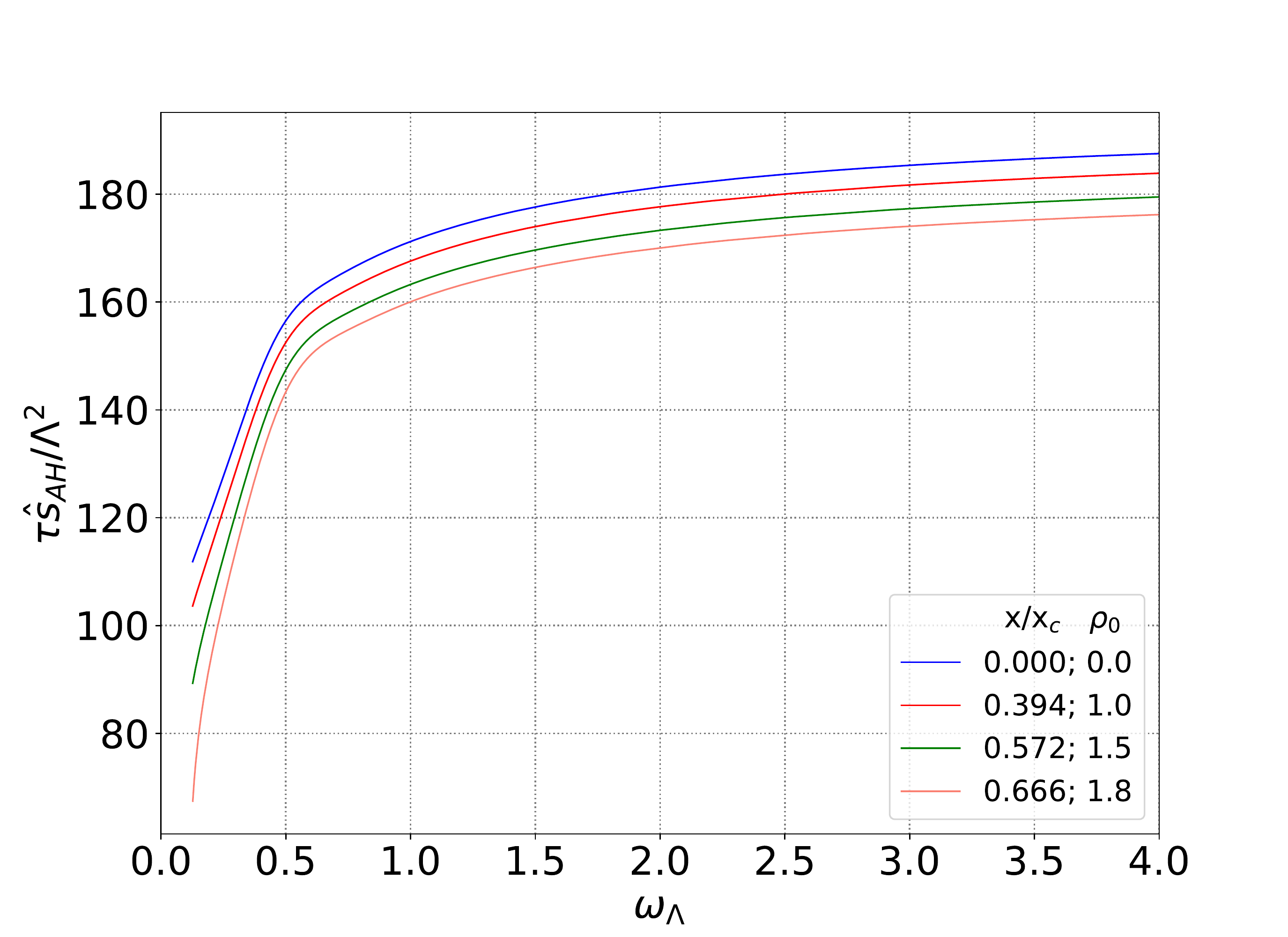}}
\subfigure[]{\includegraphics[width=0.49\textwidth]{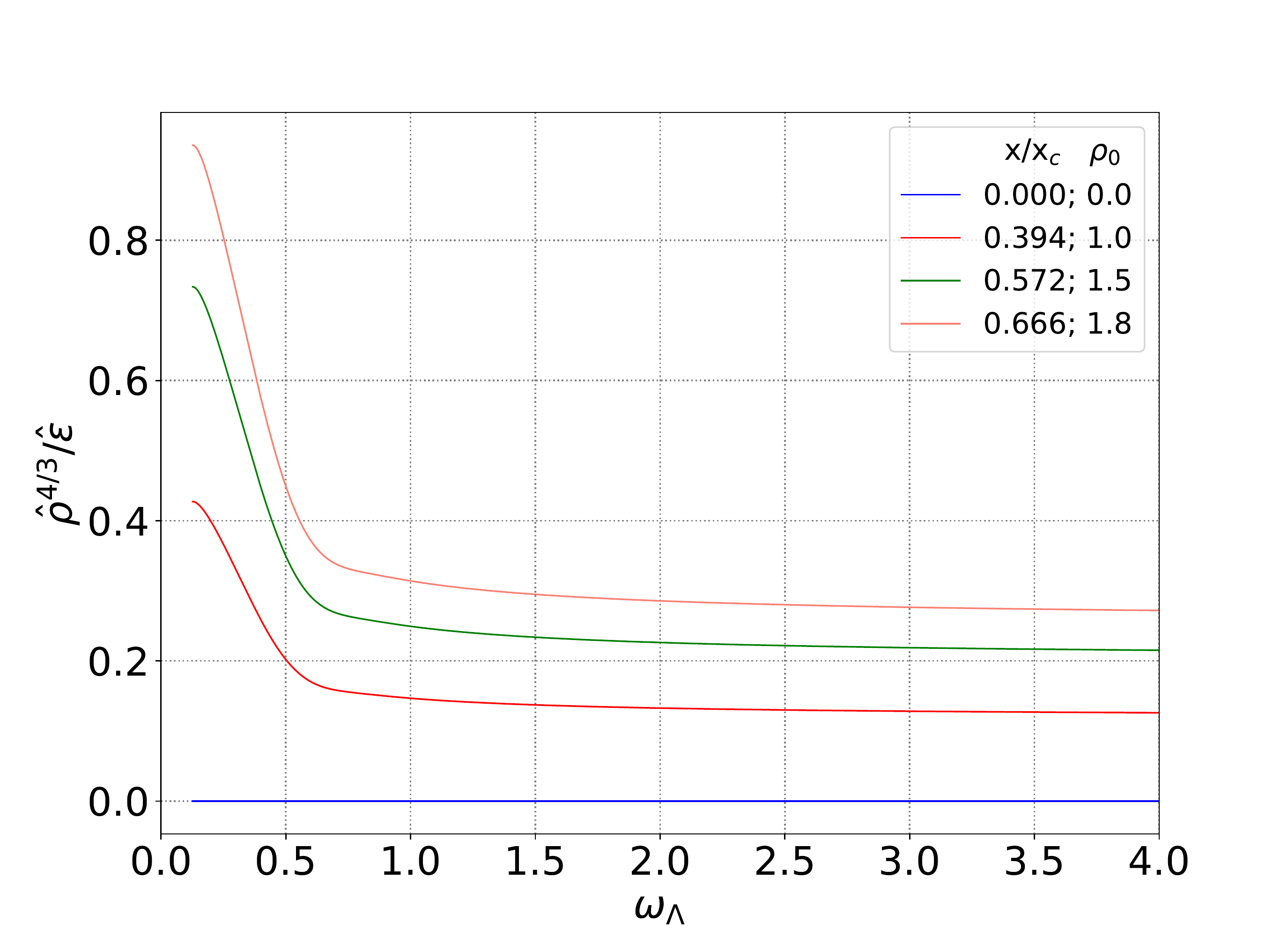}}
\subfigure[]{\includegraphics[width=0.49\textwidth]{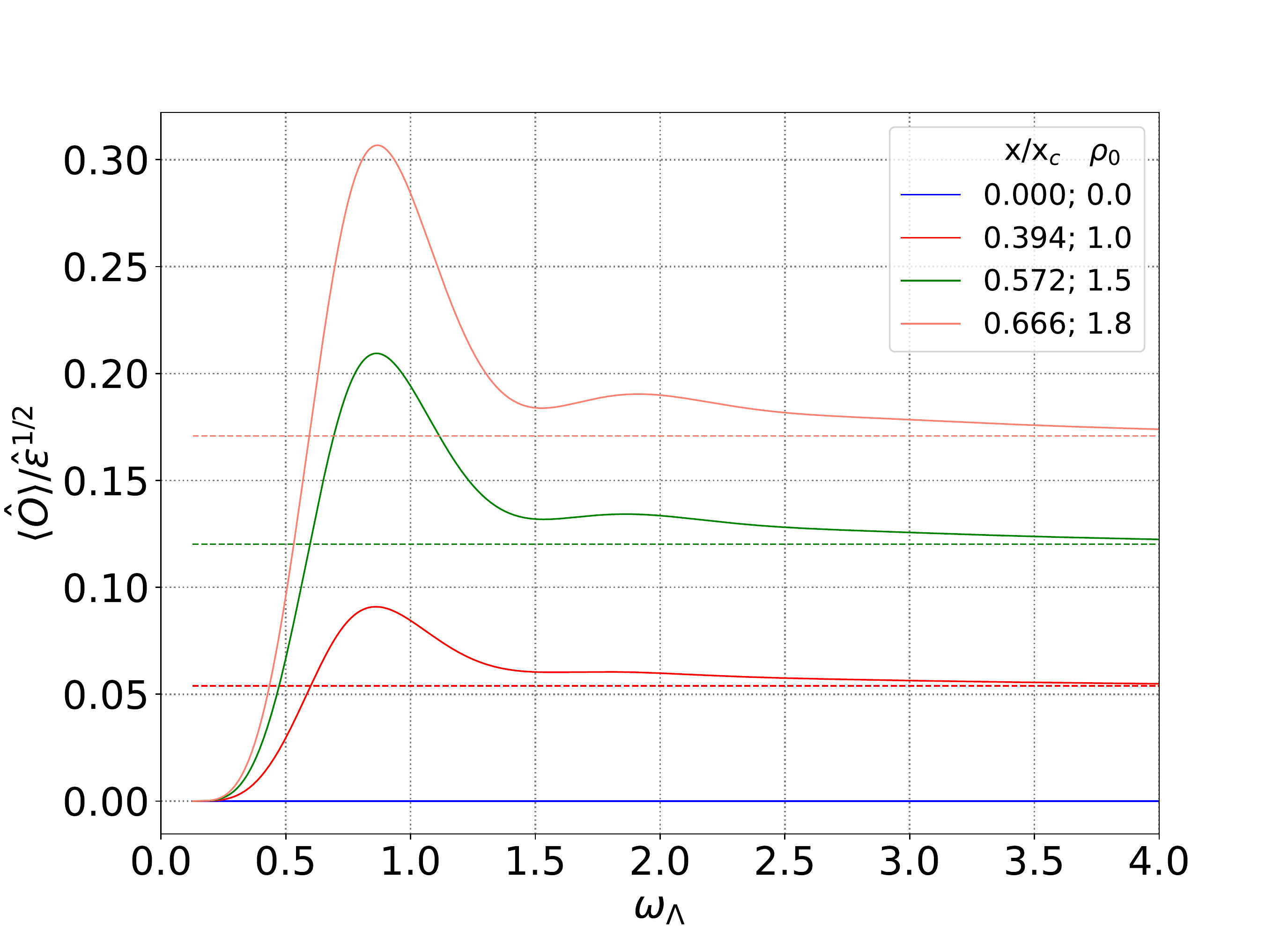}}
\caption{(a) Normalized pressure anisotropy (solid lines) and the corresponding hydrodynamic Navier-Stokes result (dashed lines), (b) normalized non-equilibrium entropy $\hat{S}_\textrm{AH}/\mathcal{A}\Lambda^2=\tau\hat{s}_\textrm{AH}/\Lambda^2$, (c) normalized charge density, and (d) normalized scalar condensate (solid lines) and the corresponding thermodynamic stable equilibrium result (dashed lines). Results obtained for variations of $\rho_0$ keeping fixed $B_s4$ in Table \ref{tabICs} with $a_2(\tau_0)=-6.67$. Note that $x_c\equiv\left(\mu/T\right)_c=\pi/\sqrt{2}$ is the critical point.}
\label{fig:result4}
\end{figure*}

\begin{figure*}%[h]
\center
\subfigure[]{\includegraphics[width=0.49\textwidth]{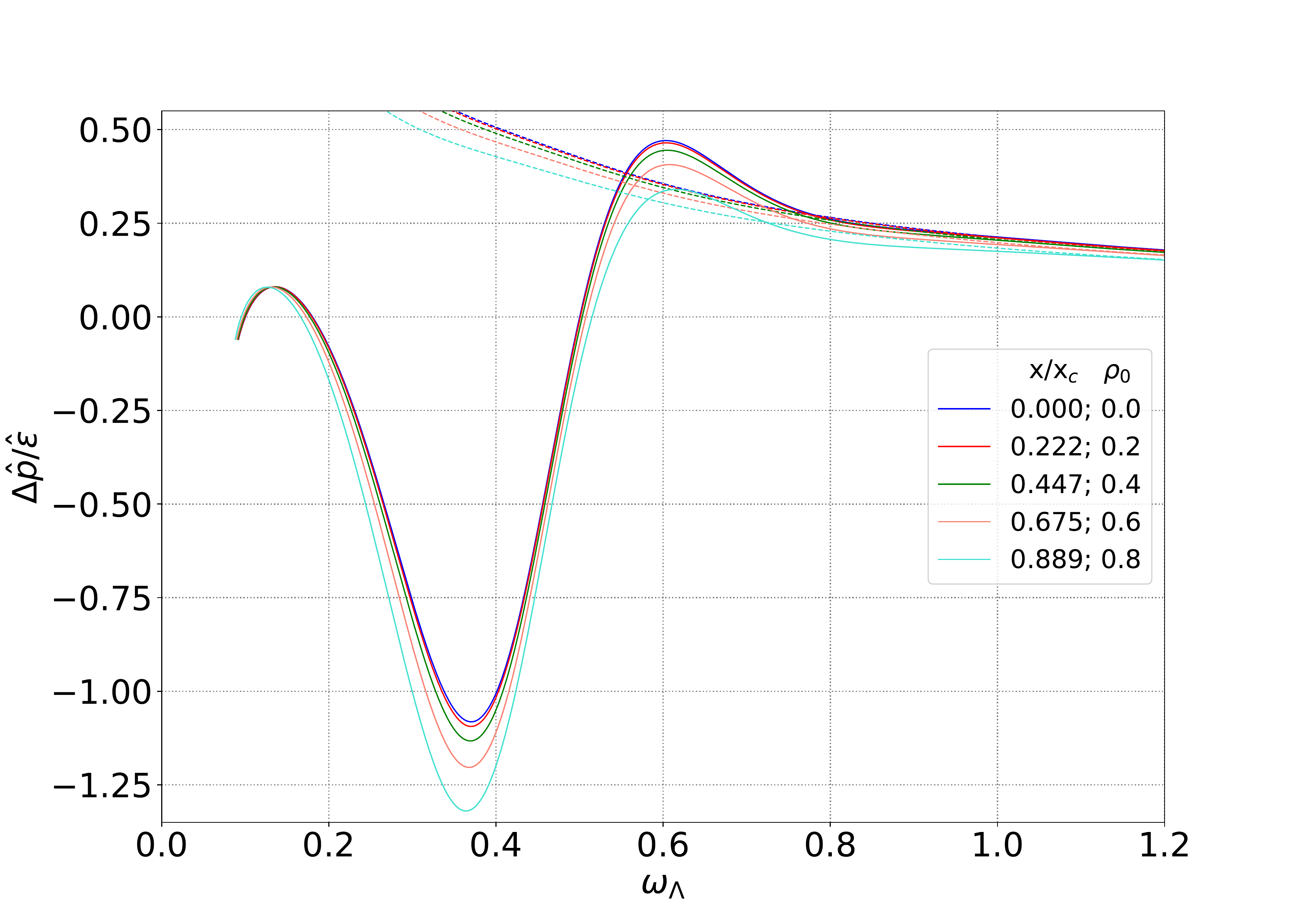}}
\subfigure[]{\includegraphics[width=0.49\textwidth]{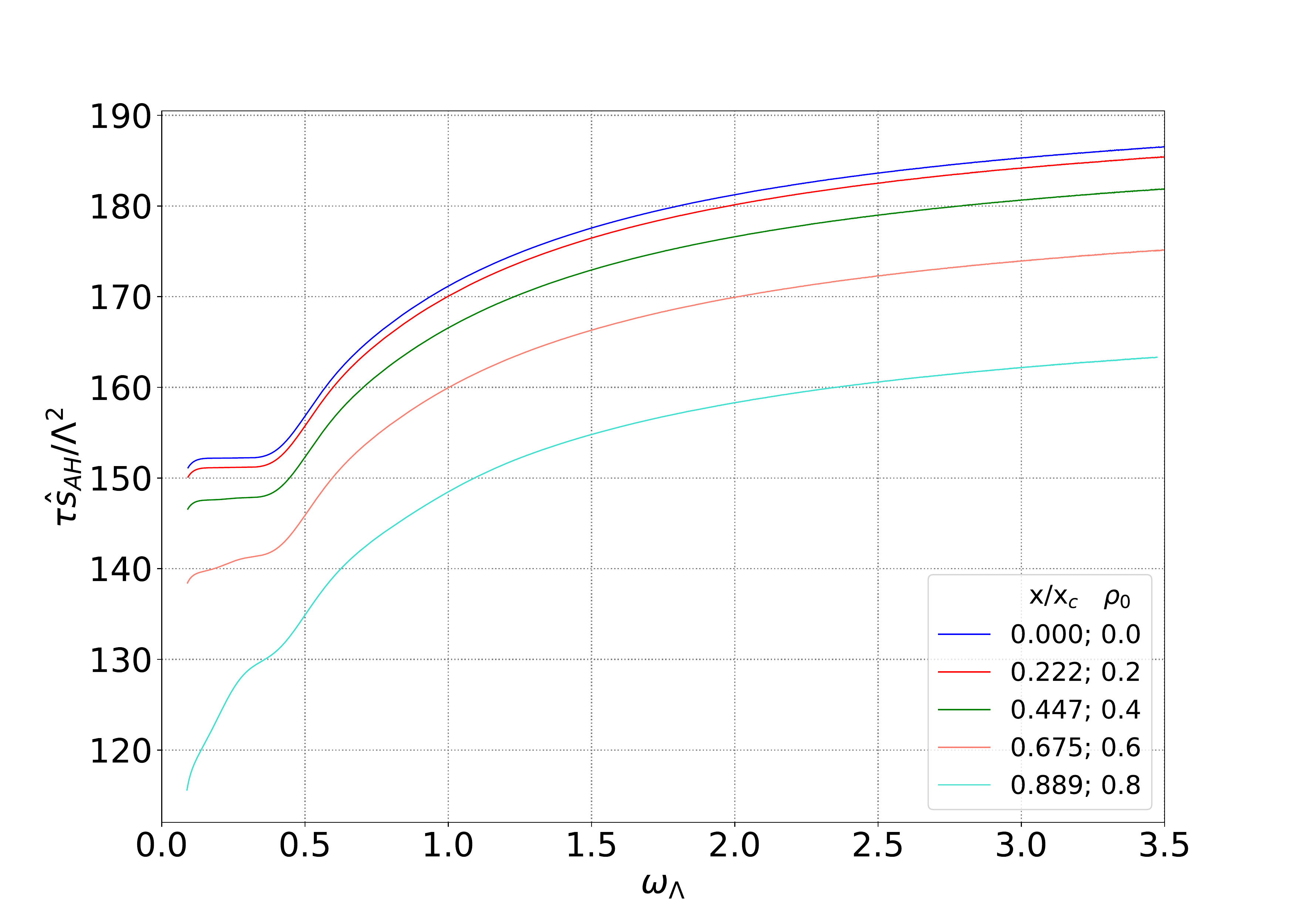}}
\subfigure[]{\includegraphics[width=0.49\textwidth]{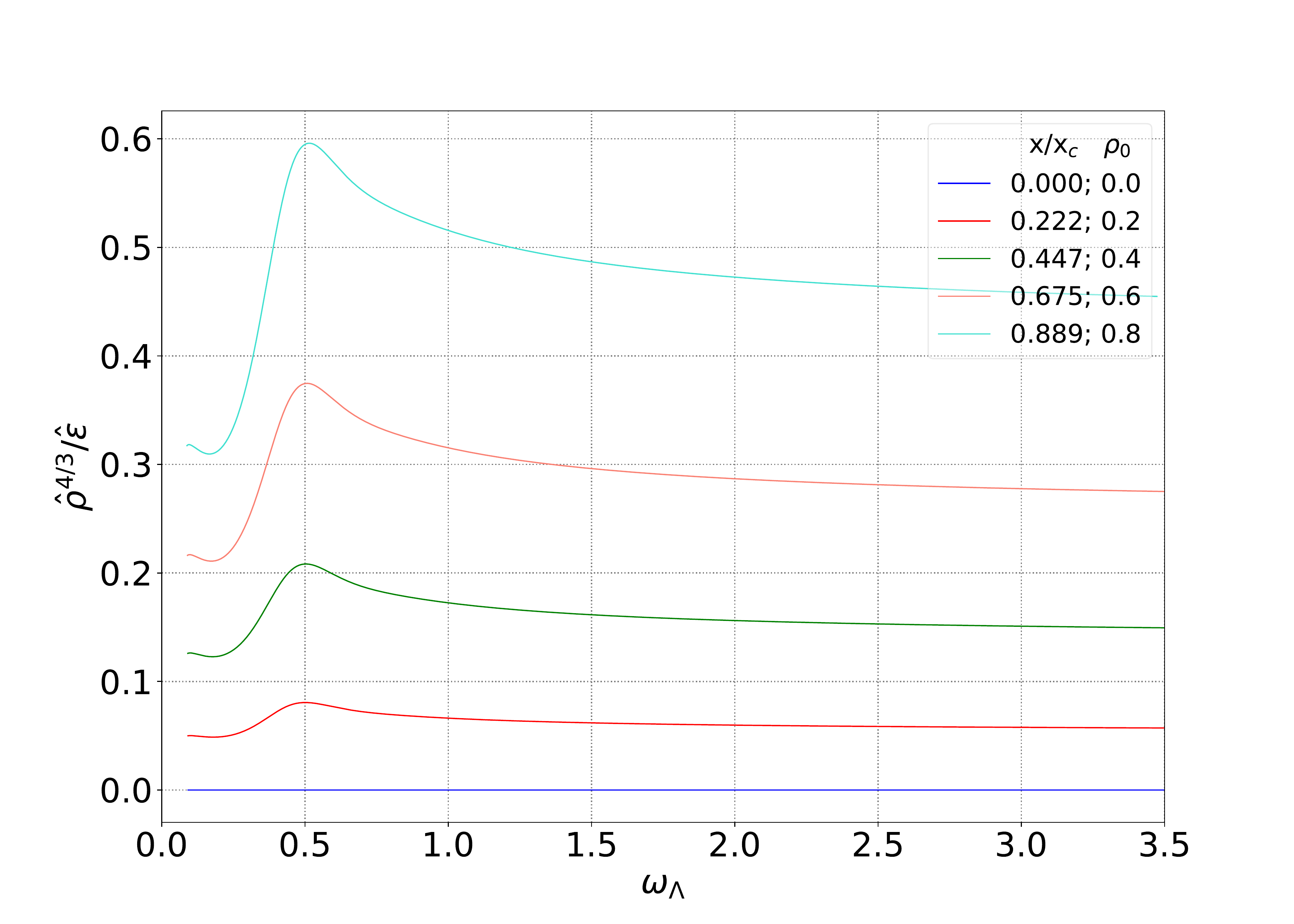}}
\subfigure[]{\includegraphics[width=0.49\textwidth]{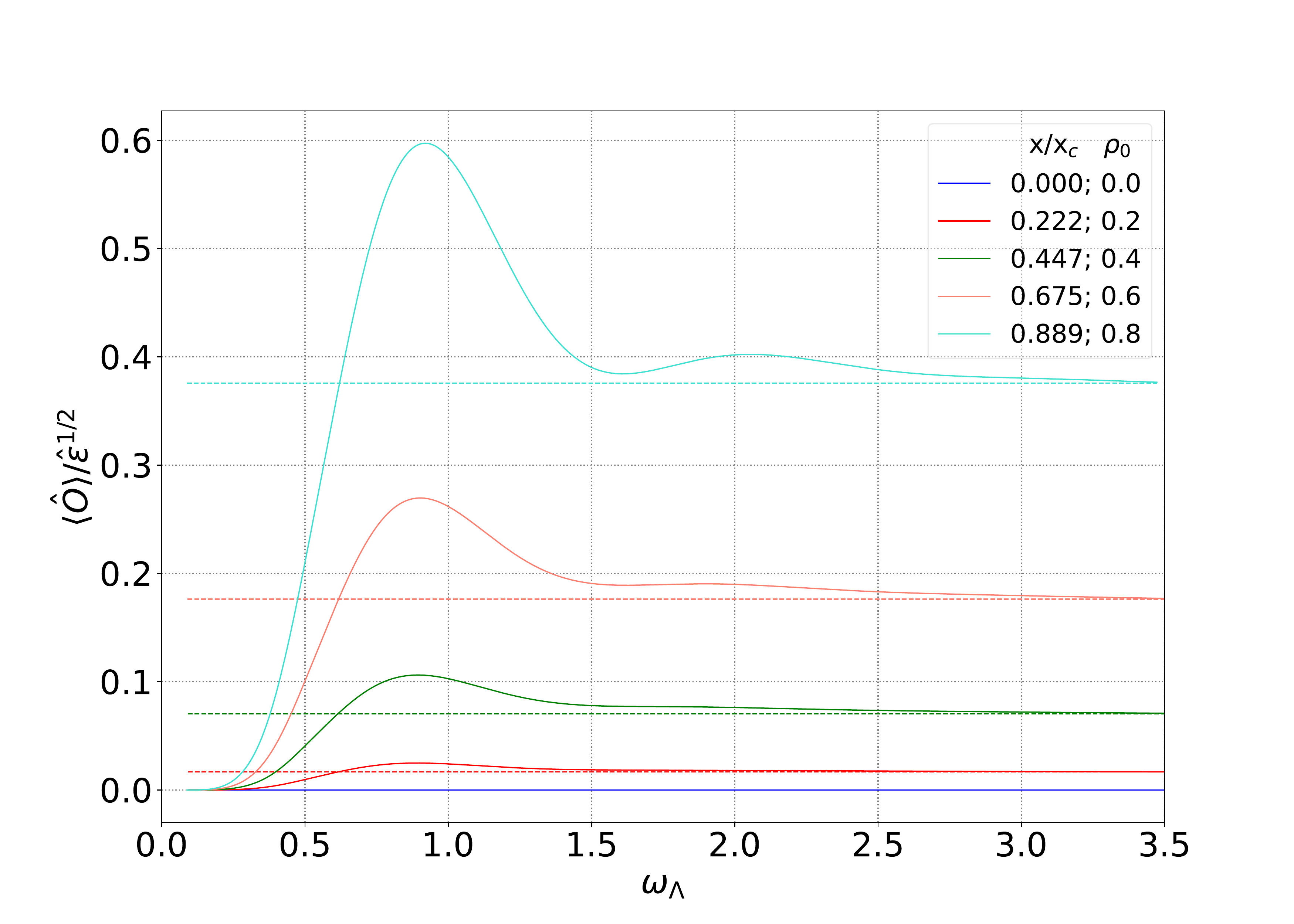}}
\caption{(a) Normalized pressure anisotropy (solid lines) and the corresponding hydrodynamic Navier-Stokes result (dashed lines), (b) normalized non-equilibrium entropy $\hat{S}_\textrm{AH}/\mathcal{A}\Lambda^2=\tau\hat{s}_\textrm{AH}/\Lambda^2$, (c) normalized charge density, and (d) normalized scalar condensate (solid lines) and the corresponding thermodynamic stable equilibrium result (dashed lines). Results obtained for variations of $\rho_0$ keeping fixed $B_s5$ in Table \ref{tabICs} with $a_2(\tau_0)=-6.67$. Note that $x_c\equiv\left(\mu/T\right)_c=\pi/\sqrt{2}$ is the critical point.}
\label{fig:result5}
\end{figure*}

\begin{figure*}%[h]
\center
\subfigure[]{\includegraphics[width=0.49\textwidth]{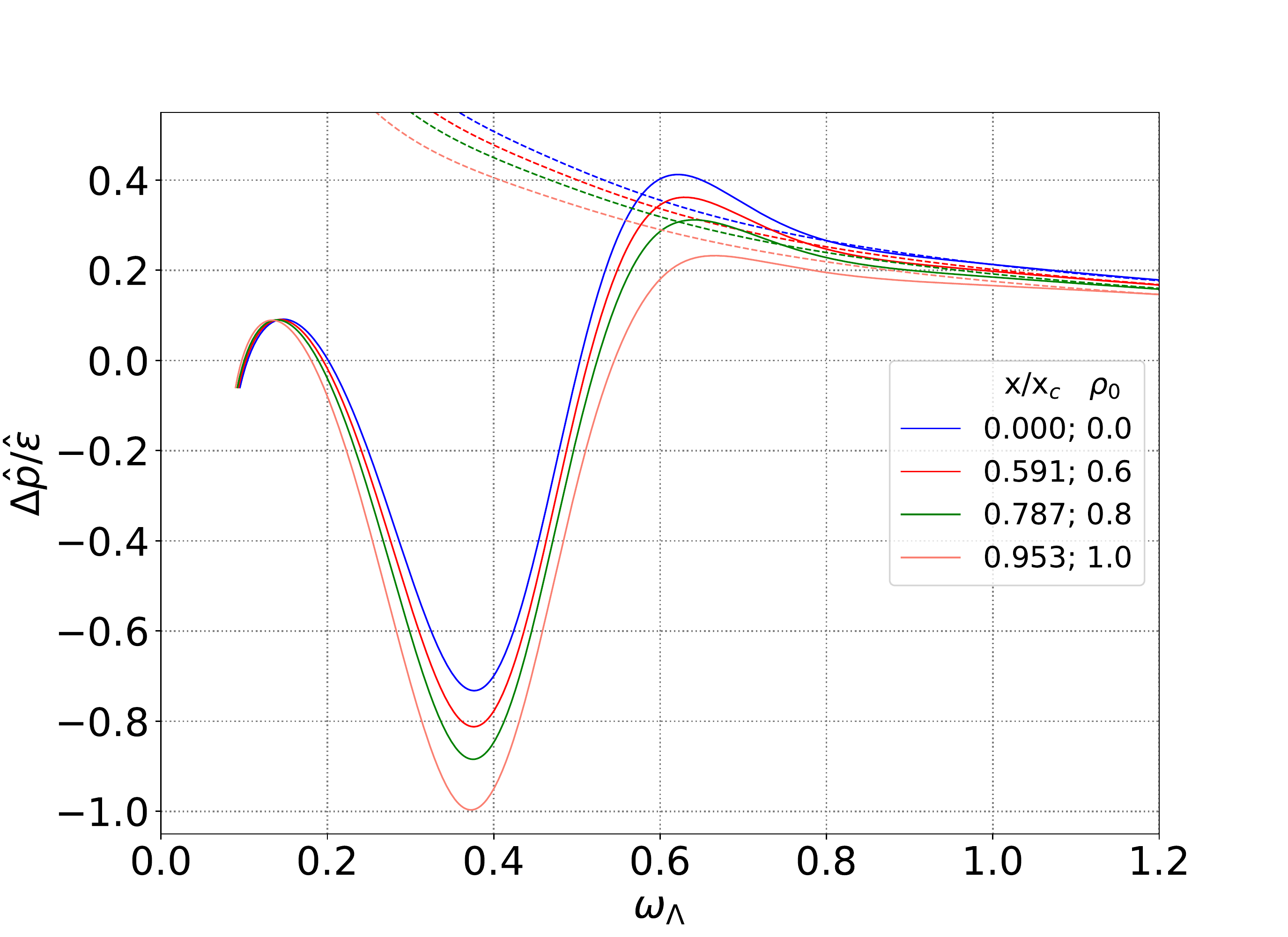}}
\subfigure[]{\includegraphics[width=0.49\textwidth]{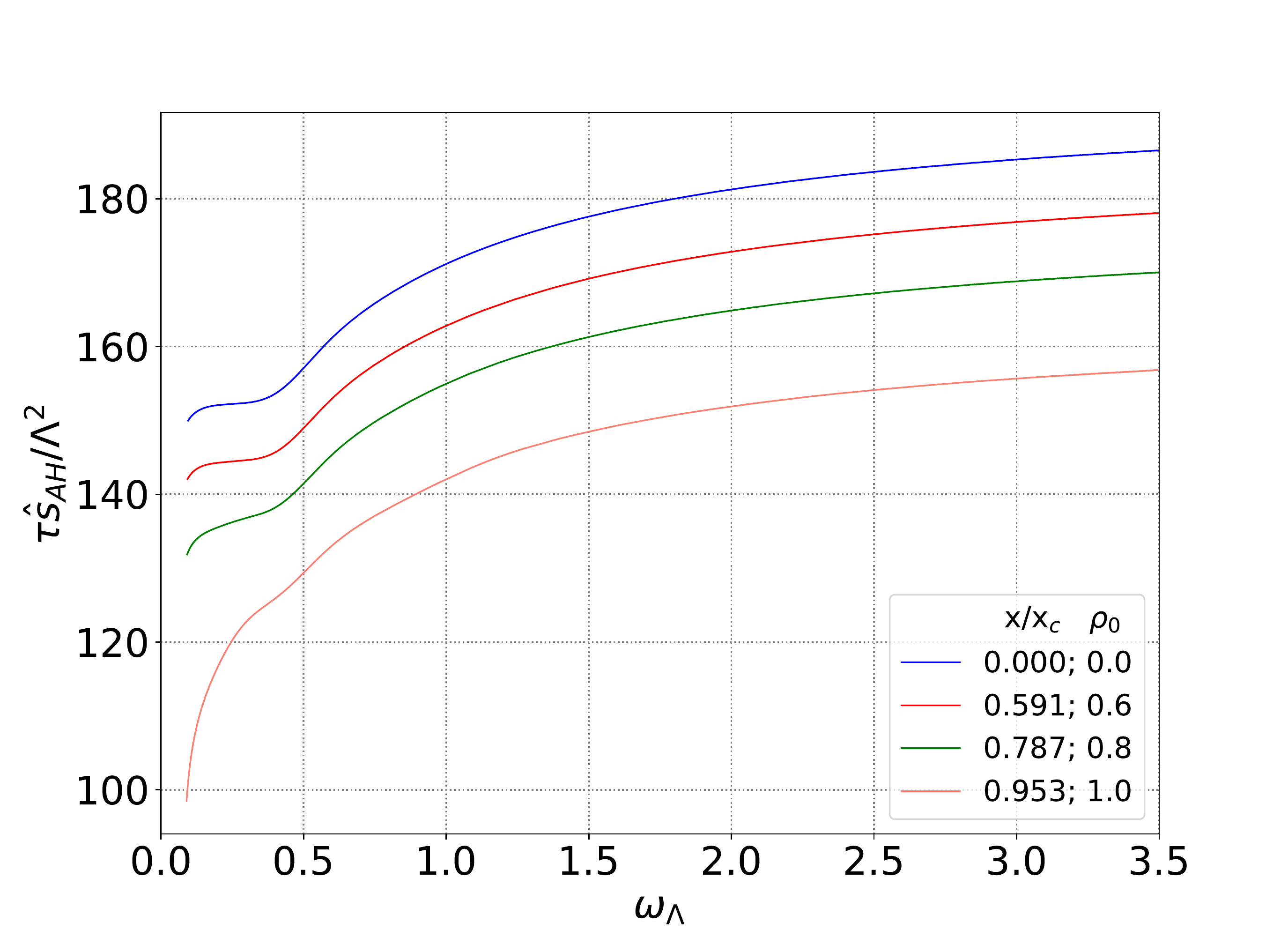}}
\subfigure[]{\includegraphics[width=0.49\textwidth]{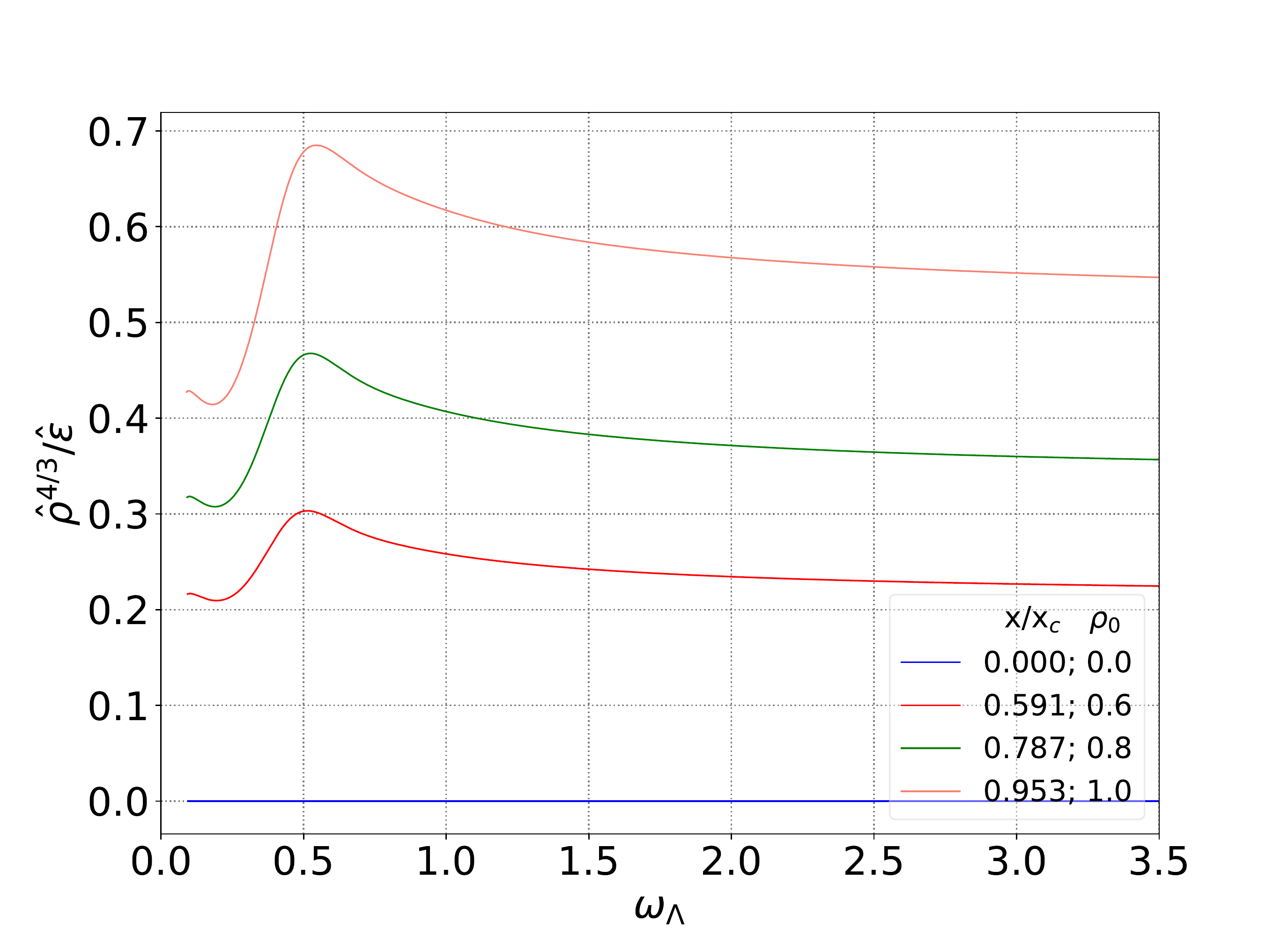}}
\subfigure[]{\includegraphics[width=0.49\textwidth]{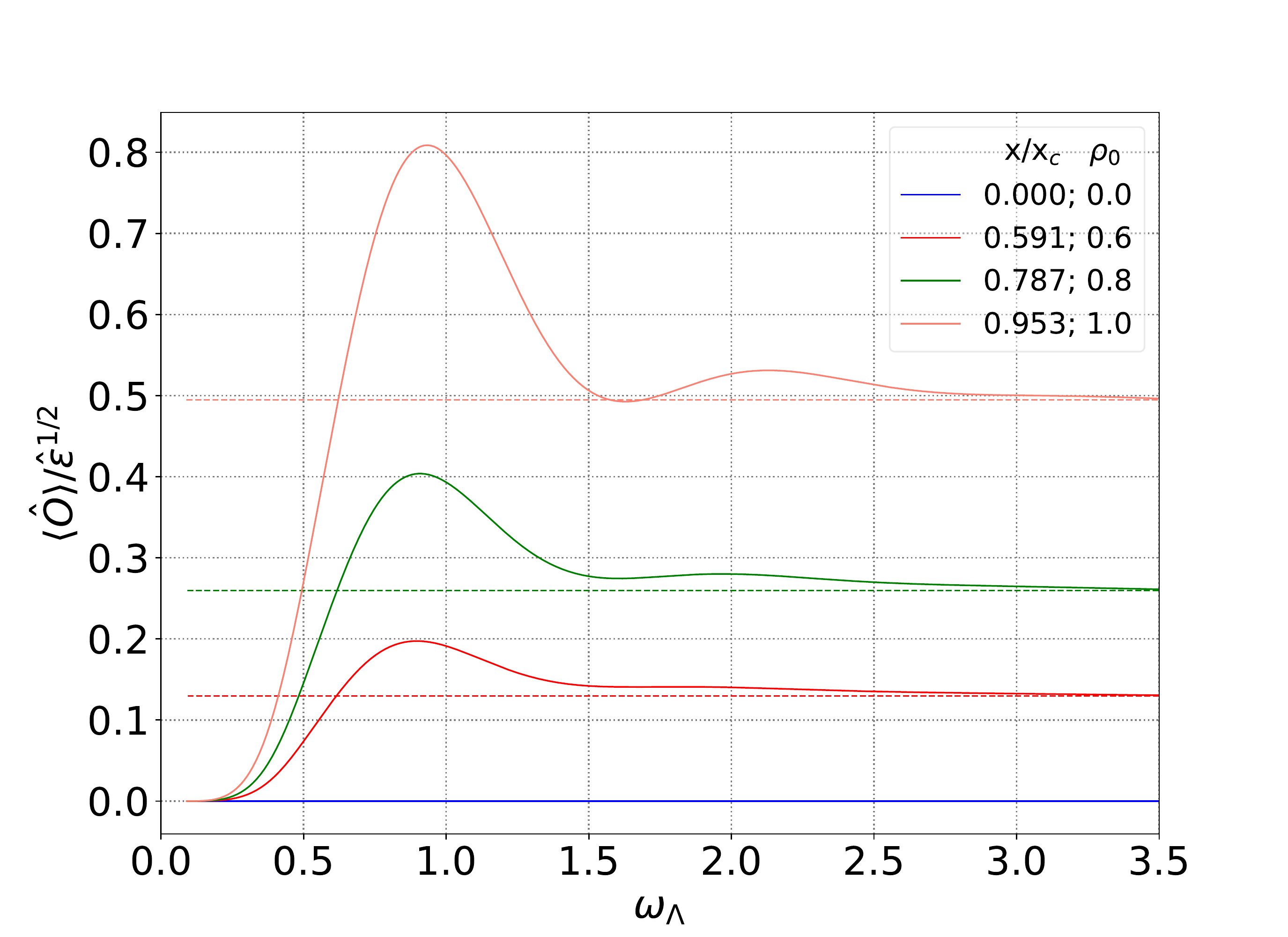}}
\caption{(a) Normalized pressure anisotropy (solid lines) and the corresponding hydrodynamic Navier-Stokes result (dashed lines), (b) normalized non-equilibrium entropy $\hat{S}_\textrm{AH}/\mathcal{A}\Lambda^2=\tau\hat{s}_\textrm{AH}/\Lambda^2$, (c) normalized charge density, and (d) normalized scalar condensate (solid lines) and the corresponding thermodynamic stable equilibrium result (dashed lines). Results obtained for variations of $\rho_0$ keeping fixed $B_s6$ in Table \ref{tabICs} with $a_2(\tau_0)=-6.67$. Note that $x_c\equiv\left(\mu/T\right)_c=\pi/\sqrt{2}$ is the critical point.}
\label{fig:result6}
\end{figure*}

\begin{figure*}%[h]
\center
\subfigure[]{\includegraphics[width=0.49\textwidth]{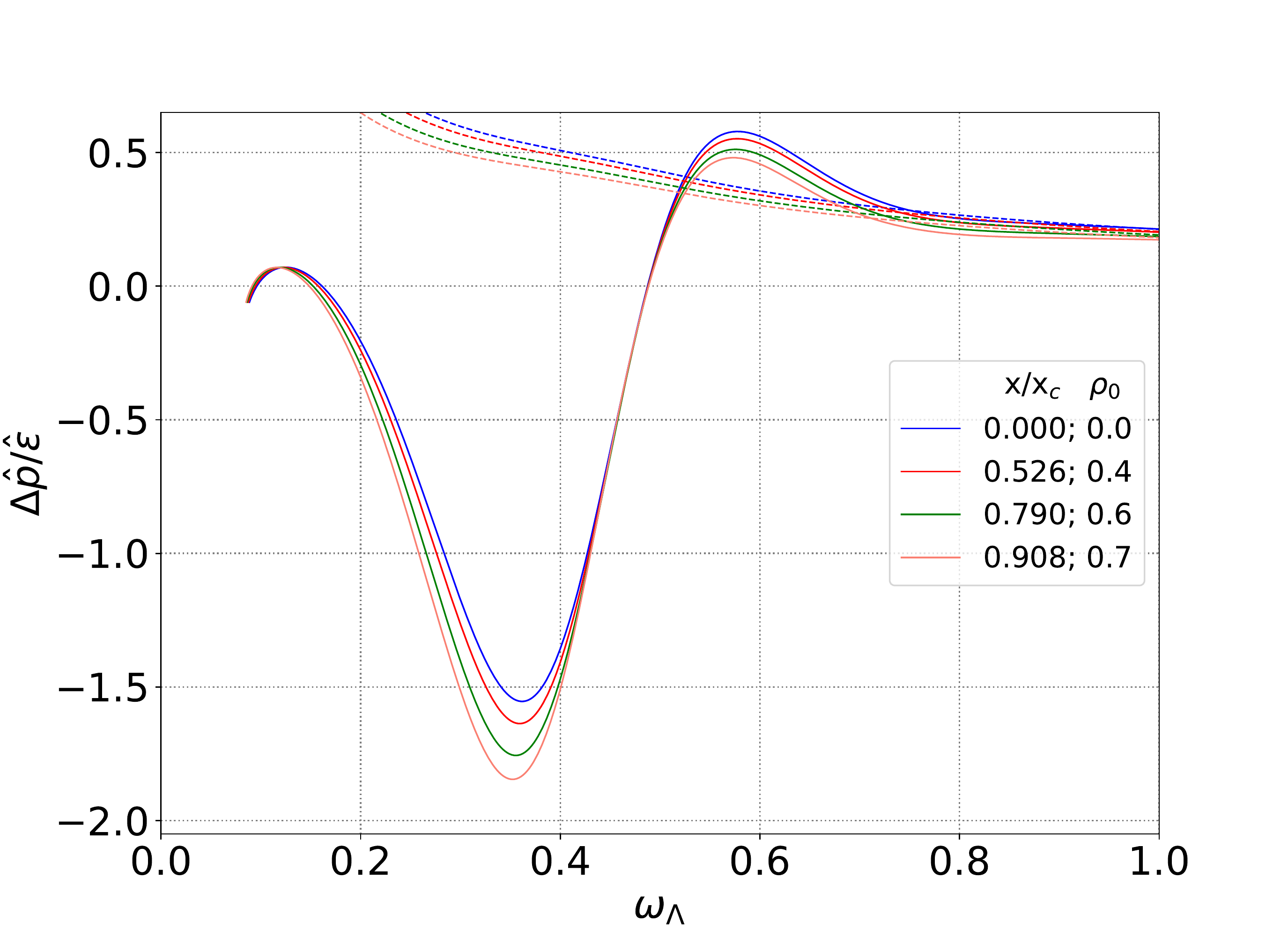}}
\subfigure[]{\includegraphics[width=0.49\textwidth]{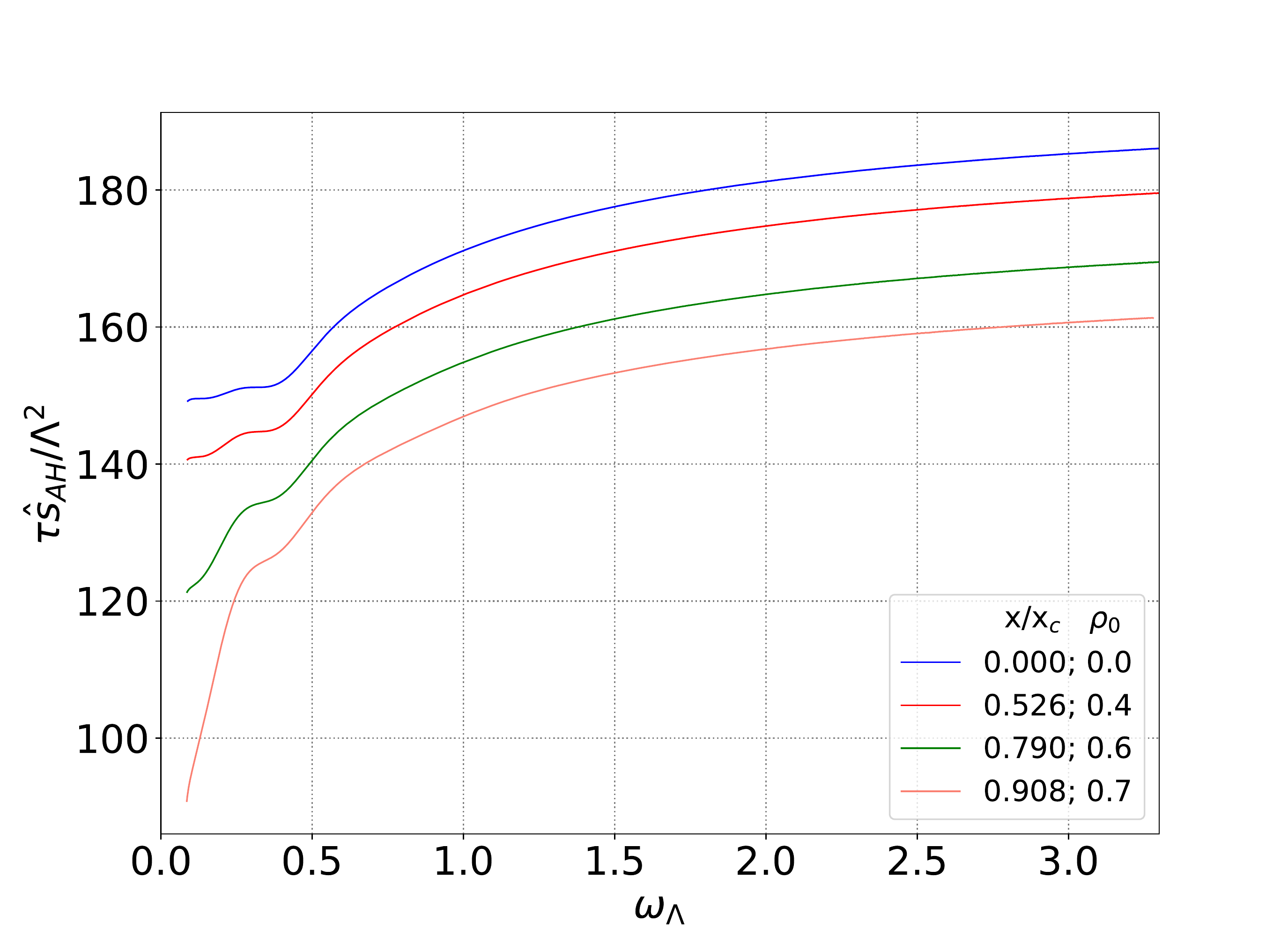}}
\subfigure[]{\includegraphics[width=0.49\textwidth]{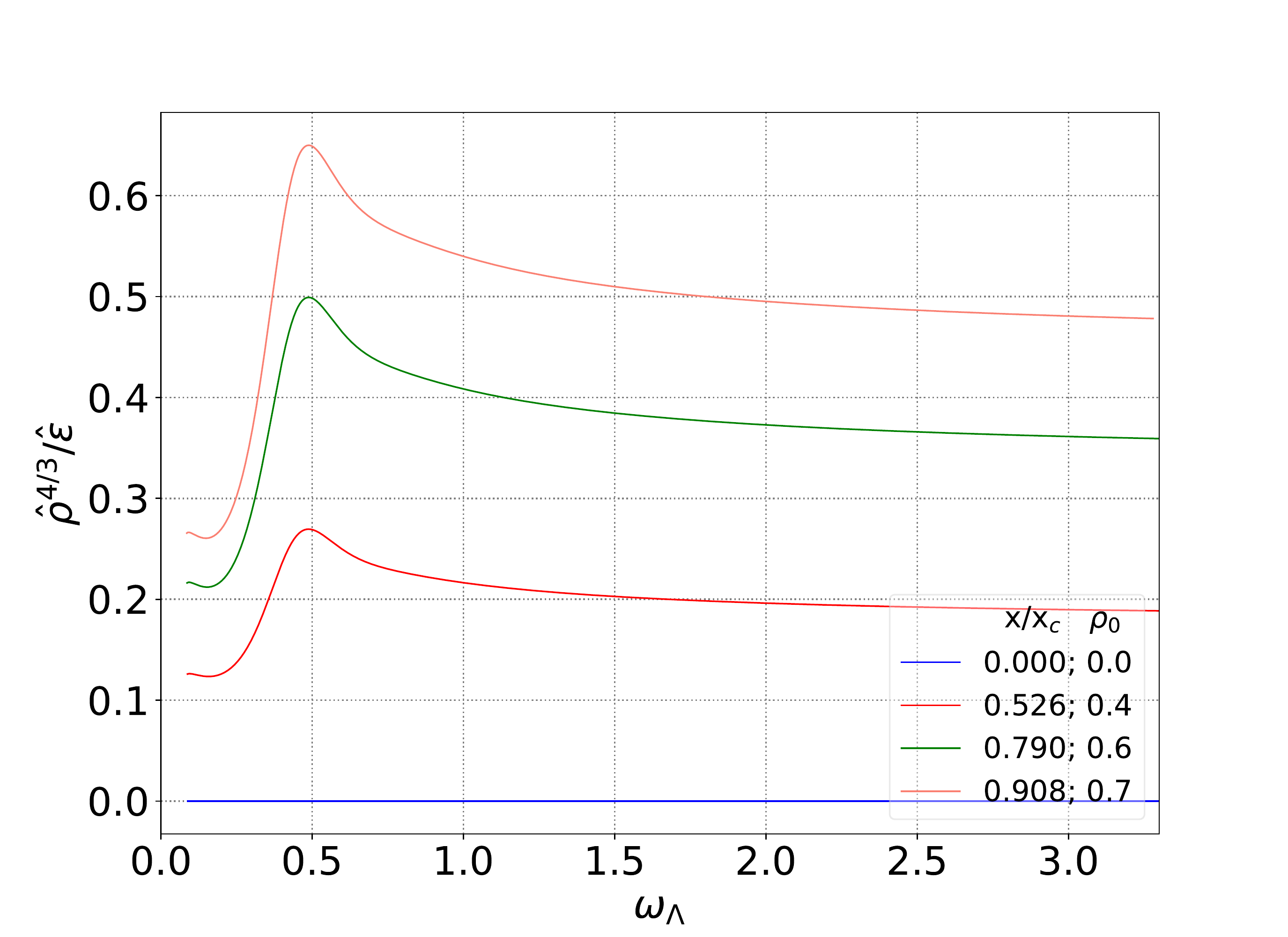}}
\subfigure[]{\includegraphics[width=0.49\textwidth]{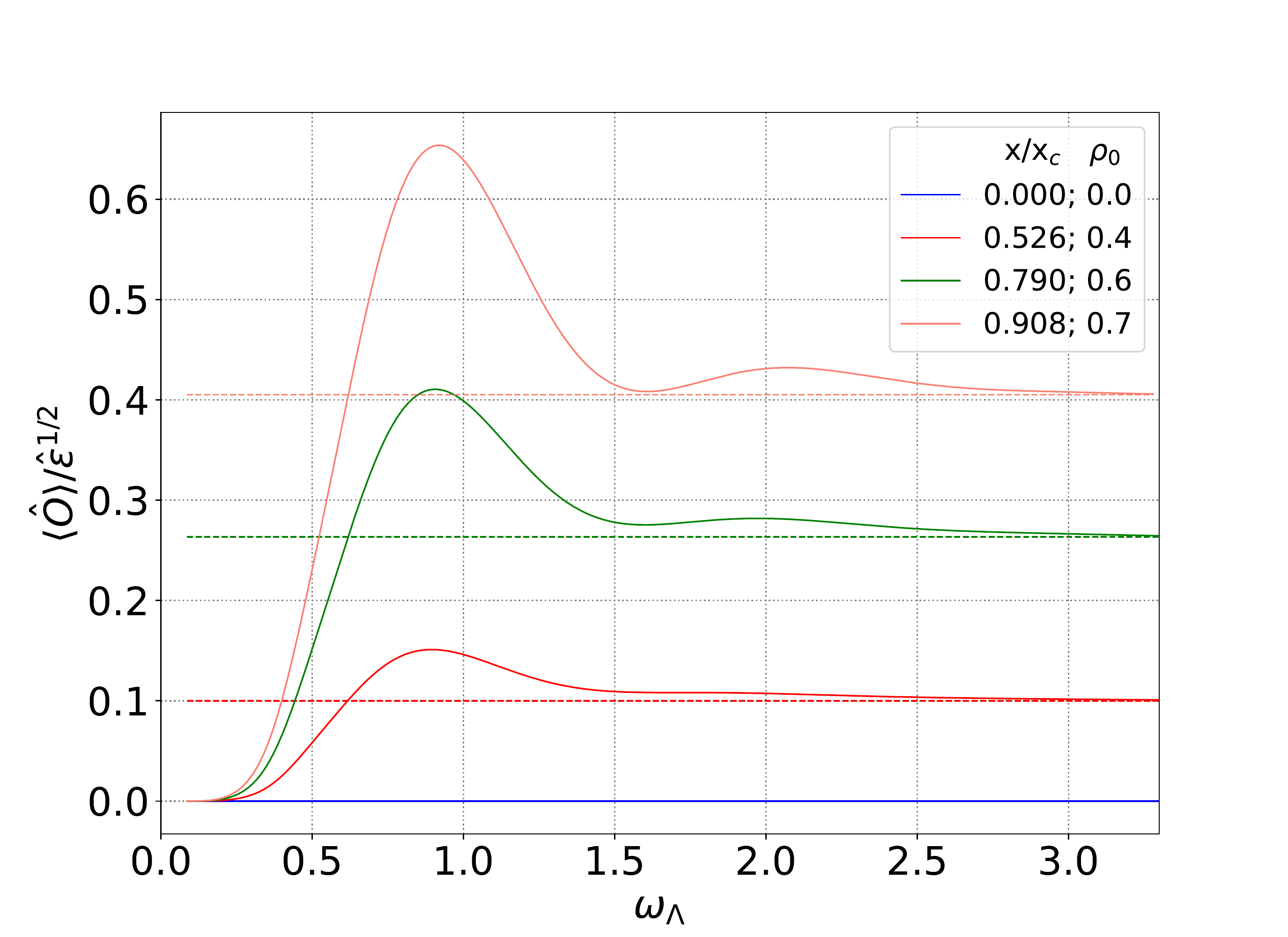}}
\caption{(a) Normalized pressure anisotropy (solid lines) and the corresponding hydrodynamic Navier-Stokes result (dashed lines), (b) normalized non-equilibrium entropy $\hat{S}_\textrm{AH}/\mathcal{A}\Lambda^2=\tau\hat{s}_\textrm{AH}/\Lambda^2$, (c) normalized charge density, and (d) normalized scalar condensate (solid lines) and the corresponding thermodynamic stable equilibrium result (dashed lines). Results obtained for variations of $\rho_0$ keeping fixed $B_s7$ in Table \ref{tabICs} with $a_2(\tau_0)=-6.67$. Note that $x_c\equiv\left(\mu/T\right)_c=\pi/\sqrt{2}$ is the critical point.}
\label{fig:result7}
\end{figure*}

\begin{figure*}%[h]
\center
\subfigure[]{\includegraphics[width=0.49\textwidth]{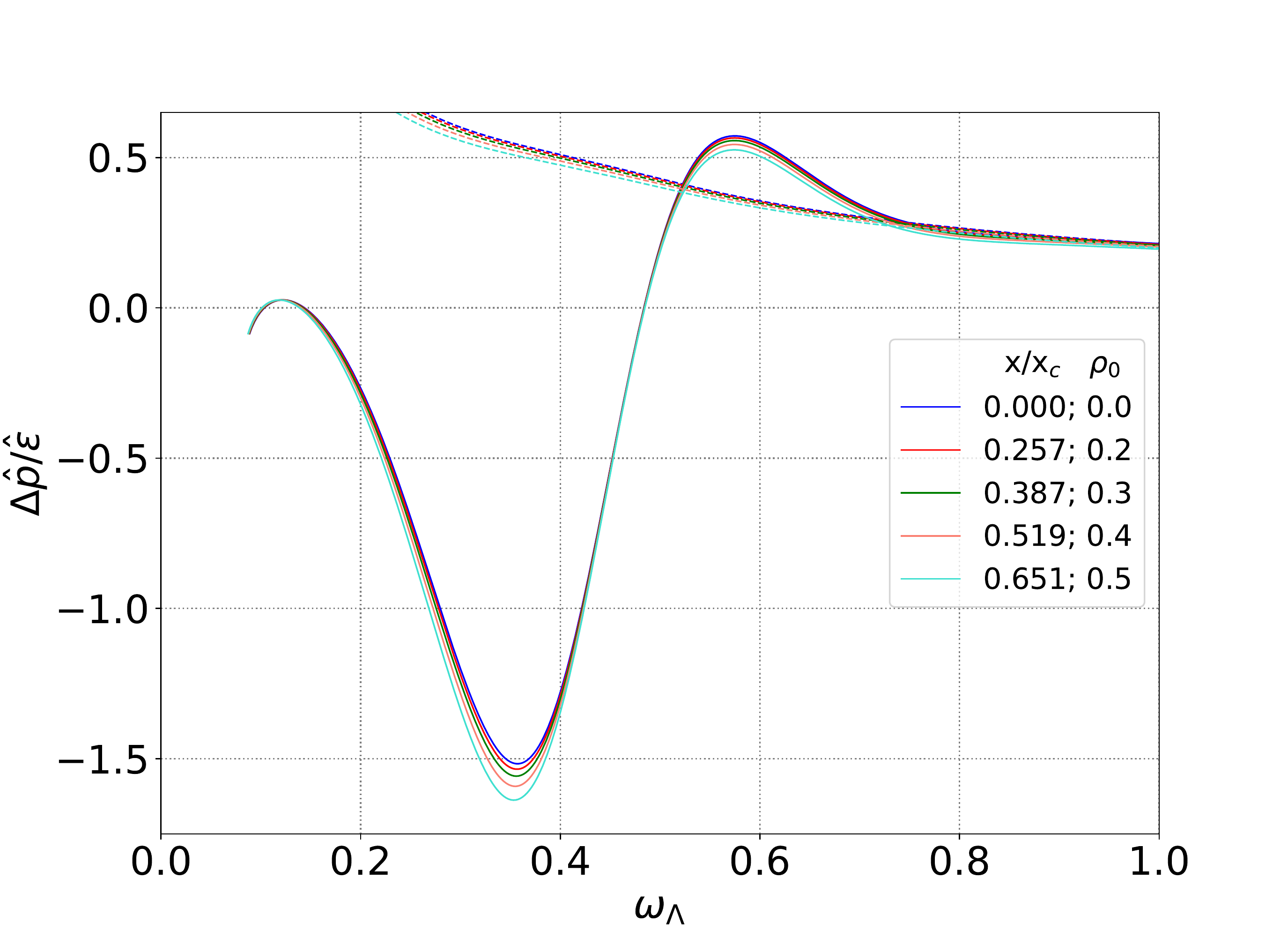}}
\subfigure[]{\includegraphics[width=0.49\textwidth]{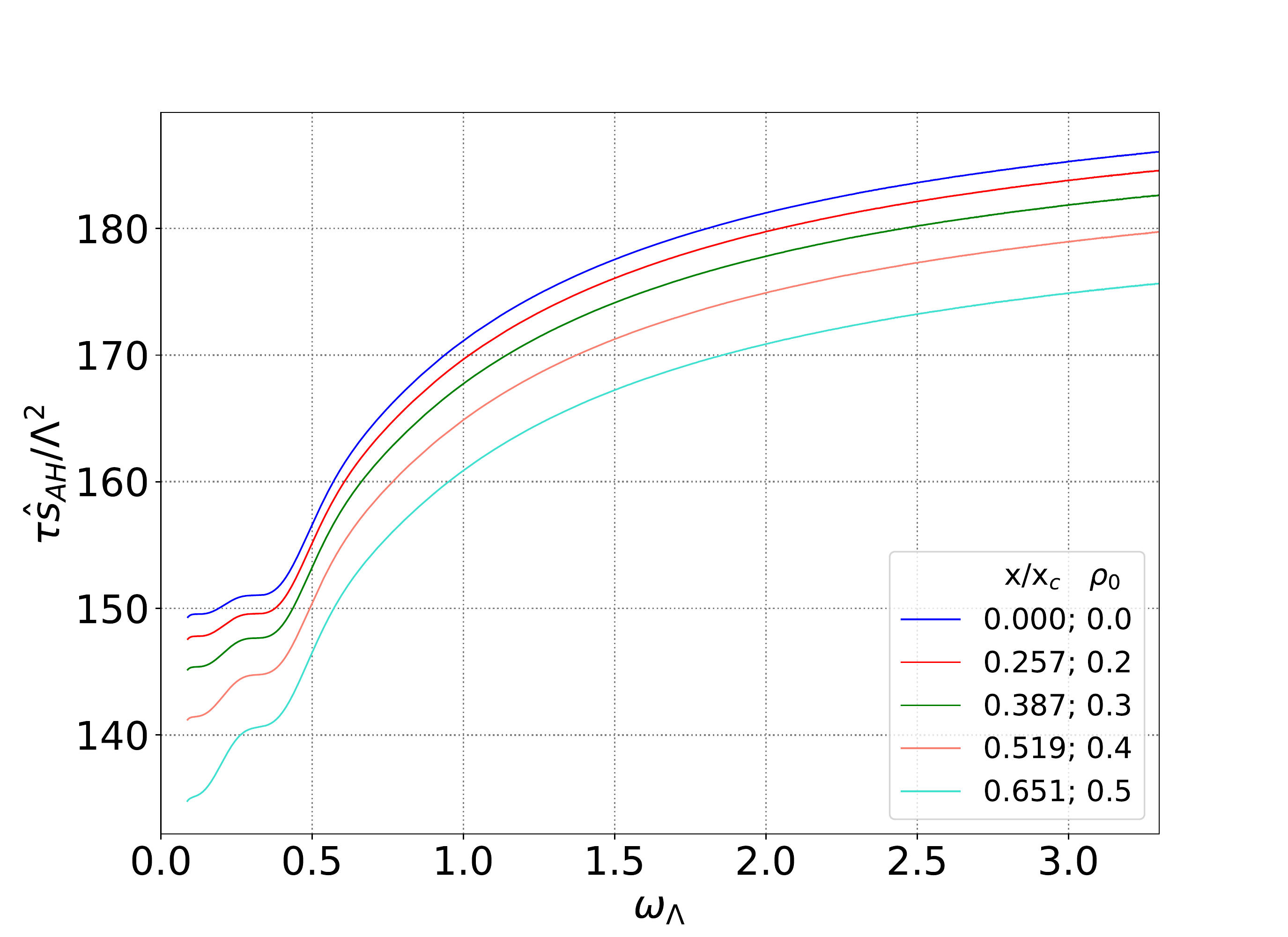}}
\subfigure[]{\includegraphics[width=0.49\textwidth]{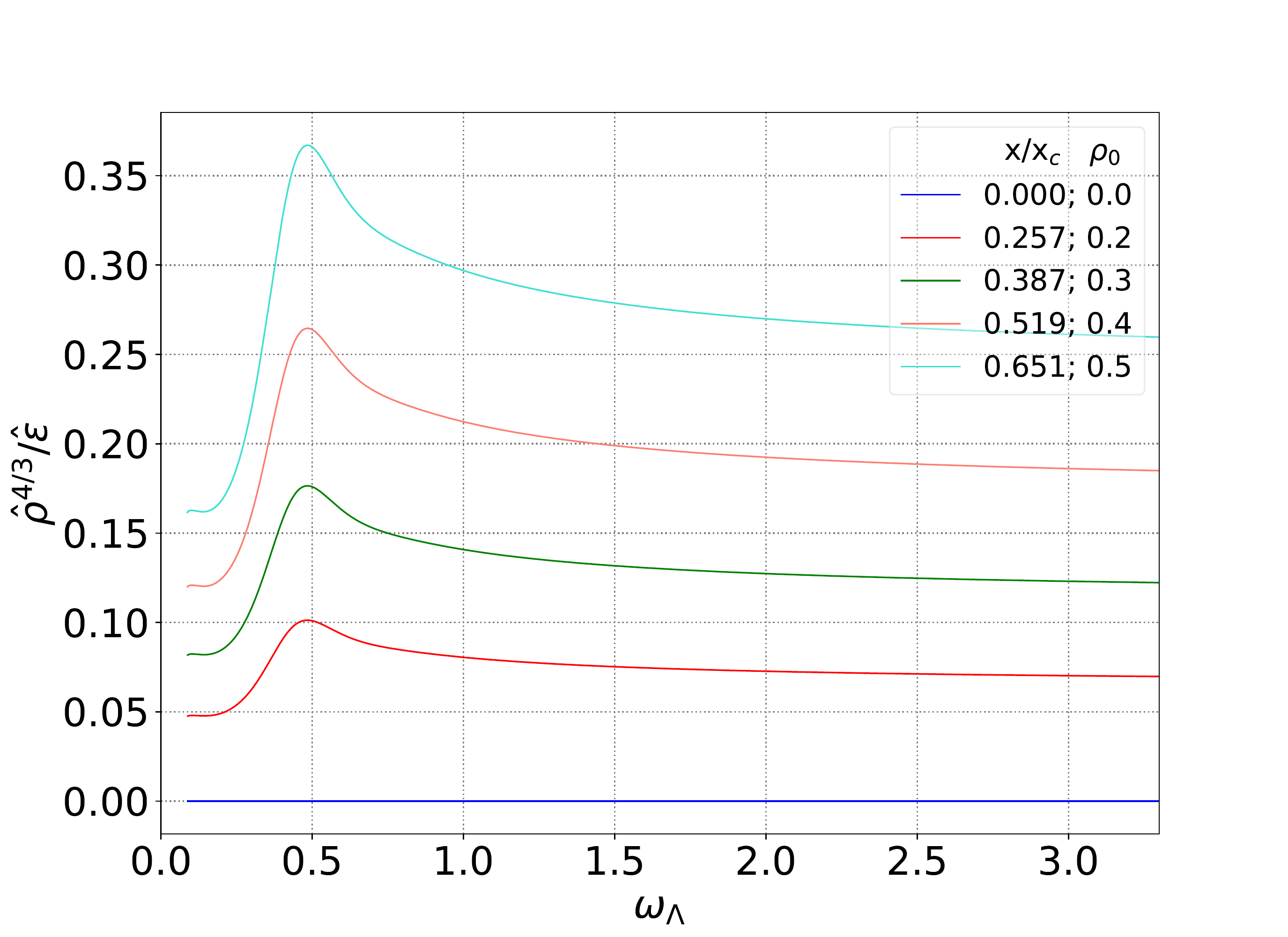}}
\subfigure[]{\includegraphics[width=0.49\textwidth]{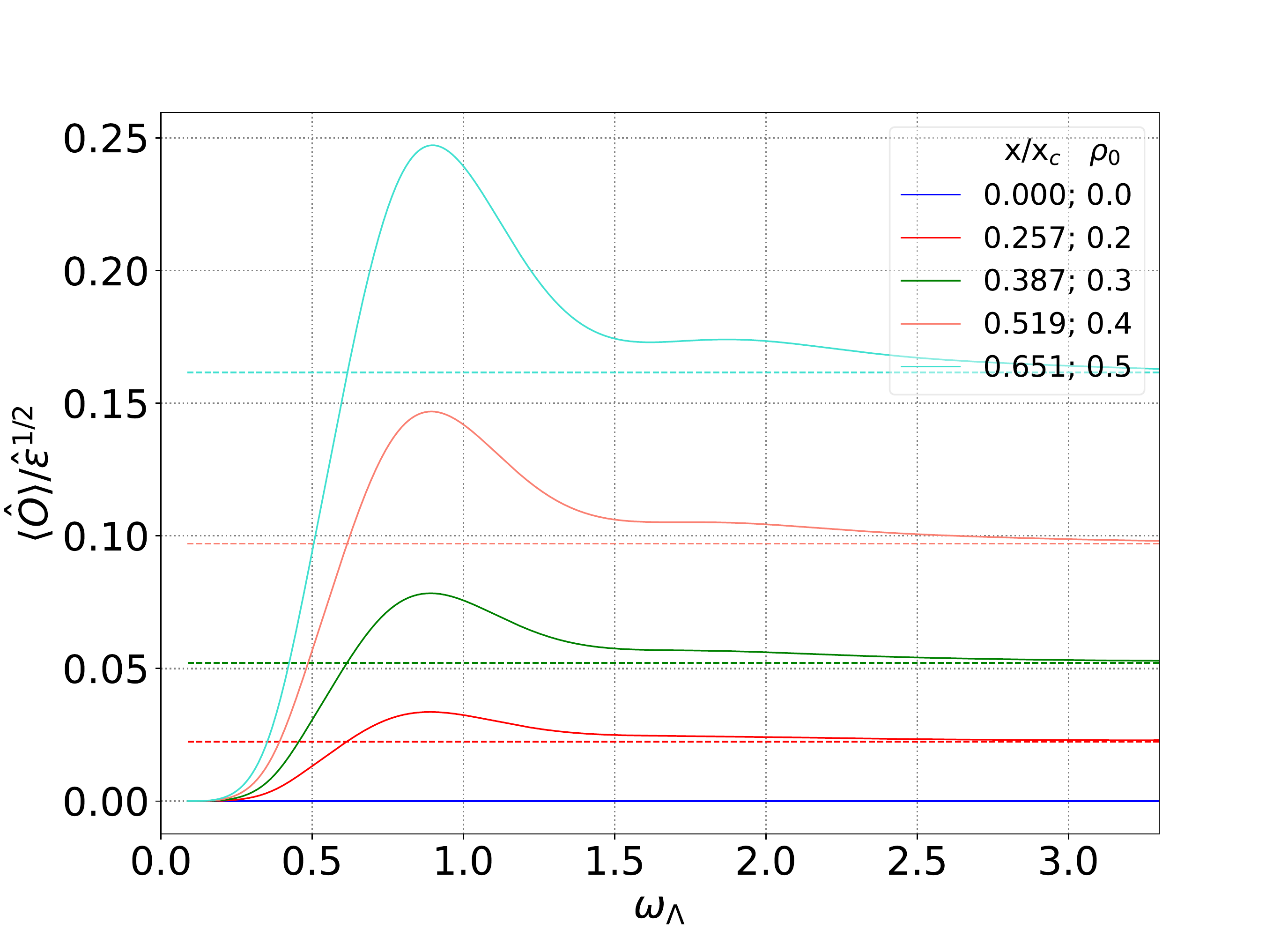}}
\caption{(a) Normalized pressure anisotropy (solid lines) and the corresponding hydrodynamic Navier-Stokes result (dashed lines), (b) normalized non-equilibrium entropy $\hat{S}_\textrm{AH}/\mathcal{A}\Lambda^2=\tau\hat{s}_\textrm{AH}/\Lambda^2$, (c) normalized charge density, and (d) normalized scalar condensate (solid lines) and the corresponding thermodynamic stable equilibrium result (dashed lines). Results obtained for variations of $\rho_0$ keeping fixed $B_s8$ in Table \ref{tabICs} with $a_2(\tau_0)=-7$. Note that $x_c\equiv\left(\mu/T\right)_c=\pi/\sqrt{2}$ is the critical point.}
\label{fig:result8}
\end{figure*}

\begin{figure*}%[h]
\center
\subfigure[]{\includegraphics[width=0.49\textwidth]{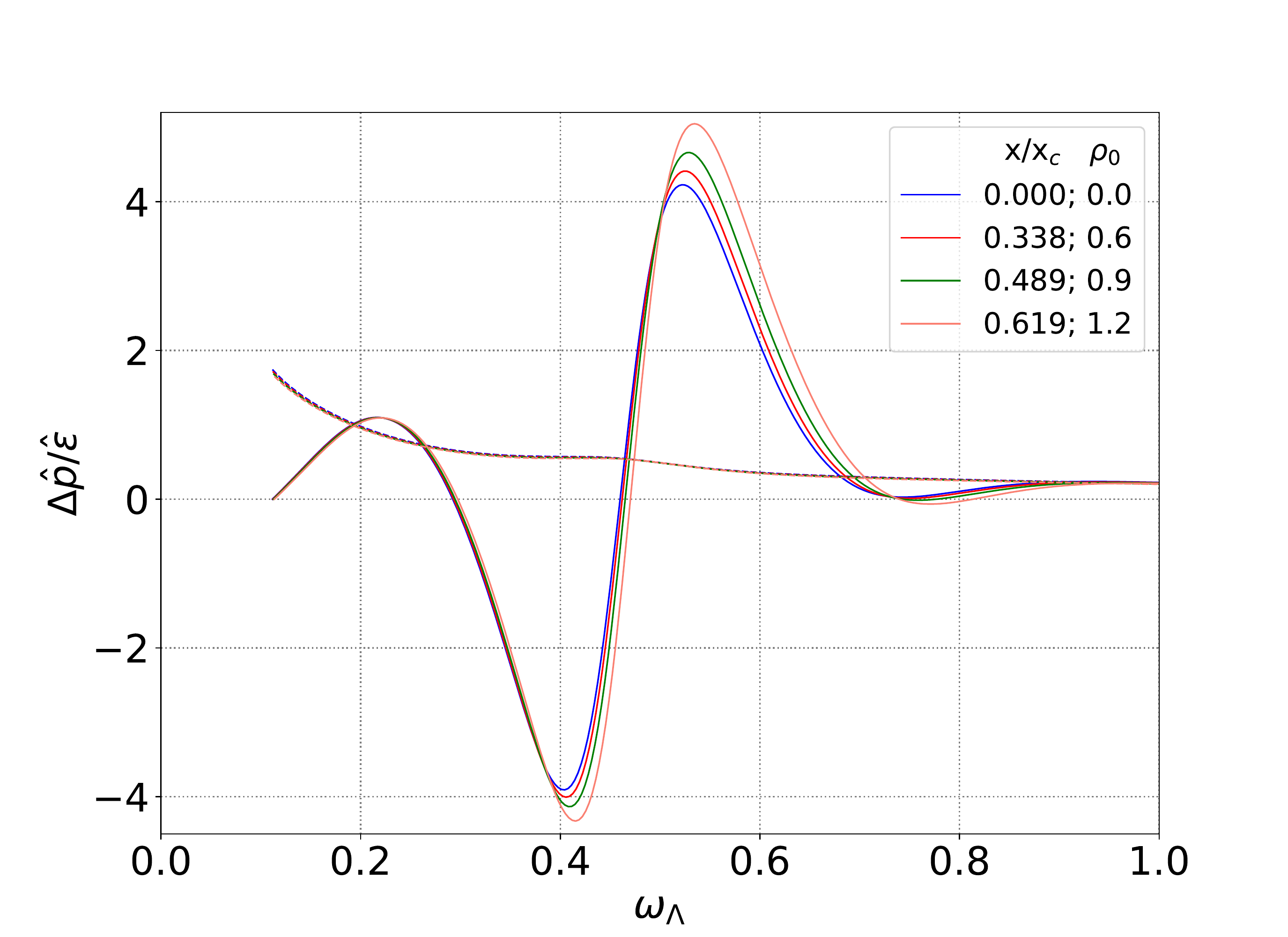}}
\subfigure[]{\includegraphics[width=0.49\textwidth]{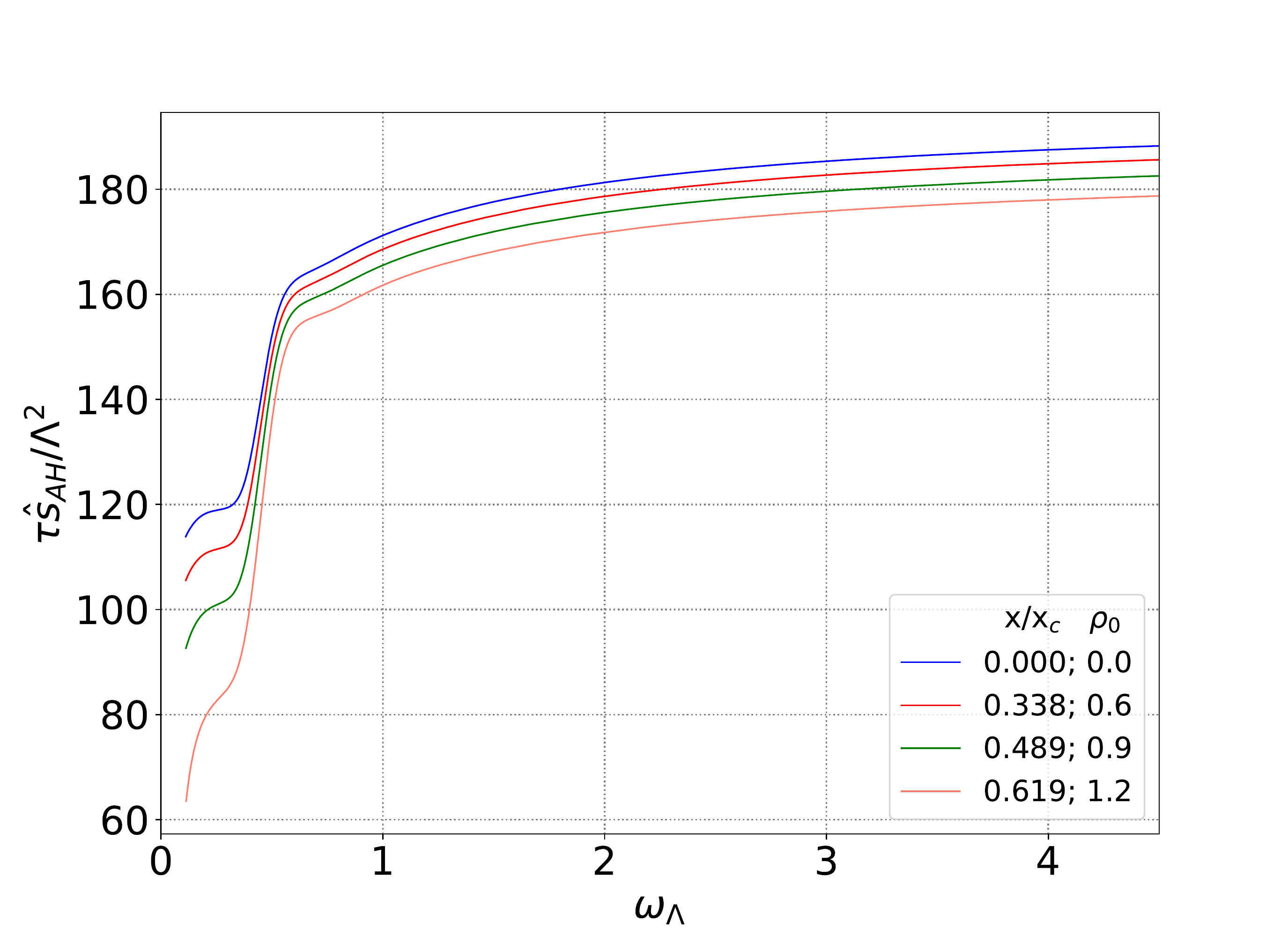}}
\subfigure[]{\includegraphics[width=0.49\textwidth]{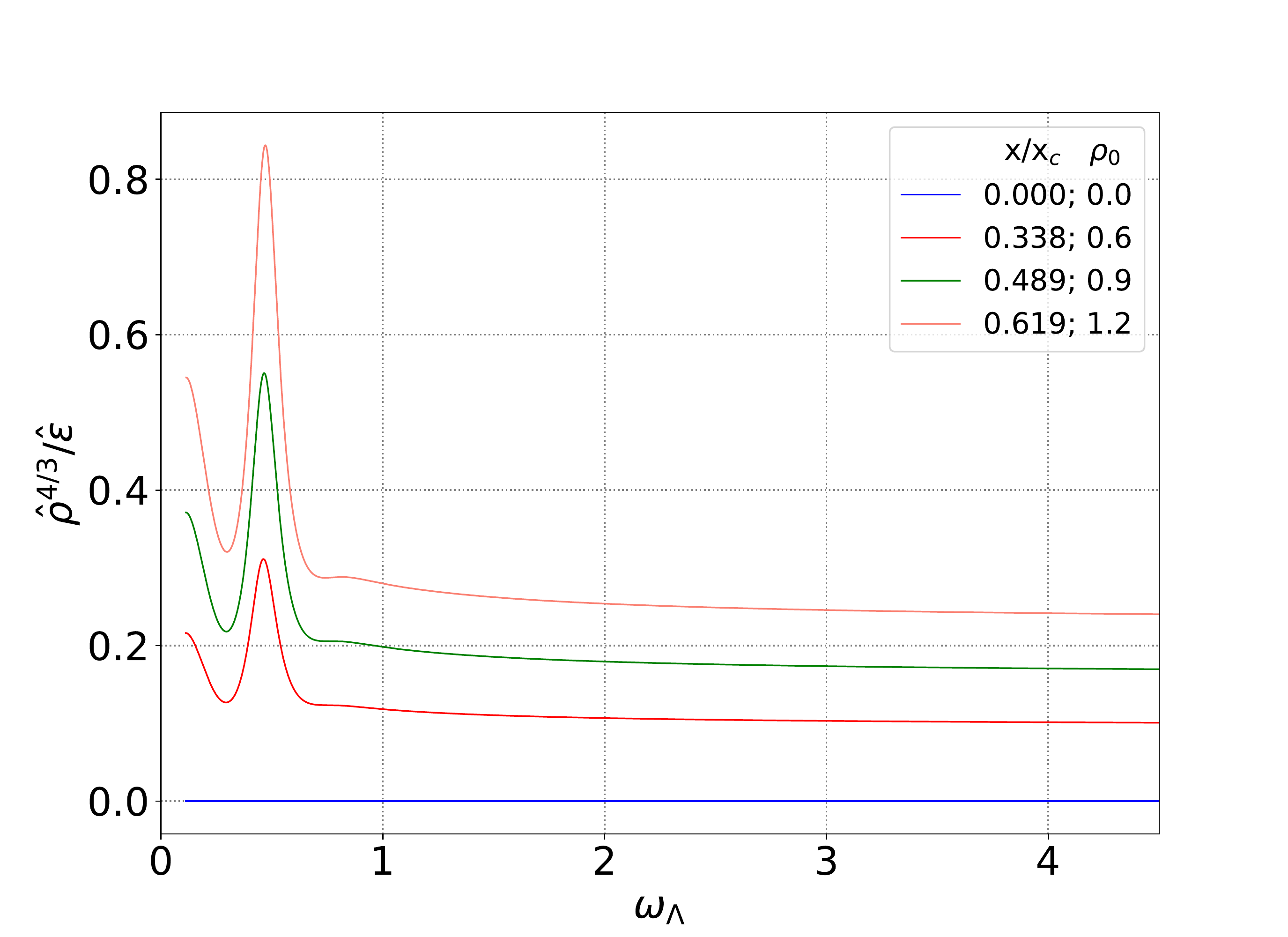}}
\subfigure[]{\includegraphics[width=0.49\textwidth]{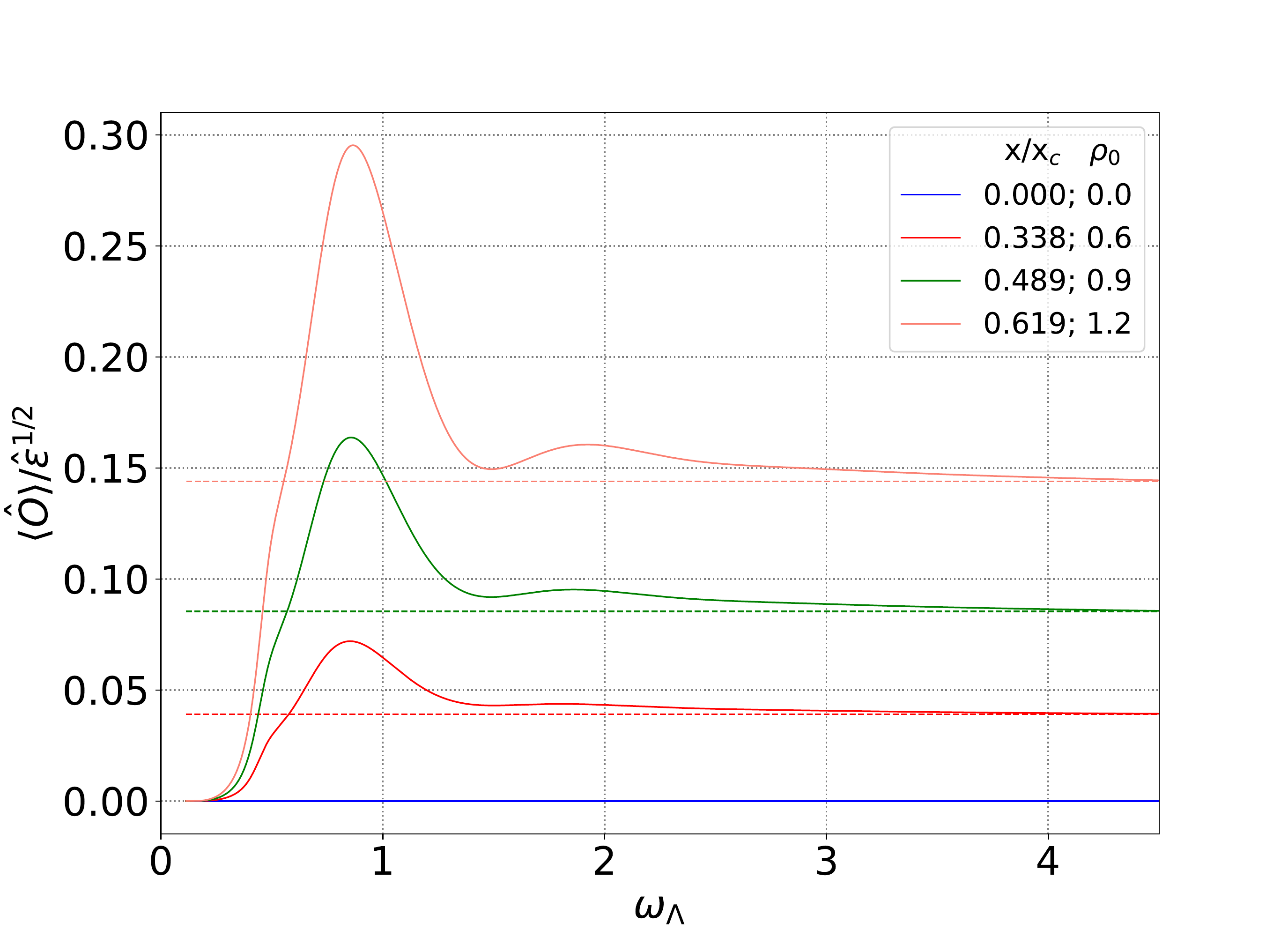}}
\caption{(a) Normalized pressure anisotropy (solid lines) and the corresponding hydrodynamic Navier-Stokes result (dashed lines), (b) normalized non-equilibrium entropy $\hat{S}_\textrm{AH}/\mathcal{A}\Lambda^2=\tau\hat{s}_\textrm{AH}/\Lambda^2$, (c) normalized charge density, and (d) normalized scalar condensate (solid lines) and the corresponding thermodynamic stable equilibrium result (dashed lines). Results obtained for variations of $\rho_0$ keeping fixed $B_s9$ in Table \ref{tabICs} with $a_2(\tau_0)=-6.67$. Note that $x_c\equiv\left(\mu/T\right)_c=\pi/\sqrt{2}$ is the critical point.}
\label{fig:result9}
\end{figure*}

\begin{figure*}%[h]
\center
\subfigure[]{\includegraphics[width=0.49\textwidth]{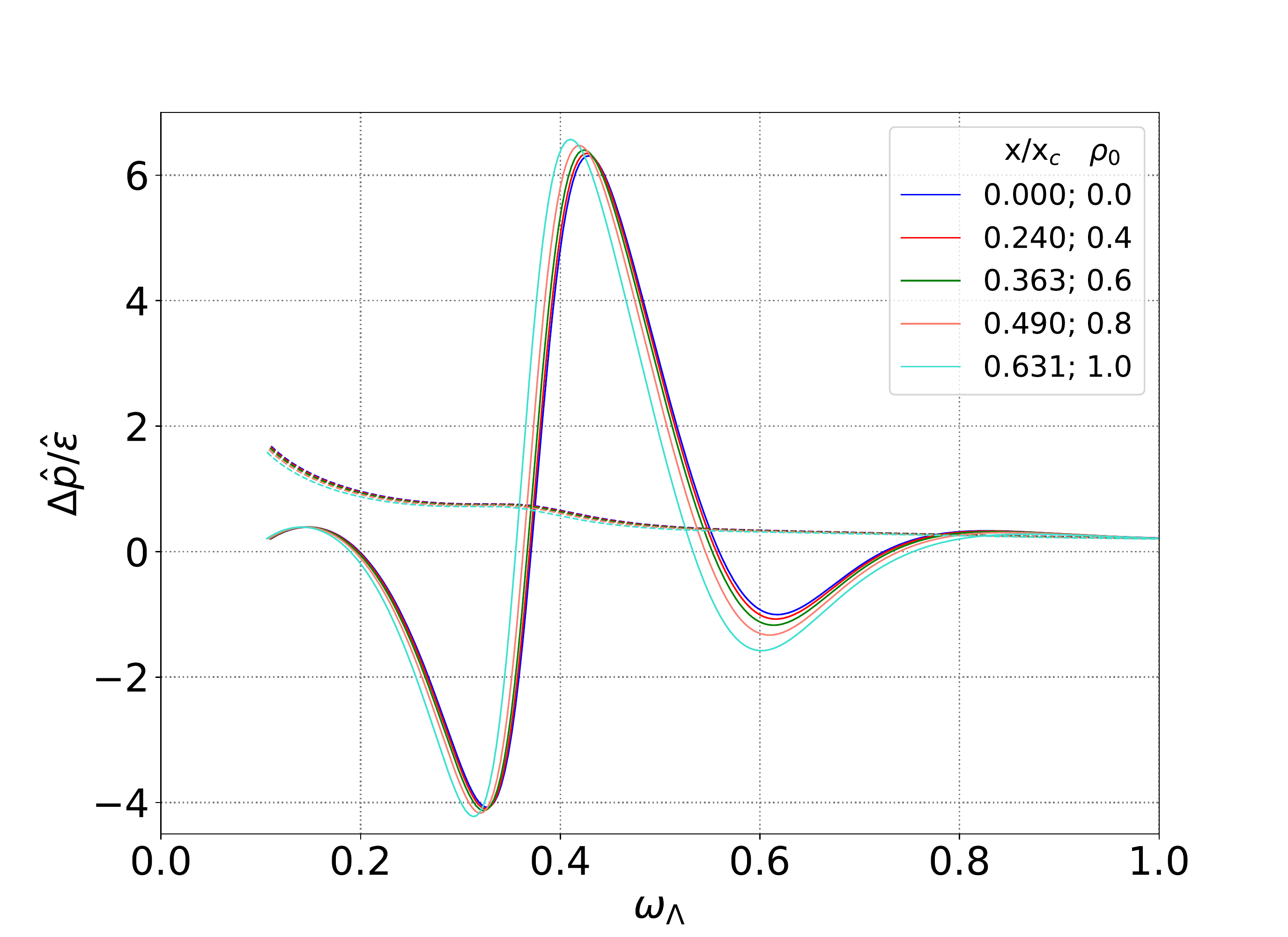}}
\subfigure[]{\includegraphics[width=0.49\textwidth]{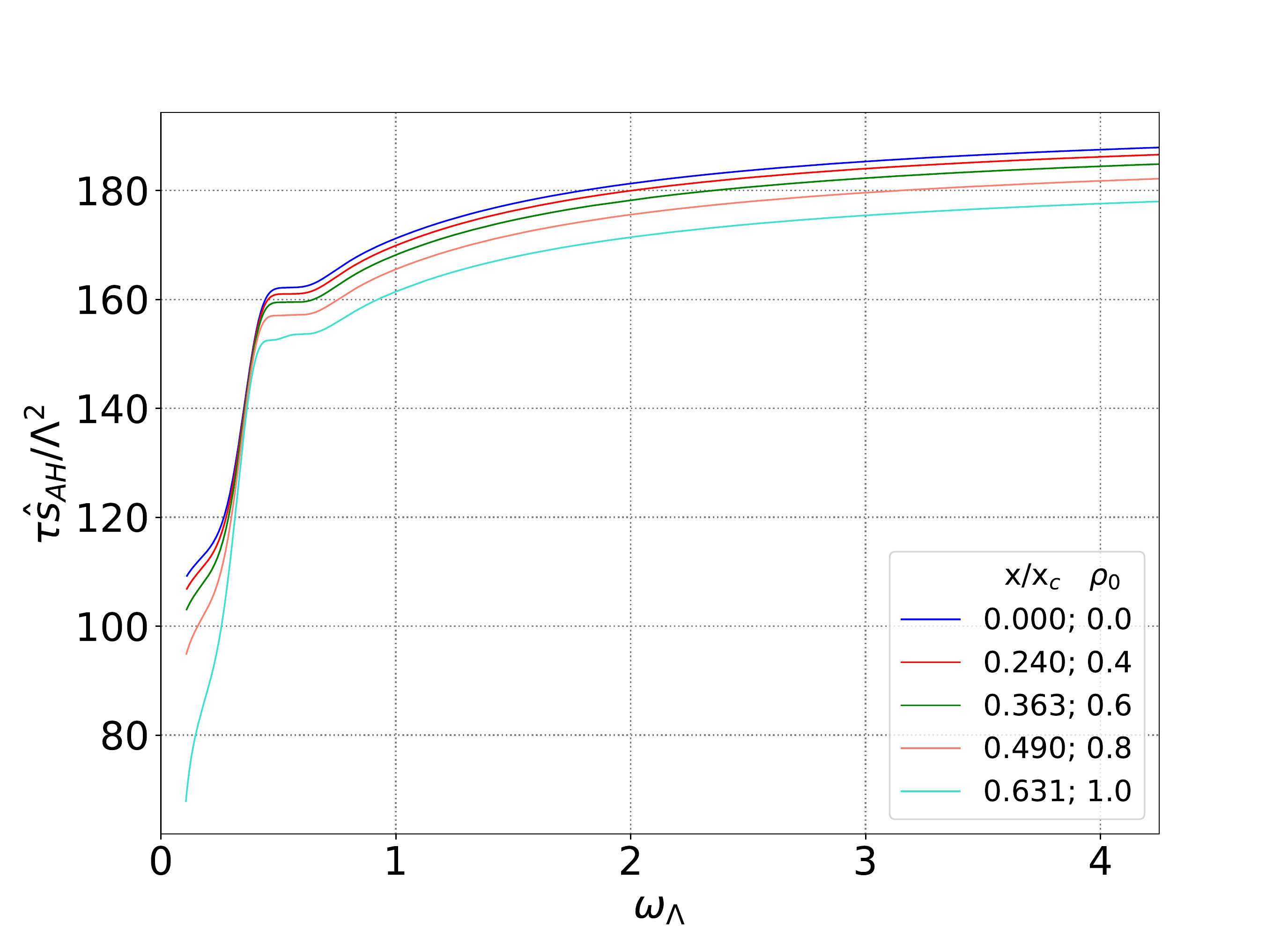}}
\subfigure[]{\includegraphics[width=0.49\textwidth]{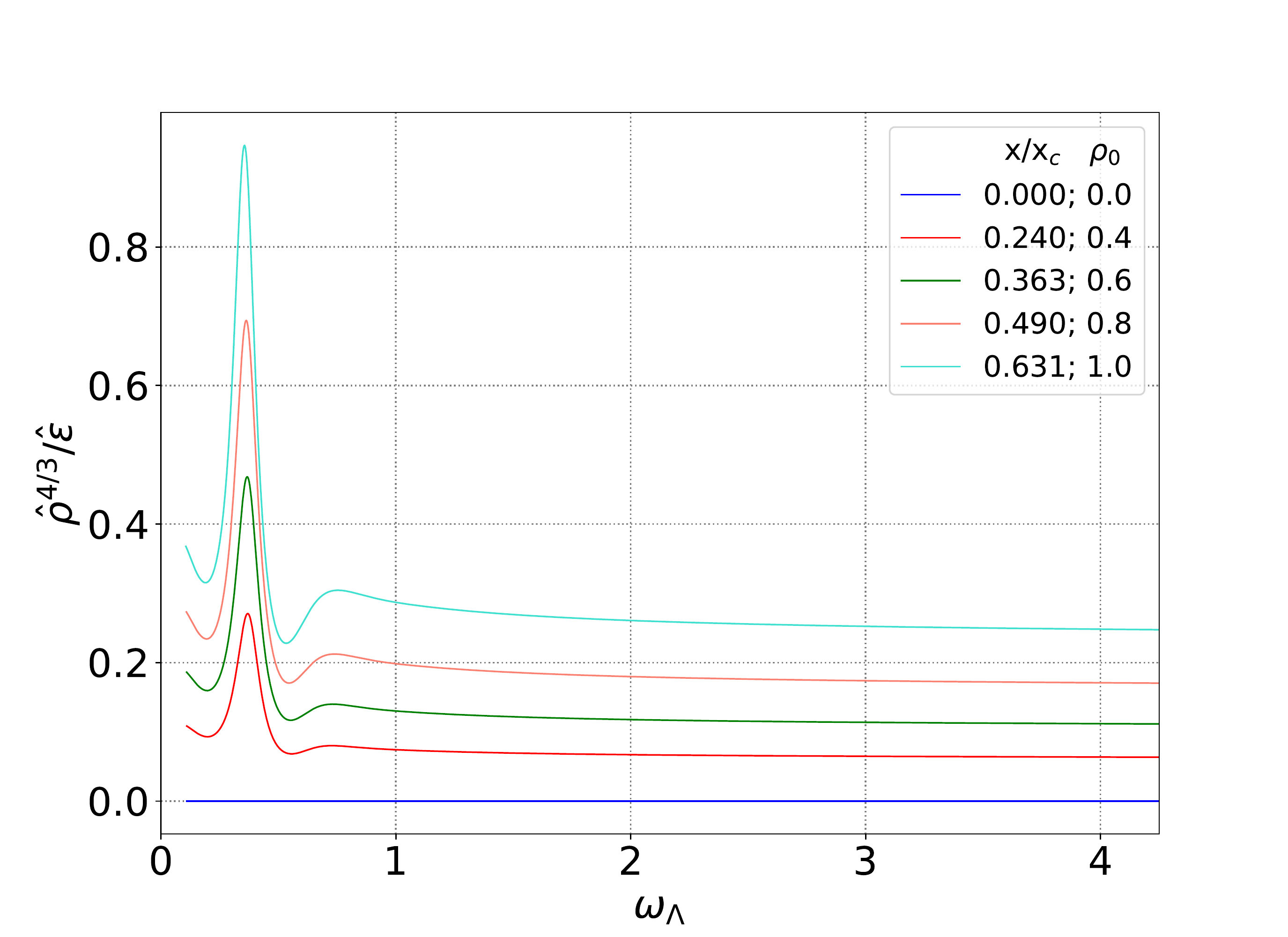}}
\subfigure[]{\includegraphics[width=0.49\textwidth]{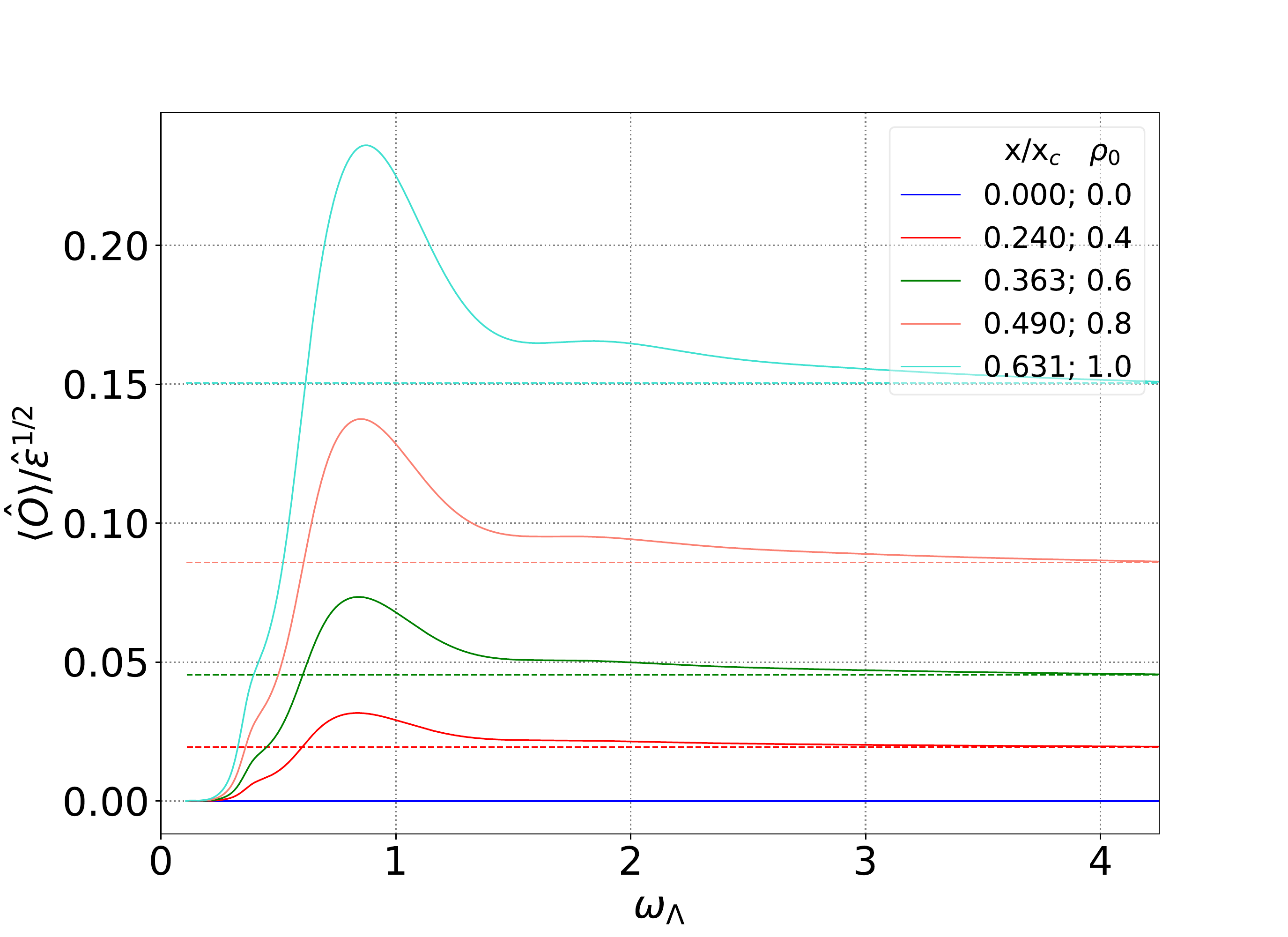}}
\caption{(a) Normalized pressure anisotropy (solid lines) and the corresponding hydrodynamic Navier-Stokes result (dashed lines), (b) normalized non-equilibrium entropy $\hat{S}_\textrm{AH}/\mathcal{A}\Lambda^2=\tau\hat{s}_\textrm{AH}/\Lambda^2$, (c) normalized charge density, and (d) normalized scalar condensate (solid lines) and the corresponding thermodynamic stable equilibrium result (dashed lines). Results obtained for variations of $\rho_0$ keeping fixed $B_s10$ in Table \ref{tabICs} with $a_2(\tau_0)=-7.75$. Note that $x_c\equiv\left(\mu/T\right)_c=\pi/\sqrt{2}$ is the critical point.}
\label{fig:result10}
\end{figure*}

\begin{figure*}%[h]
\center
\subfigure[]{\includegraphics[width=0.49\textwidth]{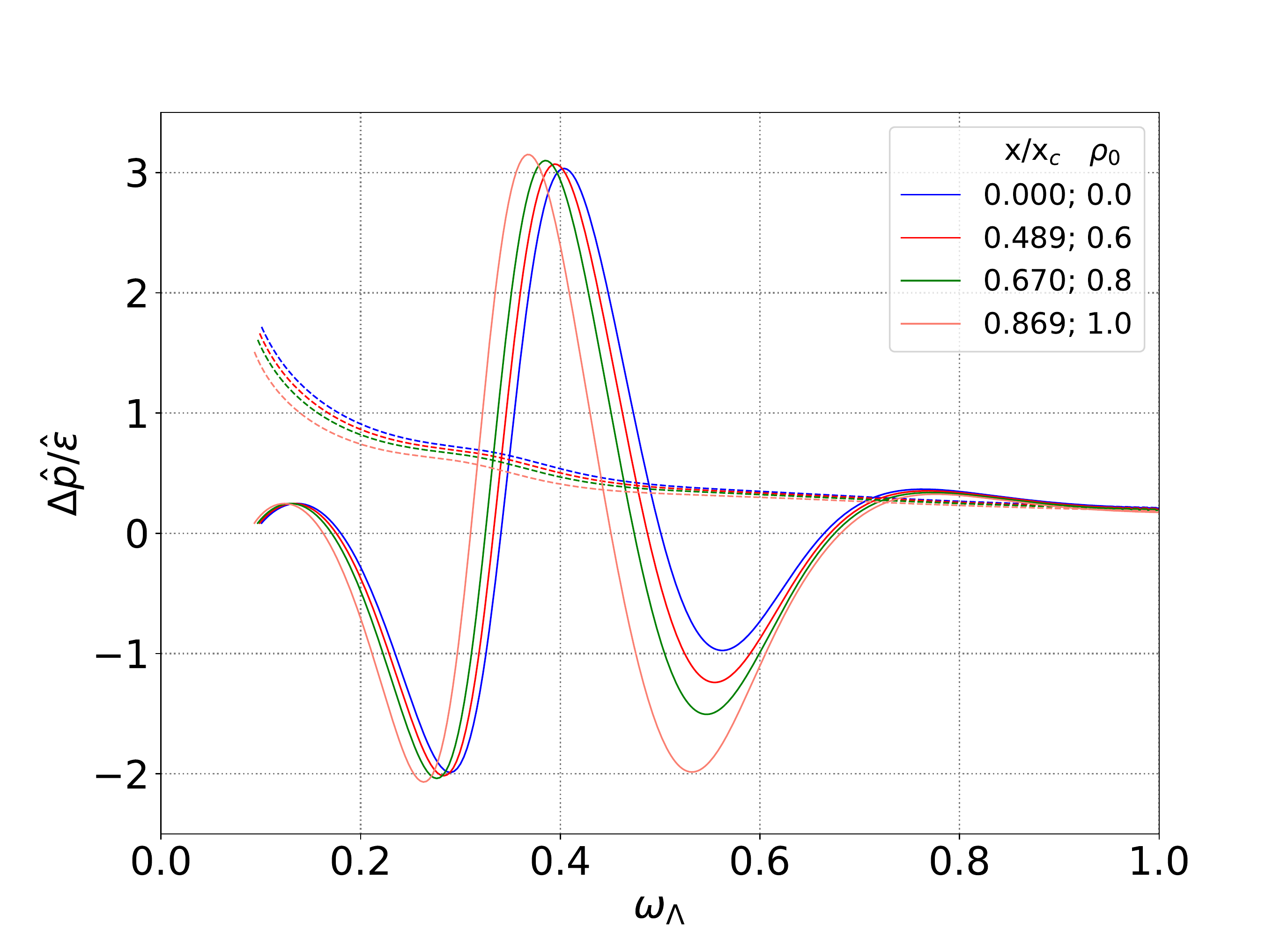}}
\subfigure[]{\includegraphics[width=0.49\textwidth]{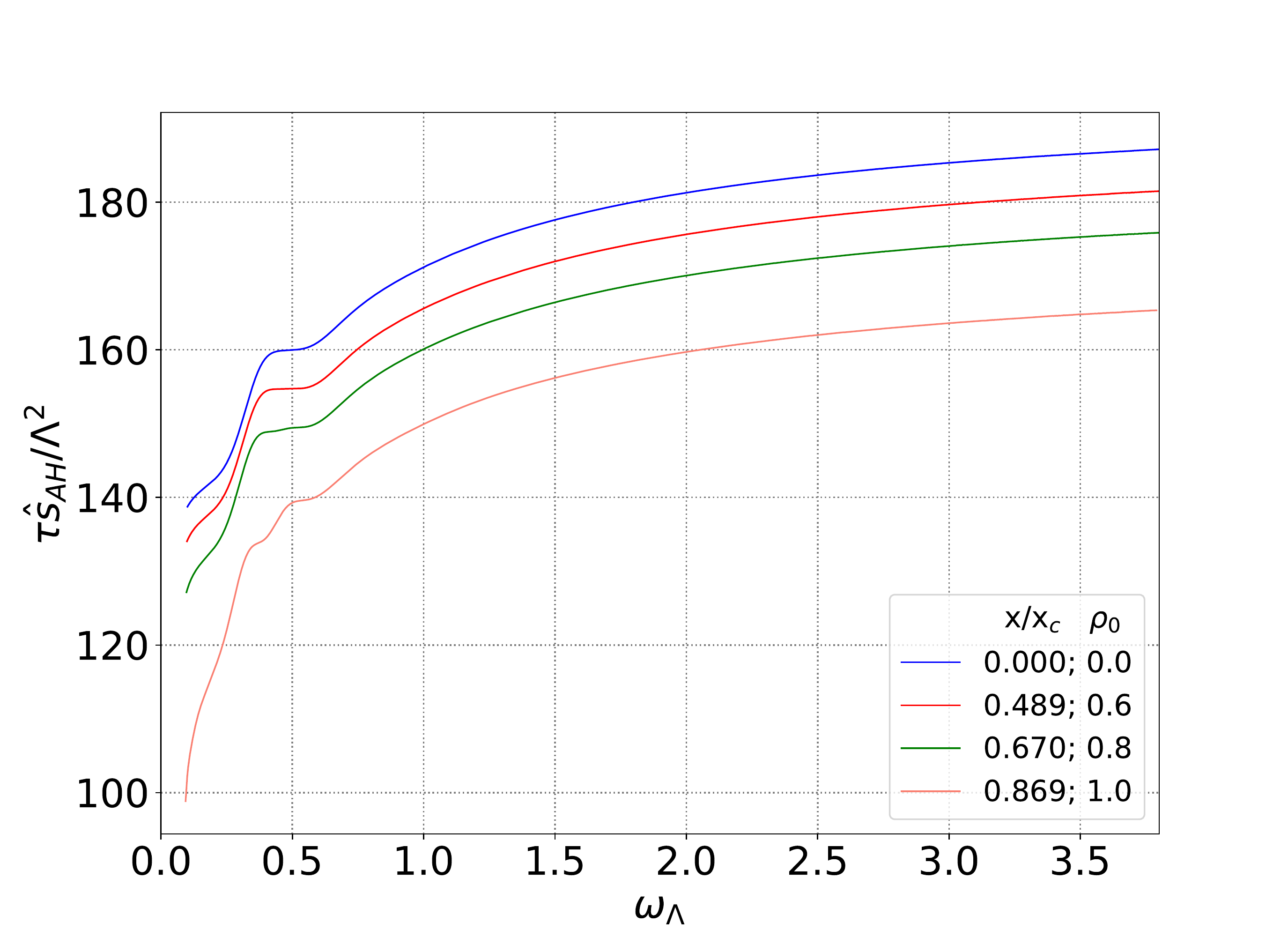}}
\subfigure[]{\includegraphics[width=0.49\textwidth]{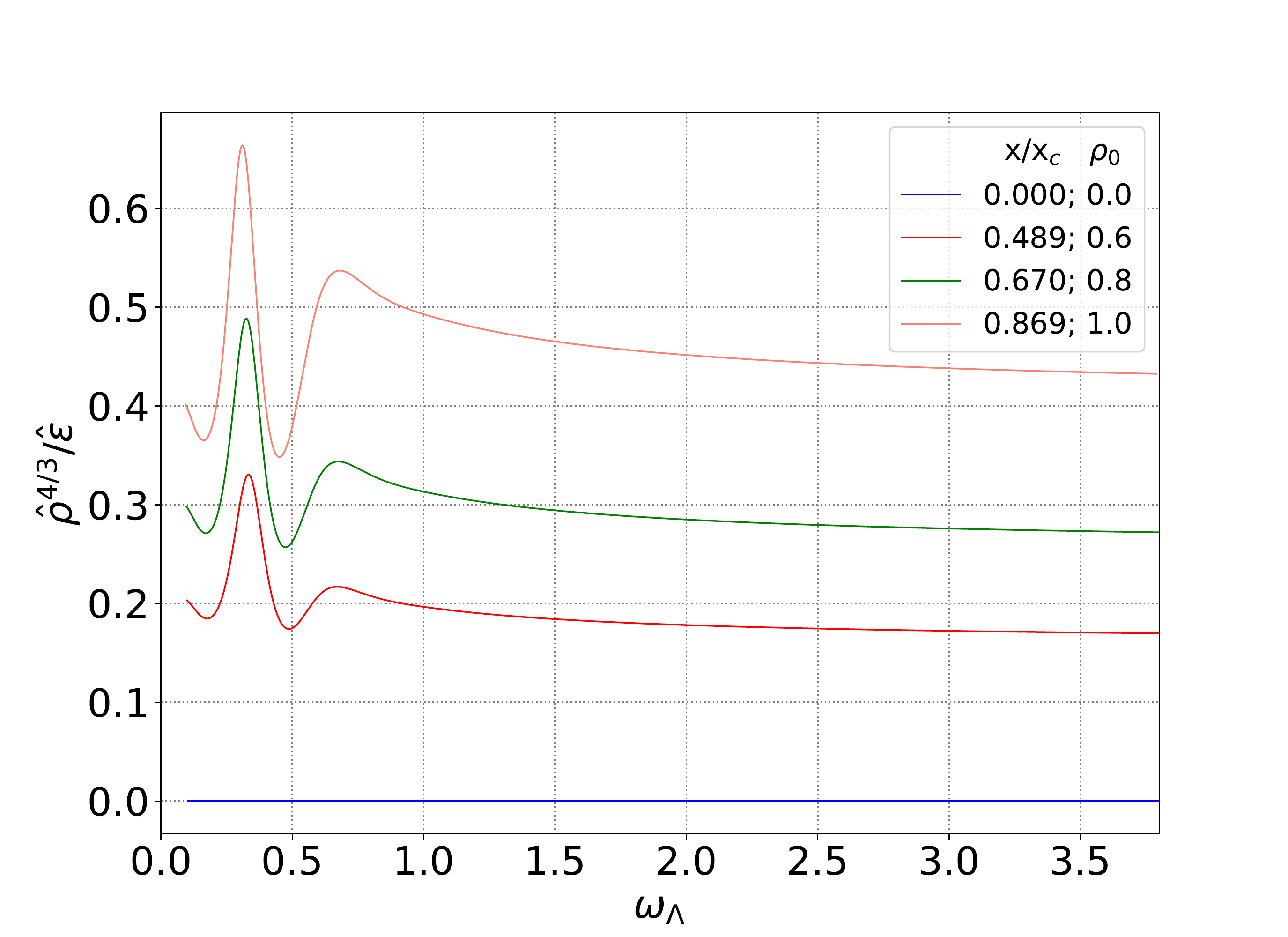}}
\subfigure[]{\includegraphics[width=0.49\textwidth]{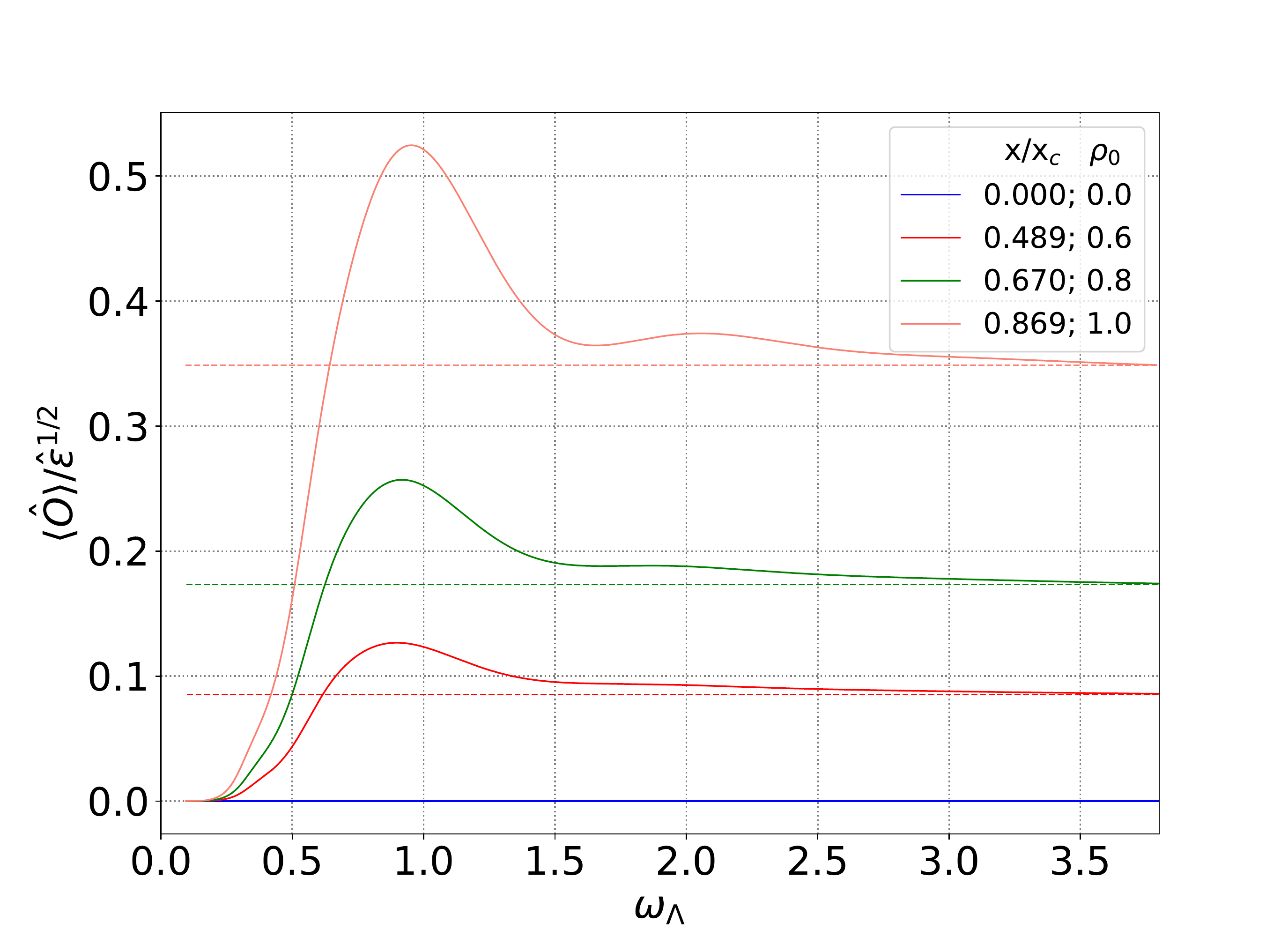}}
\caption{(a) Normalized pressure anisotropy (solid lines) and the corresponding hydrodynamic Navier-Stokes result (dashed lines), (b) normalized non-equilibrium entropy $\hat{S}_\textrm{AH}/\mathcal{A}\Lambda^2=\tau\hat{s}_\textrm{AH}/\Lambda^2$, (c) normalized charge density, and (d) normalized scalar condensate (solid lines) and the corresponding thermodynamic stable equilibrium result (dashed lines). Results obtained for variations of $\rho_0$ keeping fixed $B_s11$ in Table \ref{tabICs} with $a_2(\tau_0)=-7.1$. Note that $x_c\equiv\left(\mu/T\right)_c=\pi/\sqrt{2}$ is the critical point.}
\label{fig:result11}
\end{figure*}

In this section, we analyze the time evolution of several different far from equilibrium initial conditions of the Bjorken expanding 1RCBH plasma, where for each profile for the initial subtracted metric anisotropy specified in Eq. \eqref{eq:Bs0} and in Table \ref{tabICs}, we consider variations of the initial charge density of the medium \eqref{eq:hatrho}, $\hat{\rho}(\tau_0)=\rho_0/\tau_0$, while keeping fixed its initial energy density \eqref{eq:hatE}, $\hat{\epsilon}(\tau_0)=-3a_2(\tau_0)$ (we set $\phi_s(\tau_0,u)=0$ throughout this section). The corresponding results are shown in Figs. \ref{fig:result1} --- \ref{fig:result11}. We remark that, due to the presence of several initial data which can be independently varied in the 1RCBH plasma, in the present and in the following sections, we plot the time evolution of many initial conditions in order to explore in details several qualitatively different possibilities for the dynamic evolution of the relevant physical observables of the 1RCBH plasma undergoing Bjorken flow. The several pictures considered illustrate the main general features and observations we summarize in the text.

One notices that by increasing the initial charge density of the 1RCBH plasma (by increasing $\rho_0$), while keeping its initial energy density fixed, the value of $\mu/T$ in the medium is enhanced, producing the following effects on the physical observables analyzed as functions of the dimensionless time measure $\omega_\Lambda$:

\begin{enumerate}[a)]
\item \underline{$[\Delta\hat{p}/\hat{\epsilon}](\omega_\Lambda)$}: the hydrodynamization of the pressure anisotropy of the medium, as measured by its convergence, within small relative tolerances, to the Navier-Stokes regime, is generally delayed as $\mu/T$ increases (in line with what was reported in \cite{Critelli:2018osu}). The enhancement of $\mu/T$ also leads to different physical possibilities for the maxima and the minima of the pressure anisotropy (not always both extrema are simultaneously present): while for all the initial conditions considered which present a minimum, we observed an increase of its magnitude, for initial conditions presenting a maximum, its magnitude can either increase or decrease depending on the chosen profile for the initial subtracted metric anisotropy $B_s(\tau_0,u)$; moreover, in the cases where far from equilibrium transient violations of the DEC \eqref{eq:DEC} and of the WEC \eqref{eq:WEC} are observed, such violations increase with increasing $\mu/T$, clearly indicating that they become more and more relevant as the system approaches criticality at large charge densities.

\item \underline{$\hat{S}_\textrm{AH}(\omega_\Lambda)/\mathcal{A}\Lambda^2$}: the magnitude of this dimensionless ratio always decreases as $\mu/T$ is increased. For different choices of the profile for the initial subtracted metric anisotropy $B_s(\tau_0,u)$, one may realize different physical possibilities for the time evolution of the entropy of the medium, with the presence or absence of early time (quasi)plateau structures. \emph{When} single or double plateaus are produced in the time evolution of the entropy, indicating the existence of far from equilibrium time windows with no entropy production in the Bjorken expanding fluid, it is later observed almost or effective violations of DEC from below with $\Delta\hat{p}/\hat{\epsilon}\sim -1$ or $\Delta\hat{p}/\hat{\epsilon}<-1$, respectively (in line with the observations reported in \cite{Rougemont:2021qyk,Rougemont:2021gjm} for the particular case corresponding to the pure thermal SYM plasma with $\mu/T=0$ --- here we also observe the aforementioned correlations under much more general situations with $\mu/T \ge 0$).\footnote{We remark that, in general, such correlations do not hold in reverse order, i.e. there are evolutions for the 1RCBH plasma with transient violations of DEC which present no far from equilibrium plateaus in the entropy of the medium (see e.g. Fig. \ref{fig:result9}).} In particular, we notice the following general trend regarding the deformation of such plateau structures as $\mu/T$ is increased, leading to progressively lower dips in the pressure anisotropy: first a single plateau is formed, which then becomes more spread out in the $\omega_\Lambda$ direction, and next it is progressively deformed into double plateaus, then becoming double quasiplateaus, until the plateau structure is finally lost for sufficiently strong violations of the DEC from below (in such a progression, the lower plateau structure is undone first than the higher plateau). We also observe that it is possible to form a single quasiplateau (i.e. with a time derivative close but not actually zero) in the far from equilibrium entropy, which leads to no posterior violations of the DEC (see e.g. Fig. \ref{fig:result6}; in particular, for $\rho_0=0$, this has been already seen in \cite{Rougemont:2021qyk,Rougemont:2021gjm} for the SYM plasma at $\mu/T=0$, see the initial condition $\#20$ in those works, corresponding to the curve in cyan with the highest initial entropy). We also remark that we have not observed the final two steps in the aforementioned general trend of progressive deformations of the plateau structures in cases with $\mu/T=0$ (i.e., the progressive deformation of the double plateaus into double quasiplateaus and the final loss of the plateau structure was only observed in association with later, progressively stronger violations of the DEC from below with nonzero values of the initial charge density, which leads to a medium with $\mu/T>0$ --- this will be further illustrated in Fig. \ref{fig:app-2} in appendix \ref{sec:app1}).

\item \underline{$[\hat{\rho}^{4/3}/\hat{\epsilon}](\omega_\Lambda)$}: the magnitude of this observable always increases as $\mu/T$ is increased, but its qualitative behavior at early times may be very different depending on the chosen profile for the initial subtracted metric anisotropy $B_s(\tau_0,u)$: while in some cases this observable monotonically decreases in time, in other cases it presents the formation of extrema at early times, with the peculiar feature that the position of these extrema in the axis of the dimensionless time measure $\omega_\Lambda$ displays little to almost no variation as $\mu/T$ is enhanced.

\item \underline{$[\langle\hat{O}_\phi\rangle/\hat{\epsilon}^{1/2}](\omega_\Lambda)$}: the effective thermalization of the scalar condensate, as measured by its convergence, within small relative tolerances, to the thermodynamically stable equilibrium, is generally delayed as $\mu/T$ increases, and it only occurs for time scales much larger than the ones observed for the hydrodynamization of the pressure anisotropy. The enhancement of $\mu/T$ always leads to an increase in the magnitude of the scalar condensate, with the formation of at least one maximum as function of the dimensionless time measure $\omega_\Lambda$, with other later extrema possessing smaller magnitudes being also eventually observed for some high values of $\mu/T$ in the medium.
\end{enumerate}

We close this section with an extra technical information regarding the numerical simulations used to obtain the results displayed in Figs. \ref{fig:result1} --- \ref{fig:result11}. For almost all the initial conditions considered in those results, it is enough to use $N\sim 20$ collocations points in the radial grid in order to obtain convergent and physically reliable results for all the observables considered. However, specifically for $B_s11$ in Table \ref{tabICs} with $a_2(\tau_0)=-7.1$ and $\rho_0=1$, we noted spurious numerical oscillations at early times for the scalar condensate, with such an issue being completely fixed by increasing the number of collocation points to $N\sim 30$.

%%%%%%%%%%%%%%%%%%%%%%%%%%%%%%%%%
\section{Results for variations of the initial energy density}
\label{sec:5}

\begin{figure*}%[h]
\center
\subfigure[]{\includegraphics[width=0.49\textwidth]{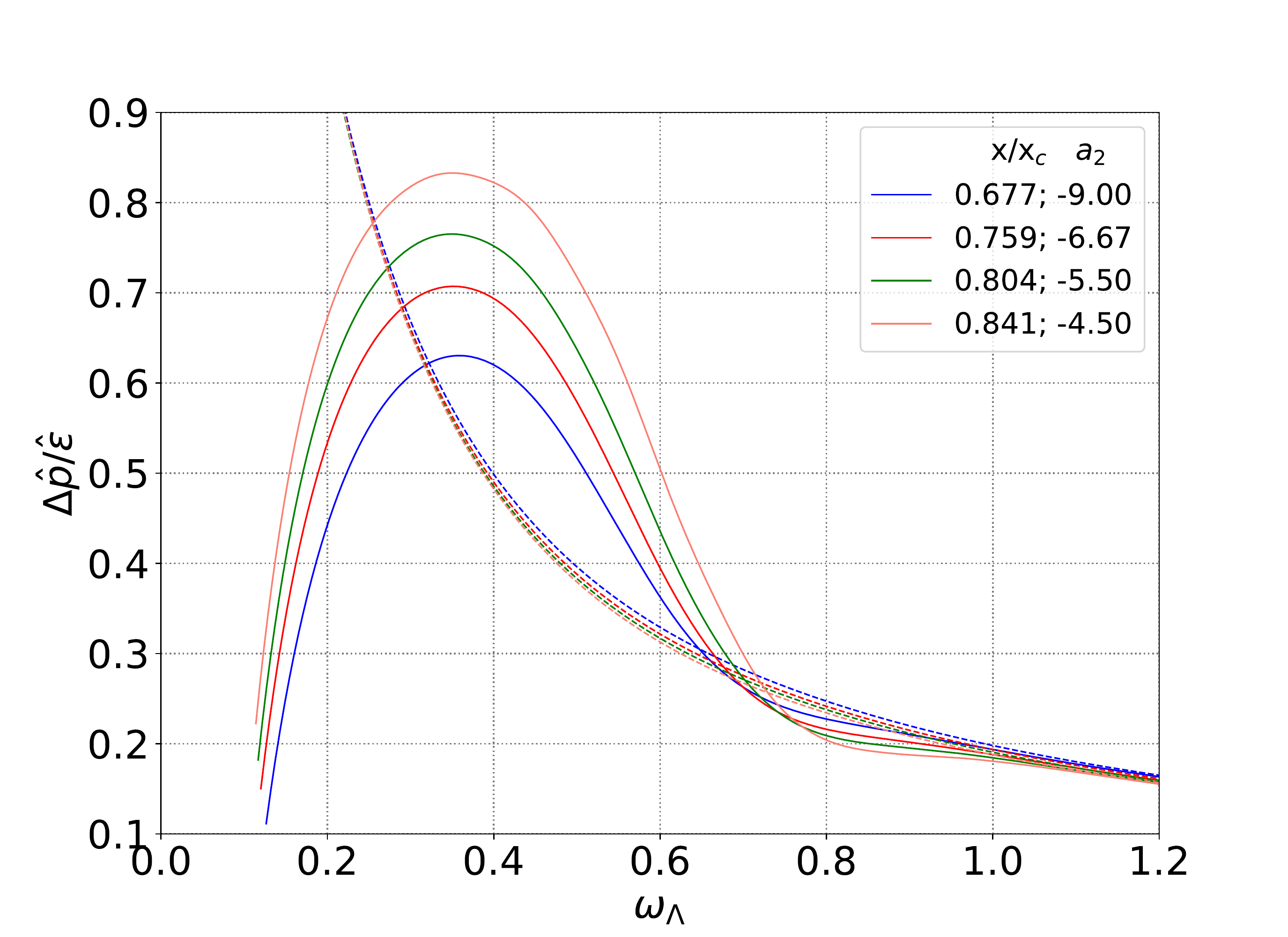}}
\subfigure[]{\includegraphics[width=0.49\textwidth]{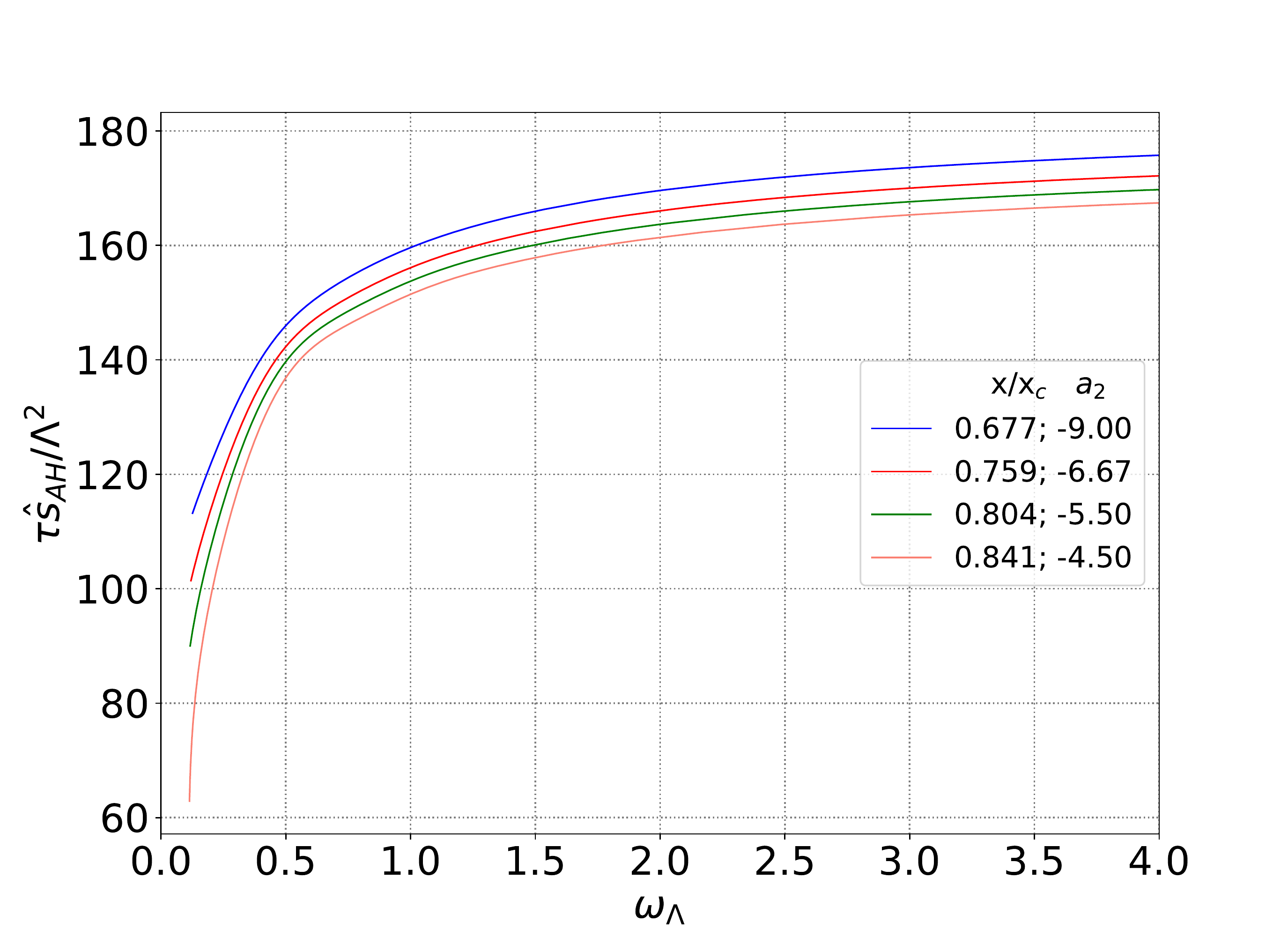}}
\subfigure[]{\includegraphics[width=0.49\textwidth]{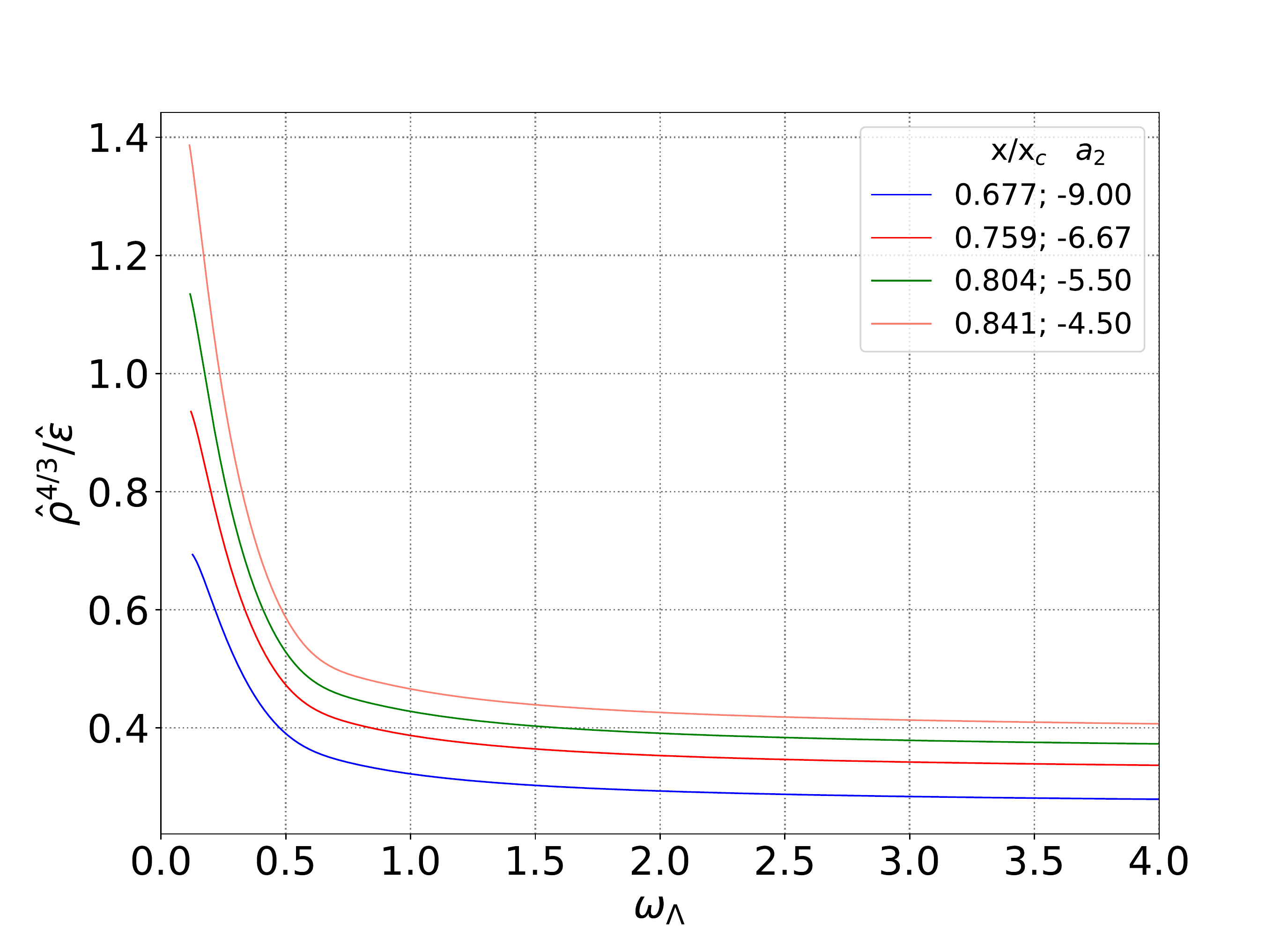}}
\subfigure[]{\includegraphics[width=0.49\textwidth]{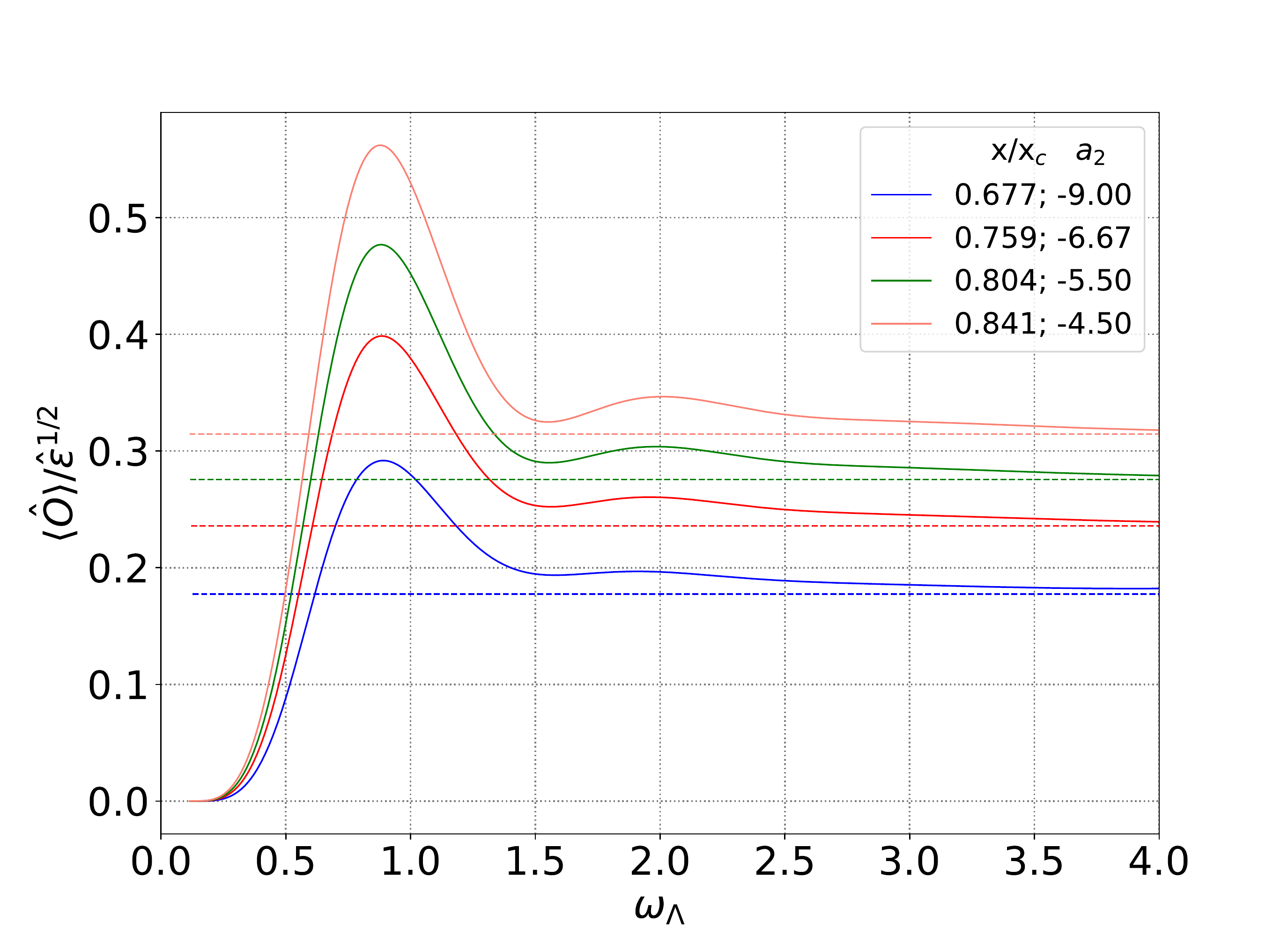}}
\caption{(a) Normalized pressure anisotropy (solid lines) and the corresponding hydrodynamic Navier-Stokes result (dashed lines), (b) normalized non-equilibrium entropy $\hat{S}_\textrm{AH}/\mathcal{A}\Lambda^2=\tau\hat{s}_\textrm{AH}/\Lambda^2$, (c) normalized charge density, and (d) normalized scalar condensate (solid lines) and the corresponding thermodynamic stable equilibrium result (dashed lines). Results obtained for variations of $a_2(\tau_0)$ keeping fixed $B_s1$ in Table \ref{tabICs} with $\rho_0=1.8$. Note that $x_c\equiv\left(\mu/T\right)_c=\pi/\sqrt{2}$ is the critical point.}
\label{fig:result12}
\end{figure*}

\begin{figure*}%[h]
\center
\subfigure[]{\includegraphics[width=0.49\textwidth]{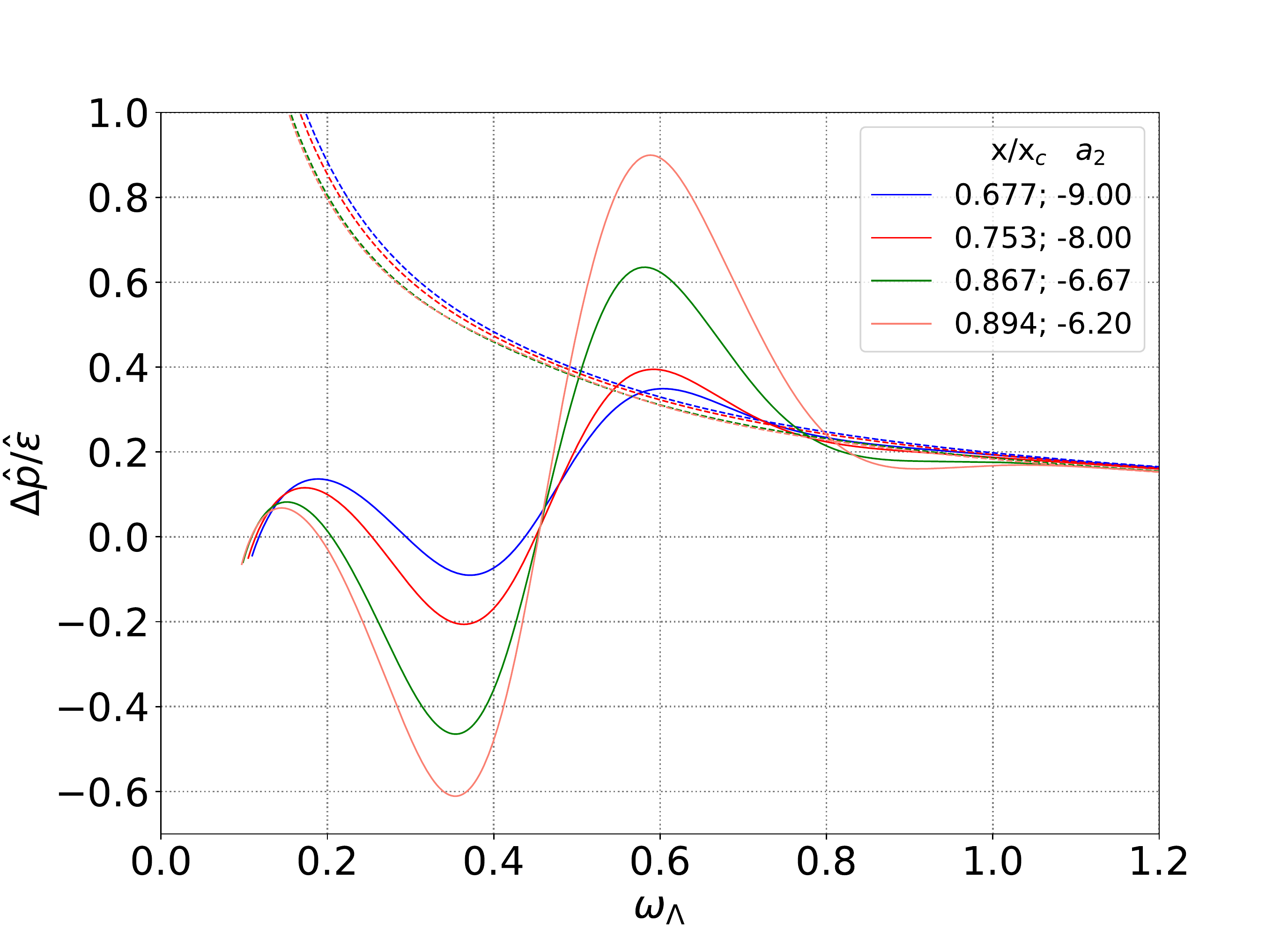}}
\subfigure[]{\includegraphics[width=0.49\textwidth]{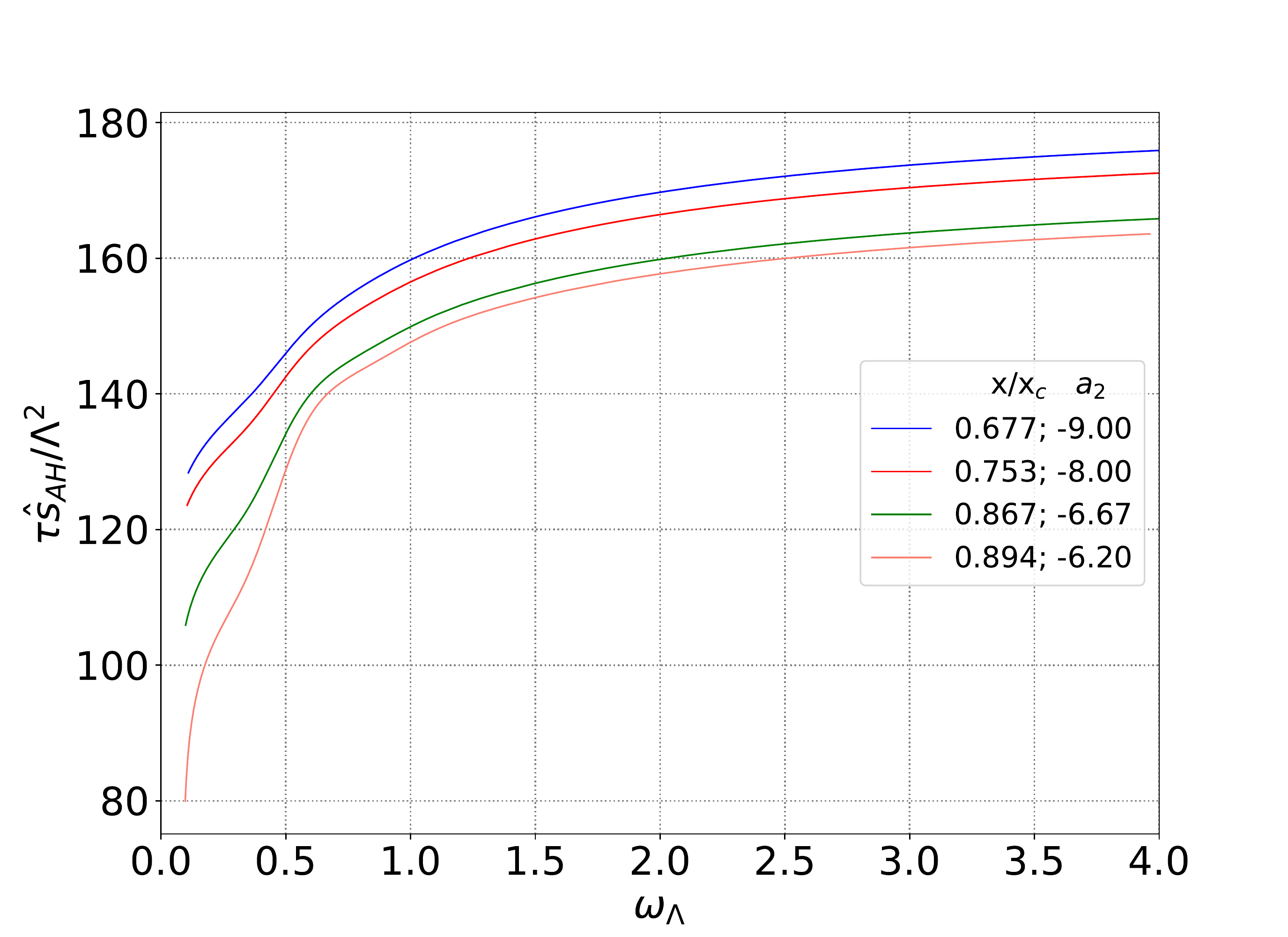}}
\subfigure[]{\includegraphics[width=0.49\textwidth]{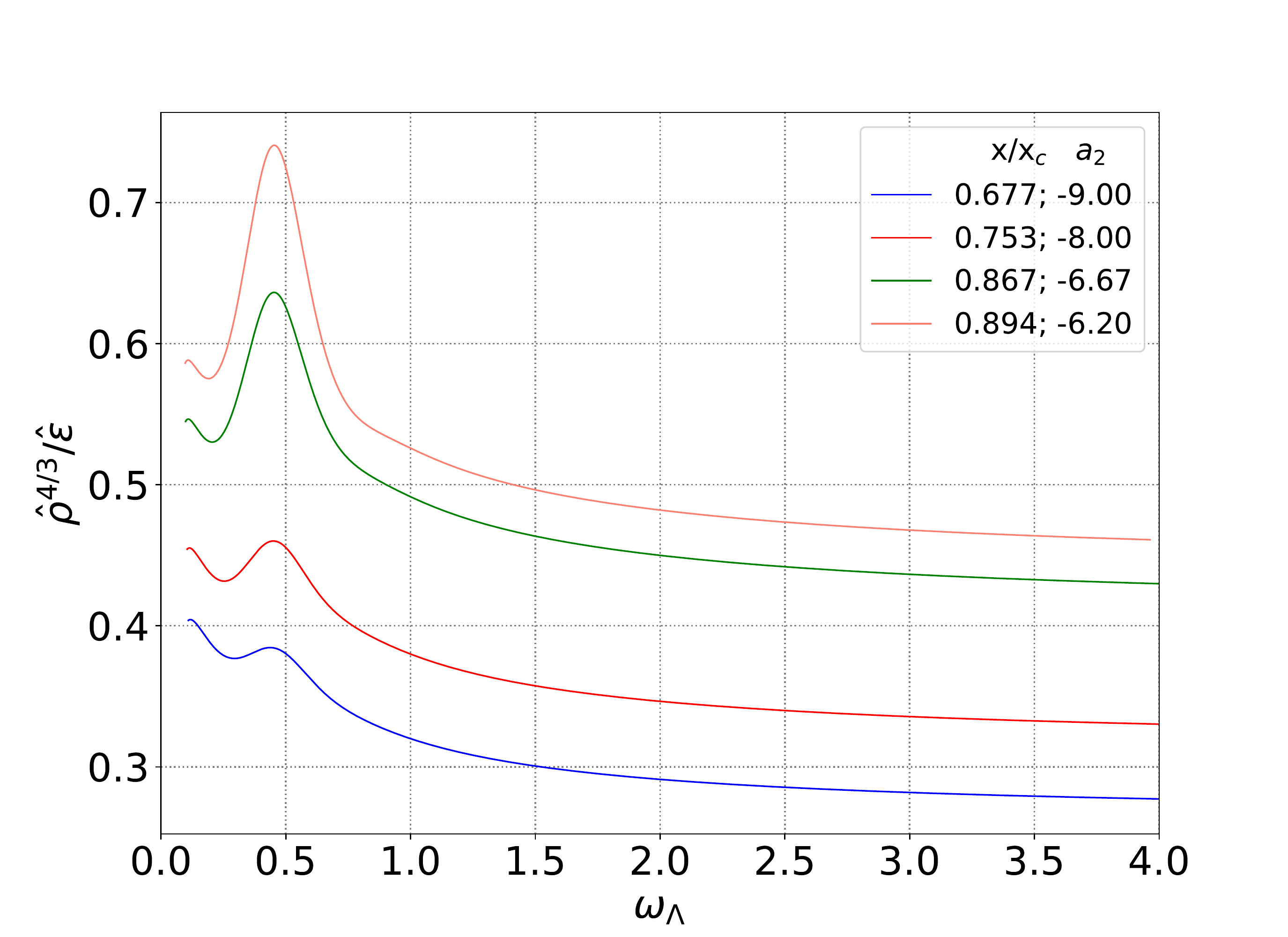}}
\subfigure[]{\includegraphics[width=0.49\textwidth]{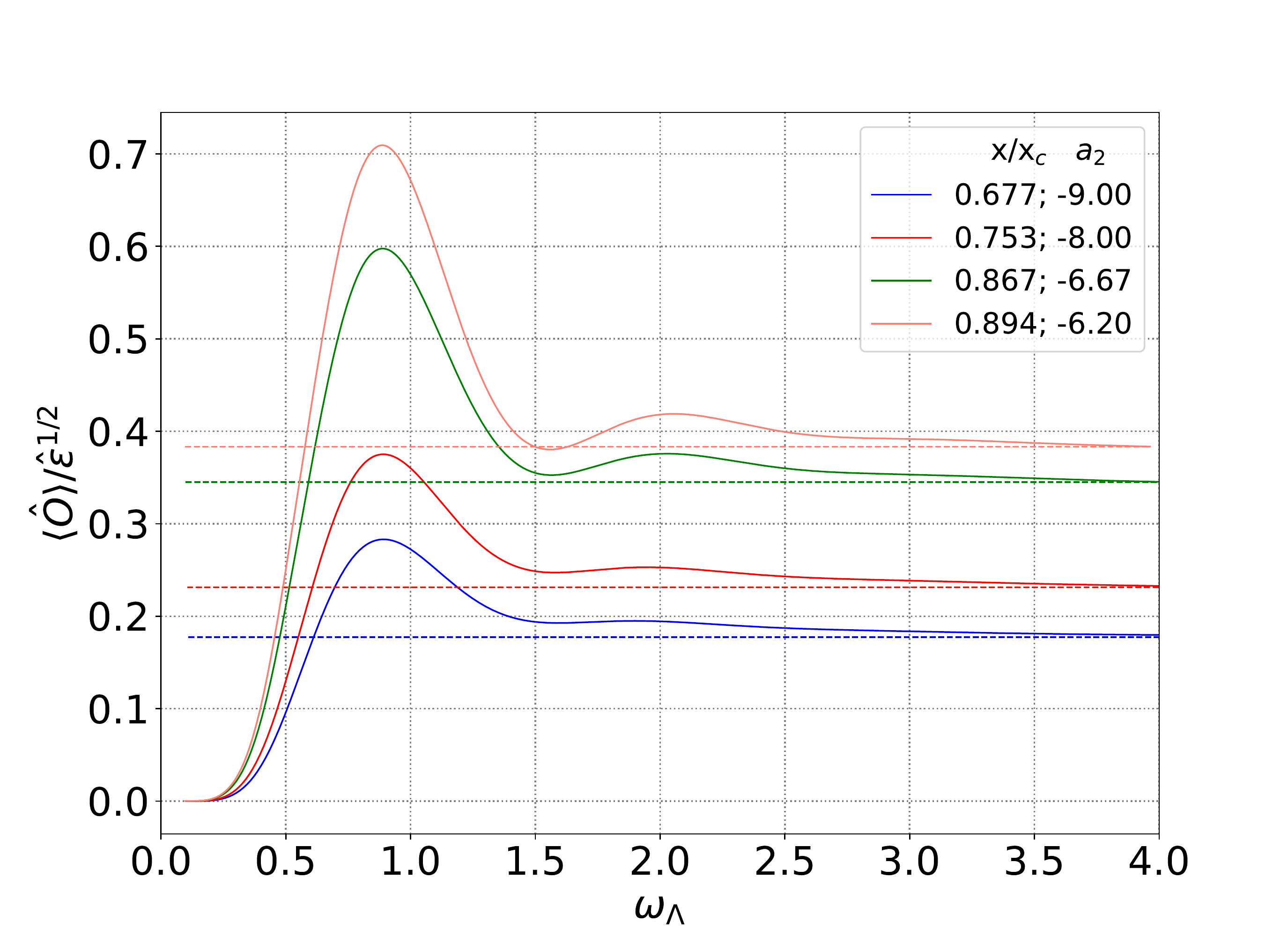}}
\caption{(a) Normalized pressure anisotropy (solid lines) and the corresponding hydrodynamic Navier-Stokes result (dashed lines), (b) normalized non-equilibrium entropy $\hat{S}_\textrm{AH}/\mathcal{A}\Lambda^2=\tau\hat{s}_\textrm{AH}/\Lambda^2$, (c) normalized charge density, and (d) normalized scalar condensate (solid lines) and the corresponding thermodynamic stable equilibrium result (dashed lines). Results obtained for variations of $a_2(\tau_0)$ keeping fixed $B_s2$ in Table \ref{tabICs} with $\rho_0=1.2$. Note that $x_c\equiv\left(\mu/T\right)_c=\pi/\sqrt{2}$ is the critical point.}
\label{fig:result13}
\end{figure*}

\begin{figure*}%[h]
\center
\subfigure[]{\includegraphics[width=0.49\textwidth]{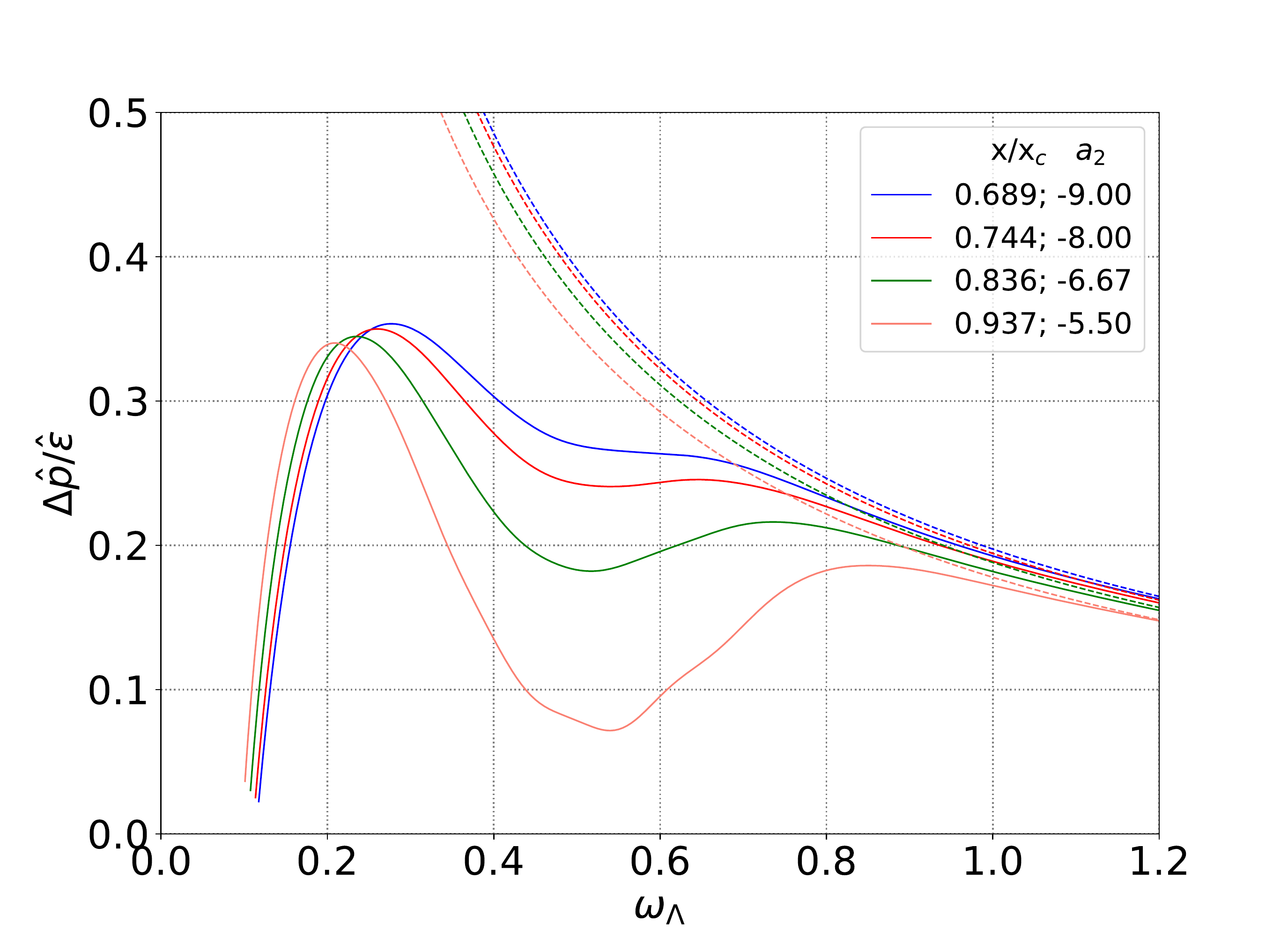}}
\subfigure[]{\includegraphics[width=0.49\textwidth]{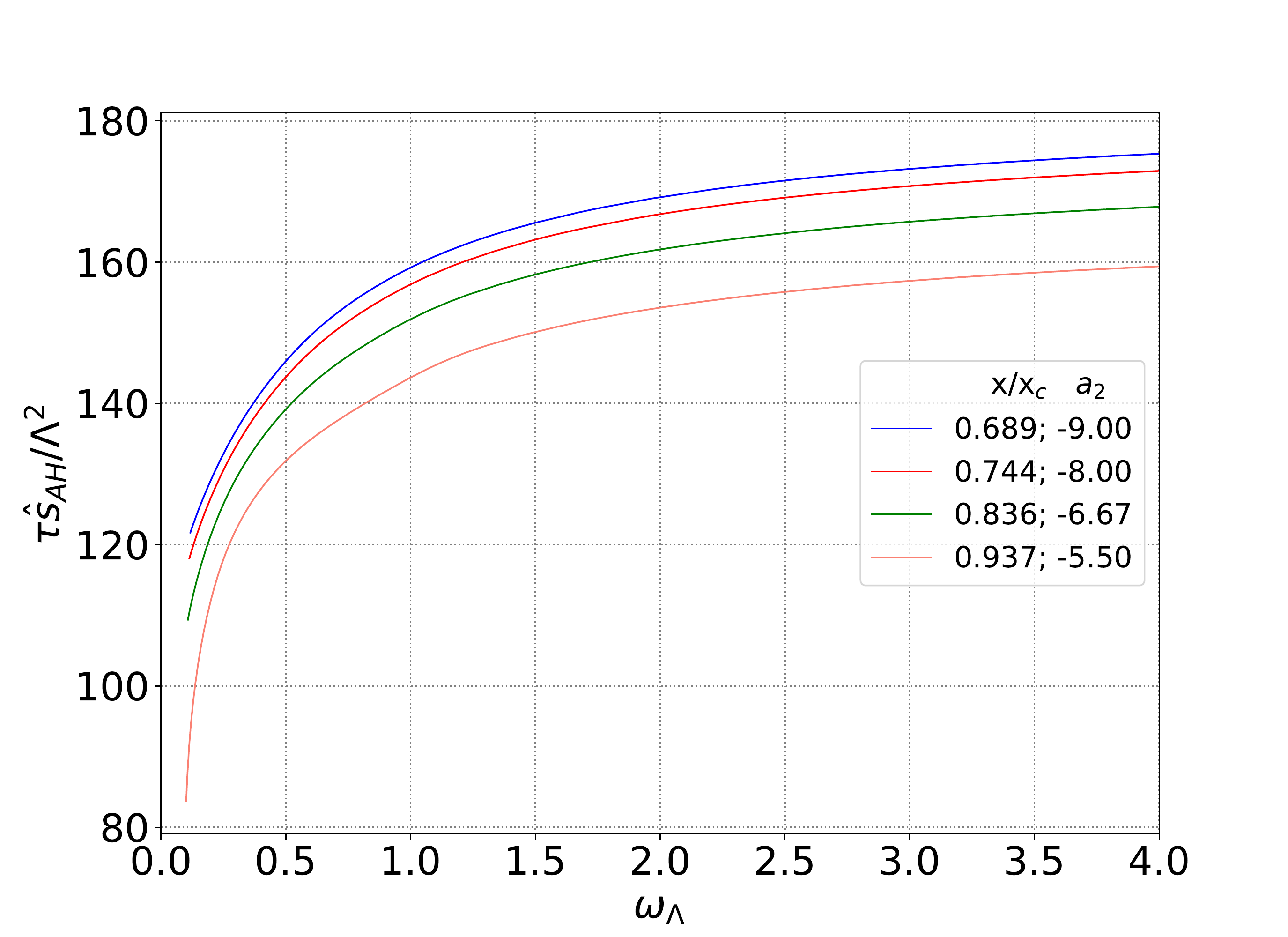}}
\subfigure[]{\includegraphics[width=0.49\textwidth]{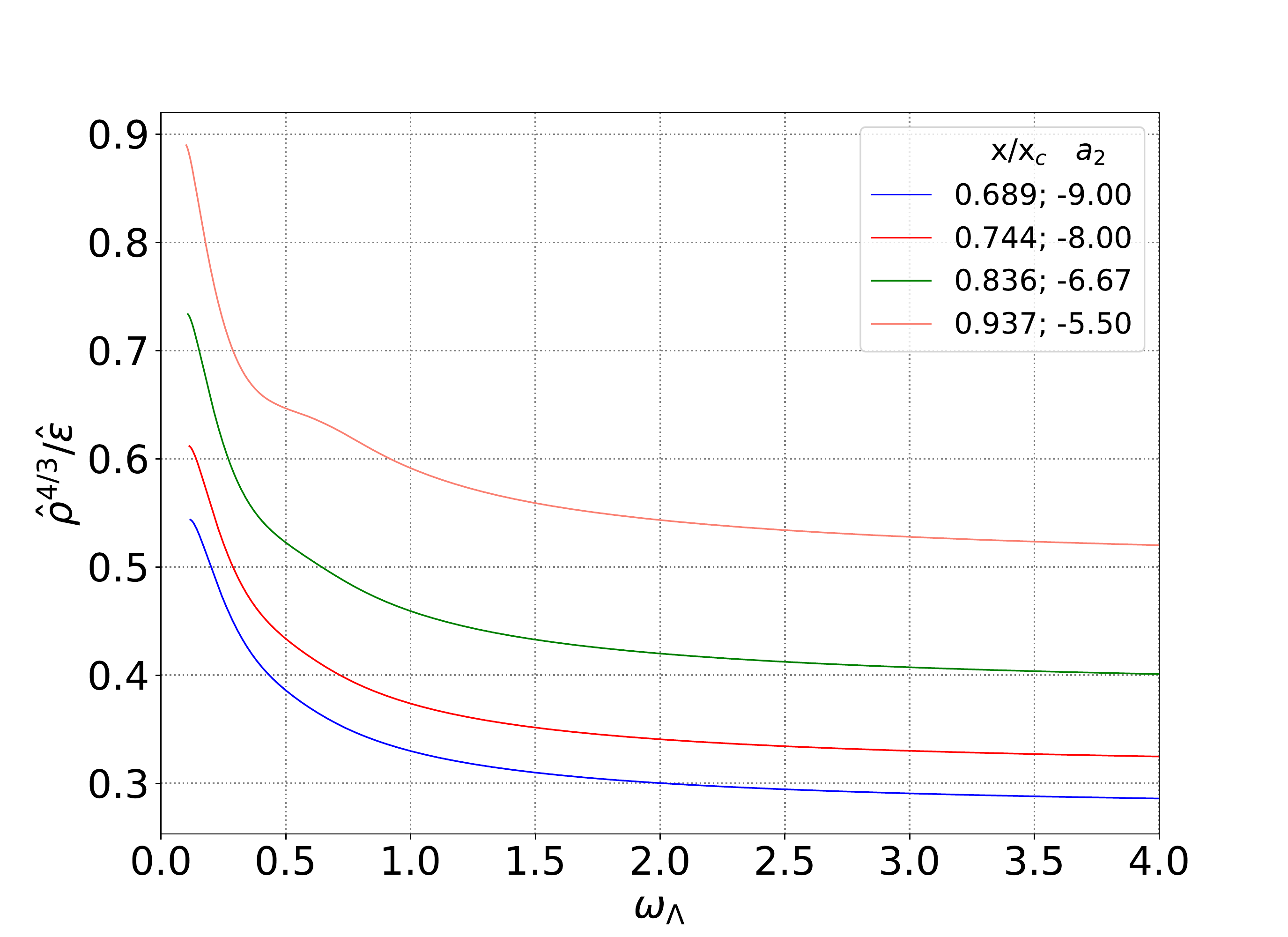}}
\subfigure[]{\includegraphics[width=0.49\textwidth]{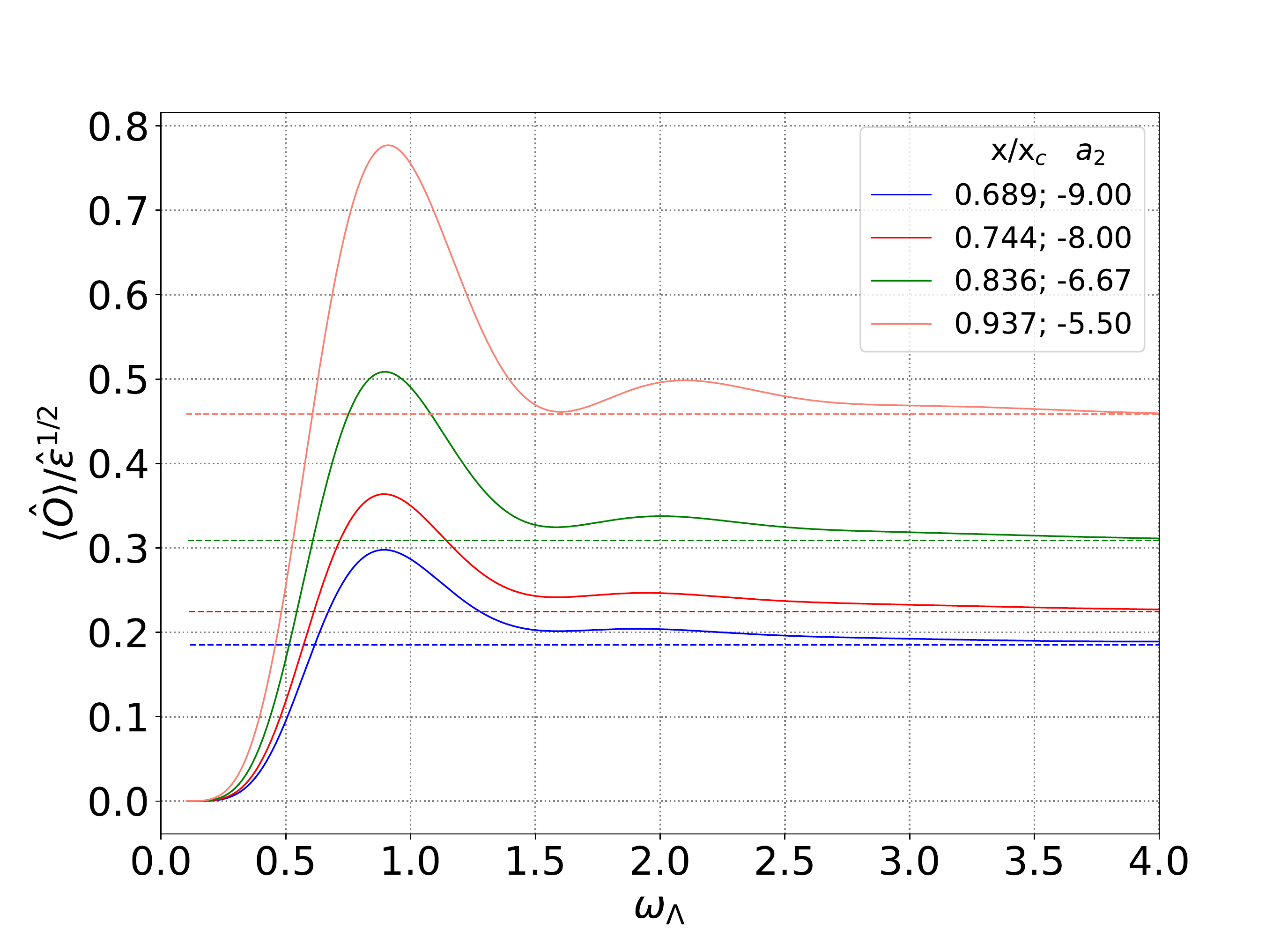}}
\caption{(a) Normalized pressure anisotropy (solid lines) and the corresponding hydrodynamic Navier-Stokes result (dashed lines), (b) normalized non-equilibrium entropy $\hat{S}_\textrm{AH}/\mathcal{A}\Lambda^2=\tau\hat{s}_\textrm{AH}/\Lambda^2$, (c) normalized charge density, and (d) normalized scalar condensate (solid lines) and the corresponding thermodynamic stable equilibrium result (dashed lines). Results obtained for variations of $a_2(\tau_0)$ keeping fixed $B_s3$ in Table \ref{tabICs} with $\rho_0=1.5$. Note that $x_c\equiv\left(\mu/T\right)_c=\pi/\sqrt{2}$ is the critical point.}
\label{fig:result14}
\end{figure*}

\begin{figure*}%[h]
\center
\subfigure[]{\includegraphics[width=0.49\textwidth]{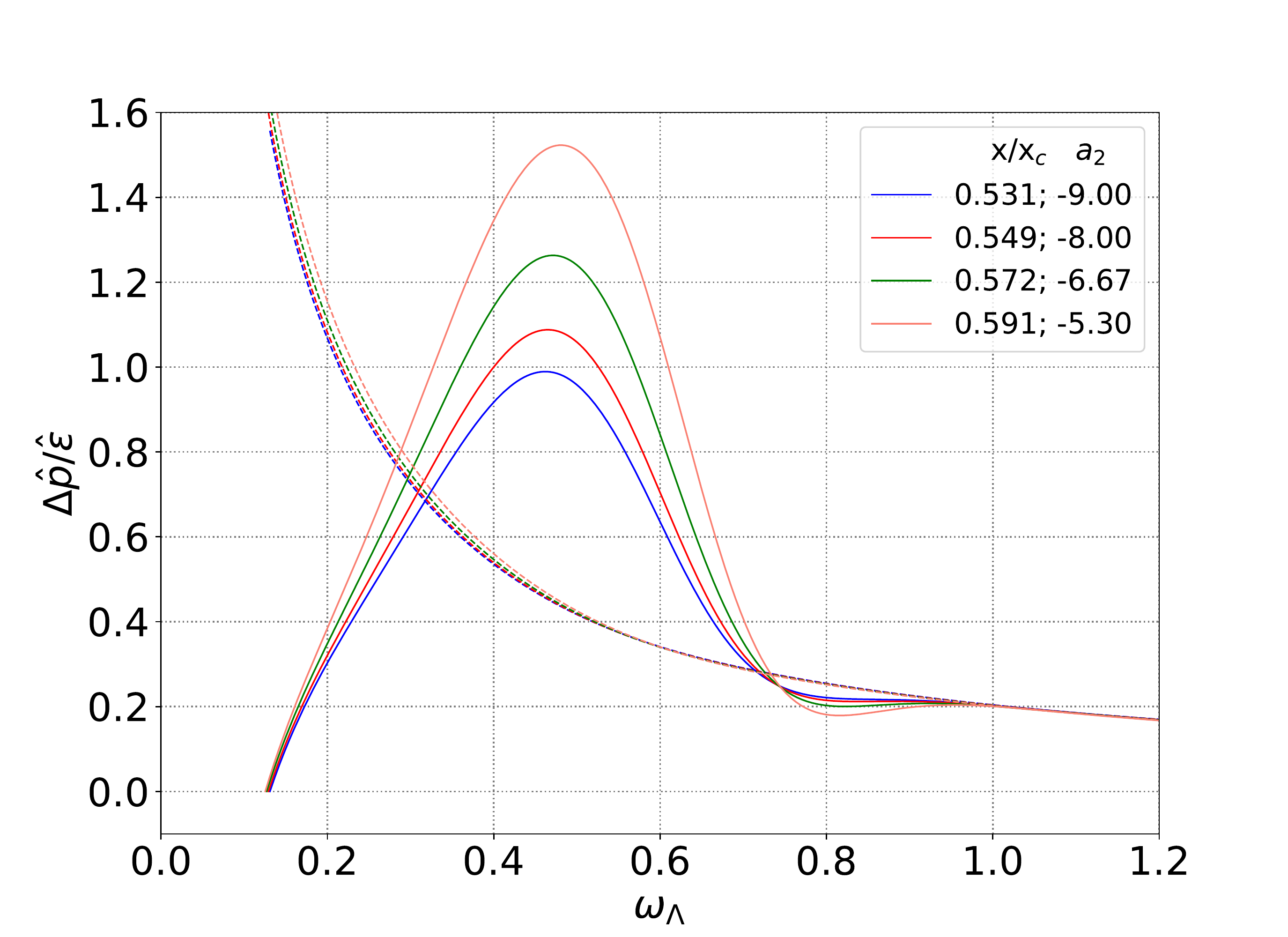}}
\subfigure[]{\includegraphics[width=0.49\textwidth]{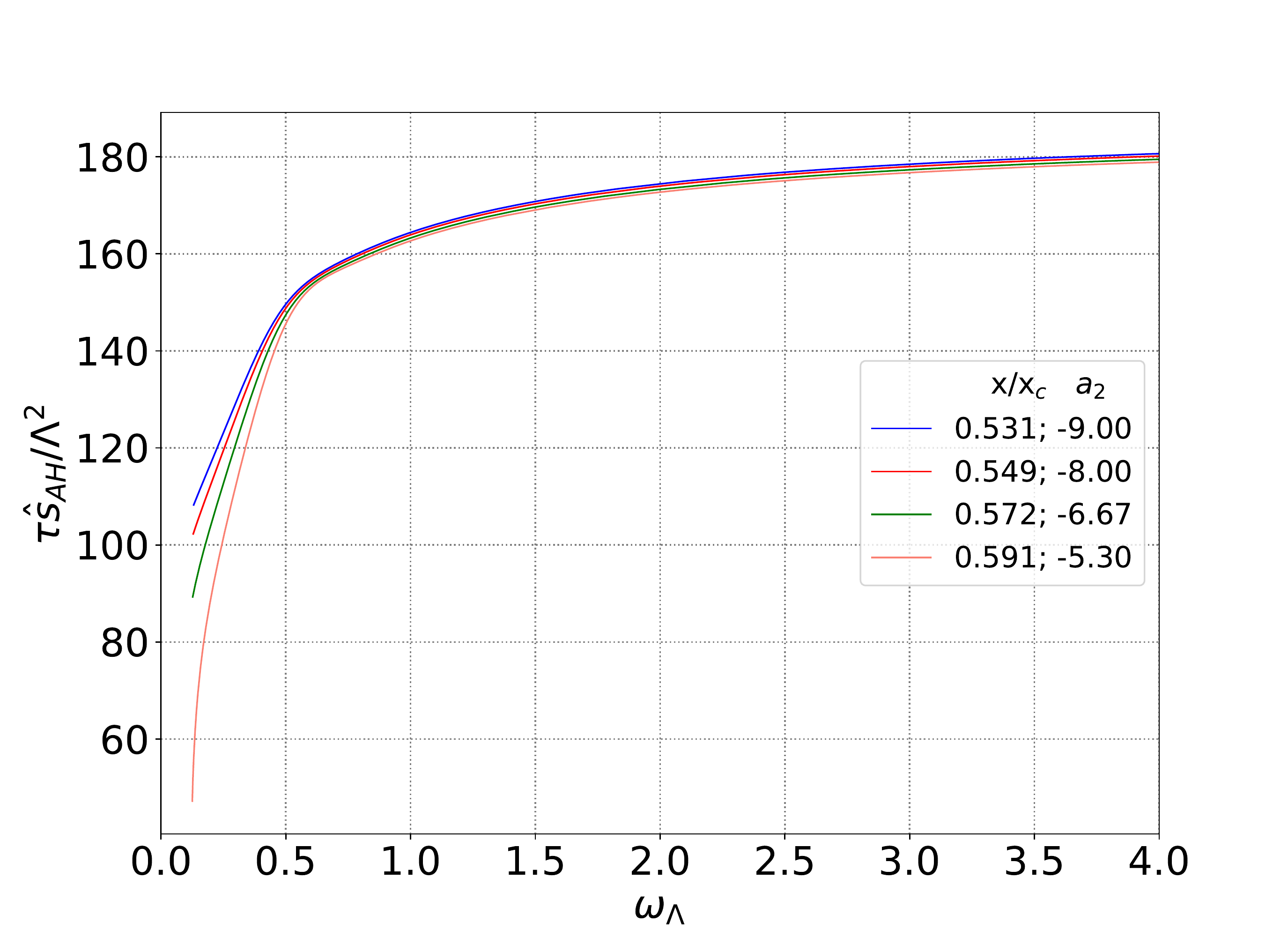}}
\subfigure[]{\includegraphics[width=0.49\textwidth]{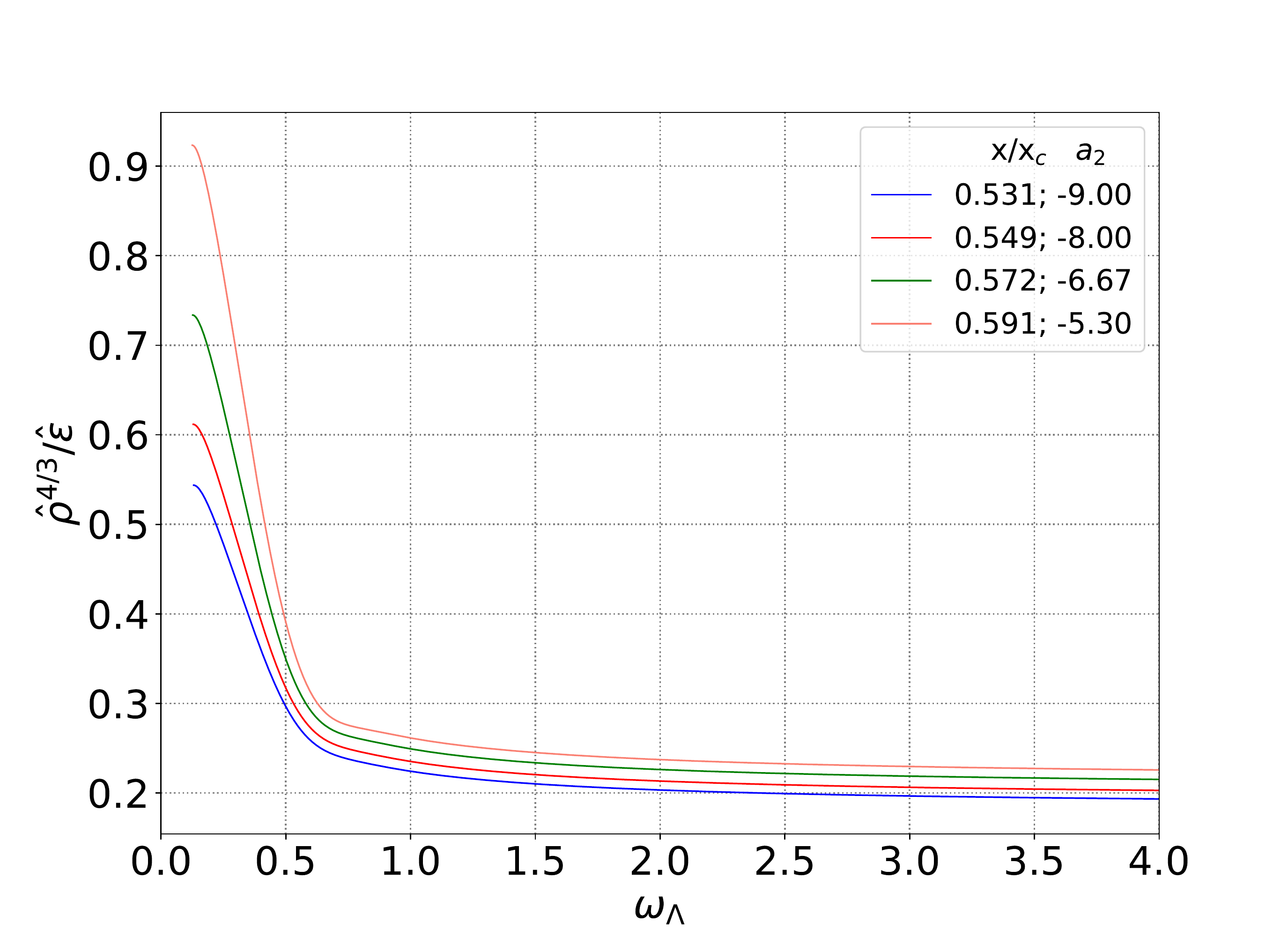}}
\subfigure[]{\includegraphics[width=0.49\textwidth]{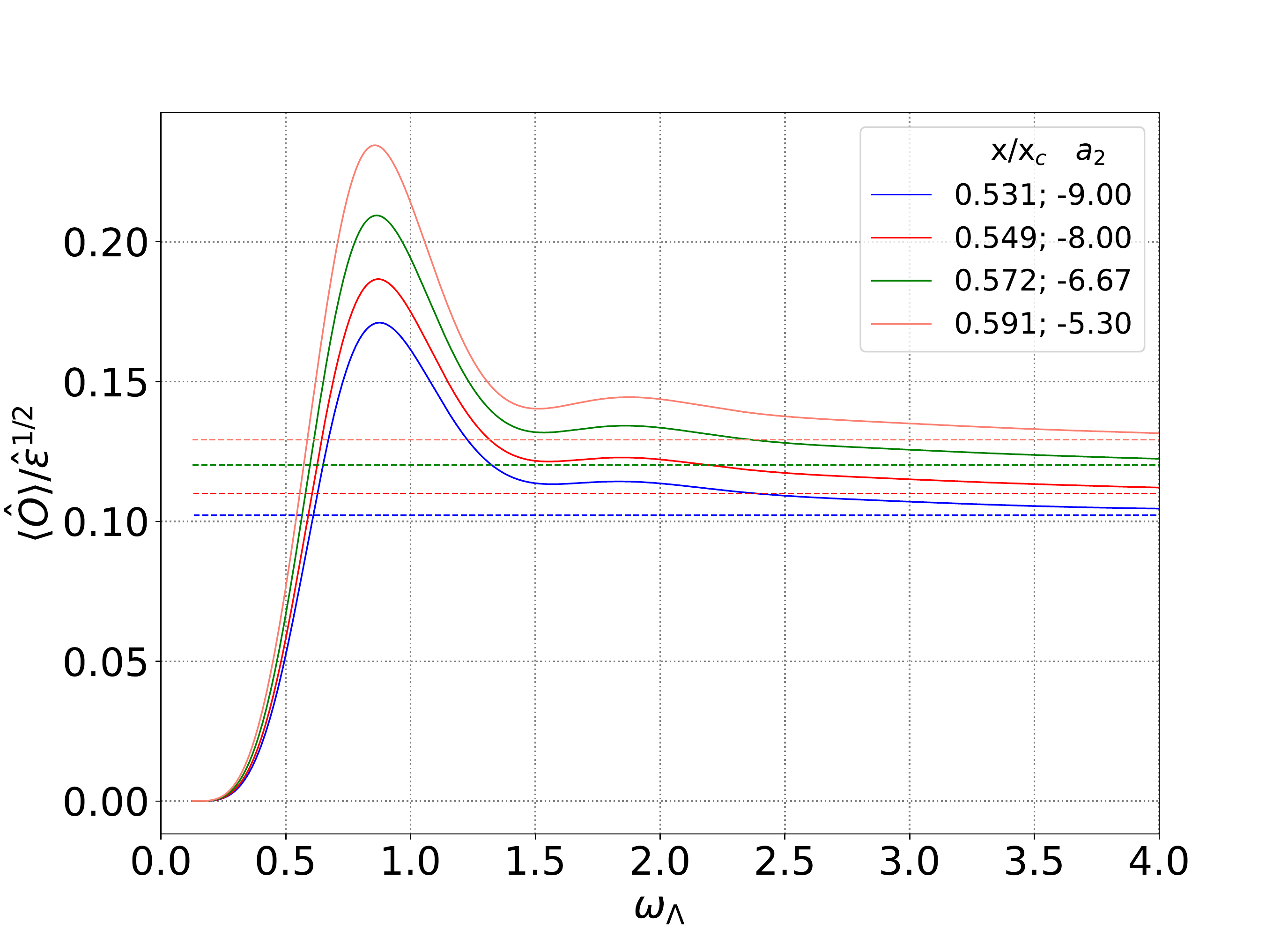}}
\caption{(a) Normalized pressure anisotropy (solid lines) and the corresponding hydrodynamic Navier-Stokes result (dashed lines), (b) normalized non-equilibrium entropy $\hat{S}_\textrm{AH}/\mathcal{A}\Lambda^2=\tau\hat{s}_\textrm{AH}/\Lambda^2$, (c) normalized charge density, and (d) normalized scalar condensate (solid lines) and the corresponding thermodynamic stable equilibrium result (dashed lines). Results obtained for variations of $a_2(\tau_0)$ keeping fixed $B_s4$ in Table \ref{tabICs} with $\rho_0=1.5$. Note that $x_c\equiv\left(\mu/T\right)_c=\pi/\sqrt{2}$ is the critical point.}
\label{fig:result15}
\end{figure*}

\begin{figure*}%[h]
\center
\subfigure[]{\includegraphics[width=0.49\textwidth]{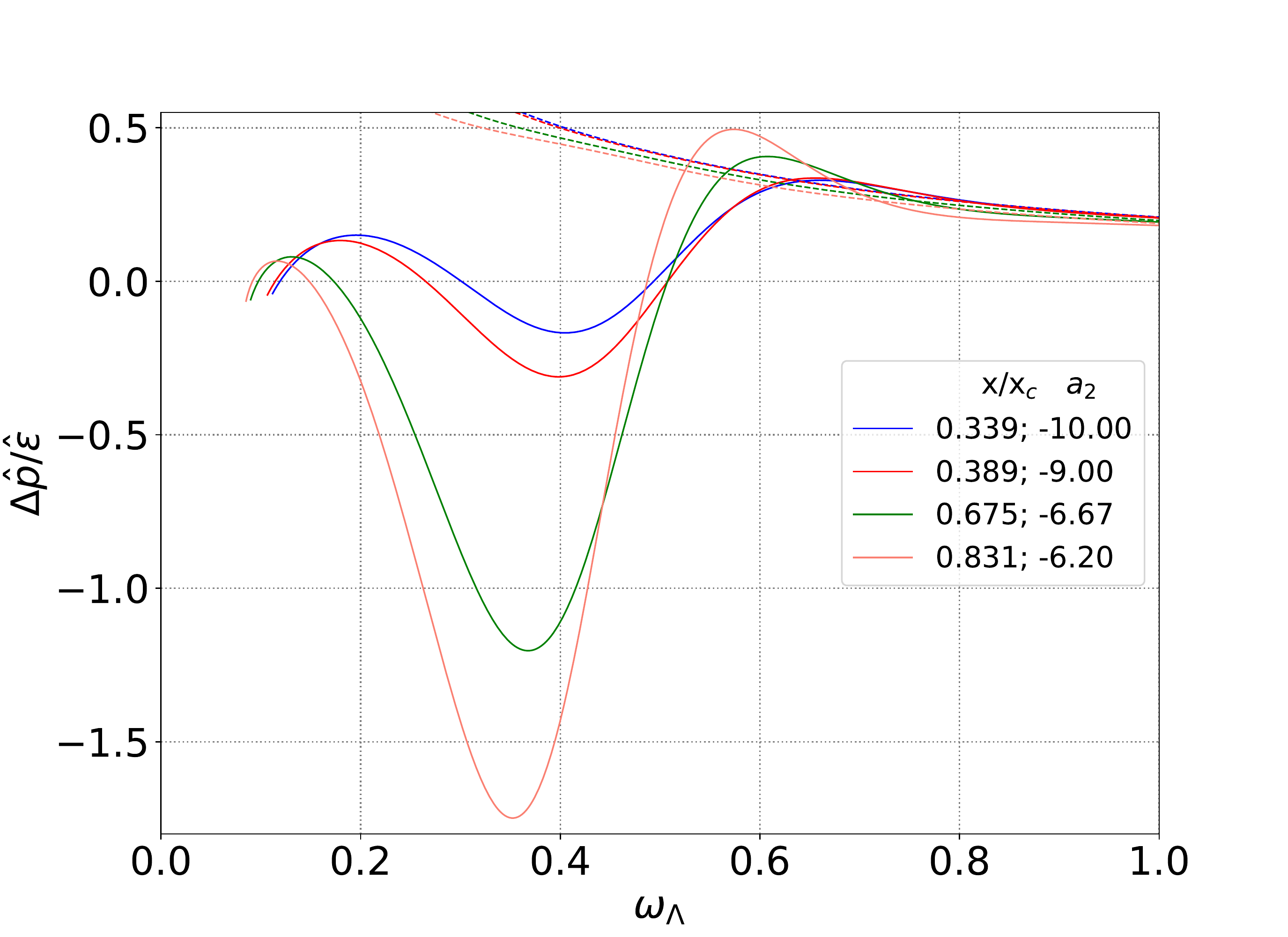}}
\subfigure[]{\includegraphics[width=0.49\textwidth]{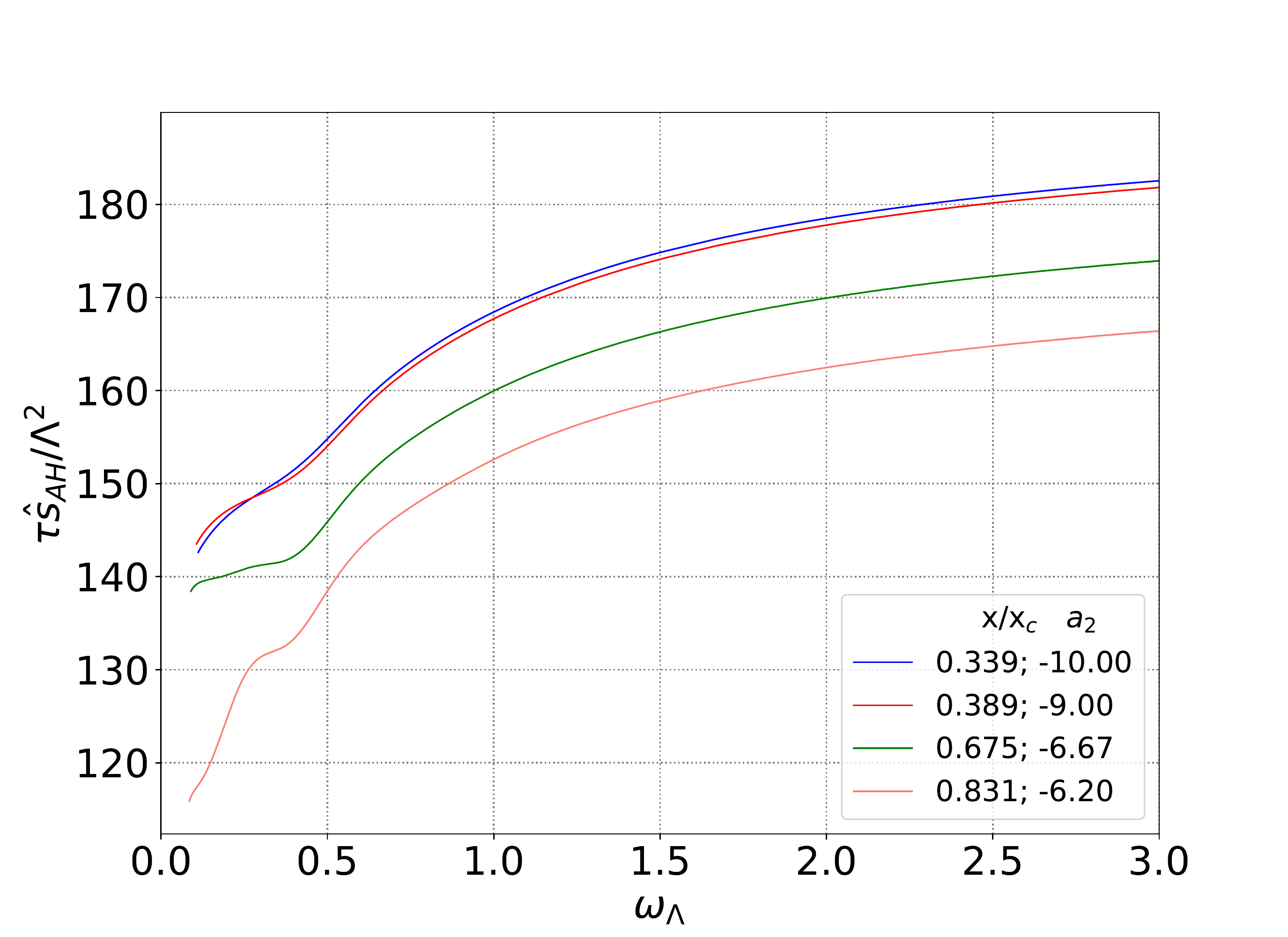}}
\subfigure[]{\includegraphics[width=0.49\textwidth]{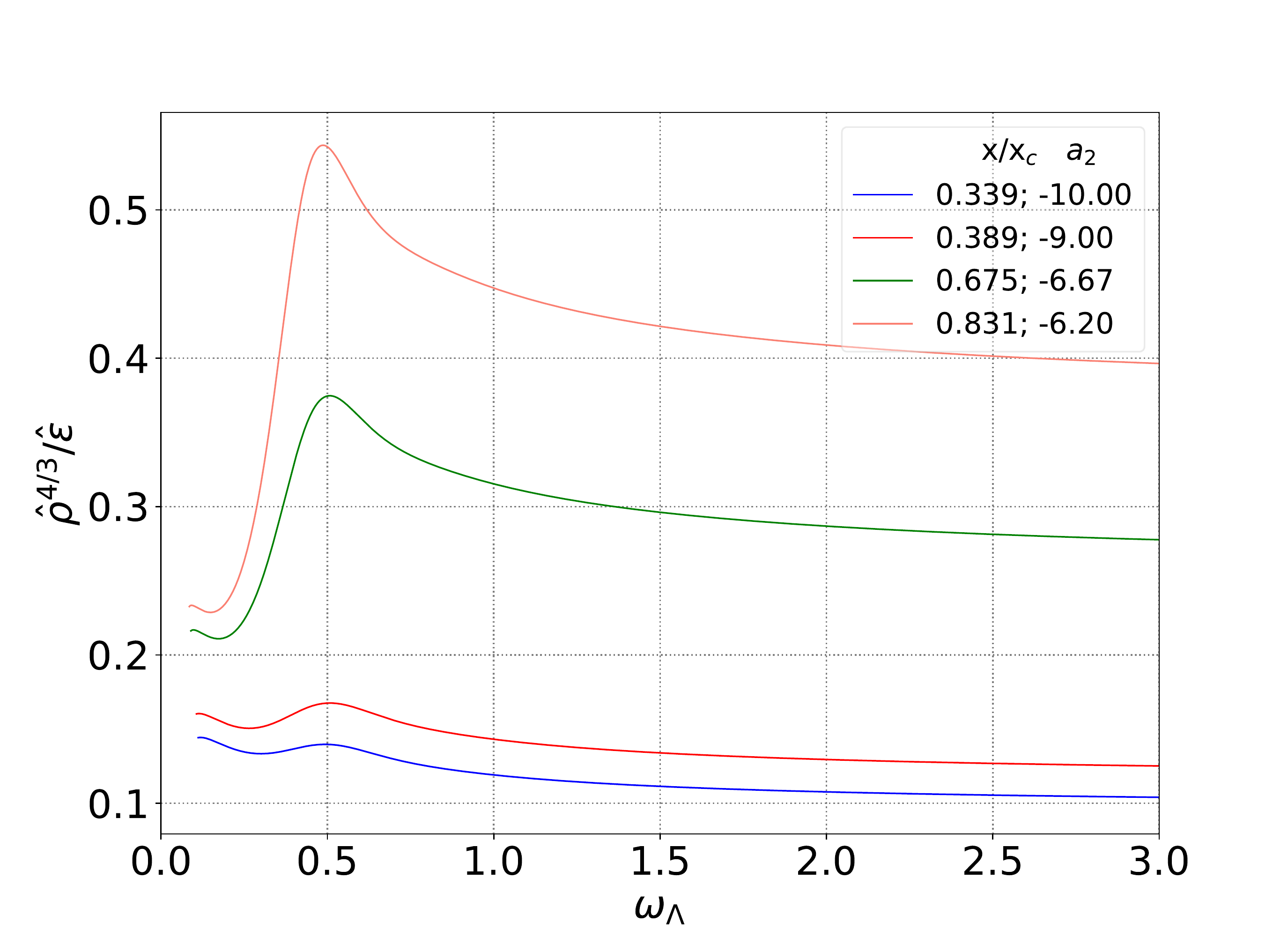}}
\subfigure[]{\includegraphics[width=0.49\textwidth]{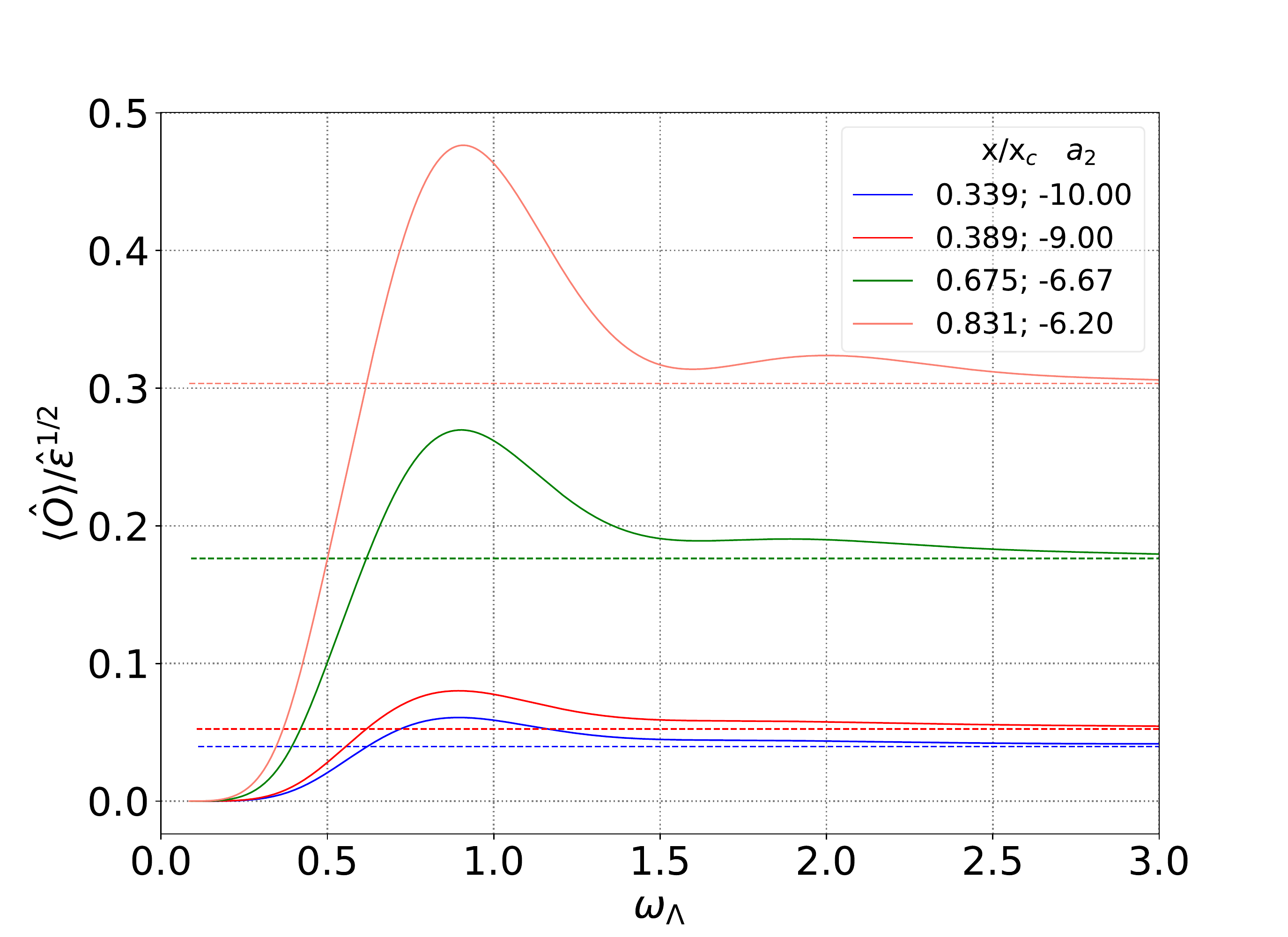}}
\caption{(a) Normalized pressure anisotropy (solid lines) and the corresponding hydrodynamic Navier-Stokes result (dashed lines), (b) normalized non-equilibrium entropy $\hat{S}_\textrm{AH}/\mathcal{A}\Lambda^2=\tau\hat{s}_\textrm{AH}/\Lambda^2$, (c) normalized charge density, and (d) normalized scalar condensate (solid lines) and the corresponding thermodynamic stable equilibrium result (dashed lines). Results obtained for variations of $a_2(\tau_0)$ keeping fixed $B_s5$ in Table \ref{tabICs} with $\rho_0=0.6$. Note that $x_c\equiv\left(\mu/T\right)_c=\pi/\sqrt{2}$ is the critical point.}
\label{fig:result16}
\end{figure*}

\begin{figure*}%[h]
\center
\subfigure[]{\includegraphics[width=0.49\textwidth]{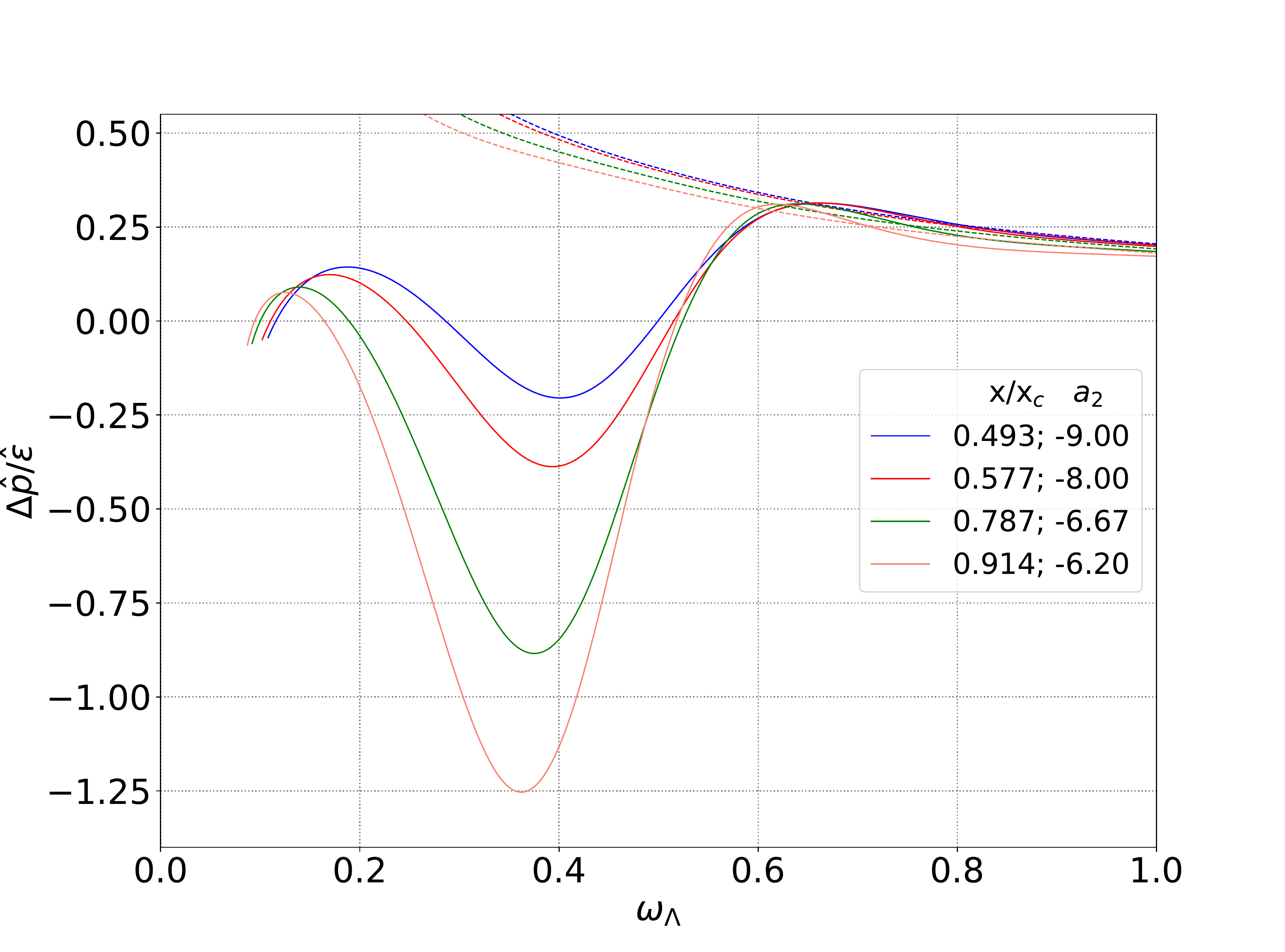}}
\subfigure[]{\includegraphics[width=0.49\textwidth]{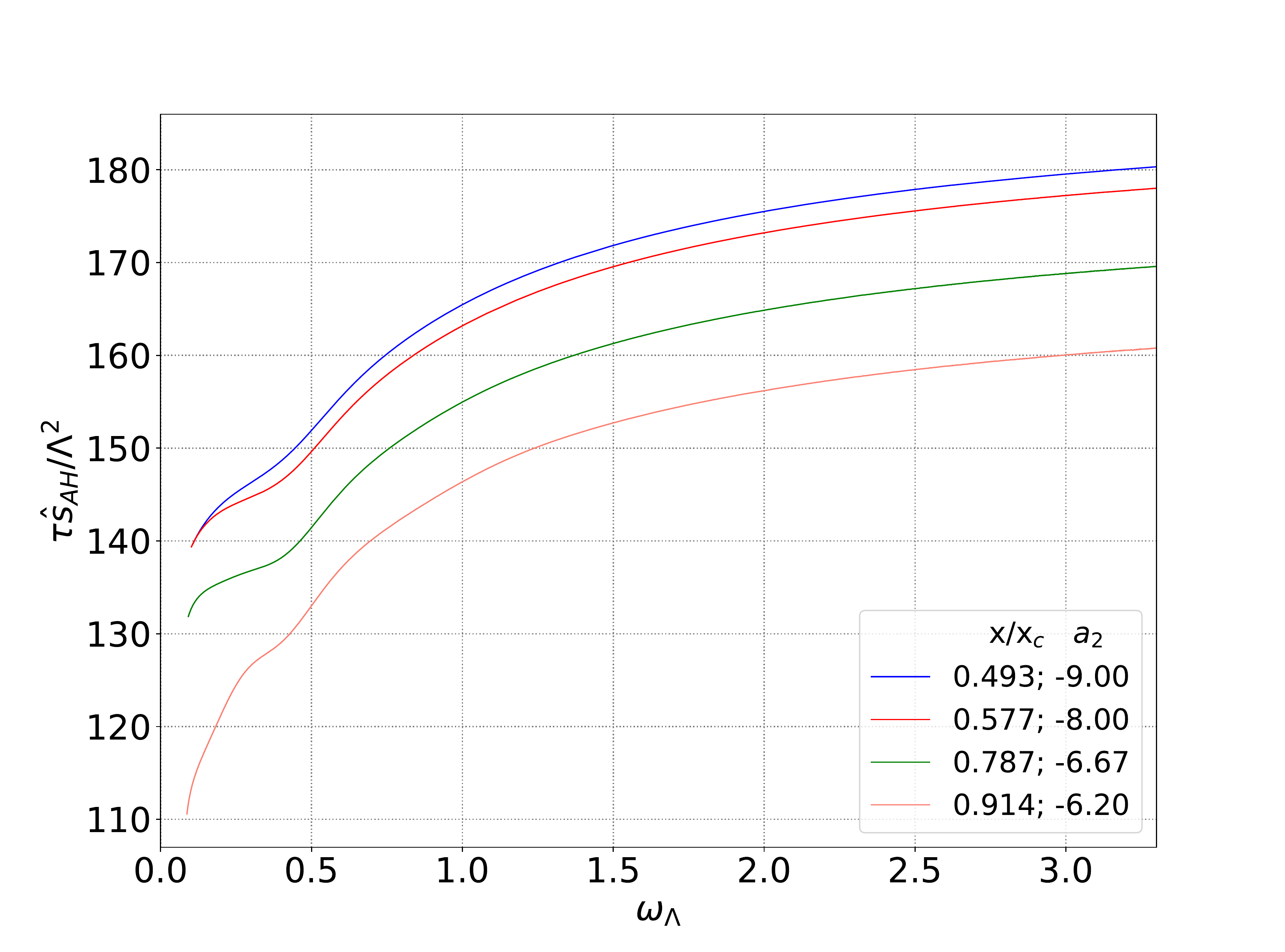}}
\subfigure[]{\includegraphics[width=0.49\textwidth]{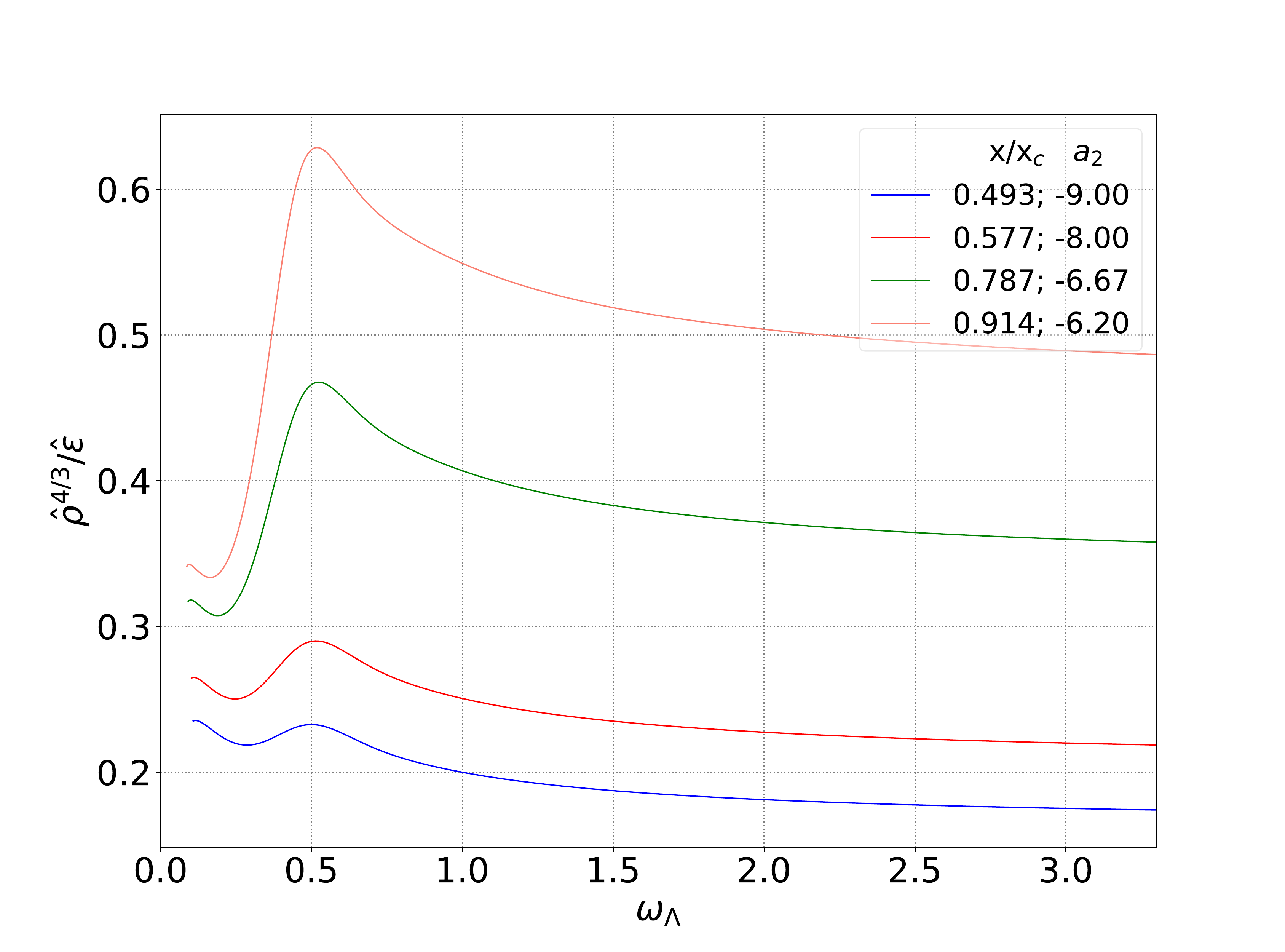}}
\subfigure[]{\includegraphics[width=0.49\textwidth]{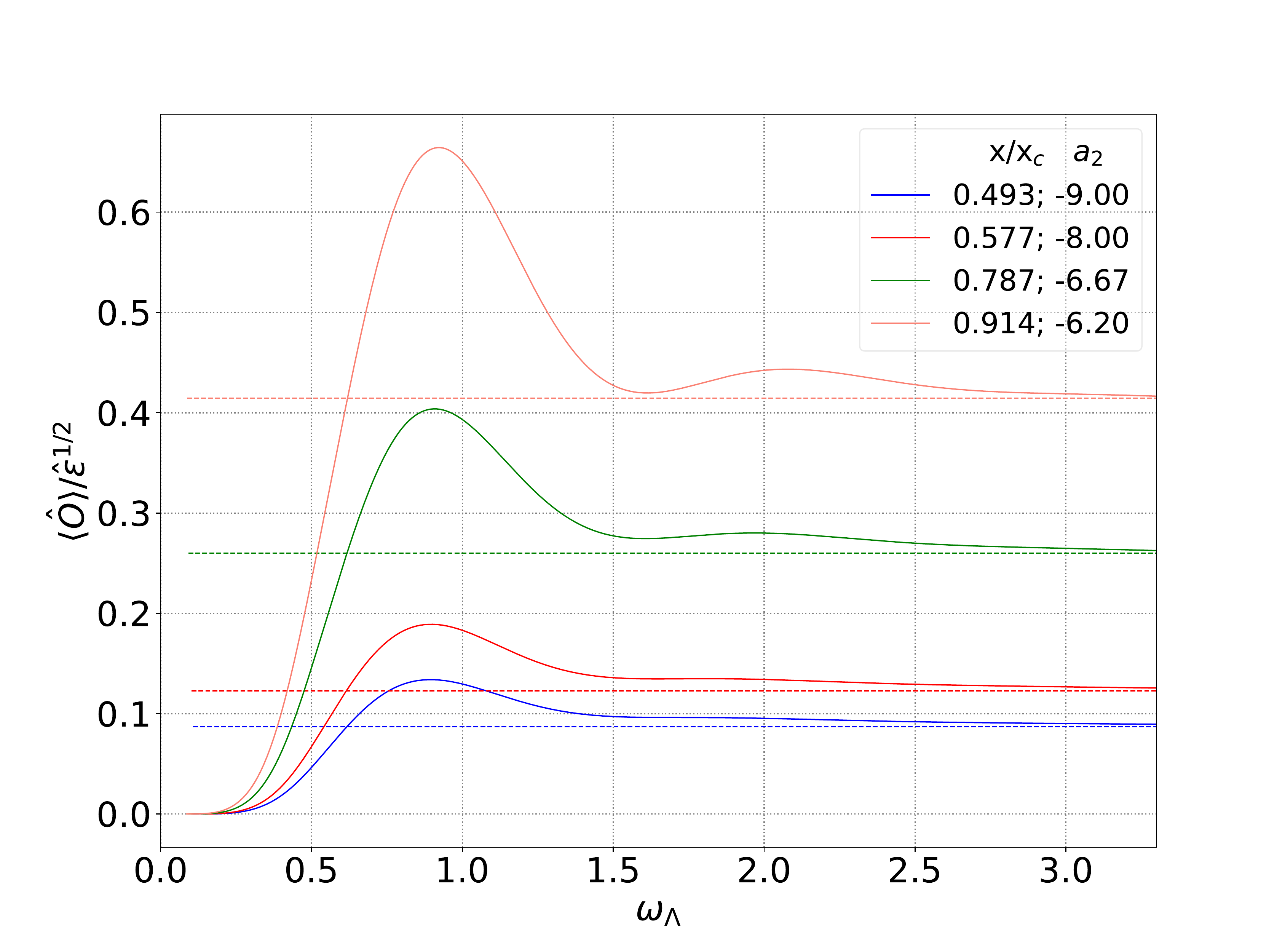}}
\caption{(a) Normalized pressure anisotropy (solid lines) and the corresponding hydrodynamic Navier-Stokes result (dashed lines), (b) normalized non-equilibrium entropy $\hat{S}_\textrm{AH}/\mathcal{A}\Lambda^2=\tau\hat{s}_\textrm{AH}/\Lambda^2$, (c) normalized charge density, and (d) normalized scalar condensate (solid lines) and the corresponding thermodynamic stable equilibrium result (dashed lines). Results obtained for variations of $a_2(\tau_0)$ keeping fixed $B_s6$ in Table \ref{tabICs} with $\rho_0=0.8$. Note that $x_c\equiv\left(\mu/T\right)_c=\pi/\sqrt{2}$ is the critical point.}
\label{fig:result17}
\end{figure*}

\begin{figure*}%[h]
\center
\subfigure[]{\includegraphics[width=0.49\textwidth]{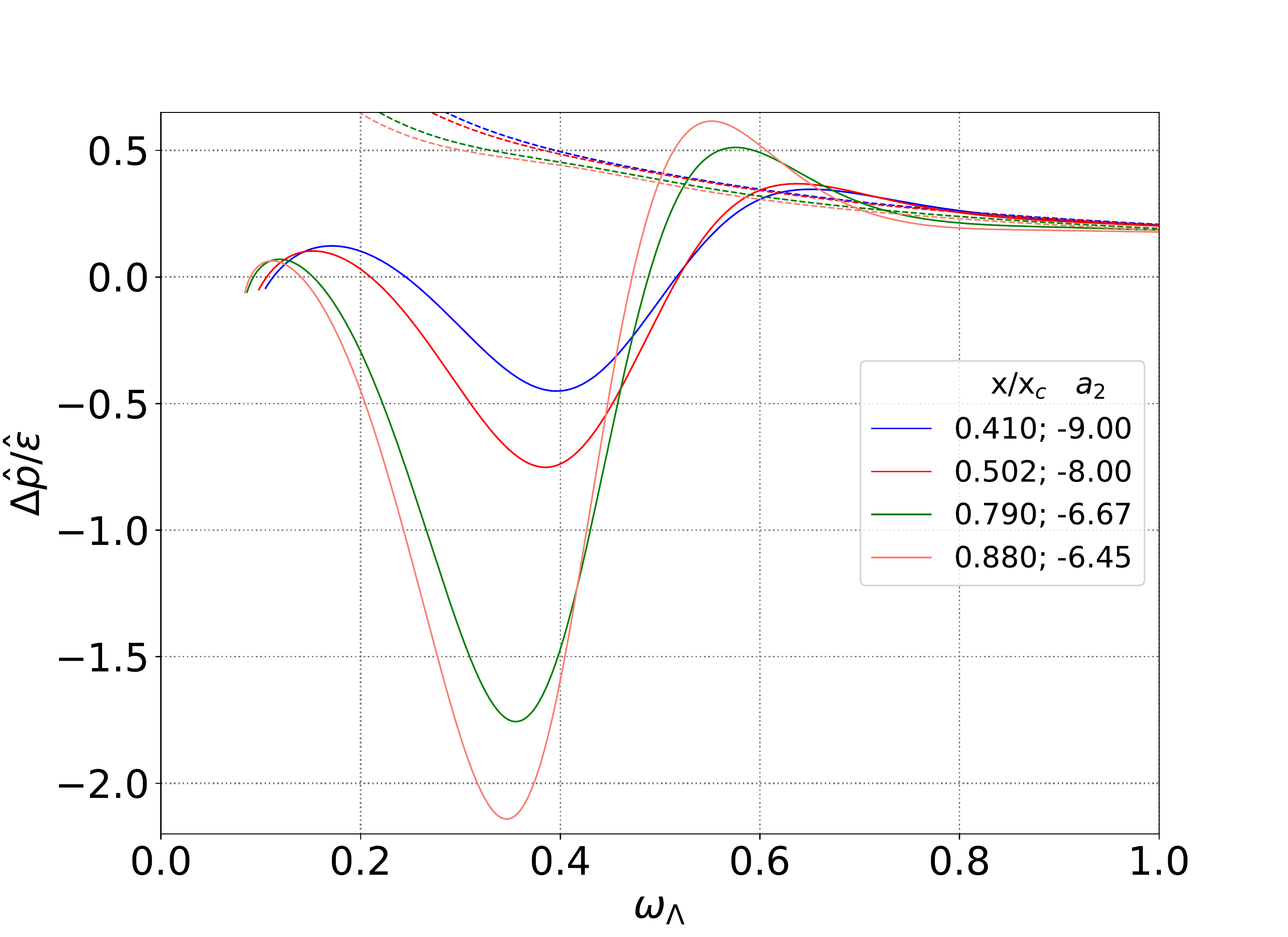}}
\subfigure[]{\includegraphics[width=0.49\textwidth]{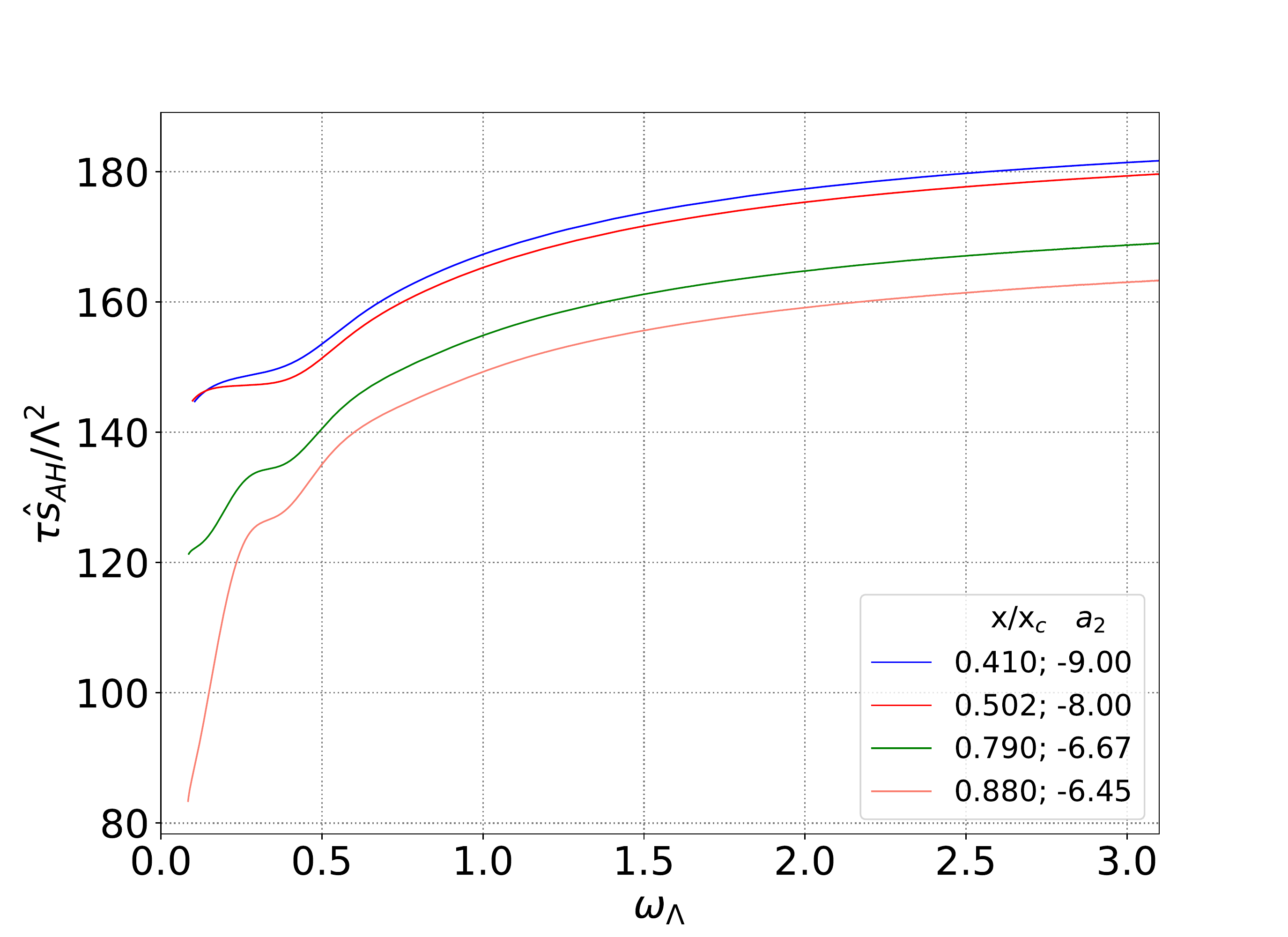}}
\subfigure[]{\includegraphics[width=0.49\textwidth]{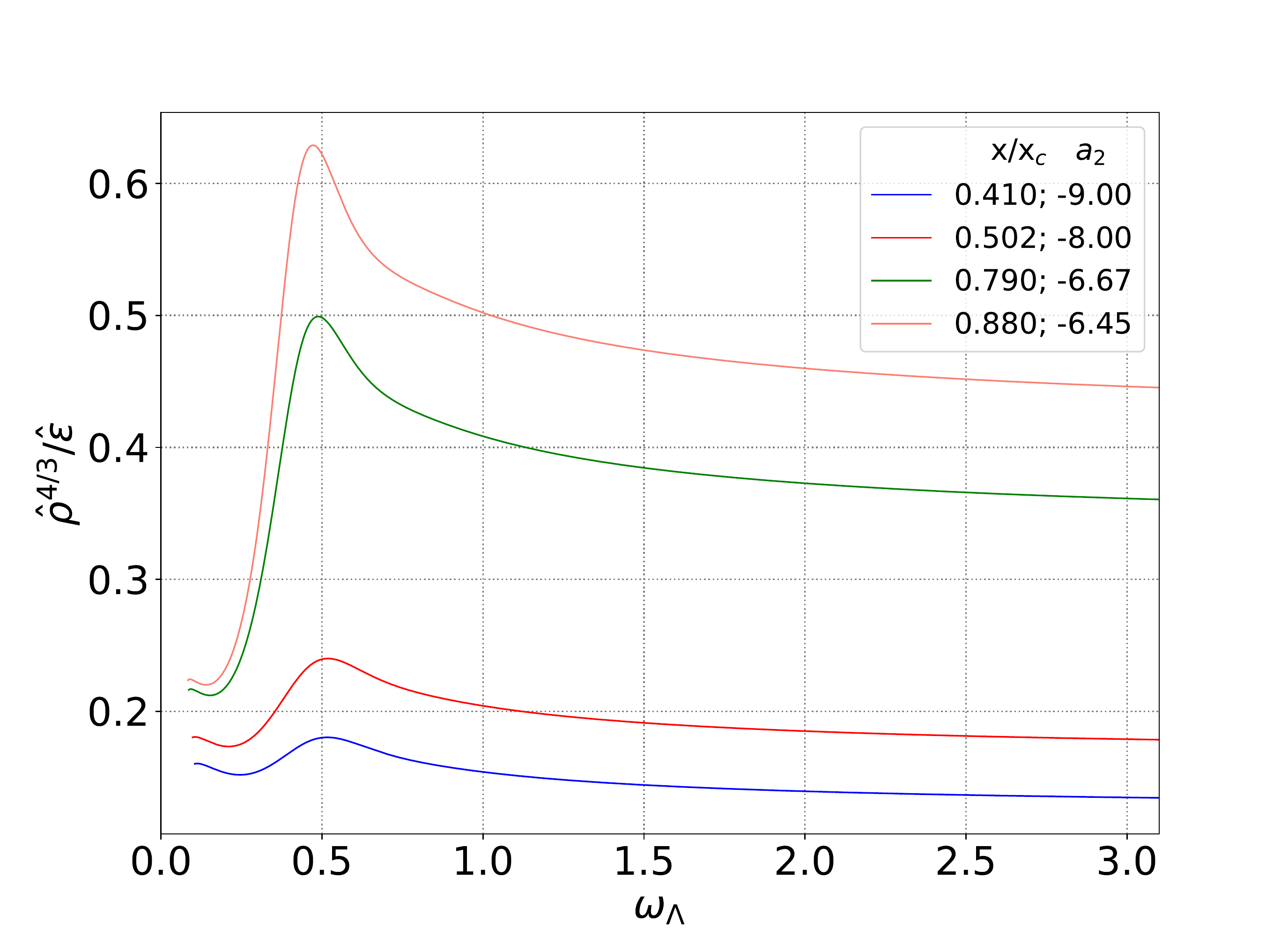}}
\subfigure[]{\includegraphics[width=0.49\textwidth]{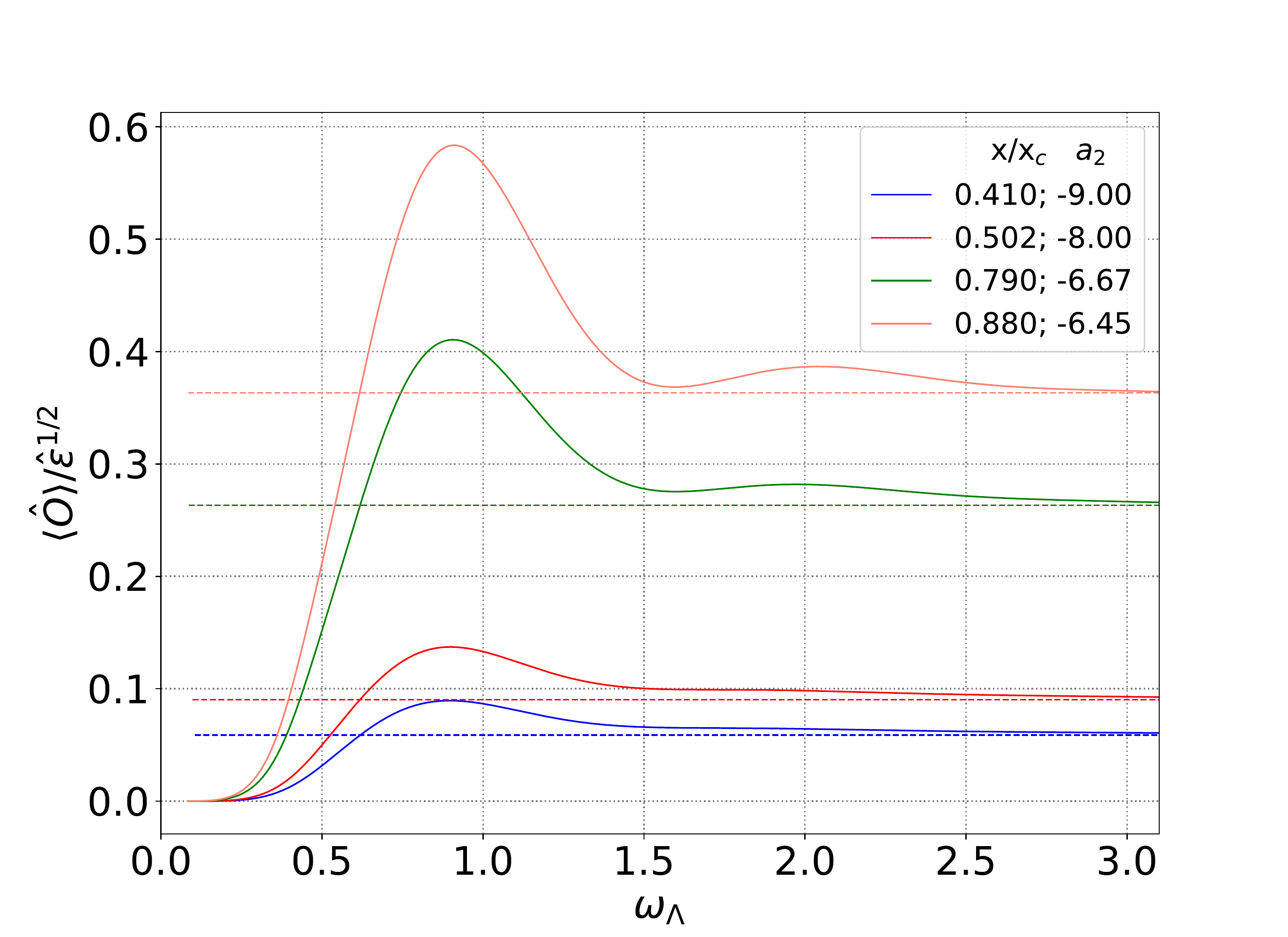}}
\caption{(a) Normalized pressure anisotropy (solid lines) and the corresponding hydrodynamic Navier-Stokes result (dashed lines), (b) normalized non-equilibrium entropy $\hat{S}_\textrm{AH}/\mathcal{A}\Lambda^2=\tau\hat{s}_\textrm{AH}/\Lambda^2$, (c) normalized charge density, and (d) normalized scalar condensate (solid lines) and the corresponding thermodynamic stable equilibrium result (dashed lines). Results obtained for variations of $a_2(\tau_0)$ keeping fixed $B_s7$ in Table \ref{tabICs} with $\rho_0=0.6$. Note that $x_c\equiv\left(\mu/T\right)_c=\pi/\sqrt{2}$ is the critical point.}
\label{fig:result18}
\end{figure*}

\begin{figure*}%[h]
\center
\subfigure[]{\includegraphics[width=0.49\textwidth]{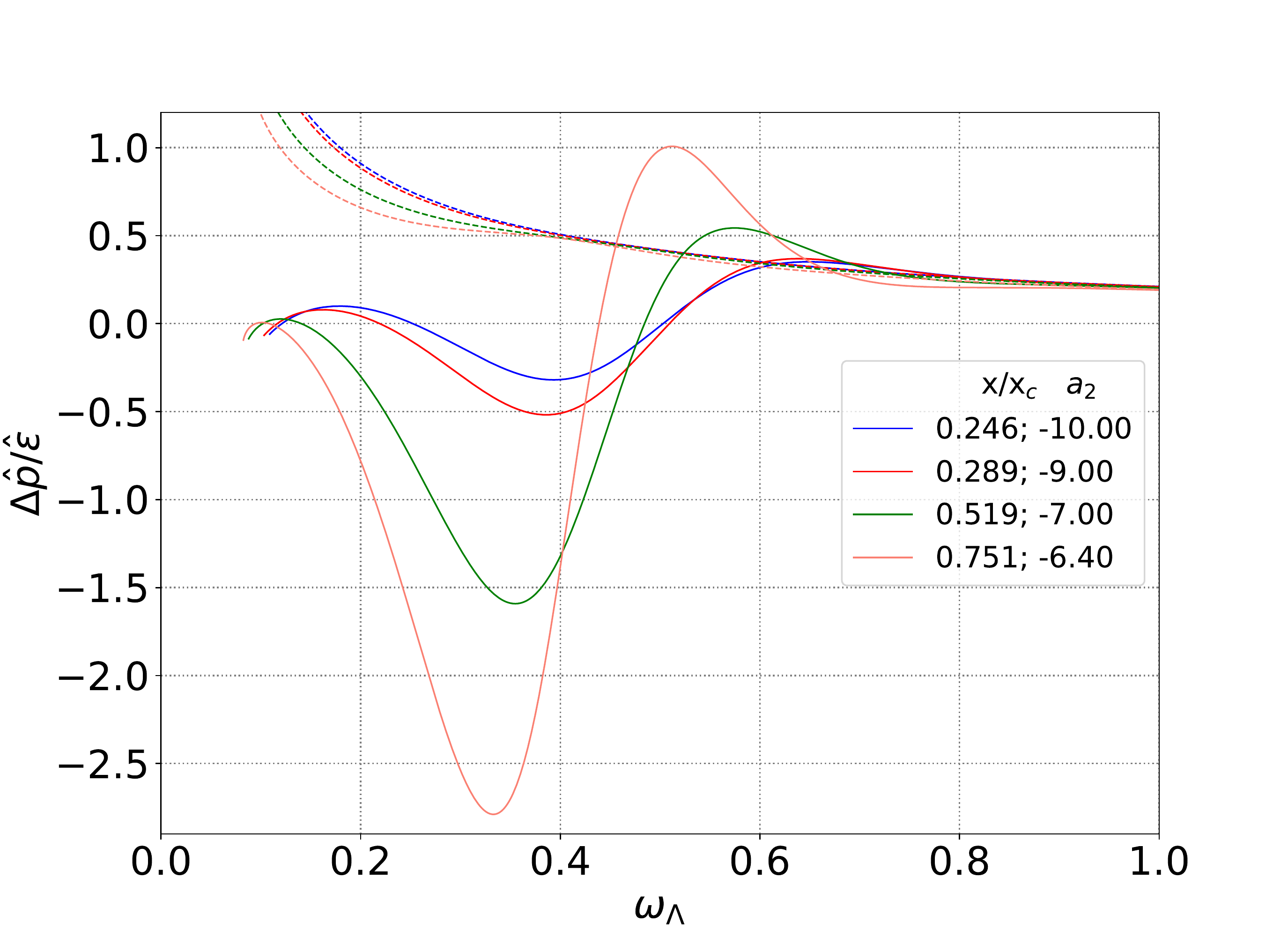}}
\subfigure[]{\includegraphics[width=0.49\textwidth]{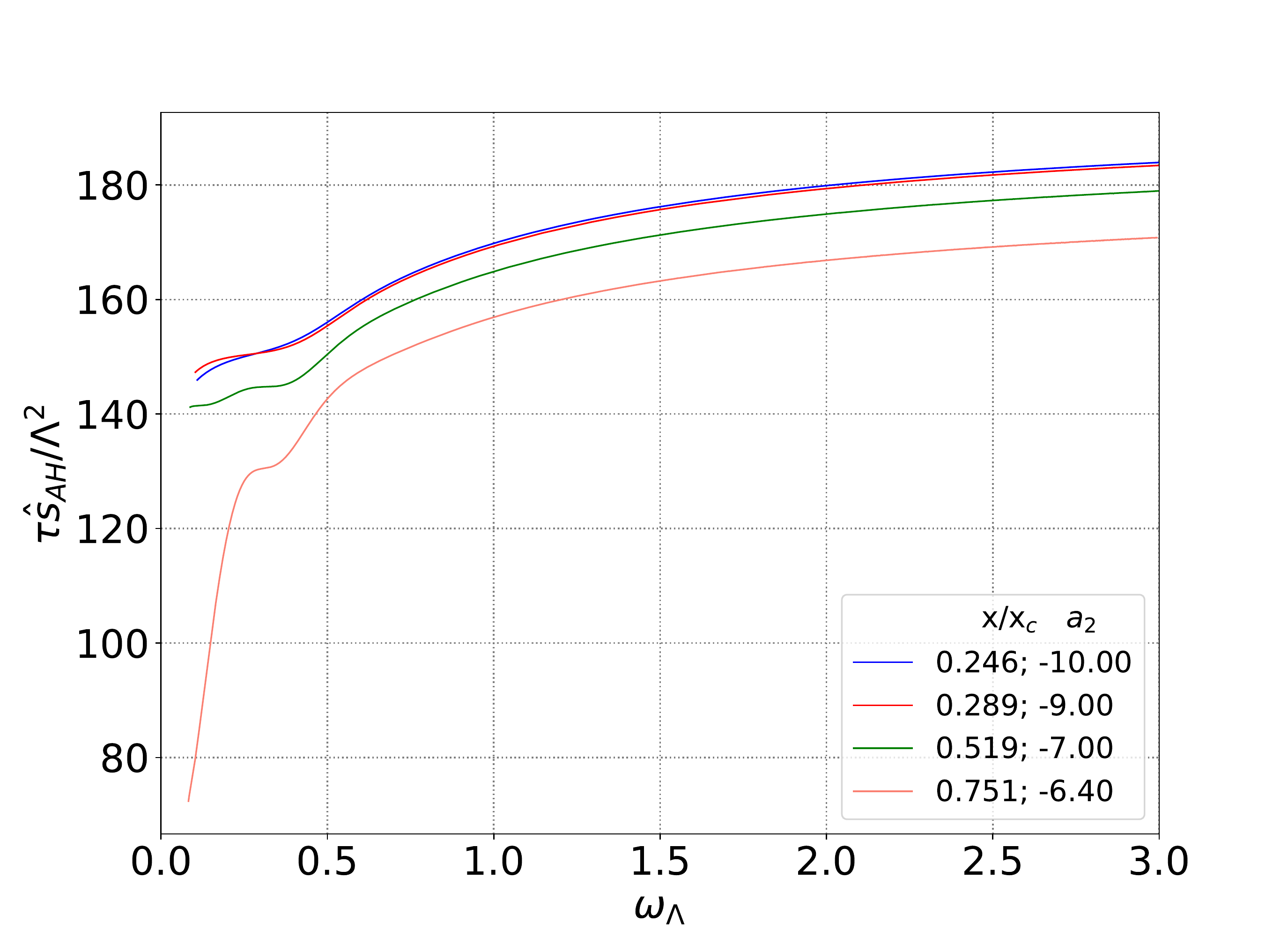}}
\subfigure[]{\includegraphics[width=0.49\textwidth]{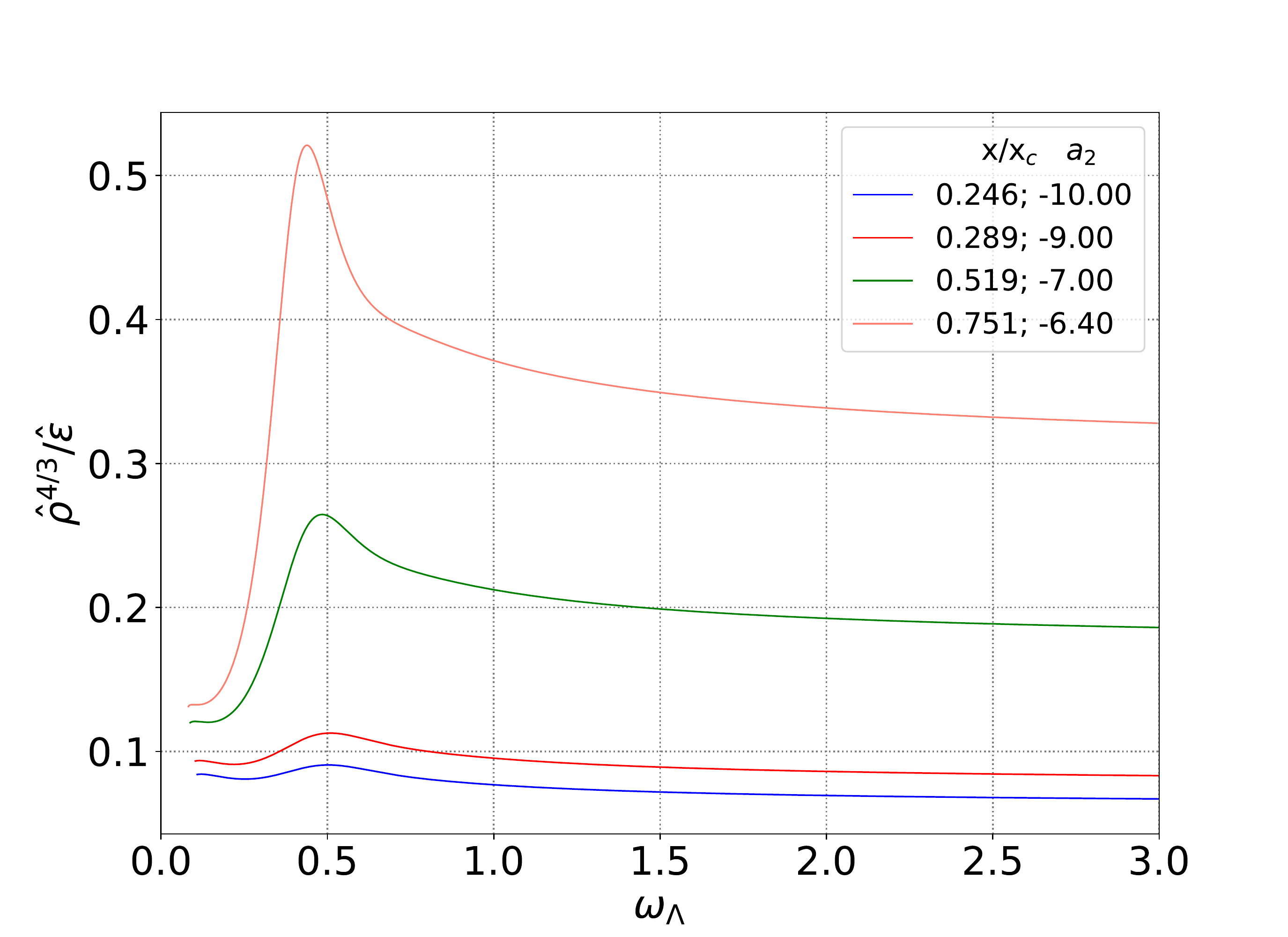}}
\subfigure[]{\includegraphics[width=0.49\textwidth]{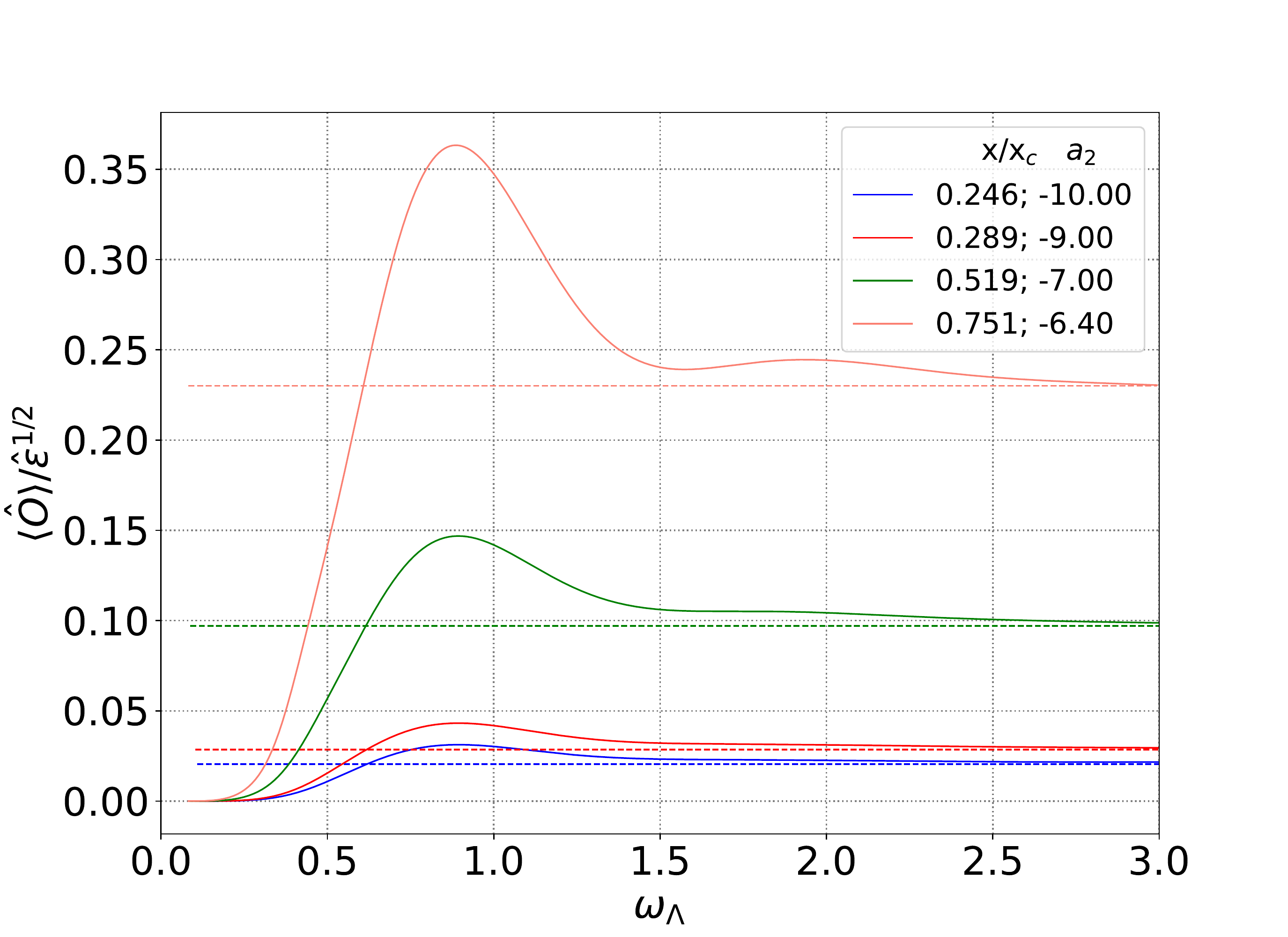}}
\caption{(a) Normalized pressure anisotropy (solid lines) and the corresponding hydrodynamic Navier-Stokes result (dashed lines), (b) normalized non-equilibrium entropy $\hat{S}_\textrm{AH}/\mathcal{A}\Lambda^2=\tau\hat{s}_\textrm{AH}/\Lambda^2$, (c) normalized charge density, and (d) normalized scalar condensate (solid lines) and the corresponding thermodynamic stable equilibrium result (dashed lines). Results obtained for variations of $a_2(\tau_0)$ keeping fixed $B_s8$ in Table \ref{tabICs} with $\rho_0=0.4$. Note that $x_c\equiv\left(\mu/T\right)_c=\pi/\sqrt{2}$ is the critical point.}
\label{fig:result19}
\end{figure*}

\begin{figure*}%[h]
\center
\subfigure[]{\includegraphics[width=0.49\textwidth]{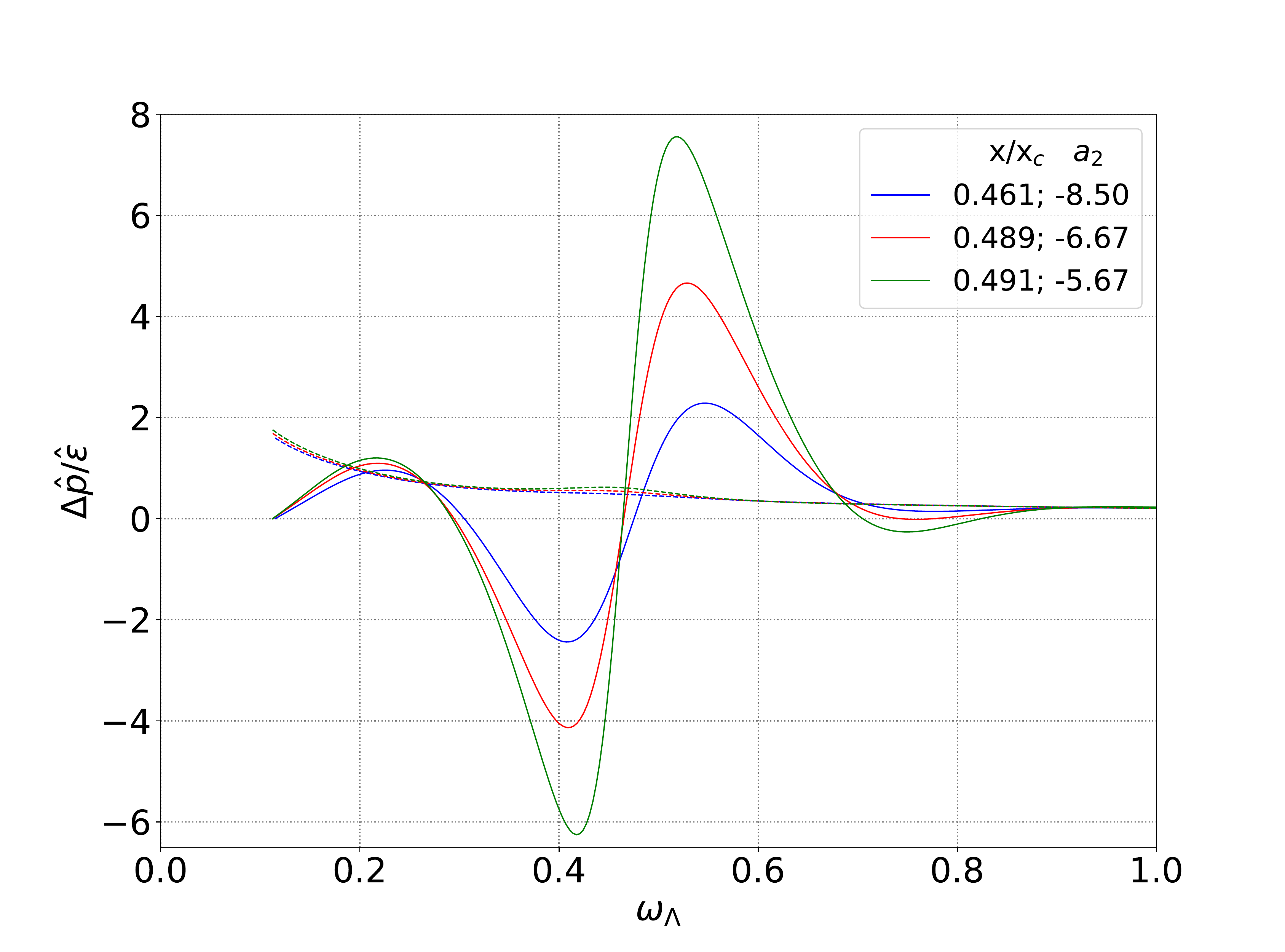}}
\subfigure[]{\includegraphics[width=0.49\textwidth]{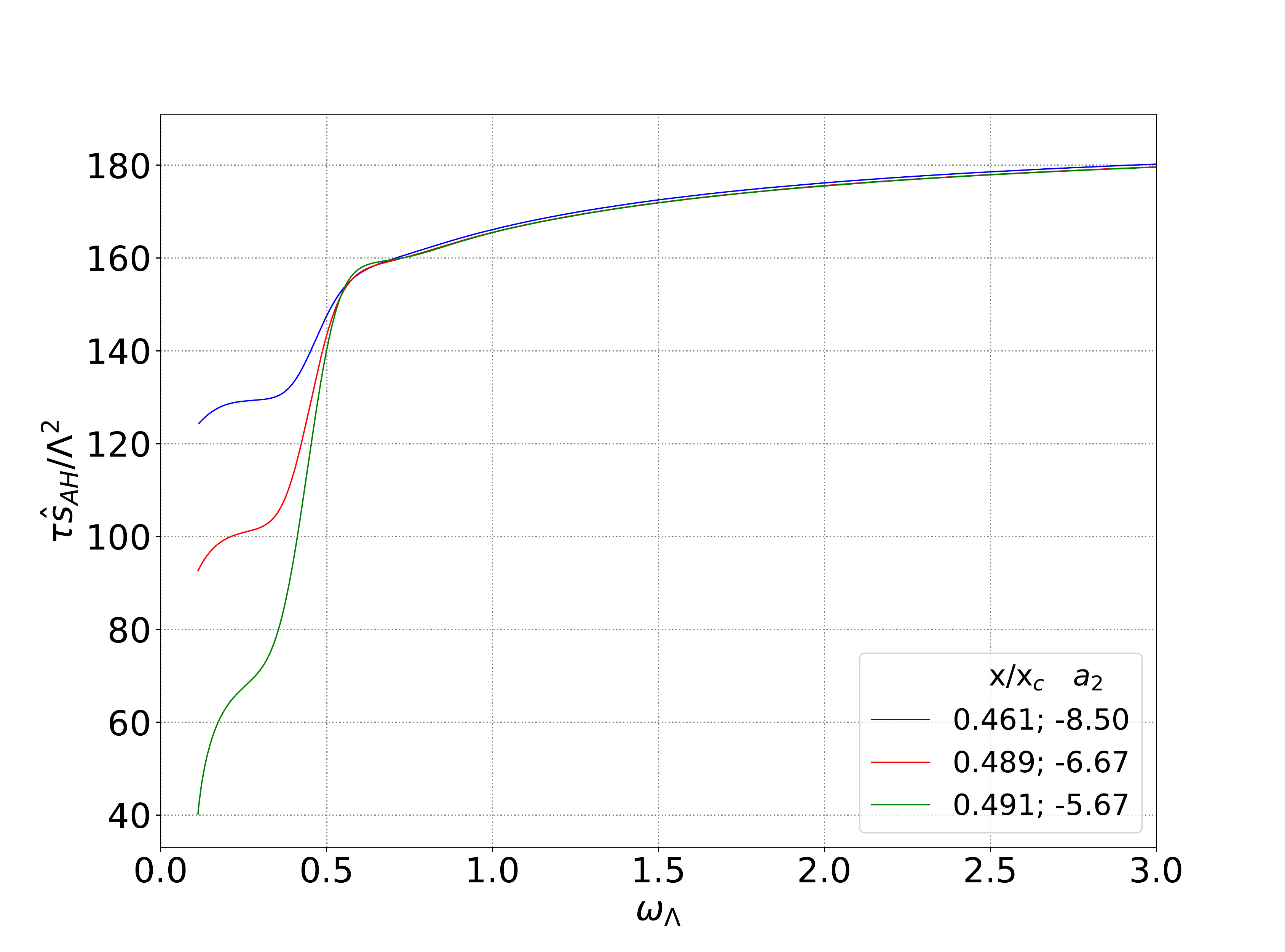}}
\subfigure[]{\includegraphics[width=0.49\textwidth]{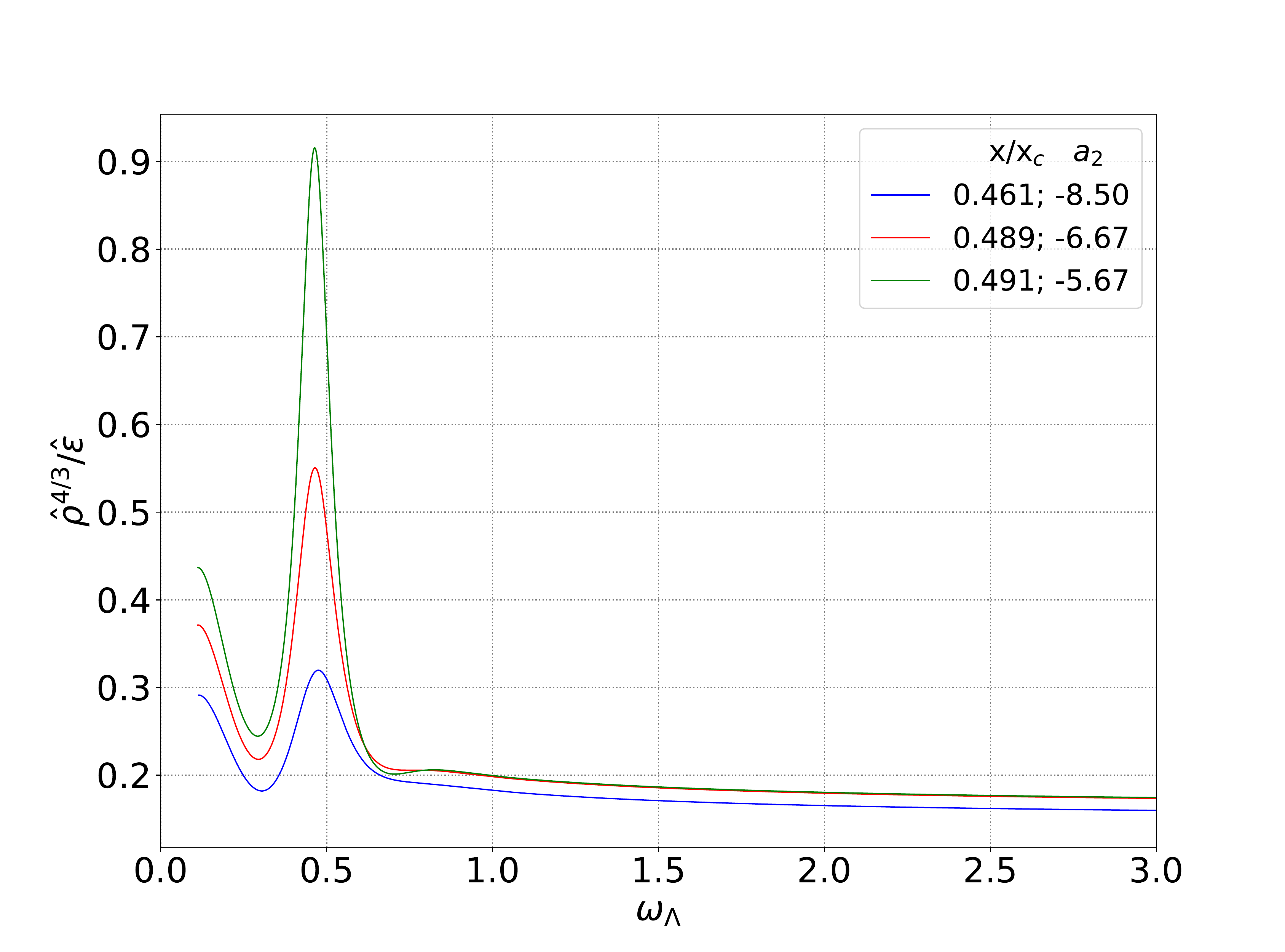}}
\subfigure[]{\includegraphics[width=0.49\textwidth]{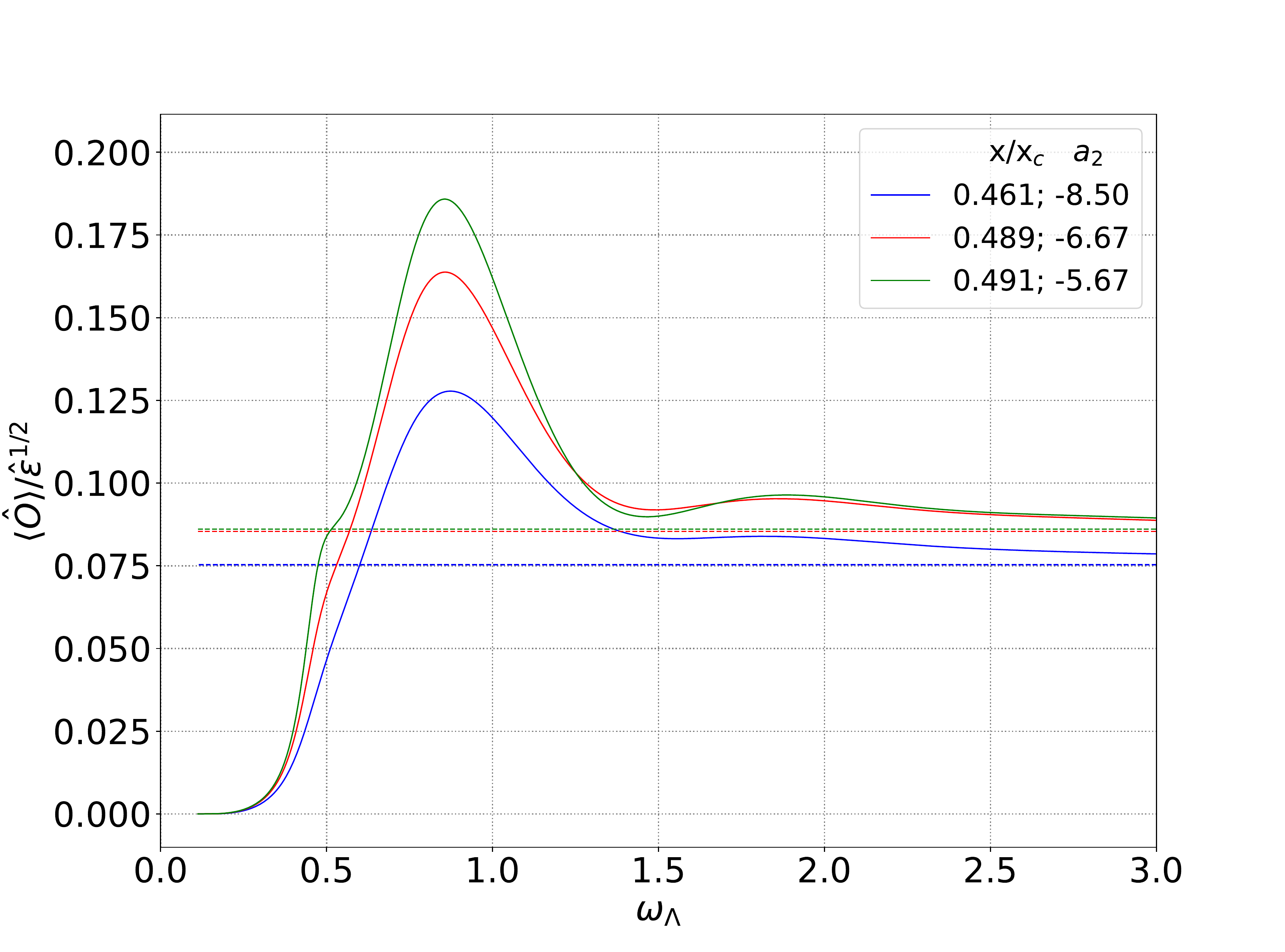}}
\caption{(a) Normalized pressure anisotropy (solid lines) and the corresponding hydrodynamic Navier-Stokes result (dashed lines), (b) normalized non-equilibrium entropy $\hat{S}_\textrm{AH}/\mathcal{A}\Lambda^2=\tau\hat{s}_\textrm{AH}/\Lambda^2$, (c) normalized charge density, and (d) normalized scalar condensate (solid lines) and the corresponding thermodynamic stable equilibrium result (dashed lines). Results obtained for variations of $a_2(\tau_0)$ keeping fixed $B_s9$ in Table \ref{tabICs} with $\rho_0=0.9$. Note that $x_c\equiv\left(\mu/T\right)_c=\pi/\sqrt{2}$ is the critical point.}
\label{fig:result20}
\end{figure*}

\begin{figure*}%[h]
\center
\subfigure[]{\includegraphics[width=0.49\textwidth]{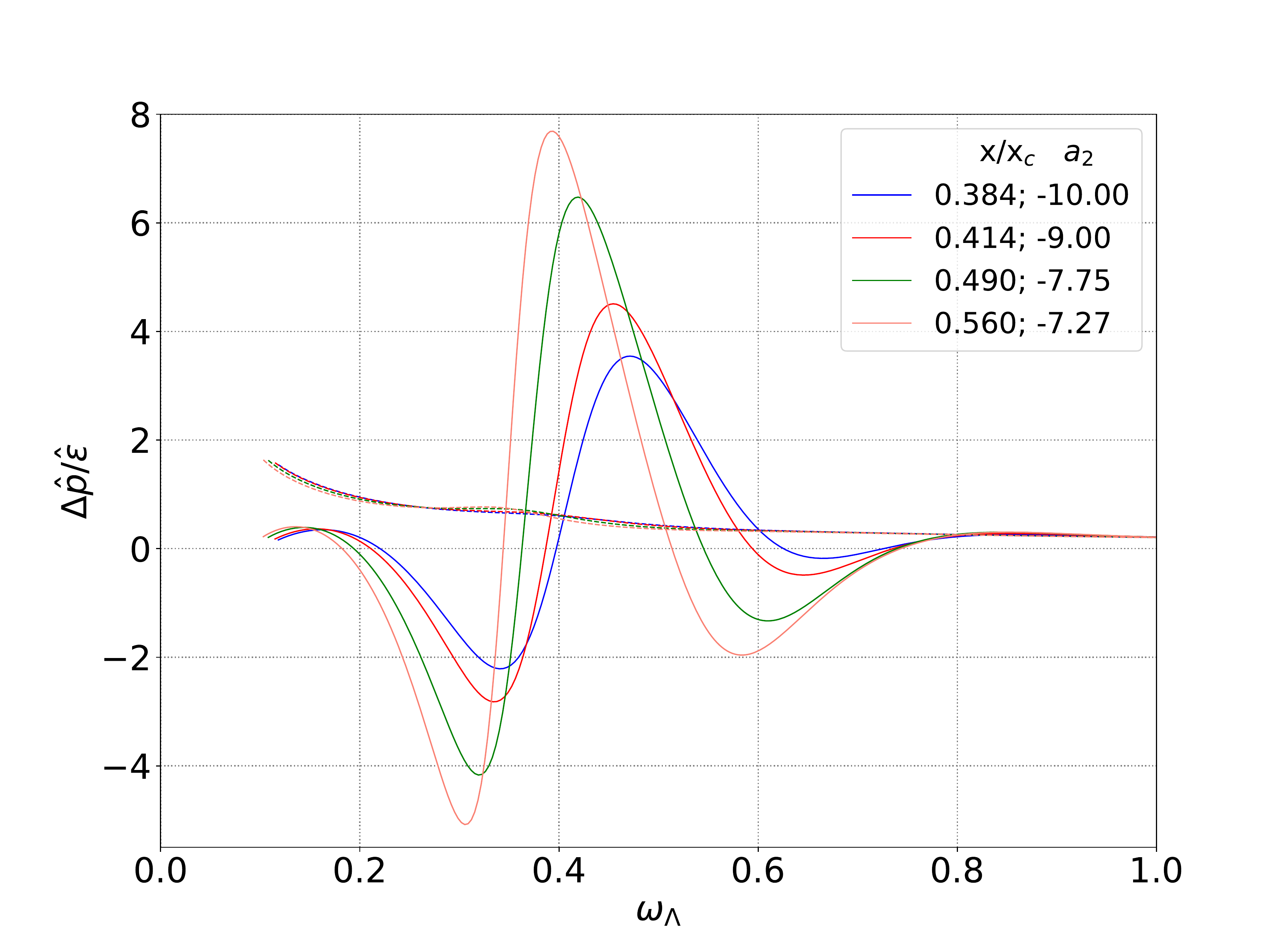}}
\subfigure[]{\includegraphics[width=0.49\textwidth]{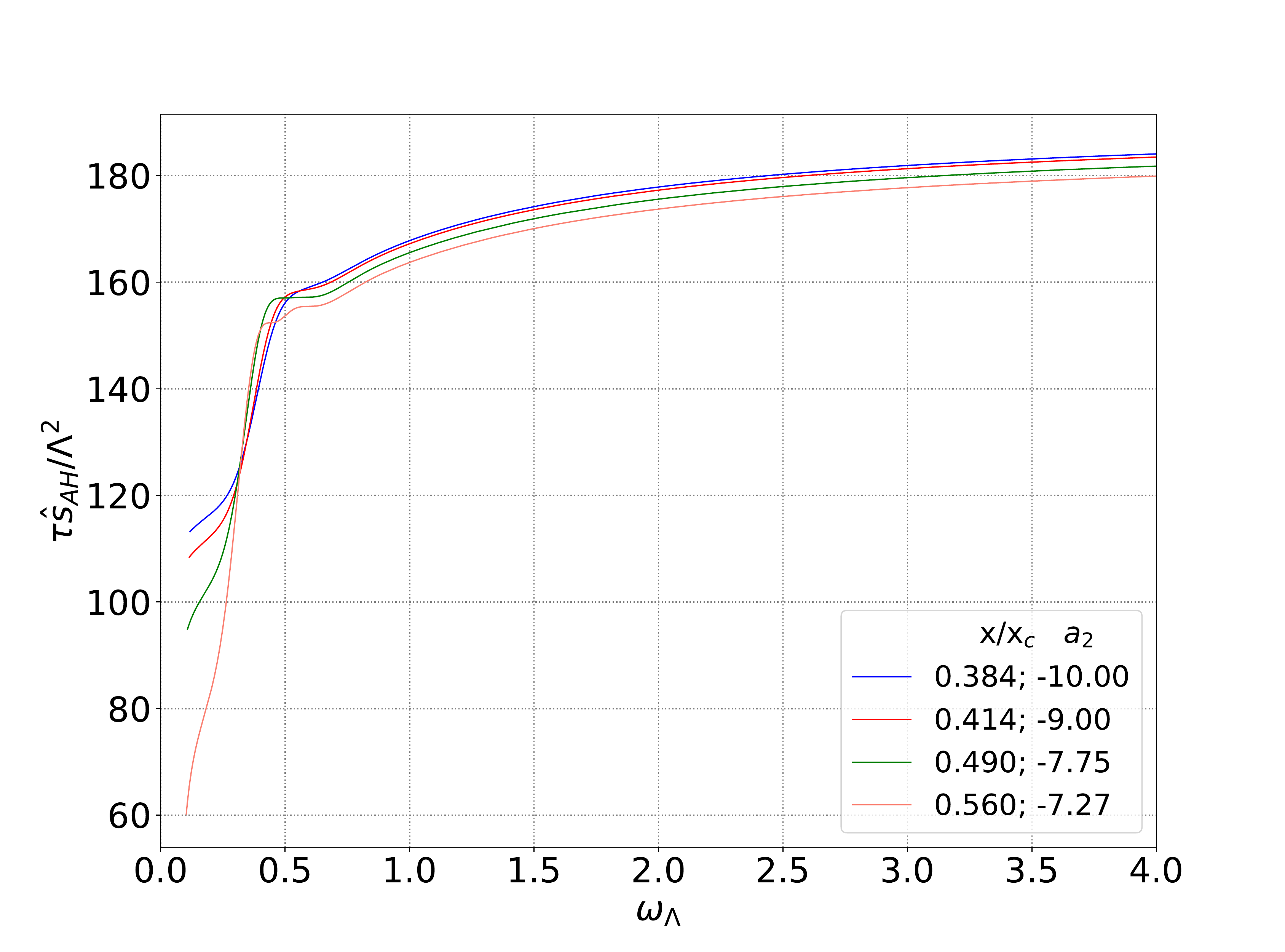}}
\subfigure[]{\includegraphics[width=0.49\textwidth]{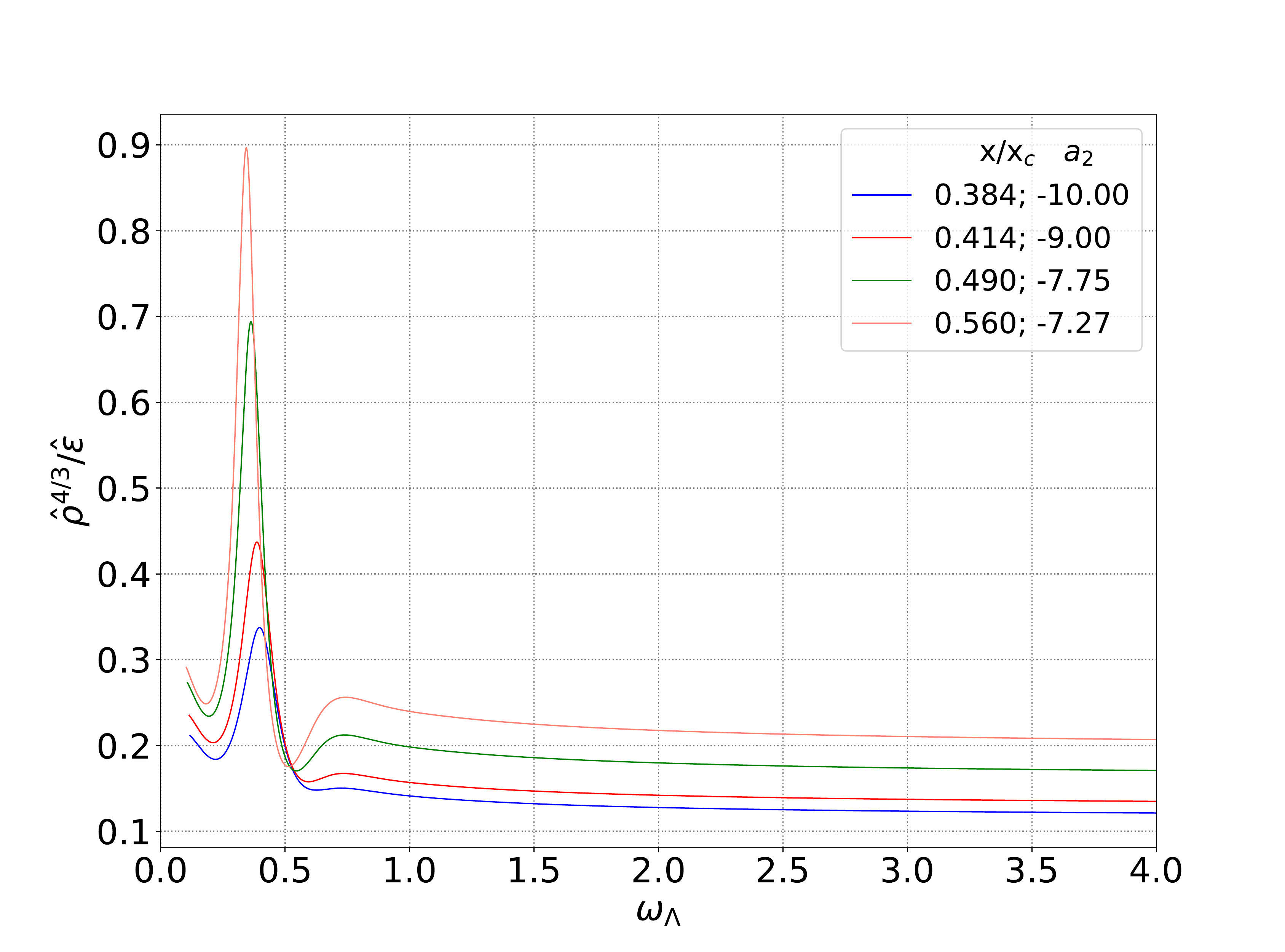}}
\subfigure[]{\includegraphics[width=0.49\textwidth]{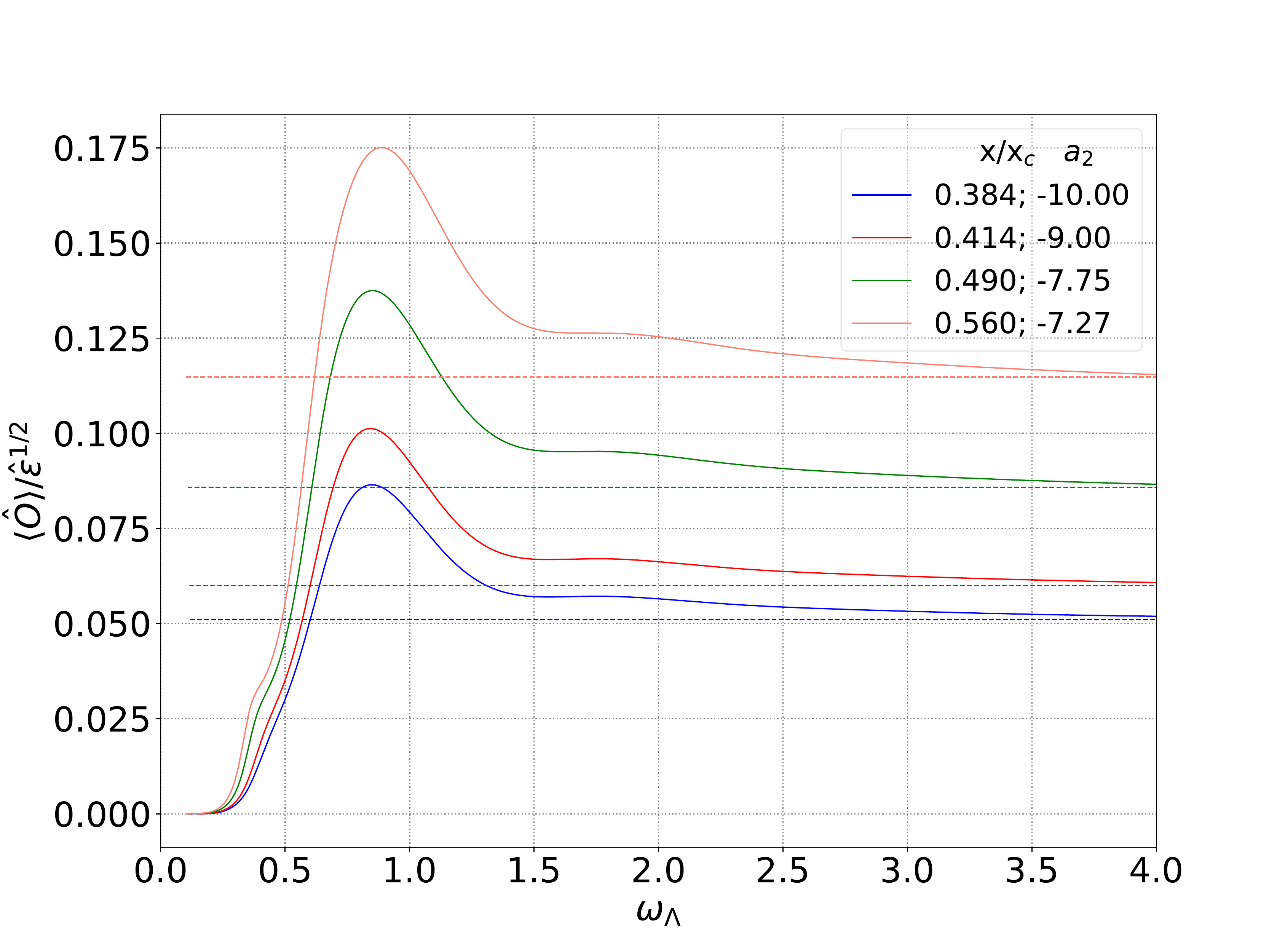}}
\caption{(a) Normalized pressure anisotropy (solid lines) and the corresponding hydrodynamic Navier-Stokes result (dashed lines), (b) normalized non-equilibrium entropy $\hat{S}_\textrm{AH}/\mathcal{A}\Lambda^2=\tau\hat{s}_\textrm{AH}/\Lambda^2$, (c) normalized charge density, and (d) normalized scalar condensate (solid lines) and the corresponding thermodynamic stable equilibrium result (dashed lines). Results obtained for variations of $a_2(\tau_0)$ keeping fixed $B_s10$ in Table \ref{tabICs} with $\rho_0=0.8$. Note that $x_c\equiv\left(\mu/T\right)_c=\pi/\sqrt{2}$ is the critical point.}
\label{fig:result21}
\end{figure*}

\begin{figure*}%[h]
\center
\subfigure[]{\includegraphics[width=0.49\textwidth]{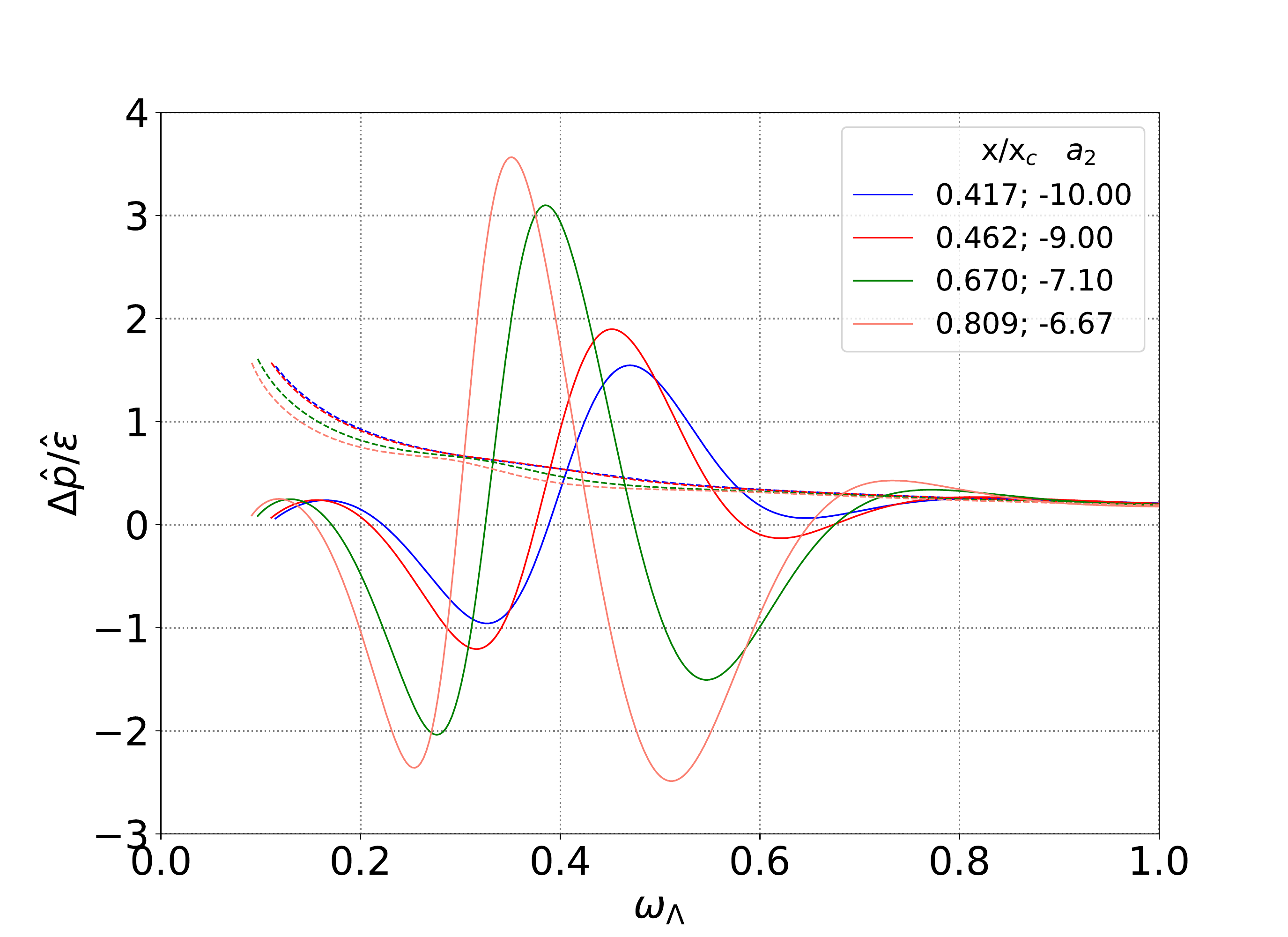}}
\subfigure[]{\includegraphics[width=0.49\textwidth]{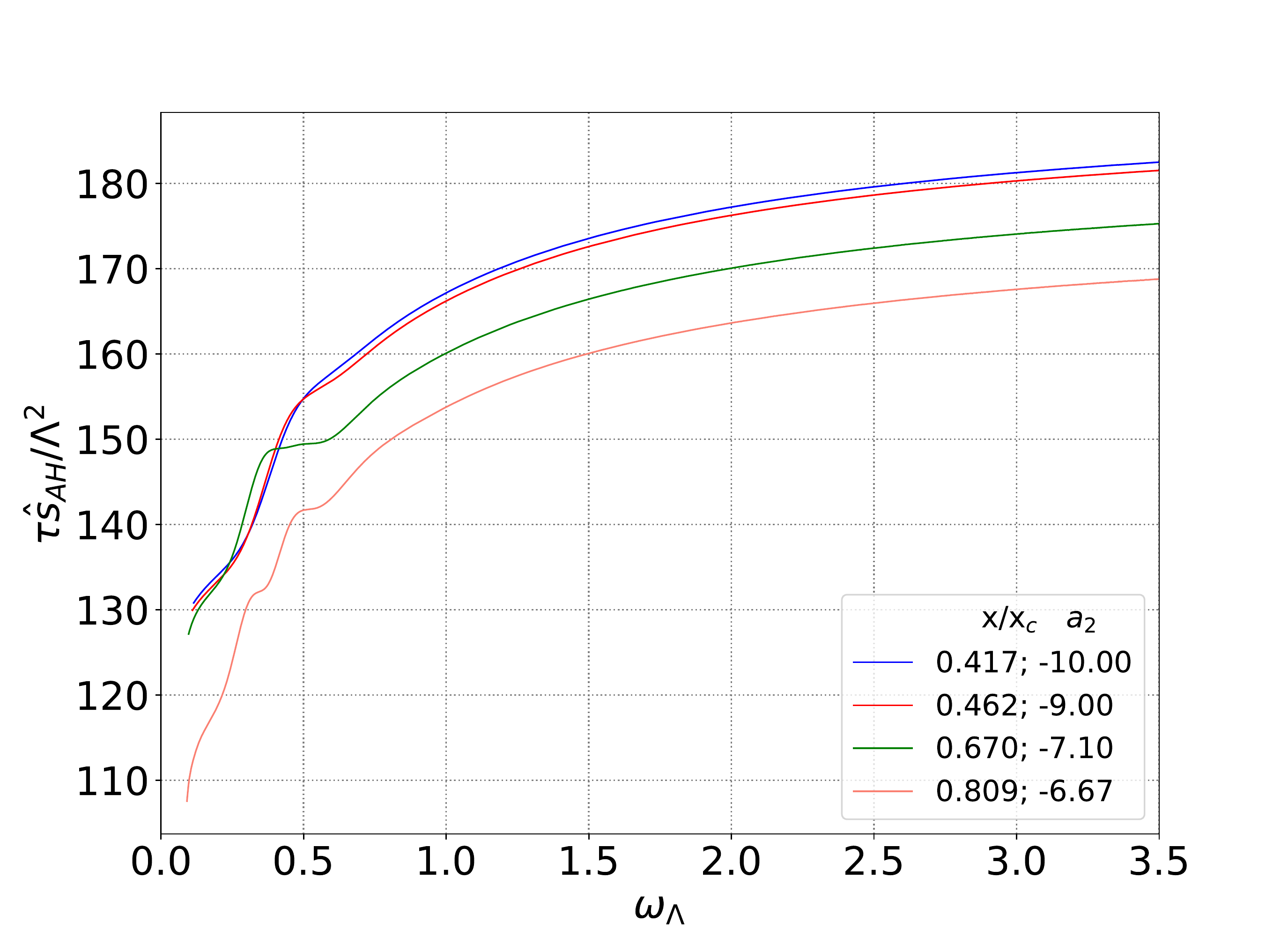}}
\subfigure[]{\includegraphics[width=0.49\textwidth]{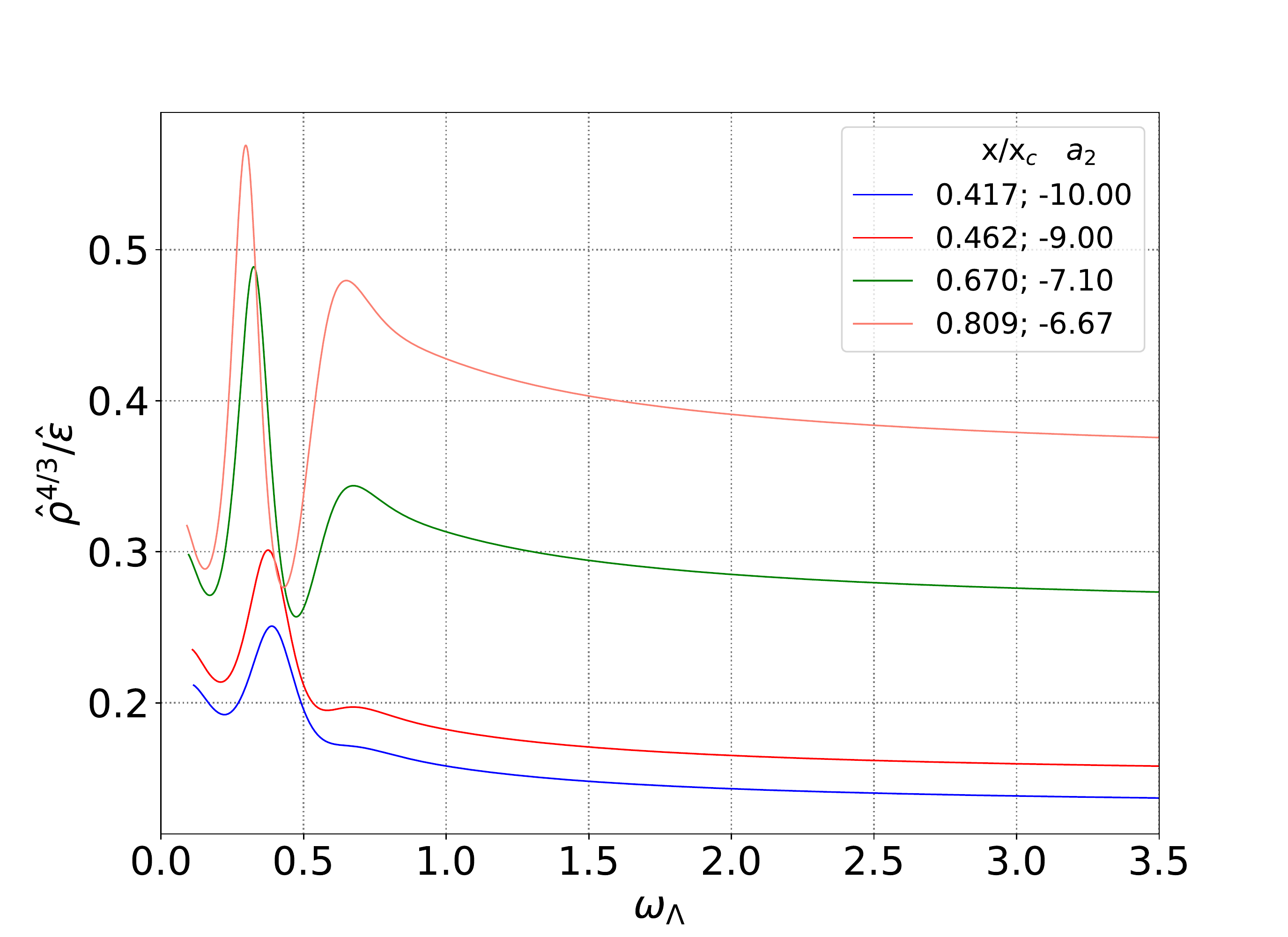}}
\subfigure[]{\includegraphics[width=0.49\textwidth]{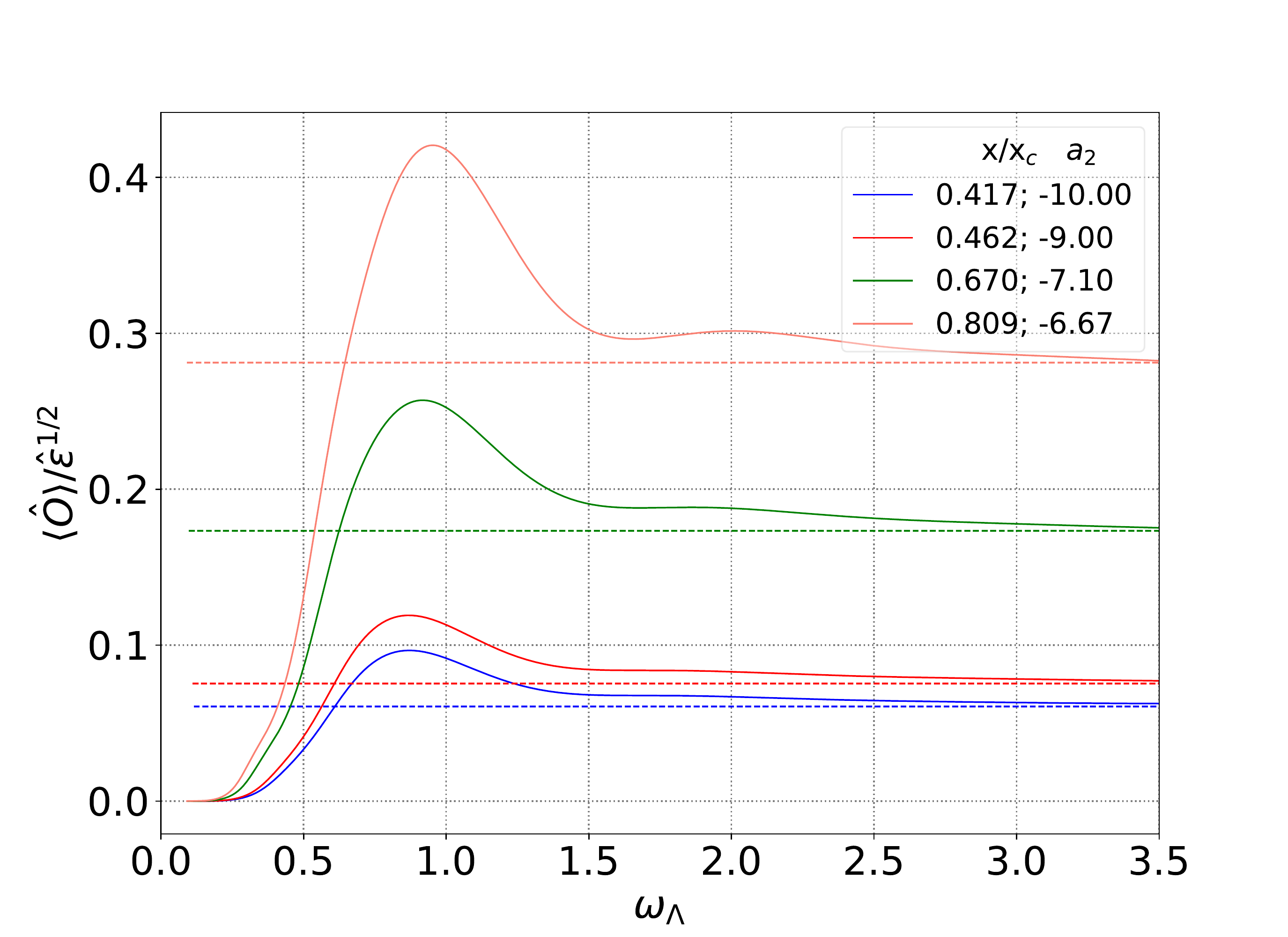}}
\caption{(a) Normalized pressure anisotropy (solid lines) and the corresponding hydrodynamic Navier-Stokes result (dashed lines), (b) normalized non-equilibrium entropy $\hat{S}_\textrm{AH}/\mathcal{A}\Lambda^2=\tau\hat{s}_\textrm{AH}/\Lambda^2$, (c) normalized charge density, and (d) normalized scalar condensate (solid lines) and the corresponding thermodynamic stable equilibrium result (dashed lines). Results obtained for variations of $a_2(\tau_0)$ keeping fixed $B_s11$ in Table \ref{tabICs} with $\rho_0=0.8$. Note that $x_c\equiv\left(\mu/T\right)_c=\pi/\sqrt{2}$ is the critical point.}
\label{fig:result22}
\end{figure*}

In this section, we analyze the time evolution of several different far from equilibrium initial conditions of the Bjorken expanding 1RCBH plasma, where for each profile for the initial subtracted metric anisotropy specified in Eq. \eqref{eq:Bs0} and in Table \ref{tabICs}, we consider variations of the initial energy density of the medium \eqref{eq:hatE}, $\hat{\epsilon}(\tau_0)=-3a_2(\tau_0)$ (we set again $\phi_s(\tau_0,u)=0$ throughout this section), while keeping fixed its initial charge density \eqref{eq:hatrho}, $\hat{\rho}(\tau_0)=\rho_0/\tau_0$. The corresponding results are shown in Figs. \ref{fig:result12} --- \ref{fig:result22}.

One notices that by decreasing the initial energy density of the 1RCBH plasma (by decreasing $|a_2(\tau_0)|$), while keeping its initial charge density fixed, the value of $\mu/T$ in the medium is enhanced, producing for the physical observables analyzed as functions of the dimensionless time measure $\omega_\Lambda$, generally the same physical possibilities as discussed in section \ref{sec:4}. However, for a given profile for the initial subtracted metric anisotropy $B_s(\tau_0,u)$, considering the enhancement of $\mu/T$ of the medium caused specifically by increasing its initial charge density, or specifically by reducing its initial energy density, it is something that may lead to qualitatively different outcomes for some physical observables. For instance, by comparing Figs. \ref{fig:result1} and \ref{fig:result12} for $B_s1$ in Table \ref{tabICs}, one notices that the peak of the normalized pressure anisotropy is reduced (increased) by increasing $\mu/T$ associated to increasing (decreasing) the initial charge (energy) density of the medium.

We close this section with an extra technical information regarding the numerical simulations used to obtain the results displayed in Figs. \ref{fig:result12} --- \ref{fig:result22}. For almost all the initial conditions considered in those results, it is enough to use $N\sim 20$ collocations points in the radial grid in order to obtain convergent and physically reliable results for all the observables considered. However, specifically for $B_s8$ in Table \ref{tabICs} with $\rho_0=0.4$ and $a_2(\tau_0)=-6.4$, and also for $B_s11$ with $\rho_0=0.8$ and $a_2(\tau_0)=-6.67$, we noted spurious numerical oscillations at early times for the scalar condensate, with such issues being completely fixed by increasing the number of collocation points to $N\sim 30$.

%%%%%%%%%%%%%%%%%%%%%%%%%%%%%%%%%
\section{Results for variations of the initial dilaton profile}
\label{sec:extra}

\begin{figure*}%[h]
\center
\subfigure[]{\includegraphics[width=0.49\textwidth]{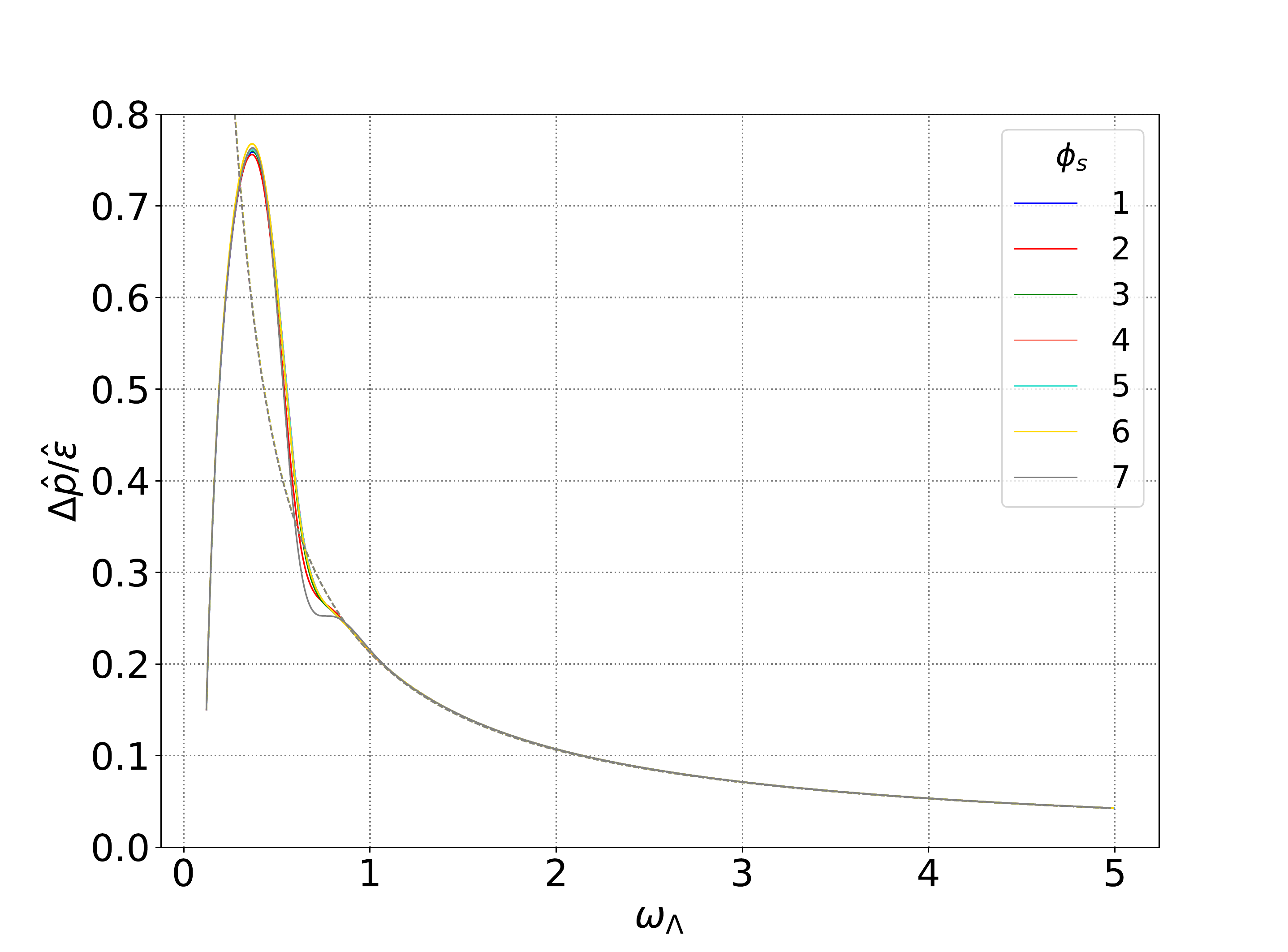}}
\subfigure[]{\includegraphics[width=0.49\textwidth]{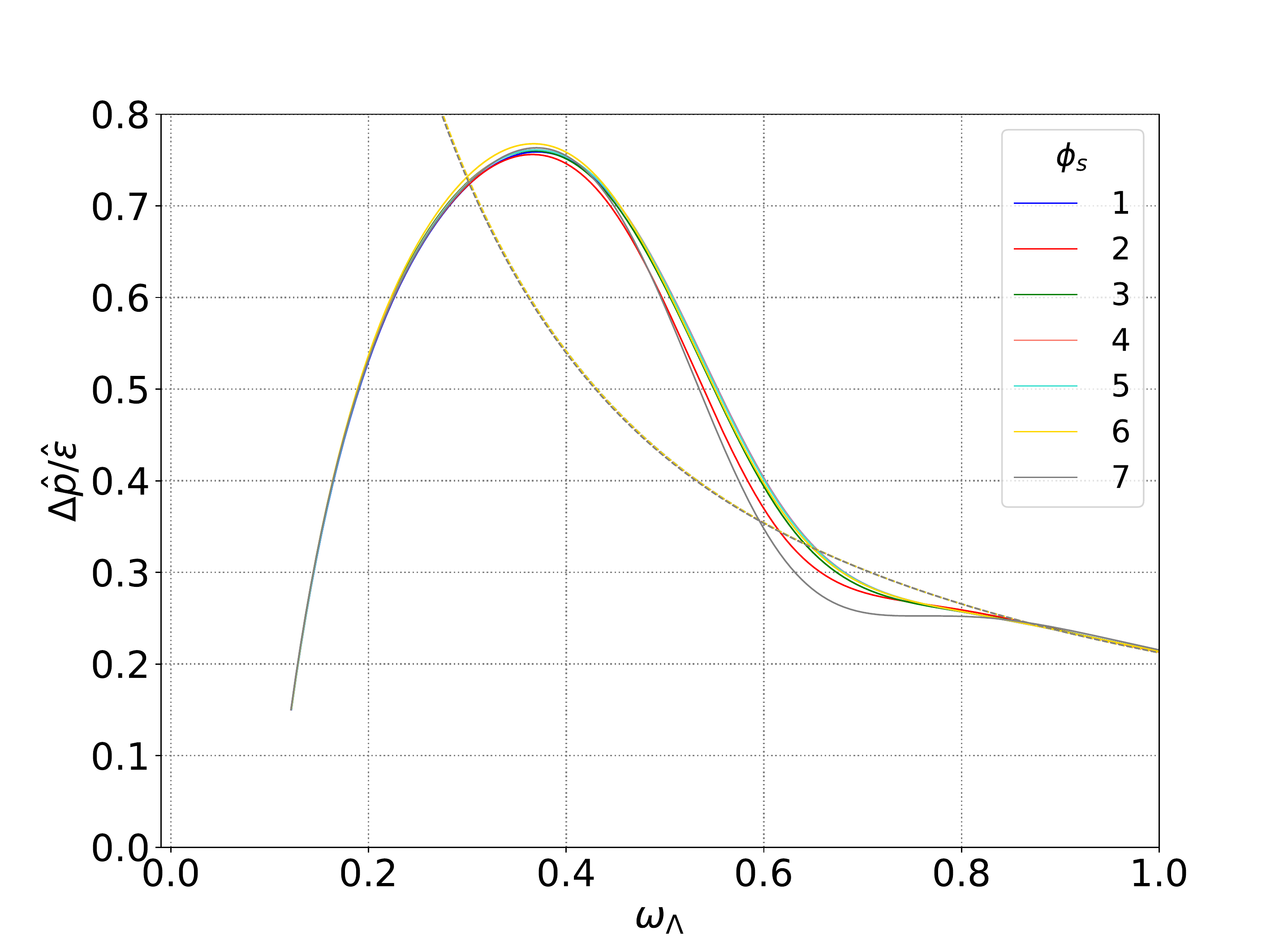}}
\subfigure[]{\includegraphics[width=0.49\textwidth]{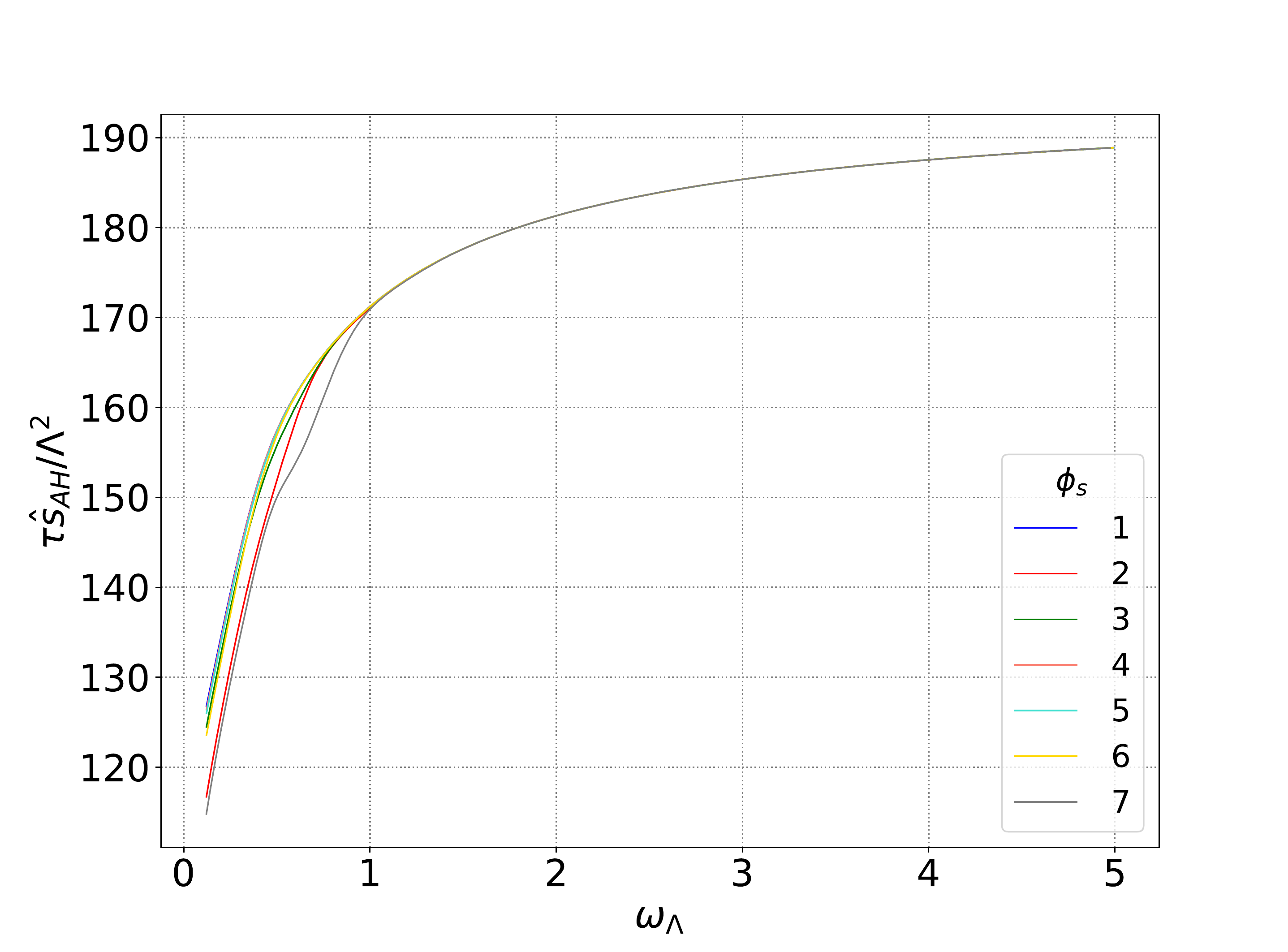}}
\subfigure[]{\includegraphics[width=0.49\textwidth]{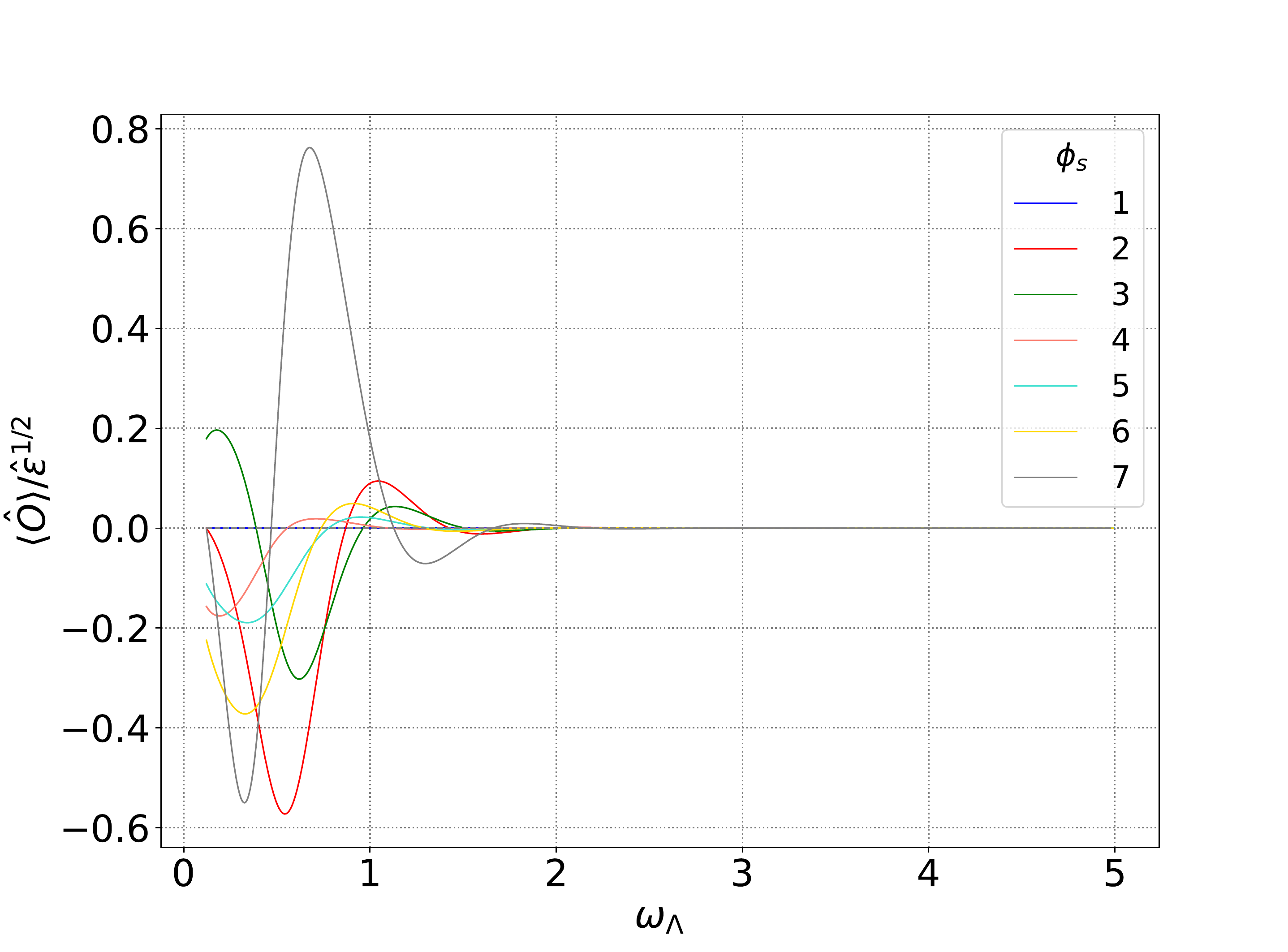}}
\caption{(a) Normalized pressure anisotropy (solid lines) and the corresponding hydrodynamic Navier-Stokes result (dashed line), (b) zoom of the pressure anisotropy up to the hydrodynamization region, (c) normalized non-equilibrium entropy $\hat{S}_\textrm{AH}/\mathcal{A}\Lambda^2=\tau\hat{s}_\textrm{AH}/\Lambda^2$, and (d) normalized scalar condensate (solid lines) and the corresponding thermodynamic stable equilibrium result (dashed line). Results obtained for variations of the initial profile for the subtracted dilaton field $\phi_s(\tau_0,u)$ in Table \ref{tabphis}, keeping fixed $B_s1$ in Table \ref{tabICs} with $a_2(\tau_0)=-6.67$ and $\rho_0=0$, which give solutions with $\mu/T=0$. Note that when $\phi_s(\tau_0,u)\neq 0$ the scalar condensate presents a nontrivial time evolution (with some impact also on other observables), different from pure thermal SYM states far from equilibrium, even though the asymptotic equilibrium state (which depends only on the value of $\mu/T$) is the same as in pure thermal SYM.}
\label{fig:extra1}
\end{figure*}

\begin{figure*}%[h]
\center
\subfigure[]{\includegraphics[width=0.49\textwidth]{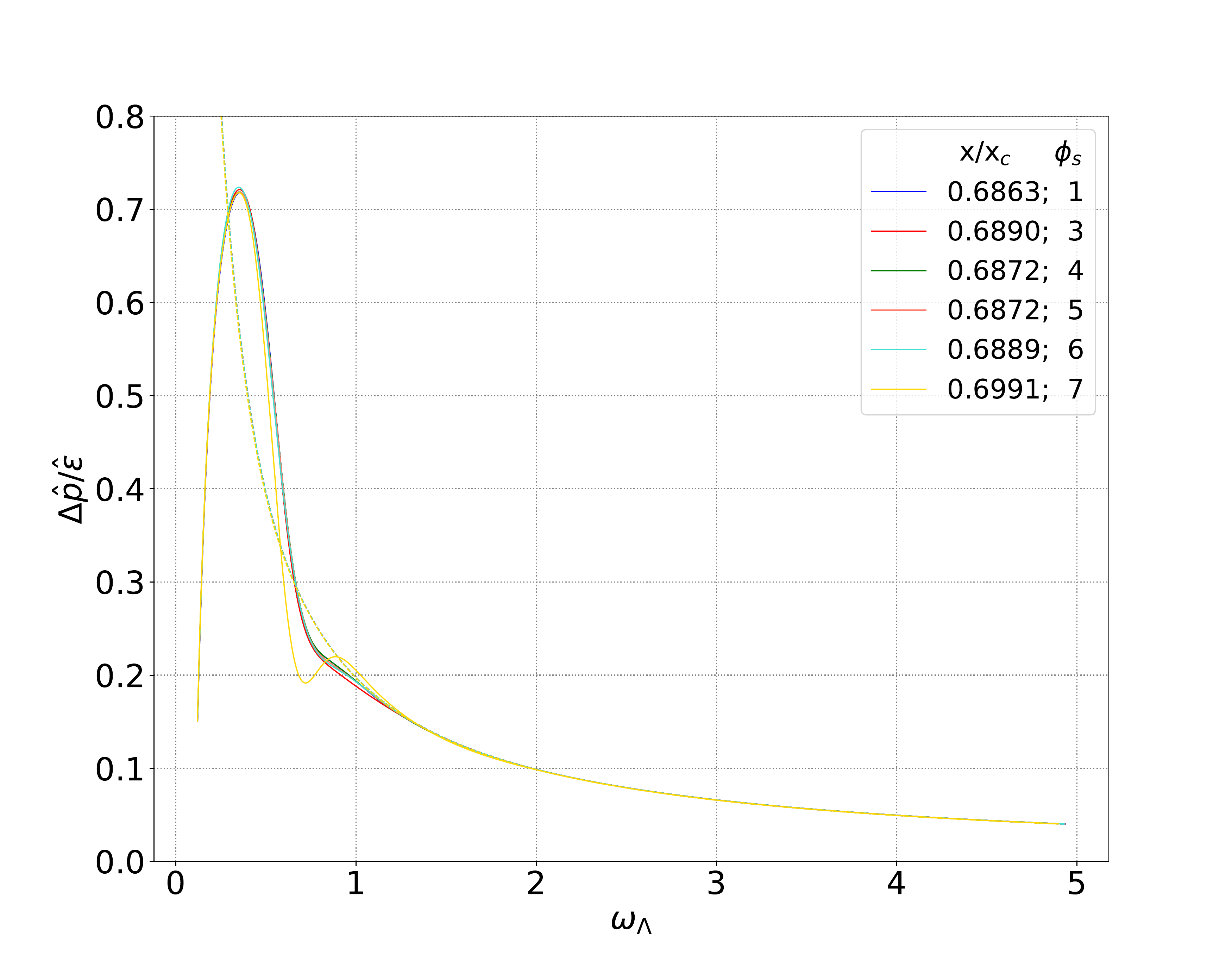}}
\subfigure[]{\includegraphics[width=0.49\textwidth]{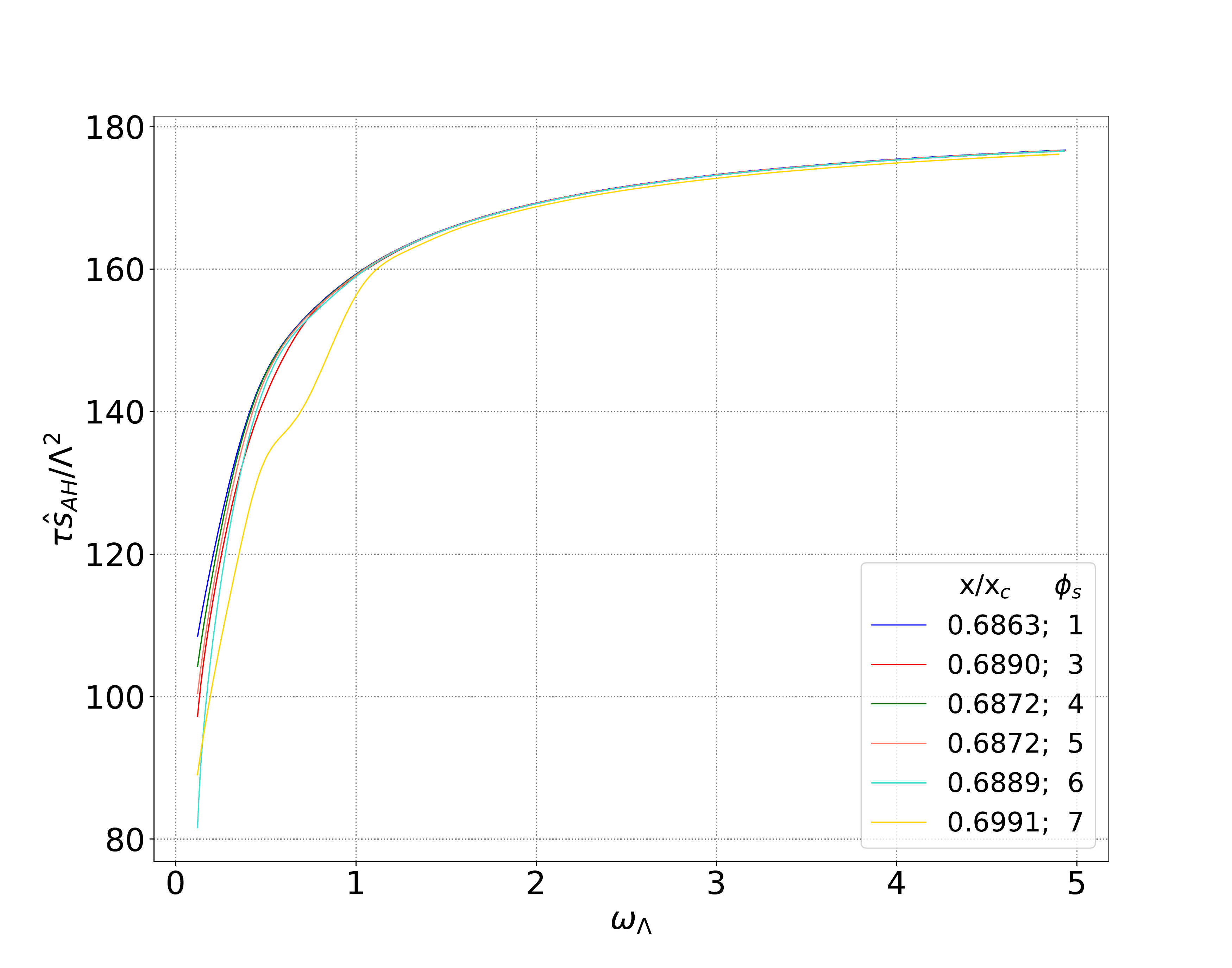}}
\subfigure[]{\includegraphics[width=0.49\textwidth]{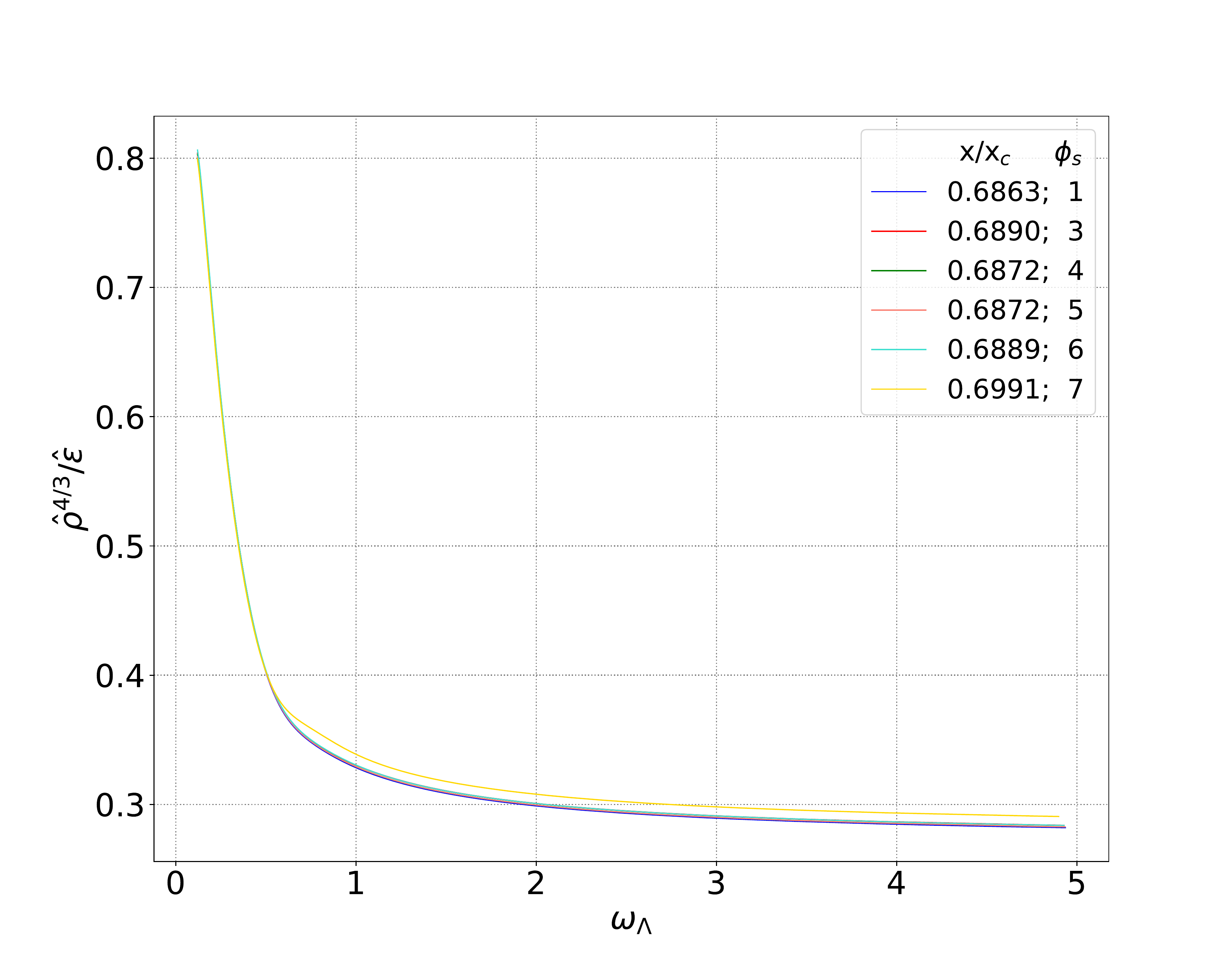}}
\subfigure[]{\includegraphics[width=0.49\textwidth]{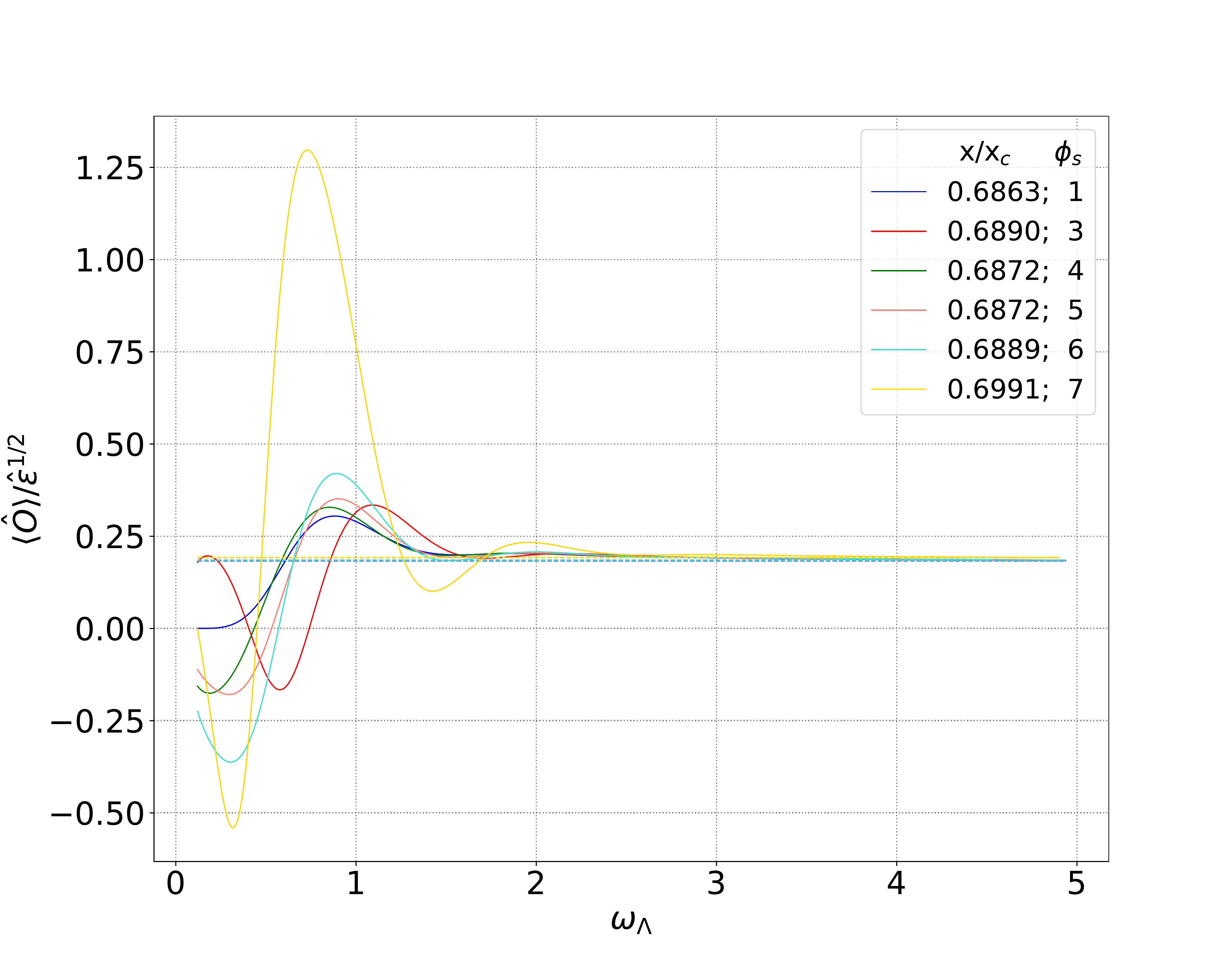}}
\caption{(a) Normalized pressure anisotropy (solid lines) and the corresponding hydrodynamic Navier-Stokes result (dashed lines), (b) normalized non-equilibrium entropy $\hat{S}_\textrm{AH}/\mathcal{A}\Lambda^2=\tau\hat{s}_\textrm{AH}/\Lambda^2$, (c) normalized charge density, and (d) normalized scalar condensate (solid lines) and the corresponding thermodynamic stable equilibrium result (dashed lines). Results obtained for variations of the initial profile for the subtracted dilaton field $\phi_s(\tau_0,u)$ in Table \ref{tabphis}, keeping fixed $B_s1$ in Table \ref{tabICs} with $a_2(\tau_0)=-6.67$ and $\rho_0=1.6$. Note that $x_c\equiv\left(\mu/T\right)_c=\pi/\sqrt{2}$ is the critical point.}
\label{fig:extra2}
\end{figure*}

In this section, we analyze the time evolution of some different far from equilibrium initial conditions of the Bjorken expanding 1RCBH plasma, where we consider variations of the initial profile for the subtracted dilaton field specified in Eq. \eqref{eq:phis0} and in Table \ref{tabphis}, while keeping fixed all the other initial data. One notices from Eq. \eqref{eq:hatE} that when the boundary value of the initial profile for the subtracted dilaton field is non-vanishing, $\phi_2(\tau_0)=\phi_s(\tau_0,u=0)\neq 0$, it contributes to the initial energy density of the medium, $\hat{\epsilon}(\tau_0)=-3a_2(\tau_0)-\phi_2(\tau_0)^2/6$. The corresponding results are shown in Figs. \ref{fig:extra1} and \ref{fig:extra2}.

In the case of time evolutions with zero charge density, as displayed in Fig. \ref{fig:extra1}, one typically has relatively small variations of the pressure anisotropy and the entropy at early times in terms of variations of the initial dilaton (keeping fixed the remaining initial data). On the other hand, the relative variations of the scalar condensate before thermalization may be large depending on the chosen profiles for the initial dilaton field. The fact that the scalar condensate may be nonzero when the 1RCBH fluid is far from equilibrium, even for time evolutions with zero charge density, implies that there are far from equilibrium solutions with zero charge density in the 1RCBH model that are different from pure thermal SYM solutions far from equilibrium. Indeed, even though the asymptotic equilibrium state is the same for all initial conditions with $\rho_0=0$, since in those cases $\mu/T=0$ and the equilibrium state in the 1RCBH model only depends on the value of $\mu/T$, at early times, when $\phi_s(\tau_0,u)\neq 0$, the time evolutions with zero charge density in the 1RCBH plasma develop transiently nontrivial profiles for the scalar condensate (which vanish when the medium thermalizes). One also notices from Figs. \ref{fig:extra1} (b) and (d) that the thermalization time associated to the effective equilibration of the scalar condensate is considerably larger than the hydrodynamization time of the pressure anisotropy of the medium.

In the case of time evolutions at finite charge density, as shown in Fig. \ref{fig:extra2}, again there are typically relatively small variations of the pressure anisotropy, entropy, and charge density at early times, while the relative variations of the scalar condensate may be large. The value of $\mu/T$ is typically almost insensitive to variations only on the initial dilaton profile.

Interestingly, and completely different from all the cases considered with $\phi_s(\tau_0,u)=0$, when the initial dilaton profile is nontrivial, the scalar condensate typically acquires negative values with pronounced dips when the medium is still far from equilibrium.

We close this section by observing in Fig. \ref{fig:extra1} that, just as it happens for the pressure anisotropy and the scalar condensate, also the normalized entropy converges to a single curve at late times when $\mu/T$ is kept fixed (in the case of Fig. \ref{fig:extra1}, it is zero). In two previous works of ours \cite{Rougemont:2021qyk,Rougemont:2021gjm}, it was not very clear that the entropy does converge to a single curve (it seemed instead that it was converging to a tiny band of values for different initial conditions at zero chemical potential) because the numerical fits done to obtain the energy scale $\Lambda$ (see below Eq. \eqref{eq:normS}) were not very precise, since they were calculated at not very late times. In the present work, since we evolved each initial condition for much longer times than in Refs. \cite{Rougemont:2021qyk,Rougemont:2021gjm}, we were able to obtain precise results for $\Lambda$, such that the normalized entropy $\hat{S}_\textrm{AH}/\mathcal{A}\Lambda^2=\tau\hat{s}_{\textrm{AH}}/\Lambda^2$ clearly converges to a single curve for different initial conditions evolving into the same value of $\mu/T$ at late times.

%%%%%%%%%%%%%%%%%%%%%%%%%%%%%%%%%
\section{Conclusions and future perspectives}
\label{sec:conc}

In this paper, we studied the time evolution of several different far from equilibrium initial states for a hot and dense strongly coupled quantum fluid expanding according to the Bjorken flow dynamics. The corresponding medium, called the 1RCBH plasma, describes a conformal $\mathcal{N}=4$ SYM plasma charged under an Abelian $U(1)$ group of the $SU(4)$ R-symmetry, having a critical point in its phase diagram. We analyzed the time evolution of the Bjorken expanding 1RCBH plasma taking into account the behavior of several physical observables, including for the first time the calculation of the holographic non-equilibrium entropy for this model.

We observed that the value of $\mu/T$ in the medium is enhanced either by increasing its initial charge density, or by decreasing its initial energy density. We found that as $\mu/T$ is enhanced towards its critical value, the hydrodynamization of the pressure anisotropy of the medium, measured by its late time convergence to the corresponding Navier-Stokes regime, is generally delayed, in line with previous work \cite{Critelli:2018osu}. We have also seen, for the first time, the effective thermalization of the scalar condensate, measured by its late time convergence to the corresponding thermodynamically stable equilibrium, which is also generally delayed as the R-charge density of the medium is increased. The thermalization of the scalar condensate only happens at much later times than the hydrodynamization of the pressure anisotropy, requiring the numerical simulations to run for much longer times than in previous works \cite{Critelli:2018osu,Rougemont:2021qyk,Rougemont:2021gjm}, which in turn demands the numerical code to be implemented using efficient programming languages for numerical calculations.

For some sets of initial data preserving all the classical energy conditions, dynamically driven transient violations of the dominant and the weak energy conditions are observed when the 1RCBH plasma is still far from the hydrodynamic regime. The far from equilibrium violations of the dominant and weak energy conditions get stronger at larger values of $\mu/T$, indicating that such violations become more and more relevant as the strongly coupled quantum fluid approaches its critical regime. For some of these energy conditions violations, it is observed a clear correlation with a previous formation of different plateau structures in the far from equilibrium entropy of the medium, indicating the presence of transient, early time windows where the Bjorken expanding plasma has zero entropy production even while being far from equilibrium.

More specifically, \emph{when} single or double plateaus are produced in the time evolution of the far from equilibrium entropy of the medium, it is later observed almost or effective violations of DEC from below with $\Delta\hat{p}/\hat{\epsilon}\sim -1$ or $\Delta\hat{p}/\hat{\epsilon}<-1$, respectively, in line with the observations in \cite{Rougemont:2021qyk,Rougemont:2021gjm} for the pure thermal SYM plasma with $\mu/T=0$, although here we also observe the aforementioned correlations under much more general situations with $\mu/T \ge 0$. In general, such correlations do not hold in reverse order, i.e. there are evolutions for the 1RCBH plasma with transient violations of DEC which present no far from equilibrium plateaus in the entropy of the medium.

In particular, we notice the following general trend regarding the deformation of plateau structures as $\mu/T$ is increased, leading to progressively lower dips in the pressure anisotropy to energy density ratio: first a single plateau is formed (near the boundary to DEC violation from below, with $\Delta\hat{p}/\hat{\epsilon}\sim -1$), which then becomes more spread out in the time direction, and next it is progressively deformed into double plateaus (with $\Delta\hat{p}/\hat{\epsilon}< -1$), then becoming double quasiplateaus, until the plateau structure is finally lost for sufficiently strong violations of the DEC from below (in such a progression, the lower plateau structure is undone first than the higher plateau). We also observe that it is possible to form a single quasiplateau (i.e. with a time derivative close but not actually zero) in the far from equilibrium entropy, which leads to no posterior violations of the DEC. However, we have not observed the final two steps in the aforementioned general trend of progressive deformations of plateau structures in cases with $\mu/T=0$ --- i.e., the progressive deformation of the double plateaus into double quasiplateaus and the final loss of the plateau structure was only observed in association with later, progressively stronger violations of the DEC from below (i.e., for $\Delta\hat{p}/\hat{\epsilon}<-1$) with nonzero values of the initial charge density, which leads to a medium with $\mu/T>0$. On the other hand, no particular correlation has been noticed between peculiar features in the far from equilibrium entropy of the medium and violations of DEC and WEC from above (i.e., for $\Delta\hat{p}/\hat{\epsilon}>2$).

We also analyzed variations on the initial profile for the subtracted dilaton field. When the initial charge density of the medium is set to zero but the initial dilaton is nontrivial, the early time evolution of the 1RCBH medium is different from pure thermal SYM far from equilibrium states, since in the case of the 1RCBH plasma the scalar condensate develops a nontrivial time dependence.\footnote{The 1RCBH model features pure thermal SYM as a particular case, which is obtained by setting both the initial charge density and the initial subtracted dilaton profile to zero.} At later times, when the 1RCBH fluid approaches equilibrium, the scalar condensate for charge neutral configurations goes to zero and all the other observables also match the pure thermal SYM equilibrium state. This asymptotic time behavior is expected, since the equilibrium state in the 1RCBH model only depends on the value of $\mu/T$, which vanishes for charge neutral configurations. When the initial charge density and the initial dilaton are both nonzero, by varying only the initial dilaton while keeping the remaining initial data fixed, one typically notices relatively small variations on the pressure anisotropy, entropy and charge density, while the relative variations on the scalar condensate may be large when the system is still far from equilibrium. By varying only the initial dilaton profile, the impact on the value of $\mu/T$ in the medium is typically negligible. Remarkably, in the cases with a nontrivial profile for the initial dilaton field, the scalar condensate typically acquires negative values and develops a pronounced dip when the medium is still far from equilibrium. This is very different from the solutions with zero initial dilaton, for which the scalar condensate has been always observed to be non-negative in the configurations generated in the present work.

%%%
Although we have not calculated in the present work the holographic entanglement entropy, whose second order functional derivative is related in Ref. \cite{Ecker:2017jdw} to the Quantum Null Energy Condition (QNEC) \cite{Bousso:2015mna}, which is a local energy condition proposed to hold for any quantum field theory, we analyzed the behavior of the second order proper time derivative of the non-equilibrium entropy associated to the area of the apparent horizon. Due to the noisy numerical data for the proper time evolution of the area of the apparent horizon at finite density (see appendix \ref{sec:app2.2}), we restricted our analysis of the second order derivative of the non-equilibrium entropy to charge neutral solutions corresponding to pure thermal SYM evolutions, which produce significantly less noisy numerical data for the proper time evolution of the area of the apparent horizon.

\begin{figure*}%[h]
\center
\subfigure[]{\includegraphics[width=0.49\textwidth]{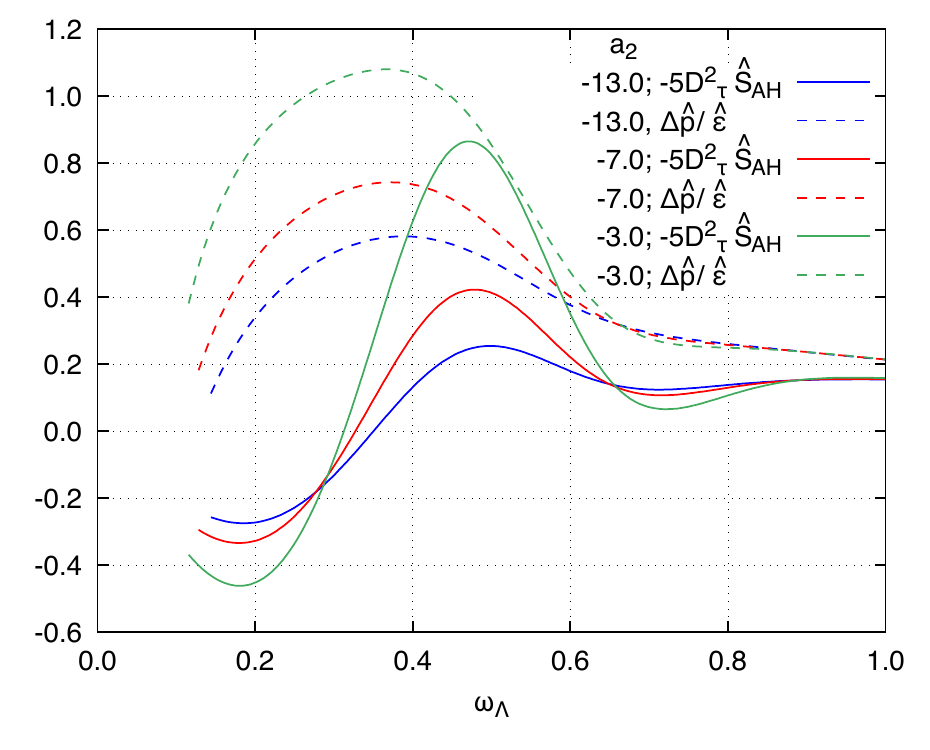}}
\subfigure[]{\includegraphics[width=0.49\textwidth]{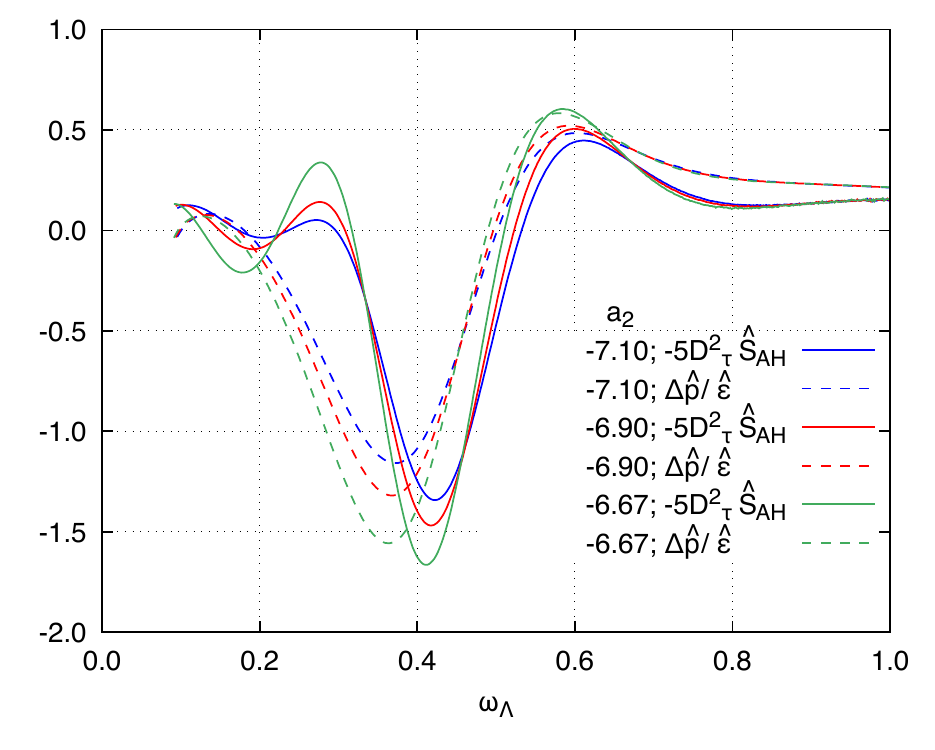}}
\subfigure[]{\includegraphics[width=0.49\textwidth]{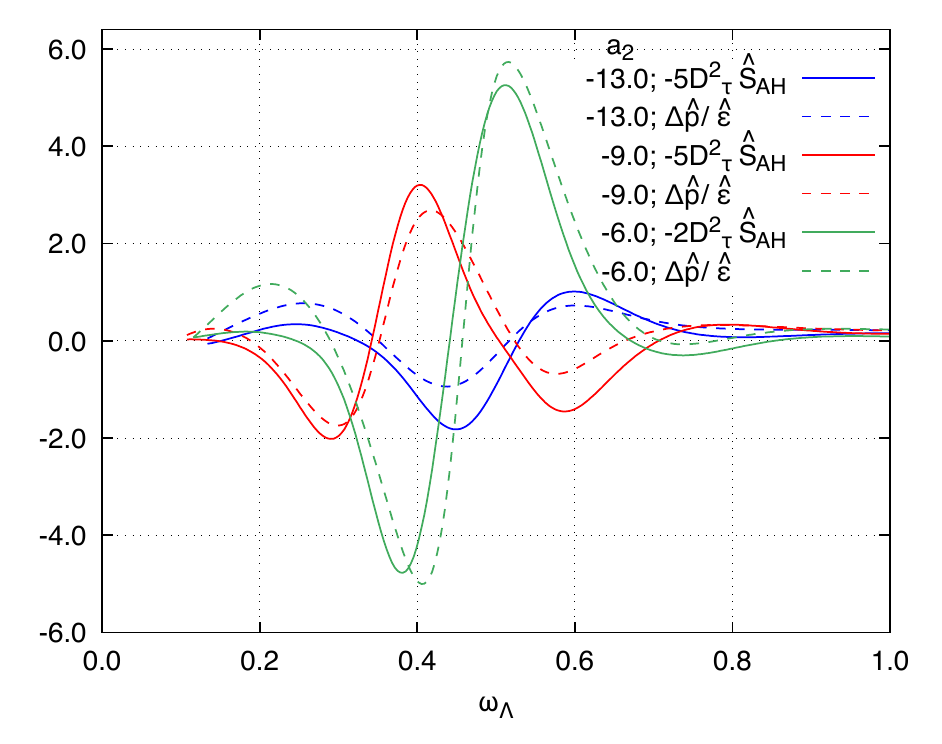}}
\subfigure[]{\includegraphics[width=0.49\textwidth]{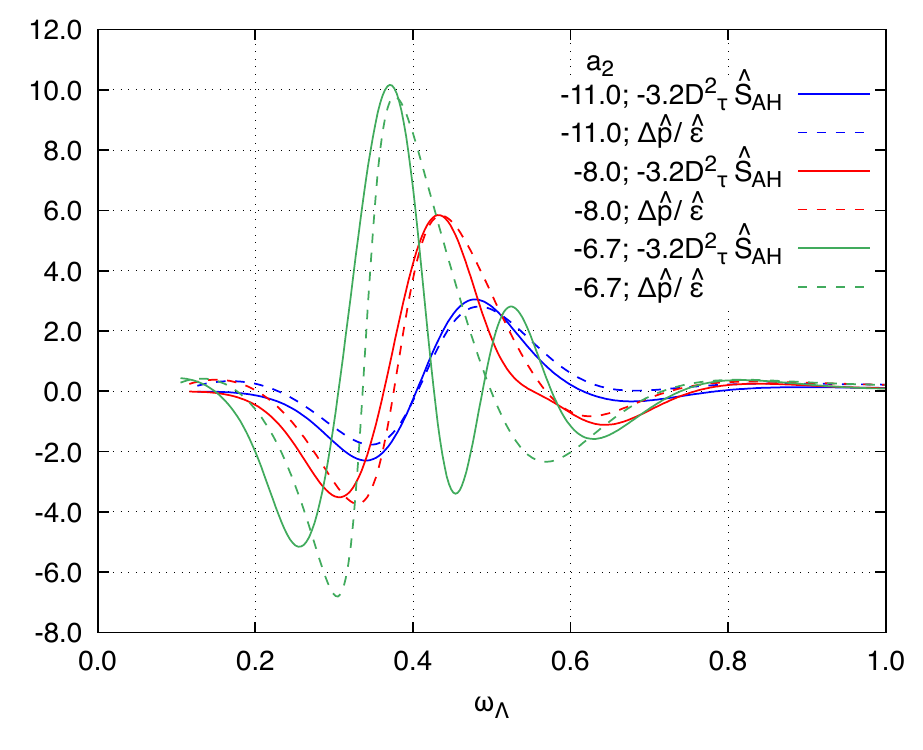}}
\caption{Comparison between the normalized pressure anisotropy (dashed lines), $\Delta\hat{p}/\hat{\varepsilon}$, and (minus) the normalized logarithmic second order derivative of the non-equilibrium entropy (solid lines), $D^2_\tau\hat{S}_\textrm{AH}\equiv\,\tau\partial_\tau\left[\tau\partial_\tau\ln\left(\hat{S}_\textrm{AH}/\mathcal{A}\Lambda^2\right)\right]$, for pure thermal SYM solutions ($\rho_0=0\Rightarrow\mu/T=0$) plotted with different values of $a_2$ and: (a) $B_s1$ in Table \ref{tabICs}, (b) $B_s7$, (c) $B_s9$, and (d) $B_s10$. The numerical factors in front of $D^2_\tau\hat{S}_\textrm{AH}$ were chosen to take the curves close to the corresponding results for the normalized pressure anisotropy, in order to facilitate graphical comparisons.}
\label{figEXTRA}
\end{figure*}

We looked at several different initial conditions and compared the normalized pressure anisotropy and the normalized logarithmic second order derivative of the non-equilibrium entropy in cases with no violations of the classical energy conditions, and also in cases with violations of DEC and WEC. The main conclusions we obtained are illustrated with the plots in Fig. \ref{figEXTRA}.

We found that when DEC or WEC/DEC are violated from below (by having, respectively, $\Delta\hat{p}/\hat{\varepsilon}<-1$ or $\Delta\hat{p}/\hat{\varepsilon}<-4$) and next WEC/DEC are violated from above (by having $\Delta\hat{p}/\hat{\varepsilon}>2$), with no further violations afterwards, then the time evolution of (minus) the normalized logarithmic second order derivative of the non-equilibrium entropy is qualitatively similar to the time evolution of the normalized pressure anisotropy, with the former being slightly delayed in time relatively to the latter. If a third violation happens afterwards (generally corresponding to a second violation of DEC from below), then this correlation is lost with $D^2_\tau\hat{S}_\textrm{AH}$ developing more extrema than $\Delta\hat{p}/\hat{\varepsilon}$ (at the boundary of such a third violation, with $\Delta\hat{p}/\hat{\varepsilon}=-1$, the extra extremum developed by $D^2_\tau\hat{S}_\textrm{AH}$ is an inflection point). When no violations of DEC or WEC are observed, or when just DEC is violated from below, $D^2_\tau\hat{S}_\textrm{AH}$ and $\Delta\hat{p}/\hat{\varepsilon}$ may generally display different behaviors depending on the chosen initial conditions.

In this regard, when DEC or WEC/DEC are violated from below and next WEC/DEC are violated from above, with no further violations afterwards, we found that the curve for $D^2_\tau\hat{S}_\textrm{AH}$ generally anticipates what will happen with the curve for $\Delta\hat{p}/\hat{\varepsilon}$. Clearly, such a inferred correlation is physically nontrivial and it can be deeply investigated elsewhere.
 
%%%

It would be also interesting to systematically investigate the violation of energy conditions in other far from equilibrium holographic models, as well as possible correlations with early time windows with zero entropy production for the out of equilibrium medium. It may be that the correlations reported here and in \cite{Rougemont:2021qyk,Rougemont:2021gjm} are general for strongly coupled quantum fluids, which is then a remarkable qualitative difference with regard to the physical possibilities which may be realized in classical weakly coupled approaches for non-equilibrium media, like kinetic theory, where violations of energy conditions are not observed.

%%%%%%%%%%%%%%%%%%%%%%%%%%%%%%%%%
\acknowledgments
W.B. is grateful to S\~{a}o Paulo Research Foundation (FAPESP) for the received support 
under grant No. 2022/02503-9. We are greatly grateful to Renato Critelli for developing an earlier version of the symbolic-algebraic and numerical code for the Bjorken flow dynamics of the 1RCBH model \cite{Critelli:2018osu}, from which we heavily based the development of our current code using Mathematica, Fortran and Python for different tasks. We are also grateful to Braudel Maqueira for developing a Docker container which we used to install and run our calculations on a Linux virtual machine.

%%%%%%%%%%%%%%%%%%%%%%%%%%%%%%%%%%
\appendix

\section{Further physical consistency checks and details}
\label{sec:app1}

In this appendix, we discuss some extra physical consistency checks of our numerical results and also provide some clear illustrations for the previously discussed behavior regarding the formation of plateau structures in the far from equilibrium entropy of the medium for some selected initial data.

\begin{figure*}%[h]
\center
\subfigure[]{\includegraphics[width=0.49\textwidth]{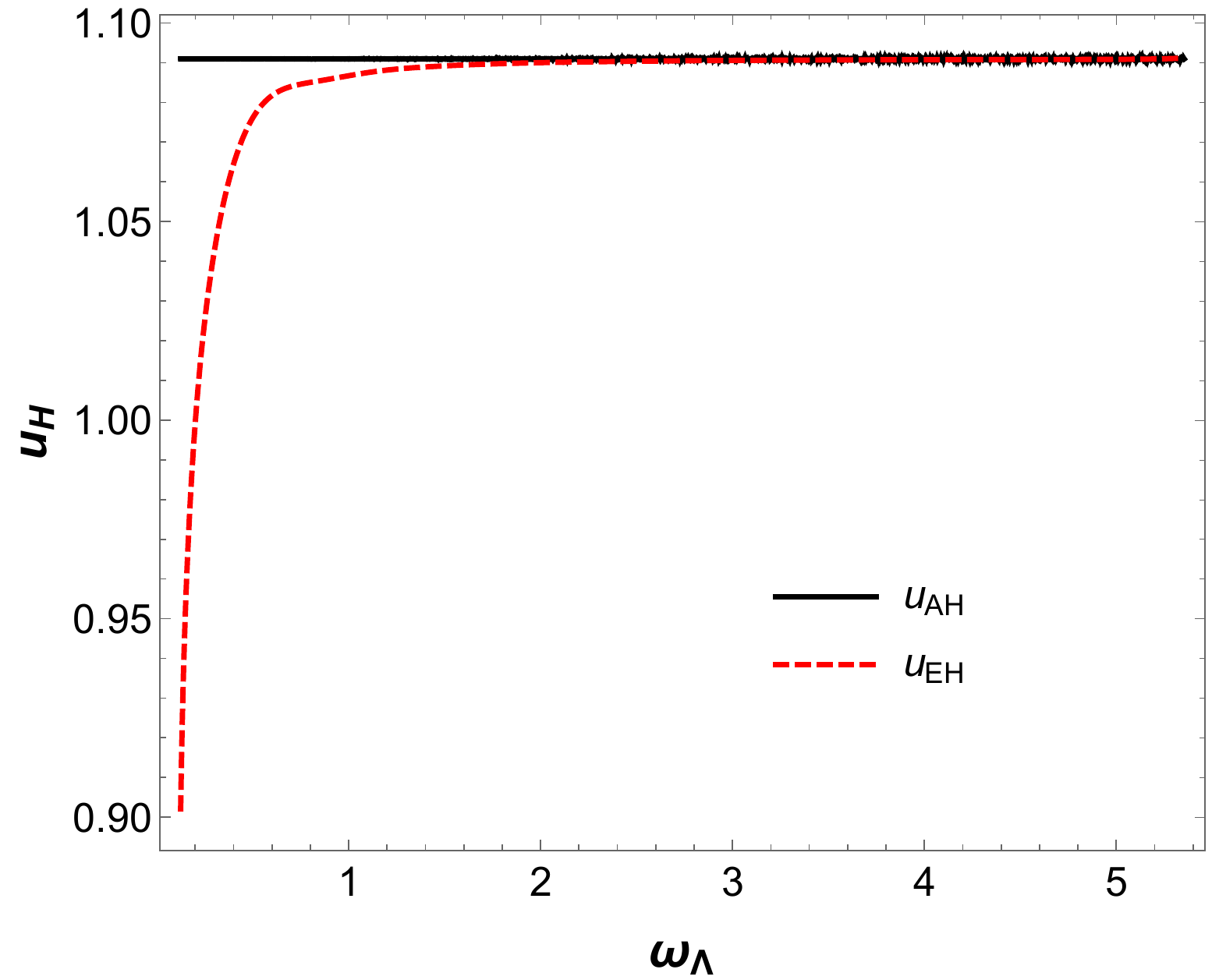}}
\subfigure[]{\includegraphics[width=0.49\textwidth]{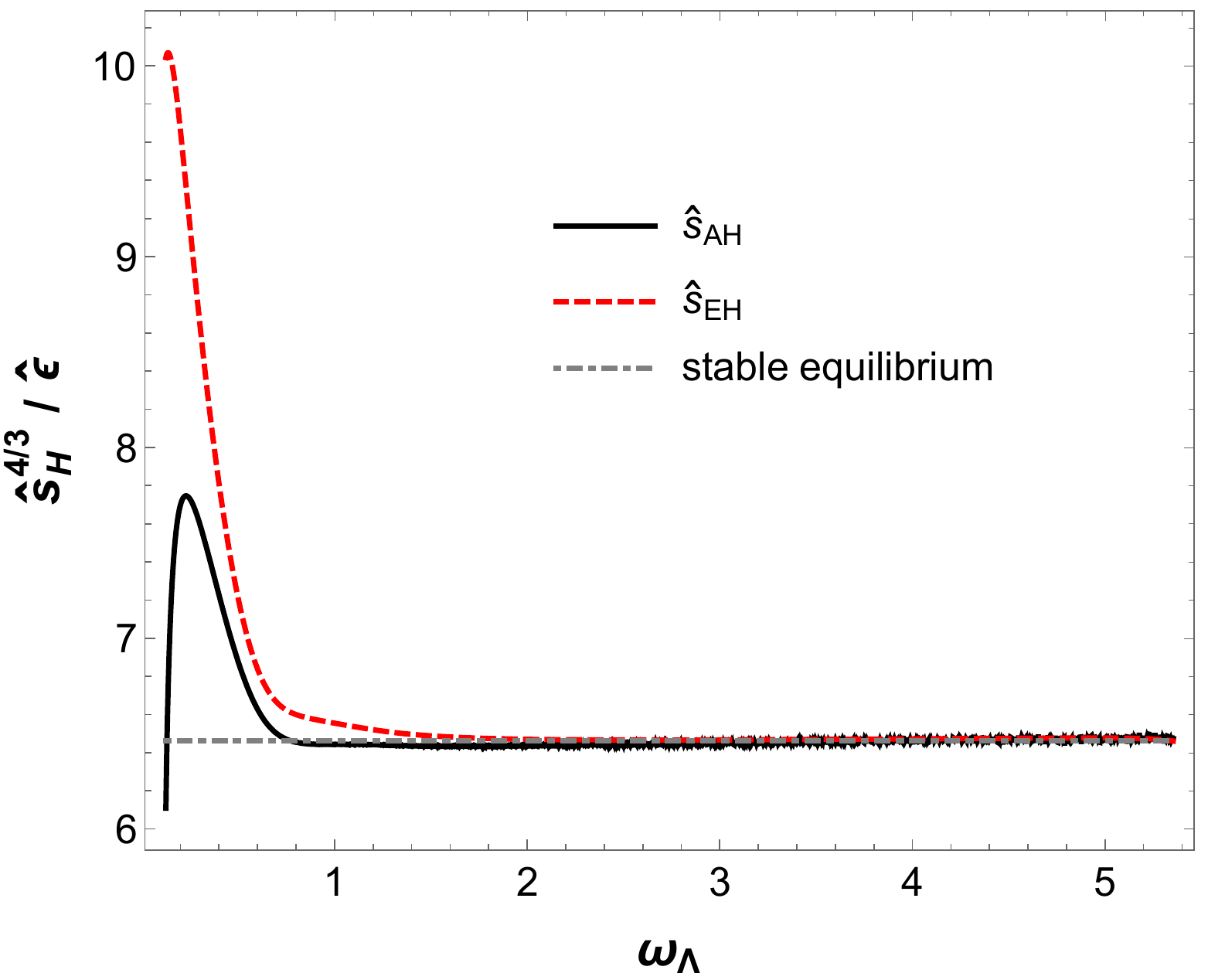}}
\caption{Time evolutions for: (a) the apparent horizon (AH) and the event horizon (EH); (b) the non-equilibrium entropy densities associated to the areas of the apparent and the event horizons, normalized by the energy density. Results obtained for $B_s1$ in Table \ref{tabICs} with $a_2(\tau_0)=-6.67$ and $\rho_0=2.2$, for which $\mu/T\sim 1.957 \rightarrow x/x_c \sim 0.881$, where $x_c\equiv\left(\mu/T\right)_c=\pi/\sqrt{2}$ is the critical point. The corresponding results for the other observables of the 1RCBH plasma are displayed in Fig. \ref{fig:result1}. The analytical stable equilibrium result for the normalized entropy density is obtained from Eq. \eqref{eqse5}.}
\label{fig:app-1}
\end{figure*}

The radial location of the event horizon is determined by the solution of the outgoing radial null geodesics equation subjected to the condition that at asymptotically large times it is given by a zero of the metric coefficient $A(\tau,r)$,
\begin{align}
\frac{dr_\textrm{EH}(\tau)}{d\tau} = A(\tau,r_\textrm{EH}(\tau)),\qquad r_\textrm{EH}(\tau\to\infty) = r_\textrm{EH}^{\textrm{(eq.)}},
\label{EventHor}
\end{align}
where $r_\textrm{EH}^{\textrm{(eq.)}}$ is the largest simple root of equation $A(\tau\to\infty,r)=0$, corresponding to the radial position of the event horizon in equilibrium. Translating the above conditions to the coordinate $u=1/r$ and making use of the subtracted field $A_s(\tau,u)$ defined by Eq. \eqref{eq:subA}, one obtains the following first order differential equation,
\begin{align}
&\frac{du_\textrm{EH}(\tau)}{d\tau} = -u^2_\textrm{EH}(\tau) A(\tau,u_\textrm{EH}(\tau)) = -u^4_\textrm{EH}(\tau) A_s(\tau,u_\textrm{EH}(\tau)) - \frac{1}{2} - u_\textrm{EH}(\tau)\lambda(\tau) - u^2_\textrm{EH}(\tau) \left(\frac{\lambda^2(\tau)}{2} -\partial_\tau\lambda(\tau)\right),\nonumber\\
&u_\textrm{EH}(\tau\to\infty) = u_\textrm{EH}^{\textrm{(eq.)}}(\mu/T),
\label{eqEH}
\end{align}
where $u_\textrm{EH}^{\textrm{(eq.)}}(\mu/T)$ is the smallest simple root of the following equation,
\begin{align}
A(\tau\to\infty,u) = u^2A_s(\tau\to\infty,u)+\frac{1}{2u^2}+\frac{\lambda(\tau\to\infty)}{u} + \frac{\lambda^2(\tau\to\infty)}{2} - \partial_\tau\lambda\biggr|_{\tau\to\infty} = 0.
\label{ueq}
\end{align}
In practice, we approximate the asymptotic limit $\tau\to\infty$ in Eq. \eqref{ueq} by evaluating it at $\tau=\tau_\textrm{end}$, where the end time of the numerical simulations needs to be sufficiently large so that the dynamical background black hole geometries are near equilibrium at $\tau=\tau_\textrm{end}$. The `alternative non-equilibrium entropy density' associated to the area of the dynamical event horizon can be calculated using Eq. \eqref{eq:hats} with the substitution $u_\textrm{AH}(\tau)\to u_\textrm{EH}(\tau)$. The corresponding results comparing the time evolution of the apparent and the event horizons, and their associated entropy densities, are shown in Fig. \ref{fig:app-1} for a given far from equilibrium initial condition. One can see that, as expected, the apparent horizon is always behind the event horizon (notice that the boundary is at $u=0$) and that they converge at late times. Moreover, at early times, the normalized entropy density $[\hat{s}_\textrm{AH}^{4/3}/\hat{\epsilon}](\tau)$ associated to the apparent horizon is always less than the corresponding result involving the event horizon, while both converge to the same stable equilibrium result at late times, as they should. In fact, this is one of the important analytical consistency checks of our numerical simulations in the late time evolution of the system, as previously discussed in section \ref{sec:3}.

\begin{figure*}%[h]
\center
\subfigure[]{\includegraphics[width=0.49\textwidth]{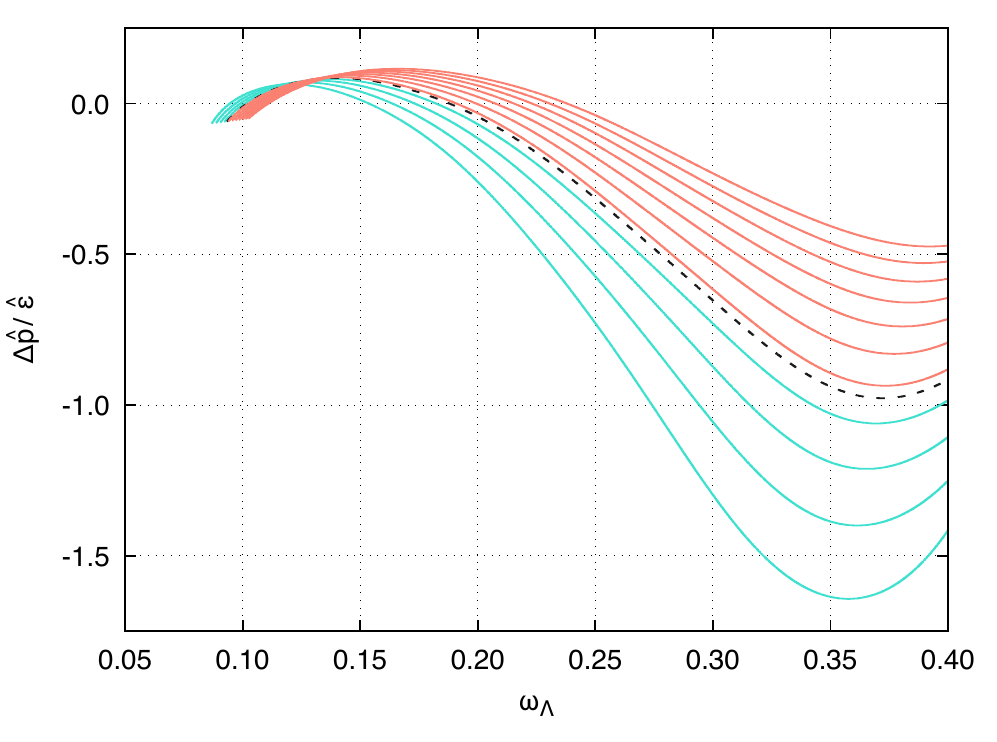}}
\subfigure[]{\includegraphics[width=0.49\textwidth]{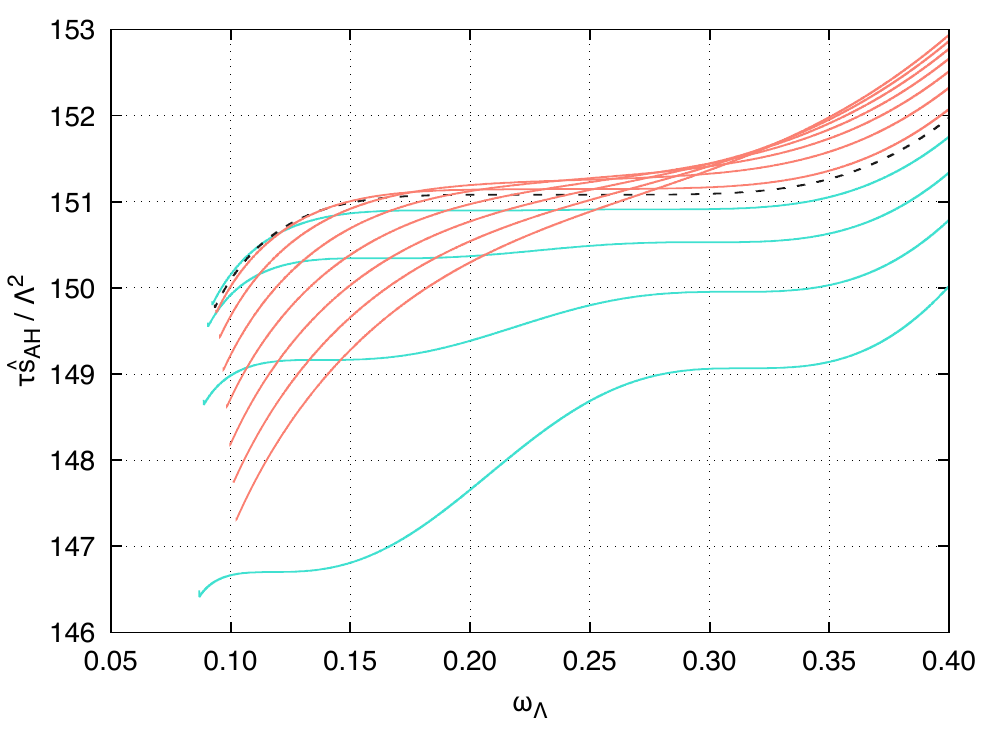}}
\subfigure[]{\includegraphics[width=0.49\textwidth]{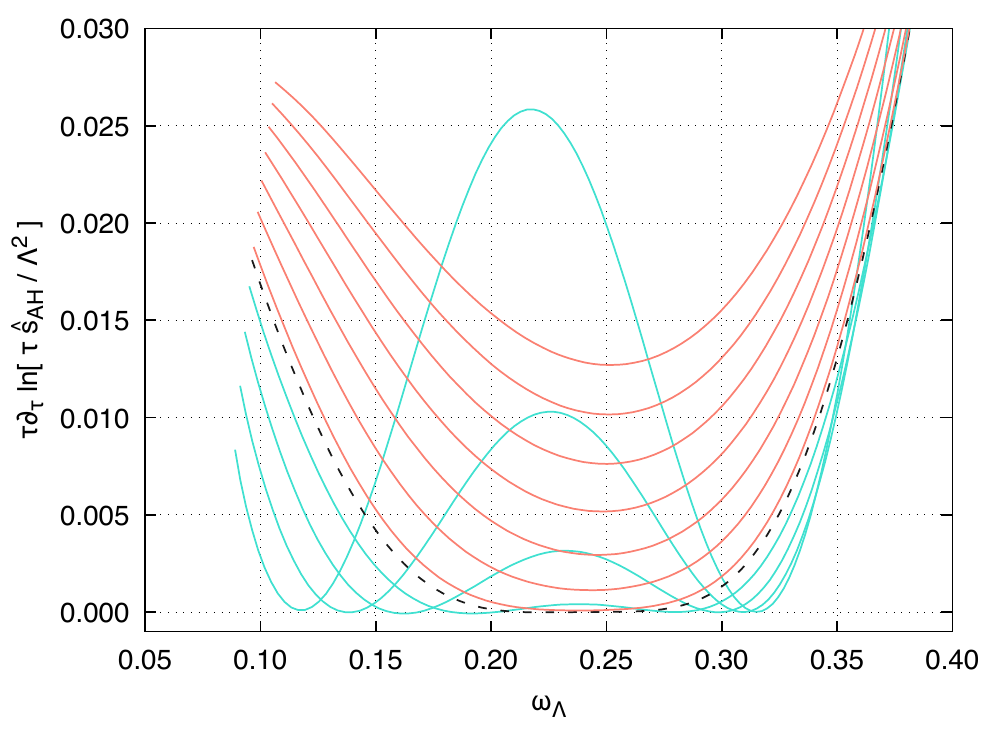}}
\subfigure[]{\includegraphics[width=0.49\textwidth]{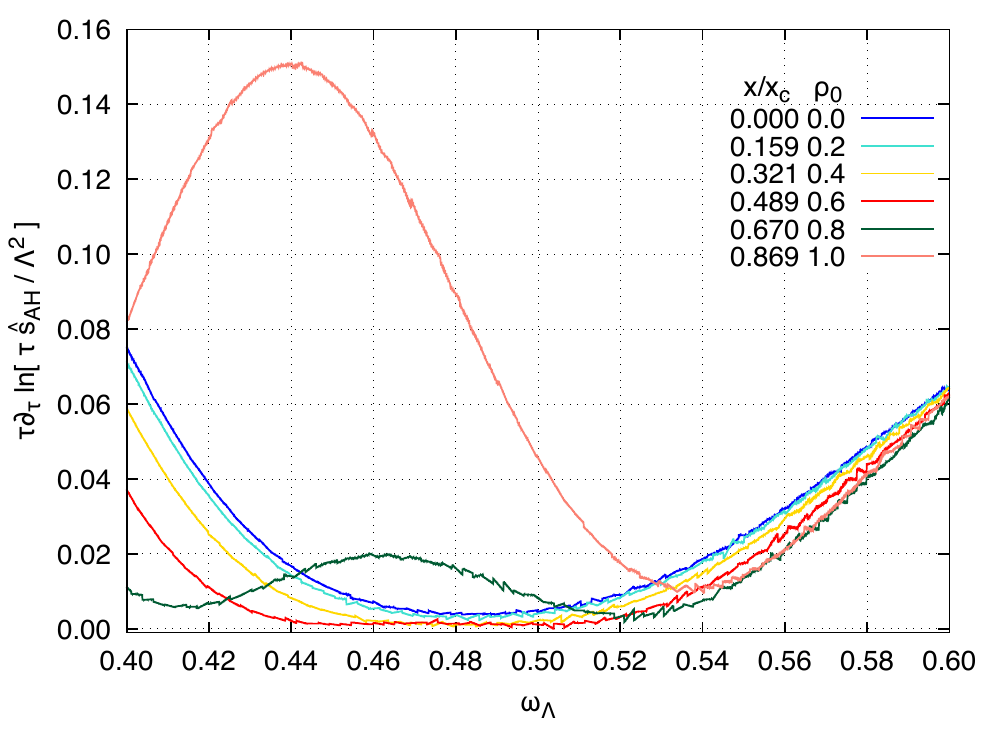}}
\caption{Zoom of the early time windows for: (a) the normalized pressure anisotropy, (b) the normalized non-equilibrium entropy $\hat{S}_\textrm{AH}/\mathcal{A}\Lambda^2=\tau\hat{s}_\textrm{AH}/\Lambda^2$, and (c) the normalized time derivative of the non-equilibrium entropy --- results obtained for variations of {$a_2(\tau_0)=\{-8.1$ (higher line in salmon with one minimum), $-7.9, -7.7, -7.5, -7.3, -7.1, -6.9, -6.83, -6.7, -6.5, -6.3, -6.1$ (higher line in turquoise with one maximum)$\}$} keeping fixed $B_s5$ in Table \ref{tabICs} with $\rho_0=0$ (these correspond to pure thermal SYM evolutions with $\mu/T=0$); the black dashed line corresponds to $a_2=-6.83$ and defines the transition between the salmon and turquoise curves. (d) Zoom of the early time windows for the normalized time derivative of the non-equilibrium entropy obtained for variations of $\rho_0$ keeping fixed $B_s11$ in Table \ref{tabICs} with $a_2(\tau_0)=-7.1$ (results obtained with the smoothing technique used to reduce numerical noise, to be discussed in appendix \ref{sec:app2}); the corresponding results for the other observables of the 1RCBH plasma are displayed in Fig. \ref{fig:result11}. Note that $x_c\equiv\left(\mu/T\right)_c=\pi/\sqrt{2}$ is the critical point.}
\label{fig:app-2}
\end{figure*}

In Fig. \ref{fig:app-2} (a), we show for $B_s5$ in Table \ref{tabICs} at $\mu/T=0$, a sequence with progressive reductions of the initial energy density of the medium, which progressively lower the dip in the pressure anisotropy leading from no violations to progressively stronger violations of the DEC \eqref{eq:DEC} from below. One notices from Fig. \ref{fig:app-2} (b) the associated behavior for the entropy of the medium, which progressively goes from an absence of plateaus, to the formation of a single plateau, which is next deformed into double plateaus, with such plateau structures being explicitly confirmed by the calculation of the time derivative of the entropy, as shown in Fig. \ref{fig:app-2} (c).

On the other hand, we show in Fig. \ref{fig:app-2} (d) the time derivative of the entropy for $B_s11$ in Table \ref{tabICs} at fixed initial energy density, considering progressive enhancements of the initial charge density of the fluid, which lead to a richer variety of steps than in the previous analyzed case at $\mu/T=0$. Indeed, in Fig. \ref{fig:app-2} (d) one notices that as $\mu/T$ is progressively increased (leading to stronger violations of the DEC from below, as previously displayed in Fig. \ref{fig:result11}), the double plateaus are further deformed into double quasiplateaus, until the plateau structure is finally lost for strong enough violations of the DEC at high values of $\mu/T$.

One also notices that the time derivative of the entropy shown in Fig. \ref{fig:app-2} (d) at finite $\mu/T$ is clearly more noisy than the results displayed in Fig. \ref{fig:app-2} (c) at $\mu/T=0$. Indeed, the results shown in Fig. \ref{fig:app-2} (d) have been already smoothed out by using a smoothing technique employed to reduce the very strong numerical noise in the original data associated to the area of the apparent horizon evaluated at finite $\mu/T$, as we discuss next in appendix \ref{sec:app2}.

%%%%%
\section{Numerical error analysis}
\label{sec:app2}

\begin{figure*}%[h]
\center
\subfigure[]{\includegraphics[width=1.\textwidth]{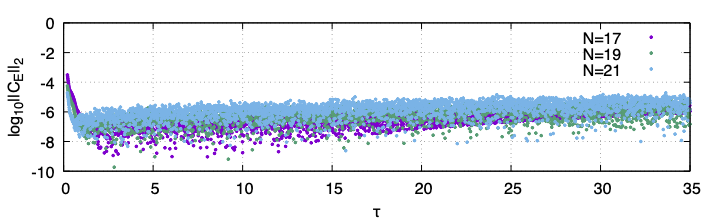}}
\subfigure[]{\includegraphics[width=1.\textwidth]{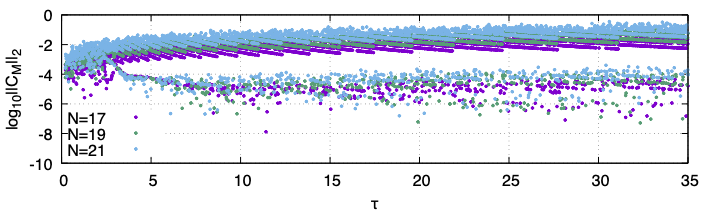}}
\caption{Time evolution of the error of the constraints $\mathcal{C}_E$ (a), and $\mathcal{C}_M$ (b), for different numbers of collocation points, $N$, evaluated with the RMS norm (\ref{eq:RMS}) for $B_s11$ in table \ref{tabICs} with $a_2(\tau_0)=-7.1$ and $\rho_0=0.6$ (corresponding to a solution with $\mu/T\sim 1.086$, which gives $x/x_c\sim 0.489$, where $x_c\equiv\left(\mu/T\right)_c=\pi/\sqrt{2}$ is the critical point). The situation is similar for the whole set of initial conditions considered in this work.}
\label{fig:error_constraints}
\end{figure*}

\begin{figure*}%[h]
\center
\subfigure[]{\includegraphics[width=0.78\textwidth]{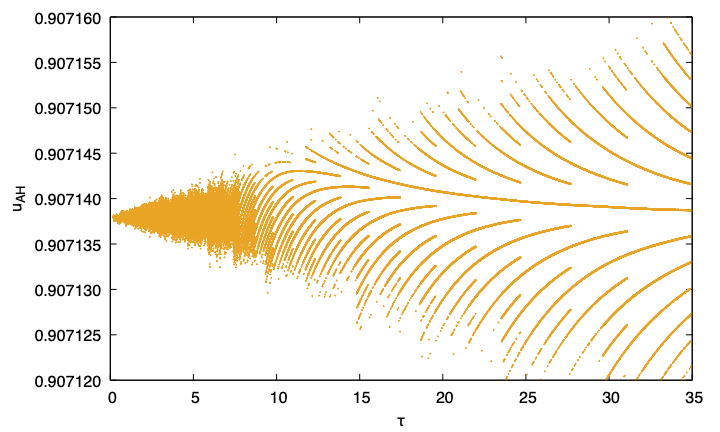}}
\subfigure[]{\includegraphics[width=0.75\textwidth]{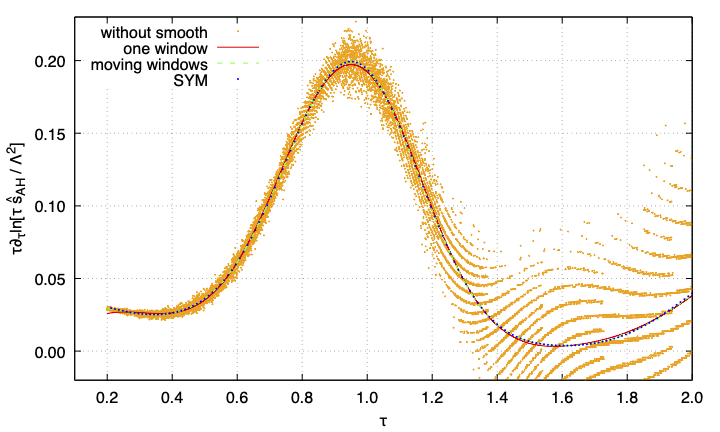}}
%\subfigure[]{\includegraphics[width=0.75\textwidth]{figure27c.pdf}}
\caption{Time evolution of the radial location of the apparent horizon $u_\textrm{AH}(\tau)$ (a), and the normalized time derivative of the non-equilibrium entropy $\tau\partial_\tau\ln[\tau\hat{s}_\textrm{AH}/\Lambda^2]$ (b), with and without smoothing for $B_s11$ in table \ref{tabICs} with $a_2(\tau_0)=-7.1$ and $\rho_0=0$ (corresponding to a pure thermal SYM solution with $\mu/T=0$).}
\label{fig:noise}
\end{figure*}

In this appendix, we briefly discuss some details regarding the numerical error and the treatment of the inherent noise in our code.

\subsection{Monitoring the convergence and the constraints}
\label{sec:app2.1}

We have two constraint equations, denoted here by $\mathcal{C}_E$ and $\mathcal{C}_M$, respectively given by Eqs. (\ref{eq:EH-Const2}) and (\ref{eq:Max-Const}), which are used to globally monitor the time evolution of the initial data. In order to accomplish such a task, we define a root mean square (RMS) norm,\footnote{We denote by $u_\textrm{IR}$ the infrared radial cutoff used as the end point of the radial integration deep into the bulk \cite{Rougemont:2021qyk,Rougemont:2021gjm}. In the present work, we selected different values for $u_\textrm{IR}$ (typically between $1.0$ and $1.2$) depending on the initial condition considered, such as to have the initial value of the radial location of the apparent horizon $u_\textrm{AH}(\tau_0)$ (calculated with $\lambda(\tau_0)=0$) within the radial grid $u\in[0,u_\textrm{IR}]$.}
\begin{equation}
||\mathcal{C}||_2=\sqrt{\frac{1}{2}\int_{0}^{u_\textrm{IR}}du\,\mathcal{C}^2}\,,
\label{eq:RMS}
\end{equation}
which is our measure of error for the time evolution of any given initial condition. The integral in Eq. \eqref{eq:RMS} is calculated using a Gauss-Lobatto quadrature off the collocation points. By doing so, we display in Fig. \ref{fig:error_constraints}, as a typical example, the time evolution of the error of the constraints for some values of the number of radial collocation points for a given initial condition (as discussed in section \ref{sec:2.4}, we use $\Delta\tau=12\times 10^{-5}$ as the time step size). The constraint $\mathcal{C}_M$, given by Eq. (\ref{eq:Max-Const}), displays an apparent high and structured error. A careful study of this error led us to efficiently remove the associated noise, which is crucial to the more accurate calculation of the entropy, and particularly its time derivative, which is used to confirm the presence of plateaus with zero entropy production.

\subsection{Treatment of numerical noise}
\label{sec:app2.2}

As shown in Fig. \ref{fig:noise} (a), in the course of time the numerically calculated radial location of the apparent horizon typically displays the same kind of structured noise as observed for the constraint $\mathcal{C}_M$. This behavior is not clearly noticed by eye in the associated entropy,\footnote{Although it does become noticeable in the entropy at large times by increasing the number of collocation points $N$.} but its time derivative does manifest it in full swing. An additional test was conducted for the calculation of the apparent horizon. We used an alternative method to the Newton-Raphson algorithm for that purpose, namely, a bisection standard method. The result was numerically the same for a similar numerical tolerance.
% The reader should note that the constraint $\mathcal{C}_M$ contains a time derivative term.

For all time evolutions the noise was monitored and, when required by the analysis of the time derivative of the entropy, we treated it within the relevant time window, as displayed e.g. in Fig. \ref{fig:app-2} (d). In the cases when it was not enough to simply select an appropriate number of radial collocation points in order to diminish the numerical noise increasing with the action of time, we proceed to smooth out the resulting data by using standard procedures and tools. As an extra reference for the particular evolutions with $\mu/T=0$, we also compared our present results with the outcomes from the numerical code developed for the pure thermal SYM plasma undergoing Bjorken flow in Refs. \cite{Rougemont:2021qyk,Rougemont:2021gjm}, which is much less noisy. We display in Fig. \ref{fig:noise} (b) a comparison for the time derivative of the entropy, where it becomes clear that our best implementation of the smoothing technique corresponds to the moving window procedure. The algorithm is quite simple: we select one window to fit a mean smooth curve which is later differentiated. Depending on the quality of the result --- comparing e.g. with the reference pure thermal SYM code \cite{Rougemont:2021qyk,Rougemont:2021gjm} --- we proceed to move the time windows considered by doing sub-samplings. For that purpose, we implemented a short script using a couple of Python libraries and functions. Particularly, the Scipy interpolating tools for a smooth spline approximation.
%This is a wrapper around the FORTRAN routines splev and splder of FITPACK.. 
Without smoothing out the results for the time derivative of the entropy it is not possible to extract useful information from this specific numerical data, as it is clear from Fig. \ref{fig:noise} (b).

%%%%%
\section{Numerical code's performance}
\label{sec:app3}

For this work, we developed a general code which makes use of different programming languages for different tasks.

All the lengthy analytical and symbolic-algebraic manipulations were implemented using Wolfram's Mathematica and the \emph{Riemannian Geometry} $\&$ \emph{Tensor Calculus} (RGTC) library developed by Sotirios Bonanos. Although Mathematica is useful for such tasks, it is extremely inefficient to deal with numerical calculations when compared to other programming languages like e.g. Fortran and C.

Indeed, for the numerical simulations considered in the present work, typically done with $N=21$ radial collocation points, a time step size of $\Delta\tau=12\times 10^{-5}$, and with the evolutions computed within the long time interval $\tau\in[\tau_0=0.2,\tau_\textrm{end}=35]$, it is practically unfeasible to numerically evolve a single initial data using Mathematica. By considering a much shorter end time for the simulations, like $\tau_\textrm{end}=7.5$, as used in \cite{Rougemont:2021qyk,Rougemont:2021gjm}, Mathematica takes several hours to run a single evolution on an Intel Core i5 1.8 GHz Dual-Core; however, $\tau_\textrm{end}=7.5$ is not large enough to allow one to see the effective thermalization of the scalar condensate as analyzed in the present work, which generally requires a much larger value for $\tau_\textrm{end}$. Indeed, as mentioned in section \ref{sec:2.4}, we used here $\tau_\textrm{end}=35$, which can only be simulated using more efficient programming languages for numerical purposes.

In this regard, we developed an integrated numerical code combining Fortran and Python to automate each run, including all the post-processing and plotting. The performance of this numerical code for each run with the same general configurations as aforementioned, including the calculation of $\Lambda$ and $\mu/T$ in the post-processing, is about $20$ minutes on an Intel Core i5 1.8 GHz Dual-Core.

In order to standardize the runs between different operating systems, we used a Docker container to configure a virtual Linux machine and run the initial data to produce all the results analyze in this work.

%%%%%%%%%%%%%%%%%%%%%%%%%%%%%%%%%%
\bibliographystyle{apsrev4-2}
\bibliography{Bibliography} % name of the bibtex file (in the same directory as the main tex file)

\end{document}